\documentclass[bibyear]{aa} 

\usepackage{lscape}
\usepackage{float}
\usepackage{graphicx}
\usepackage{txfonts}
\begin{document} 

\title{What did the seahorse swallow? APEX 170 GHz observations of the chemical conditions in the Seahorse infrared dark cloud
\thanks{Based on observations with the Atacama Pathfinder EXperiment (APEX) telescope under programme {\tt 0104.C-0023(A)}. APEX is a collaboration between the Max-Planck-Institut f{\"u}r Radioastronomie, the European Southern Observatory, and the Onsala Space Observatory.}}

   \author{O.~Miettinen}

   \institute{Academy of Finland, Hakaniemenranta 6, P.O. Box 131, FI-00531 Helsinki, Finland \\ \email{oskari.miettinen@aka.fi}}

   \date{Received ; accepted}

\authorrunning{Miettinen}
\titlerunning{Chemistry in the Seahorse IRDC: A 170~GHz molecular line survey with APEX}

\abstract{Infrared dark clouds (IRDCs) are useful target sources for the studies of molecular cloud substructure evolution and early stages 
of star formation. Determining the chemical composition of IRDCs helps to constrain the initial conditions and timescales (via chemical clocks) of star formation in these often filamentary, dense interstellar clouds.}{We aim to determine the fractional abundances of multiple different molecular species in the filamentary IRDC G304.74+01.32, nicknamed the Seahorse IRDC, and to search for relationships between the abundances and potential evolutionary trends.}{We used the Atacama Pathfinder EXperiment (APEX) telescope to observe spectral lines occurring at about 170~GHz frequency towards 14 positions along the full extent of the Seahorse filament. The sample is composed of five clumps that appear dark in the mid-IR, eight clumps that are associated with mid-IR sources, and one clump that is already hosting an \ion{H}{ii} region and is, hence, likely to be in the most advanced stage of evolution of all the target sources. We also employed our previous 870~$\mu$m dust continuum imaging data of the Seahorse.}{Six spectral line transitions were detected ($\geq3\sigma$) altogether, namely, SO$(N_J=4_4-3_3)$, H$^{13}$CN$(J=2-1)$, H$^{13}$CO$^+(J=2-1)$, SiO$(J=4-3)$, HN$^{13}$C$(J=2-1)$, and C$_2$H$(N=2-1)$. While SO, H$^{13}$CO$^+$, and HN$^{13}$C were detected in every source, the detection rates for C$_2$H and H$^{13}$CN were 92.9\% and 85.7\%, respectively. Only one source (SMM~3) showed detectable SiO emission (7.1\% detection rate). Three clumps (SMM~5, 6, and 7) showed the SO, H$^{13}$CN, H$^{13}$CO$^+$, HN$^{13}$C, and C$_2$H lines in absorption. Of the detected species, C$_2$H was found to be the most abundant one with respect to H$_2$ (a few times $10^{-9}$ on average), while HN$^{13}$C was found to be the least abundant species (a few times $10^{-11}$). We found three positive correlations among the derived molecular abundances, of which those between C$_2$H and HN$^{13}$C and HN$^{13}$C and H$^{13}$CO$^+$ are the most significant (correlation coefficient $r\simeq0.9$). The statistically most significant evolutionary trends we uncovered are the drops in the C$_2$H abundance and in the $[{\rm HN^{13}C}]/[{\rm H^{13}CN}]$ ratio as the clump evolves from an IR dark stage to an IR bright stage and then to an \ion{H}{ii} region.}{The absorption lines detected towards SMM~6 and SMM~7 could arise from continuum radiation from an embedded young stellar object and an extragalactic object seen along the line of sight. However, the cause of absorption lines in the IR dark clump SMM~5 remains unclear. The correlations we found between the different molecular abundances can be understood as arising from the gas-phase electron (ionisation degree) and atomic carbon abundances. With the exception of H$^{13}$CN and H$^{13}$CO$^+$, the fractional abundances of the detected molecules in the Seahorse are relatively low compared to those in other IRDC sources. The [C$_2$H] evolutionary indicator we found is in agreement with previous studies, and can be explained by the conversion of C$_2$H to other species (e.g. CO) when the clump temperature rises, especially after the ignition of a hot molecular core in the clump. The decrease of $[{\rm HN^{13}C}]/[{\rm H^{13}CN}]$ as the clump evolves is also likely to reflect the increase in the clump temperature, which leads to an enhanced formation of HCN and its $^{13}$C isotopologue. Both single-dish and high-resolution interferometric imaging of molecular line emission (or absorption) of the Seahorse filament are required to understand the large-scale spatial distribution of the gas and to search for possible hot, high-mass star-forming cores in the cloud.}


   \maketitle
%

\section{Introduction}

Some of the Galactic interstellar clouds are so dense that when they are 
seen against the Galactic mid-infrared (mid-IR) background radiation
field, they manifest themselves as dark absorption features 
(\cite{perault1996}; \cite{egan1998}; \cite{simon2006}; \cite{peretto2009}). 
These clouds are known as infrared dark clouds (IRDCs). 

Ever since the IRDCs were uncovered almost a quarter-century ago, we have seen a growing interest in 
learning more about them. Studies of IRDCs have addressed a wide range of their properties, 
all the way from their formation (e.g. \cite{jimenezserra2010}; \cite{henshaw2013}), 
Galactic distribution (\cite{jackson2008}; \cite{finn2013}; \cite{giannetti2015}), magnetic fields (e.g. \cite{hoq2017}; \cite{soam2019}), fragmentation (e.g. \cite{jackson2010}; \cite{wang2011}; \cite{busquet2016}; \cite{henshaw2016}; \cite{tang2019}), 
substructure properties (e.g. \cite{rathborne2010}; \cite{ragan2013}), star formation (e.g. \cite{rathborne2005}, 2006; \cite{chambers2009}; \cite{kauffmann2010}), maser emission (e.g. \cite{wang2006}), and dust properties (e.g. \cite{juvela2018}) up to their chemistry 
(e.g. \cite{vasyunina2011}; \cite{sanhueza2012}; \cite{vasyunina2014}; \cite{gerner2014}; \cite{zeng2017}; \cite{colzi2018}; \cite{taniguchi2019}). 

Observations of IRDCs have shown that they often appear to be filamentary structures, which could be the results of 
converging atomic gas flows or collisions between precursor interstellar clouds that led to the formation of the IRDC filament (e.g. \cite{jimenezserra2010}; cf.~\cite{heitsch2006}). Moreover, the IRDC filaments appear to be fragmented along their longest axis, potentially via the so-called sausage-type fluid instability in cylindrical geometry, and the substructures thus formed, known as clumps, are found to be hierarchically fragmented down to smaller units, the dense cores (e.g. \cite{jackson2010}; \cite{wang2011}; \cite{beuther2015}; \cite{mattern2018}; \cite{svoboda2019}). Some of the dense substructures of IRDCs are found to be associated with early stages of high-mass star formation (e.g. \cite{rathborne2006}; \cite{beuther2007}; \cite{chambers2009}; \cite{battersby2010}). This makes IRDCs very useful targets for the star formation research because the process(es) of high-mass star formation still remains inconclusive (e.g. \cite{motte2018} for a review; \cite{padoan2020}).

\begin{figure}[!htb]
\centering
\resizebox{\hsize}{!}{\includegraphics{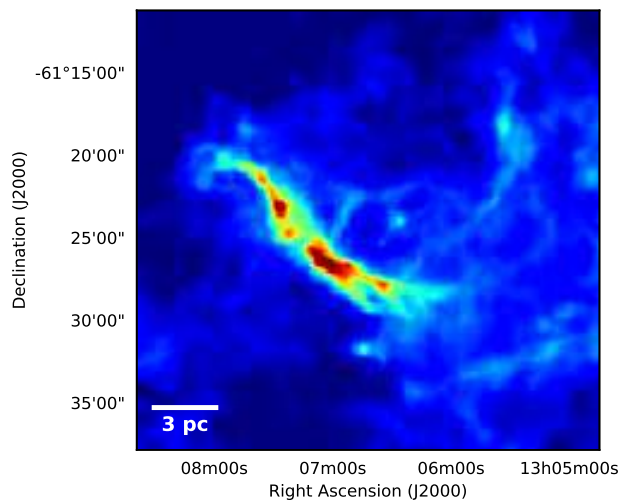}}
\caption{\textit{Herschel}/SPIRE (Spectral and Photometric Imaging REceiver) 350~$\mu$m image towards the IRDC G304.74+01.32 (the Seahorse IRDC) and its surroundings. The image is displayed using a non-linear (logarithmic) stretch to improve the visibility of the faint, extended dust emission. There appear to be dusty striations on the western side of the Seahorse filament. A scale bar of 3~pc projected length is shown in the bottom left corner.}
\label{figure:spire}
\end{figure}

In this paper, we present a study of the chemical properties of the filamentary IRDC G304.74+01.32. This IRDC has been the target of submillimetre and millimetre dust continuum studies (\cite{beltran2006}; \cite{miettinenharju2010}; \cite{miettinen2018}), and molecular spectral line observations (\cite{miettinen2012}). 
Most recently, Miettinen (2018) found that 36\% of the clumps along G304.74+01.32 are fragmented into smaller cores, and that 65\% of those cores are associated with IR sources detected with the \textit{Wide-field Infrared Survey Explorer} (\textit{WISE}; \cite{wright2010}). Those results strongly point to a hierarchical fragmentation of the filament. Miettinen (2018) also found that 
G304.74+01.32 has a seahorse-like morphology in the \textit{Herschel}\footnote{\textit{Herschel} is an ESA space observatory with science instruments pro- vided by European-led Principal Investigator consortia and with important participation from NASA.} (\cite{pilbratt2010}) far-IR and submillimetre images (Fig.~\ref{figure:spire}; cf.~Fig.~2 in \cite{miettinen2018}) and, therefore, proposed the nickname Seahorse Nebula for the cloud. Moreover, there appear to be elongated, dusty striations that are roughly perpendicularly projected with respect to the main Seahorse filament. The striations are also seen in absorption in the \textit{WISE} mid-IR images (Fig.~\ref{figure:map}; cf.~Fig.~3 in \cite{miettinen2018}). Such features could be an indication that the filament is still accreting gas from its surrounding medium, but this hypothesis needs to be tested via spectral line imaging observations. 

Only one of the clumps in the Seahorse IRDC appears to be associated with high-mass star formation, namely the one associated with the IR source IRAS~13039-6108. S\'anchez-Monge et al. (2013) found that IRAS~13039-6108 shows radio continuum emission characteristic of an optically thin \ion{H}{ii} region. The IRAS~13039-6108 region is also surrounded by extended and diffuse IR emission, most notably seen in the \textit{WISE} 12~$\mu$m image (Fig.~\ref{figure:map}), which suggests that the \ion{H}{ii} region is surrounded by a photodissociation region (PDR).

The present follow-up study of the chemistry in the Seahorse IRDC is based on observations done with the the Atacama Pathfinder EXperiment (APEX\footnote{\url{http://www.apex-telescope.org/}}; \cite{gusten2006}). The observations and data reduction are described in Sect.~2. The analysis and the corresponding results are presented in Sect.~3. In Sect.~4, we discuss the results and in Sect.~5 we summarise our results and present our main conclusions. Throughout this paper, we adopt a distance of $d=2.54$~kpc to the Seahorse IRDC (\cite{miettinen2012}, 2018).

\section{Observations and data reduction}

\begin{figure*}[!htb]
\begin{center}
\includegraphics[scale=0.75]{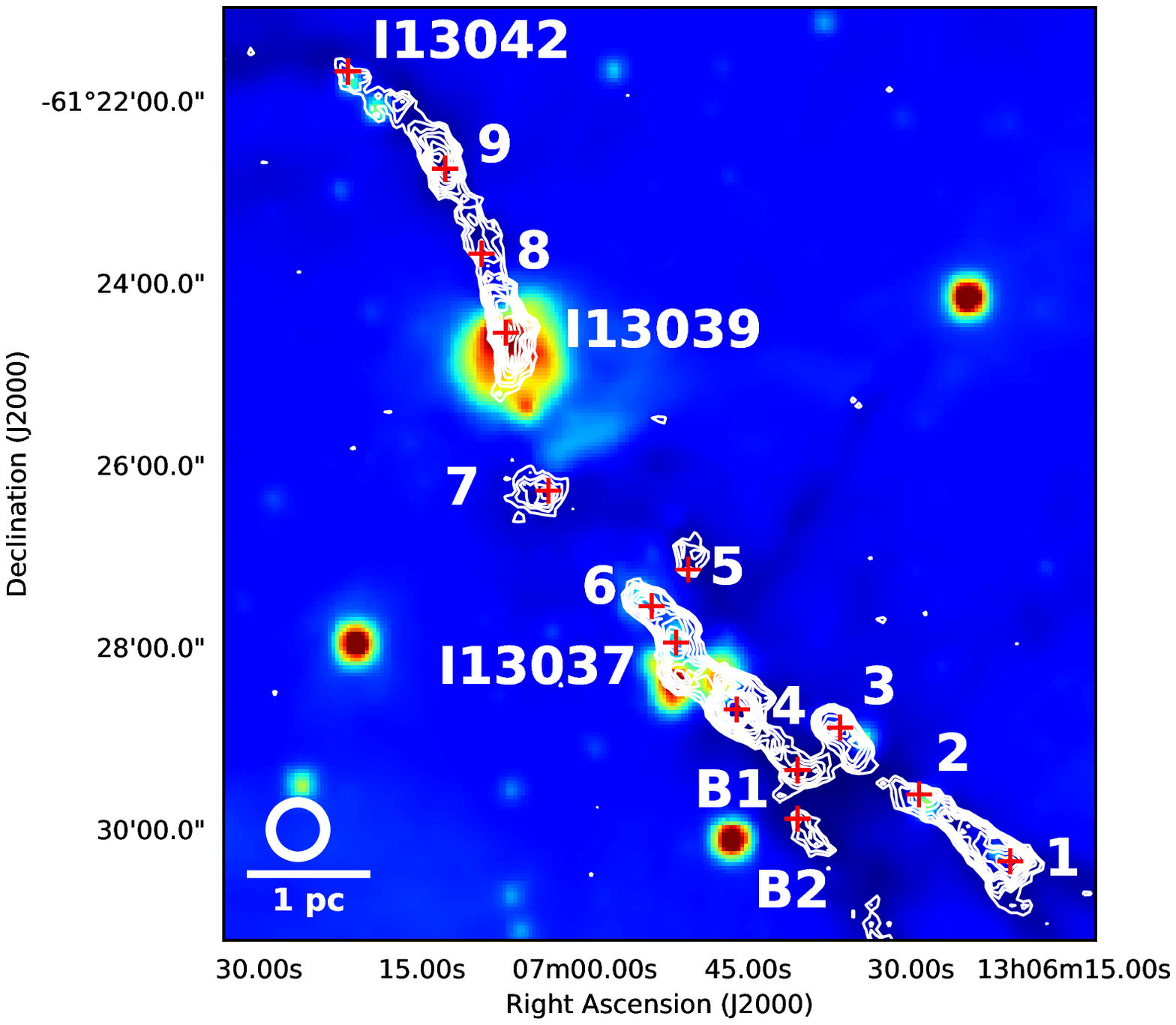}
\caption{\textit{WISE} 12~$\mu$m image towards the IRDC G304.74+01.32 (the Seahorse IRDC). The image is displayed using a logarithmic stretch
to improve the colour contrast. The overlaid contours represent the LABOCA 870~$\mu$m emission (\cite{miettinenharju2010}; \cite{miettinen2018}); the contours start at $3\sigma$, and increase in steps of $3\sigma$, where $3\sigma=120$~mJy~beam$^{-1}$. The target clumps are labelled so that the numbers refer to the SMM IDs (e.g. 1 refers to SMM~1), while the sources I13037, I13039, and I13042 are the
three \textit{IRAS} sources in the filament. The sources BLOB~1 and BLOB~2 are labelled as B1 and B2. The red plus signs indicate the target positions of the present spectral line observations (i.e. LABOCA 870~$\mu$m emission peaks of the clumps). A scale bar of 1~pc projected length, and the beam size at the tuned frequency of the present spectral line observations ($35\farcs9$ HPBW) are shown in the bottom left corner.}
\label{figure:map}
\end{center}
\end{figure*}

Using the Large APEX BOlometer CAmera (LABOCA; \cite{siringo2009}), the 870~$\mu$m peak positions of the clumps in the Seahorse IRDC were  
observed with the 12~m APEX telescope at a frequnecy of $\nu=173.68831$~GHz, that is, at the frequency of the SiO$(J=4-3)$ rotational transition. The target positions are shown in Fig.~\ref{figure:map}, and the source list is given in Table~\ref{table:sources}.

The observations were carried out on 2 September and 15-16 December 2019, and the amount of precipitable water vapour
(PWV) during the observations was measured to be between 0.5~mm and 1.2~mm, which corresponds to a zenith atmospheric transmission 
range of about 96\%--91\% at the observed frequency\footnote{The APEX atmospheric transmission calculator is available at {\tt http://www.apex-telescope.org/sites/chajnantor/ \\atmosphere/transpwv/}.}.

As a front end, we used Band~5 of the Swedish ESO PI receiver for APEX (SEPIA; \cite{belitsky2018}). The SEPIA Band~5, or SEPIA180, 
is a sideband separating (2SB) dual polarisation receiver, which operates in the frequency range 159--211~GHz.
The backend was a Fast Fourier Transform fourth-Generation spectrometer (FFTS4G) that consists of two sidebands. Each sideband has two spectral windows of 4~GHz bandwidth, which provides both orthogonal polarisations and leads to a total bandwidth of 8~GHz. The observed frequency $\nu=173.68831$~GHz was tuned in the upper sideband (USB). This setting allowed us to probe the frequency ranges 159.69-–163.69~GHz in the lower sideband (LSB) and 171.69-–175.69~GHz in the USB. The 65\,536 channels of the spectrometer yielded a 
spectral resolution of 61~kHz, which corresponds to 105.4~m~s$^{-1}$ at the observed frequency of 173.68831~GHz (and ranges from 104.2~m~s$^{-1}$ to 114.6~m~s$^{-1}$ over the observed frequency range). The beam size (half-power beam width; HPBW) at the observed frequency of 173.68831~GHz is $35\farcs9$ ($35\farcs5-39\farcs1$ over the observed frequency range), which corresponds to a physical resolution of 0.44~pc, which in turn is comparable to the diameter of the 870~$\mu$m clumps in the filament ($\sim0.4$~pc on average; \cite{miettinen2018}).

The observations were performed in the wobbler-switching mode with a $50\arcsec$ azimuthal throw between two positions on sky (symmetric offsets), and a chopping rate of $R = 0.5$~Hz. The total on-source integration time (excluding overheads) was about 6.4~min for each target source. The telescope focus and pointing were optimised and checked at regular intervals on the red supergiant star VY Canis Majoris, the carbon stars IRC +10216 (CW Leonis), 07454-7112 (RAFGL 4078), IRAS~15194-5115, and X Trianguli Australis, and the variable star of Mira Cet type R Carinae. The pointing was found to be accurate to $\sim3\arcsec$. The system temperatures during the observations were within
the range of $T_{\rm sys}\simeq 77-88$~K. Calibration was made by means of the chopper-wheel technique, and the output intensity scale given by the system is the antenna temperature corrected for the atmospheric attenuation ($T_{\rm A}^{\star}$). The observed intensities were converted to the main-beam brightness temperature scale by $T_{\rm MB}=T_{\rm A}^{\star}/\eta_{\rm MB}$, 
where $\eta_{\rm MB}$ is the main-beam efficiency. The value of the latter efficiency was derived as follows. 
The aperture efficiency at the central frequency of SEPIA180 ($\eta_{\rm a}=0.67$ when estimated from observations towards Uranus) was first scaled to that at the observed frequency using the Ruze formula (\cite{ruze1952}), and then converted to 
the value of $\eta_{\rm MB}$ using the approximation formula $\eta_{\rm MB}\simeq1.2182 \times \eta_{\rm a}$. This yielded 
the values $\eta_{\rm a}=0.67$ and $\eta_{\rm MB}=0.82$.\footnote{The APEX antenna efficiences can be found at {\tt http://www.apex-telescope.org/telescope/efficiency/ \\index.php}.} The absolute calibration uncertainty was estimated to
be about 10\% (e.g. \cite{dumke2010}).

The spectra were reduced using the {\tt CLASS90} of the GILDAS software package\footnote{Grenoble Image and Line Data Analysis Software (GILDAS) is provided and actively developed by Institut de Radioastronomie Millim\'etrique (IRAM), and is available at \url{http://www.iram.fr/IRAMFR/GILDAS}.} (version mar19b). The individual spectra or sub-scans of 9.6~s in duration (on-source time) were averaged with weights proportional to the integration time, divided by the square of the system temperature ($w_i \propto t_{\rm int}/T_{\rm sys}^2$). The resulting spectra were smoothed using the Hann window function to a velocity resolution of 210.8~m~s$^{-1}$ (value at 173.68831~GHz) to improve the signal-to-noise ratio. Linear (first-order) baselines were determined from the velocity ranges free of spectral line features, and then subtracted from the spectra. The resulting $1\sigma$ rms noise levels at the 0.21~km~s$^{-1}$ velocity resolution were about 9--17~mK on a $T_{\rm MB}$ scale. The observational parameters are provided in Table~\ref{table:observations}.

\begin{table}[H]
\renewcommand{\footnoterule}{}
\caption{Source sample.}
{\scriptsize
\begin{minipage}{1\columnwidth}
\centering
\label{table:sources}
\begin{tabular}{c c c c c c}
\hline\hline 
Source & $\alpha_{2000.0}$ & $\delta_{2000.0}$ & $T_{\rm dust}$ & $n({\rm H_2})$ & Type\\
       & [h:m:s] & [$\degr$:$\arcmin$:$\arcsec$] & [K] & [$10^4$ cm$^{-3}$] & \\ 
\hline 
SMM 1 & 13 06 20.89	& -61 30 20.51 & 14.0 & $2.1\pm1.1$ & IR bright\tablefootmark{a}\\
SMM 2 &	13 06 29.29	& -61 29 36.67 & 12.5 & $3.4\pm1.8$ & IR bright \\ 
SMM 3 &	13 06 36.56	& -61 28 52.79 & 13.7 & $5.1\pm2.7$ & IR bright \\ 
BLOB 2 & 13 06 40.44 & -61 29 52.84 & 14.9 & $5.1\pm2.8$ & IR dark \\ 
BLOB 1 & 13 06 40.46 & -61 29 20.84 & 11.5 & $>2.5$ & IR bright \\ 
SMM 4 &	13 06 46.06	& -61 28 40.90 & 13.2 & $5.4\pm2.8$ & IR dark \\ 
SMM 5 &	13 06 50.54	& -61 27 08.94 & 14.0 & $4.9\pm2.3$ & IR dark \\ 
IRAS 13037 & 13 06 51.65 & -61 27 56.95 & 20.0 & $4.9\pm2.5$ & IR bright \\ 
-6112      &             &              &      &            &   \\
SMM 6 &	13 06 53.88	& -61 27 32.96 & 14.1 & $>2.1$ & IR bright\\ 
SMM 7 &	13 07 03.38	& -61 26 17.00 & 14.5 & $4.1\pm2.1$ & IR dark\tablefootmark{b}\\ 
IRAS 13039 & 13 07 07.28 & -61 24 33.00 & 22.2 & $3.0\pm1.6$ & IR bright/ \\ 
-6108      &             &              &      &             & \ion{H}{ii} \\
SMM 8 &	13 07 09.51	& -61 23 41.00 & 13.9 & $>1.5$ & IR bright\tablefootmark{a}\\ 
SMM 9 &	13 07 12.85	& -61 22 44.99 & 16.3 & $6.3\pm3.4$ & IR dark\\ 
IRAS 13042 & 13 07 21.74 & -61 21 40.95 & 15.1 & $>0.9$ & IR bright\\
-6105      &             &              &      &             & \\  	
\hline
\end{tabular} 
\tablefoot{The coordinates refer to the LABOCA 870~$\mu$m peak positions of the sources (\cite{miettinen2018}). The dust temperatures and volume-averaged H$_2$ number densities are adopted from Miettinen (2018) but the latter are scaled upward by a factor of 1.41 (see Sect.~3.3). The source type in the last column is based on the source appearance in the \textit{WISE} IR images (\cite{miettinen2018}).\tablefoottext{a}{The clump is associated with a \textit{WISE} IR source that appears to be a shock emission knot.}\tablefoottext{b}{The clump appears to be associated with a weak \textit{WISE} IR source but it is likely an extragalactic object (see Sect.~4.1).} }
\end{minipage} }
\end{table}

\begin{table}[H]
\renewcommand{\footnoterule}{}
\caption{Observational parameters.}
{\normalsize
\begin{minipage}{1\columnwidth}
\centering
\label{table:observations}
\begin{tabular}{c c c c c}
\hline\hline 
Source & $t_{\rm ON}$\tablefootmark{a} & PWV & $T_{\rm sys}$\tablefootmark{b} & $1\sigma$ rms\tablefootmark{b} \\
       & [min] & [mm] & [K] & [mK] \\ 
\hline 
SMM 1 & 6.4 & 0.5 & 94 & 13.1 \\ 
SMM 2 & 6.4 & 0.5 & 94 & 13.1 \\ 
SMM 3 &	6.4 & 0.5 & 95 & 12.8 \\
BLOB 2 & 6.4 & 0.5 \& 1.2\tablefootmark{c} & 106 & 11.9 \\
BLOB 1 & 6.4 & 0.6 & 100 & 14.4 \\
SMM 4 &	6.4 & 0.6 & 98 & 12.9 \\
SMM 5 &	6.4 & 0.6 & 97 & 8.7 \\
IRAS 13037-6112 & 6.4 & 0.6 & 97 & 11.8 \\      
SMM 6 &	6.5 & 0.6 & 96 & 14.5 \\
SMM 7 &	6.4 & 0.6 & 96 & 12.8 \\
IRAS 13039-6108 & 6.4 & 0.6 & 95 & 15.0 \\     
SMM 8 &	6.5 & 0.6 & 95 & 12.7 \\ 
SMM 9 &	6.4 & 0.6 & 95 & 13.1 \\
IRAS 13042-6105 & 6.4 & 0.6 \& 1.2\tablefootmark{c} & 107 & 16.7 \\
\hline
\end{tabular} 
\tablefoot{\tablefoottext{a}{On-source integration time.}\tablefoottext{b}{The system temperature and $1\sigma$ rms noise level (at a velocity resolution of 0.21~km~s$^{-1}$) are given in the main-beam brightness temperature scale.}\tablefoottext{c}{The source was observed on two different days (the higher PWV occurred on 16 December 2019).} }
\end{minipage} }
\end{table}

\section{Analysis and results}

\subsection{Line identification and spectral line parameters}

To identify the spectral lines in the observed frequency bands, 
we used the {\tt Weeds} interface, which is an
extension of {\tt CLASS90} (\cite{maret2011}), to access the Cologne Database for Molecular Spectroscopy (CDMS\footnote{\url{http://www.astro.uni-koeln.de/cdms}}; \cite{muller2005}). The identified lines and selected spectroscopic parameters 
are listed in Table~\ref{table:lines}. As for the detection threshold, we used three times the rms noise level ($3\sigma$) in the spectrum.
We note that all the lines were detected in the USB, and no lines were visible in the LSB, although there are lines such as 
C$_2$S$(N_J=6_5-5_6)$ at 161.16~GHz and C$_2$S$(N_J=0_1-1_2)$ at 166.75~GHz that one could have expected to be detectable owing to the fairly low upper-state energies of the transitions (19.2~K and 9.4~K, respectively).

The spectra are shown in Figs.~\ref{figure:so}--\ref{figure:c2h}. The SO and SiO lines were fitted by single Gaussian functions using {\tt CLASS90}. All the other detected species, that is H$^{13}$CN, H$^{13}$CO$^+$, HN$^{13}$C, and C$_2$H exhibit  
hyperfine splitting of their rotational transitions. Hence, with the exception of HN$^{13}$C, the corresponding lines were fit using the hyperfine structure method of {\tt CLASS90}.

The $J=2-1$ transition of H$^{13}$CN is split into six hyperfine components owing to the nuclear quadrupole interaction by the N nucleus (e.g. \cite{fuchs2004}). The hyperfine component frequencies were taken from the Jet Propulsion Laboratory (JPL) spectroscopic database\footnote{\url{http://spec.jpl.nasa.gov/}} (\cite{pickett1998}). The central frequency was taken to be 172.677959~GHz, which corresponds to that of the strongest hyperfine component $F=3-2$ with a relative intensity of $R_i = 7/15$. 
The H$^{13}$CO$^+(J=2-1)$ line has eight hyperfine components owing to the presence of H and $^{13}$C nuclei (\cite{schmid2004}). To fit this hyperfine structure, we used the component frequencies from the CDMS (the strongest component, $F_1,\,F = 5/2,\,3 - 3/2,\,2$, with $R_i = 7/20$ has a frequency of 173.5067081~GHz). Although the rotational lines of HN$^{13}$C are known to exhibit hyperfine splitting (e.g. \cite{frerking1979}; \cite{turner2001}; \cite{hirota2003}; \cite{vandertak2009}; \cite{padovani2011}),  
to our knowledge the frequencies and relative intensities of the hyperfine components 
of the HN$^{13}$C$(J=2-1)$ transition have not been published or are not available at the CDMS or the JPL database. Hence, the HN$^{13}$C$(J=2-1)$ lines were fitted using single Gaussian profiles, which follows an approach used in previous studies and is justified by the heavy blending of the hyperfine components (e.g. \cite{hilyblant2010}; \cite{jin2015}; \cite{saral2018}). In the C$_2$H molecule, the coupling of angular momentum and spin of the H nucleus leads to the hyperfine splitting of its rotational levels (e.g. \cite{reitblat1980}; \cite{padovani2009}). We fit the detected hyperfine structure of the $N=2-1$ line of C$_2$H using the frequencies and relative intensities of the seven components from Padovani et al. (2009; Table~5 therein). The reference hyperfine component was taken to be the strongest component $J,\,F = 5/2,\,3 - 3/2,\,2$ with $R_i=7/20$.

The derived line parameters are listed in Columns~3--6 in Table~\ref{table:parameters}. Besides the formal $1\sigma$ fitting errors output by {\tt CLASS90}, the errors in the peak intensity ($T_{\rm MB}$) and the integrated intensity of the line ($\int T_{\rm MB} {\rm d}v$) also include the 10\% calibration uncertainty. These two sources of uncertainty were added in quadrature.

\begin{table*}
\caption{Detected spectral lines and selected spectroscopic parameters.}
\begin{minipage}{2\columnwidth}
\centering
\renewcommand{\footnoterule}{}
\label{table:lines}
\begin{tabular}{c c c c c c c c}
\hline\hline
Transition & $\nu$ & $E_{\rm u}/k_{\rm B}$ & $\mu$ & $B$ & $Z_{\rm rot}^{\rm CDMS}$ & $Z_{\rm rot}^{\rm JPL}$ & $Z_{\rm rot}^{\rm Eq.~(\ref{eqn:part})}$\\ 
           & [MHz] & [K] & [D] & [MHz] & & & \\
\hline 
SO$(N_J=4_4-3_3)$ & 172\,181.40340 & 33.77 & 1.535 & 21\,523.556 & 15.903 & 15.904 & 9.409\\
H$^{13}$CN$(J=2-1)$ & 172\,677.95900\tablefootmark{a} & 12.43 & 2.9852 & 43\,170.127 & 14.621 & 14.622 & 4.858 \\
H$^{13}$CO$^+(J=2-1)$ & 173\,506.70810\tablefootmark{b} & 12.49 & 3.90 & 3\,377.30 & 4.852 & 4.852 & 58.173 \\
SiO$(J=4-3)$ & 173\,688.31 & 20.84 & 3.098 & 21\,711.96 & 9.338 & 9.339 & 9.330\\
HN$^{13}$C$(J=2-1)$\tablefootmark{c} & 174\,179.411 & 12.54 & 2.699 & 43\,545.61 & \ldots & 4.835 & 4.819 \\
C$_2$H$(N=2-1)$ & 174\,663.199\tablefootmark{d} & 12.58 & 0.769 & 3\,674.52 & 19.284 & 19.284 & 53.495 \\
\hline
\end{tabular} 
\tablefoot{The spectroscopic data were compiled from the CDMS database unless otherwise stated. In Columns~2--5 we list the line rest frequency, upper-state energy divided by the Boltzmann constant, permanent electric dipole moment, and the rotational constant of the molecule. In Columns~6--8 we tabulate the partition function values at $T_{\rm ex}=9.375$~K taken from the CDMS and JPL databases and calculated using Eq.~(\ref{eqn:part}), respectively (Sect.~3.3). \tablefoottext{a}{Frequency of the strongest hyperfine component $F = 3 - 2$ (relative intensity $R_i=7/15$) taken from the JPL database.}\tablefoottext{b}{Frequency of the strongest hyperfine component $F_1,\,F = 5/2,\,3 - 3/2,\,2$ ($R_i=7/20$).}\tablefoottext{c}{The quoted parameters for HN$^{13}$C$(J=2-1)$ were taken from the JPL database (not available in the CDMS).}\tablefoottext{d}{Frequency of the strongest hyperfine component $J,\,F = 5/2,\,3 - 3/2,\,2$ ($R_i=7/20$; \cite{padovani2009}, Table~5 therein).}}
\end{minipage} 
\end{table*}

\subsection{Line optical thicknesses and excitation temperatures}

For spectral lines split into hyperfine components the optical thickness, $\tau$, can be derived via 
the relative strengths of the different components. However, the hyperfine structure in the detected 
H$^{13}$CO$^+(J=2-1)$ (and HN$^{13}$C$(J=2-1)$) lines could not be resolved to 
reliably derive the corresponding optical thicknesses. In the case of H$^{13}$CN$(J=2-1)$, 
the six hyperfine components were resolved into three separate groups, and hence the optical thickness 
could be derived via the {\tt CLASS90} hyperfine structure fitting method. However, the derived values 
have relatively large uncertainties owing to the partial blending of the hyperfine lines. In the case of
C$_2$H$(N=2-1)$, we could clearly resolve seven hyperfine components, and hence determine the optical thickness via 
the aforementioned method of {\tt CLASS90}. The excitation temperatures of the H$^{13}$CN$(J=2-1)$ and 
C$_2$H$(N=2-1)$ transitions were then calculated by using the formula: 

\begin{equation}
\label{eqn:Tex}
T_{\rm ex}= \frac{h \nu}{k_{\rm B}} \left \{ \ln \left[\frac{1}{\frac{T_{\rm MB}}{1-e^{-\tau}}\frac{k_{\rm B}}{h\nu}+F(T_{\rm bg} )}+1\right]\right \}^{-1}\,,
\end{equation}
where $h$ is the Planck constant and the function $F(T_{\rm bg})$ is defined as $F(T_{\rm bg})=\left(e^{h\nu /k_{\rm B}T_{\rm bg}}-1\right)^{-1}$. We assumed that the background temperature, $T_{\rm bg}$, is equal to that of the cosmic microwave background (CMB) radiation, that is $T_{\rm bg}=T_{\rm CMB}=2.725$~K (e.g. \cite{fixsen2009}).

For all the other lines, we assumed that the rotational transition is thermalised at the dust temperature of the clump, that is 
we assumed that $T_{\rm ex}=T_{\rm dust}$ (see Column~4 in Table~\ref{table:sources}). The corresponding line optical 
thicknesses were calculated by solving Eq.~(\ref{eqn:Tex}) for $\tau$. The values of $\tau$ and $T_{\rm ex}$ are listed in Columns~7 and 8 in Table~\ref{table:parameters}. 

\begin{table*}
\caption{Statistics of the molecular abundances.}
{\footnotesize
\begin{minipage}{1\columnwidth}
\centering
\renewcommand{\footnoterule}{}
\label{table:statistics}
\begin{tabular}{c c c c c c}
\hline\hline
 & SO & H$^{13}$CN & H$^{13}$CO$^+$ & HN$^{13}$C & C$_2$H \\ 
 & [$10^{-10}$] & [$10^{-10}$] & [$10^{-10}$] & [$10^{-11}$] & [$10^{-9}$]\\
\hline
& \multicolumn{5}{c}{IR dark clumps}\\
Mean & $1.7\pm0.5$ & $2.7\pm2.2$ & $7.4\pm2.0$ & $2.6\pm0.6$ & $2.7\pm0.4$  \\
Median & 1.5 & 0.7 & 8.3 & 2.6 & 2.7\\
\hline
& \multicolumn{5}{c}{IR bright clumps}\\
Mean & $1.6\pm0.4$ & $4.0\pm1.8$ & $5.5\pm0.9$ & $2.2\pm0.3$ & $2.2\pm0.3$ \\
Median & 1.1 & 1.0 & 5.7 & 2.3 & 2.4\\
\hline
& \multicolumn{5}{c}{\ion{H}{ii} region\tablefootmark{a}}\\
 & $0.9\pm0.2$ & $4.1\pm2.4$ & $6.8\pm0.8$ & $2.0\pm0.2$ & $1.4\pm0.2$ \\
\hline 
 & \multicolumn{5}{c}{Full sample}\\
Mean & $1.6\pm0.3$ & $3.5\pm1.3$& $6.2\pm1.0$ & $2.4\pm0.3$ & $2.3\pm0.3$\\
Median & 1.2 & 1.0 & 6.1 & 2.5 & 2.5\\
\hline
\end{tabular} 
\tablefoot{SiO was detected in only one source (SMM~3), so it is excluded from this table. A K-M survival analysis was used to calculate the mean and median abundances whenever the data set contained left-censored values or upper limits. The uncertainty quoted in the mean values is the standard error of the mean (i.e. standard deviation divided by the square root of the sample size).\tablefoottext{a}{The present sample contains only one \ion{H}{ii} region (IRAS~13039-6108), so the reported values refer to the abundances in that source.}}
\end{minipage} }
\end{table*}

\subsection{Molecular column densities and fractional abundances}

To calculate the beam-averaged column densities of the molecules, $N$, we made  
an assumption of local thermodynamic equilibrium (LTE). Under this assumption, the molecular column 
density is given by (e.g. \cite{turner1991}, Appendix~therein):

\begin{equation}
\label{eqn:CD}
N=\frac{3h \epsilon_0}{2\pi^2}\frac{1}{\mu^2S}\frac{Z_{\rm rot}(T_{\rm ex})}{g_Kg_I}e^{E_{\rm u}/k_{\rm B}T_{\rm ex}}F(T_{\rm ex})\int \tau(v){\rm d}v\, ,
\end{equation}
where $\epsilon_0$ is the vacuum permittivity, $S$ is the line strength, $Z_{\rm rot}$ is the partition function, 
$g_K$ is the $K$-level degeneracy, $g_I$ is the reduced nuclear spin degeneracy, and the function $F(T_{\rm ex})$ is defined in a similar way as $F(T_{\rm bg})$ in Sect.~3.2. For a Gaussian line profile, the last integral term in Eq.~(\ref{eqn:CD}) can be expressed as: 

\begin{equation}
\label{eqn:tau}
\int \tau(v){\rm d}v = \frac{\sqrt{\pi}}{2\sqrt{\ln 2}}\Delta v \tau_0 \simeq 1.064\Delta v \tau_0\, ,
\end{equation}
where $\tau_0$ is the peak optical thickness of the line. The total optical thickness of the lines split into hyperfine components was 
calculated by scaling the peak value by the relative strength of the strongest component that was always used as the reference line. 

The column densities of SO, H$^{13}$CN, H$^{13}$CO$^+$, SiO, and C$_2$H were calculated from the line optical thickness as outlined above, but HN$^{13}$C was treated in a different way. As mentioned in Sect.~3.1, the hyperfine component data for the HN$^{13}$C$(J=2-1)$ transition are not published. For this reason, we assumed that the line is optically thin ($\tau \ll 1$), which is supported by the values of the derived peak optical thicknesses (Table~\ref{table:parameters}, Column~7), and calculated the column density from the integrated line intensity (this approach for HN$^{13}$C was also used by e.g. Hily-Blant et al. (2010)). 
When $\tau \ll 1$, Eq.~(\ref{eqn:Tex}) can be solved for the velocity-integrated optical thickness as:

\begin{equation}
\label{eqn:tau2}
\int \tau(v){\rm d}v \simeq \frac{k_{\rm B}}{h \nu} \left[F(T_{\rm ex})-F(T_{\rm bg})\right]^{-1}\int T_{\rm MB}{\rm d}v\,. 
\end{equation}

The values of the product $\mu^2S$ appearing in Eq.~(\ref{eqn:CD}) were taken from the Splatalogue database\footnote{\url{http://www.cv.nrao.edu/php/splat/}}. All the detected molecular species are linear molecules, which allowed 
us to approximate the partition function as (e.g. \cite{mangum2015}; Eq.~(52) therein):

\begin{equation}
\label{eqn:part}
Z_{\rm rot} \simeq \frac{k_{\rm B}T_{\rm ex}}{hB}+\frac{1}{3}\,.
\end{equation}
Equation~(\ref{eqn:part}) is valid at high temperatures defined as $k_{\rm B}T_{\rm ex}\gg hB$ (see also \cite{turner1991}). As shown in the last three columns in Table~\ref{table:lines}, Eq.~(\ref{eqn:part}) yields very similar partition function values for SiO and HN$^{13}$C as those given in the CDMS and JPL databases. We note that at a reference $T_{\rm ex}$ of 9.375~K, the $k_{\rm B}T_{\rm ex}/hB$ ratios for the present species range from 4.5 to 57.8 and, hence, the aforementioned high-temperature threshold where Eq.~(\ref{eqn:part}) is valid is expected to be fulfilled. However, for SO and H$^{13}$CN, Eq.~(\ref{eqn:part}) yields lower values than those in the CDMS and JPL databases (by factors of 1.7 and 3 at 9.375~K, respectively). On the other hand, for H$^{13}$CO$^+$ and C$_2$H, Eq.~(\ref{eqn:part}) gives $Z_{\rm rot}({\rm 9.375~K})$ values that are 12 and 2.8 times larger than in the aforementioned two databases. In the case of H$^{13}$CN and C$_2$H, the discrepancy could arise from the fact that the partition function values reported in the CDMS and JPL catalogues take the vibrational states (and potentially electronic corrections) into account (cf.~\cite{favre2014}). Hence, a comparison between the rotational partition functions calculated using Eq.~(\ref{eqn:part}) and the partition function values reported in the CDMS and JPL catalogues is not always straightforward, and care should be taken when comparing the molecular column densities and abundances between different studies where $Z_{\rm rot}$ might have been determined in different ways. Because linear molecules do not have any $K$-level degeneracy and the observed species do not have identical nuclei that could be interchanged while preserving the symmetry, both $g_K$ and $g_I$ are equal to unity (e.g. \cite{turner1991}). 

The fractional abundances of the molecules were calculated by dividing the column density 
by the H$_2$ column density, that is $x = N/N({\rm H_2})$. The $N({\rm H_2})$ values were derived
from our LABOCA dust continuum data re-reduced by Miettinen (2018). For a proper comparison with the present line observations, the LABOCA map was smoothed to the angular resolution of the line observations, where the beam size was about 1.8 times larger than in our LABOCA observations. Under the assumption of optically thin dust emission, which is valid at the wavelength probed by our LABOCA observations, $N({\rm H_2})$ can be calculated from the dust peak surface brightness, $I_{\nu}^{\rm dust}$, as (e.g. \cite{kauffmann2008}):

\begin{equation}
N({\rm H_2}) = \frac{I_{\nu}^{\rm dust}}{B_{\nu}(T_{\rm dust})\mu_{\rm H_2}m_{\rm H}\kappa_{\nu}R_{\rm dg}}\,,
\end{equation}
where $B_{\nu}(T_{\rm dust})$ is the Planck function, $\mu_{\rm H_2}$ is the mean molecular weight per H$_2$ molecule (taken to be 2.82, which corresponds to a hydrogen mass percentage of 71\% (i.e. $X=0.71$; \cite{kauffmann2008}, Appendix~A.1 therein)), 
$m_{\rm H}$ is the mass of a hydrogen atom, $\kappa_{\nu}$ is the dust opacity, which was assumed to be 1.38~cm$^2$~g$^{-1}$ at 870~$\mu$m (based on the Ossenkopf \& Henning (1994) dust model of graphite-silicate dust grains that have coagulated and accreted thin ice mantles over a period of $10^5$~yr at a gas density of $n_{\rm H}=10^5$~cm$^{-3}$), and $R_{\rm dg}$ is the dust-to-gas mass ratio that we assumed to be 1/141, that is 1.41 times lower than the canonical dust-to-hydrogen mass ratio of 1/100, where the factor 1.41 
($\simeq1/X$) is the ratio of total mass (H+He+metals) to hydrogen mass.

\begin{figure}[!htb]
\centering
\resizebox{0.98\hsize}{!}{\includegraphics{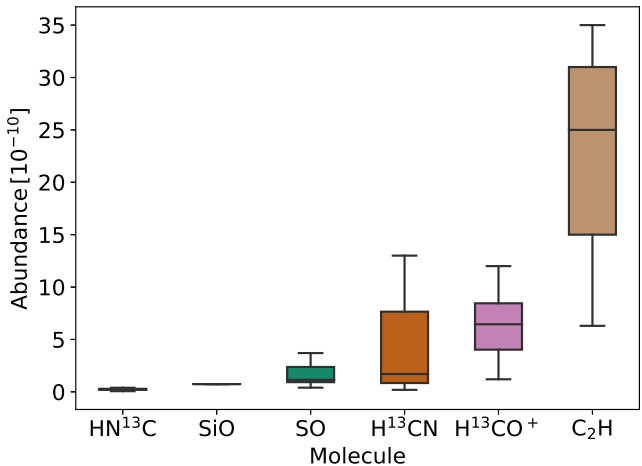}}
\caption{Box plot (or box-and-whisker plot) showing the distributions of the derived molecular fractional abundances. The boxes are 
arranged in ascending order according to the median value. The height of the box represents the interquartile range ${\rm IQR}\equiv[Q_1,\,Q_3]$ (i.e. from the 25th to 75th percentile). A horizontal line that goes through the box shows the median abundance ($Q_2$ or the 50th percentile). The whiskers are defined to extend from $Q_1-1.5\times {\rm IQR}$ to $Q_3+1.5\times {\rm IQR}$ unless the minimum and maximum values of the distribution are encountered before that (in which case the whiskers extend to the minimum and maximum values). We note that only the uncensored data are used to plot the distributions (i.e. the upper limits are ignored), and SiO was detected in only one of the target sources (SMM~3). }
\label{figure:boxall}
\end{figure}

The beam-averaged column densities and abundances with respect to H$_2$ calculated in this section are listed in the last two columns in Table~\ref{table:parameters}. Statistics of the derived fractional abundances are given in Table~\ref{table:statistics}. 
To calculate the statistical parameters presented in Table~\ref{table:statistics}, we applied survival analysis to take the left-censored data or upper limits into account. To estimate the survival curve, we used a non-parametric method called the Kaplan-Meier (K-M) estimator (\cite{kaplan1958}). The method was implemented in {\tt R} (version 3.6.2; \cite{R2019}) using the package Nondetects And Data Analysis for environmental data ({\tt NADA}; \cite{helsel2005}; \cite{lee2017}). The distributions of the derived fractional abundances are visualised using box plots in Fig.~\ref{figure:boxall}. Finally, we also calculated the HN$^{13}$C/H$^{13}$CN and HN$^{13}$C/H$^{13}$CO$^+$ abundance ratios from the corresponding column densities, and these values are listed in Table~\ref{table:ratios}.

\begin{table}[H]
\renewcommand{\footnoterule}{}
\caption{HN$^{13}$C/H$^{13}$CN and HN$^{13}$C/H$^{13}$CO$^+$ abundance ratios.}
{\small
\begin{minipage}{1\columnwidth}
\centering
\label{table:ratios}
\begin{tabular}{c c c}
\hline\hline 
Source & $[{\rm HN^{13}C}]/[{\rm H^{13}CN}]$ & $[{\rm HN^{13}C}]/[{\rm H^{13}CO^+}]$  \\
\hline 
SMM 1 & $0.63\pm0.21$ & $0.03\pm0.002$ \\ 
SMM 2 & $0.21\pm0.09$ & $0.06\pm0.01$ \\ 
SMM 3 &	$0.19\pm0.16$ & $0.04\pm0.01$  \\
BLOB 2 & $0.54\pm0.12$ & $0.04\pm0.003$  \\
BLOB 1 & $0.05\pm0.02$ & $0.06\pm0.002$  \\
SMM 4 &	$0.61\pm0.10$ & $0.05\pm0.02$ \\
SMM 5 &	$0.04\pm0.02$ & $0.03\pm0.002$  \\
IRAS 13037-6112 & $0.015\pm0.004$ & $0.02\pm0.002$  \\      
SMM 6 &	$>0.15$ & $0.09\pm0.02$  \\
SMM 7 &	$>0.09$ & $0.05\pm0.01$  \\
IRAS 13039-6108 & $0.05\pm0.03$ & $0.03\pm0.002$  \\     
SMM 8 &	$0.03\pm0.02$ & $0.03\pm0.002$  \\ 
SMM 9 &	$0.26\pm0.06$ & $0.03\pm0.001$  \\
IRAS 13042-6105 & $>0.26$ & $0.05\pm0.01$  \\
\hline
& \multicolumn{2}{c}{IR dark clumps}\\
Mean & $0.38\pm0.11$ & $0.04\pm0.004$ \\
Median & 0.54 & 0.04\\
\hline
& \multicolumn{2}{c}{IR bright clumps\tablefootmark{a}}\\
Mean & $0.27\pm0.10$ & $0.05\pm0.008$\\
Median & 0.19 & 0.05\\
\hline
 & \multicolumn{2}{c}{Full sample}\\
Mean & $0.30\pm0.07$ & $0.04\pm0.005$\\
Median & 0.21 & 0.04 \\
\hline
\end{tabular} 
\tablefoot{The abundance ratios were calculated from the molecular column densities. The quoted lower limits take the derived uncertainties into account. Survival analysis based on the K-M method was used to calculate the mean and median value of the $[{\rm HN^{13}C}]/[{\rm H^{13}CN}]$ ratio to take the right-censored values (i.e. lower limits) into account.\tablefoottext{a}{The source IRAS~13039-6108 was not included in the sub-sample of IR bright sources because it is associated with an \ion{H}{ii} region and represents a third type of clumps included in the present study.}}
\end{minipage} }
\end{table}

\subsection{Correlation analysis}

\begin{figure*}
\begin{center}
\includegraphics[scale=0.33]{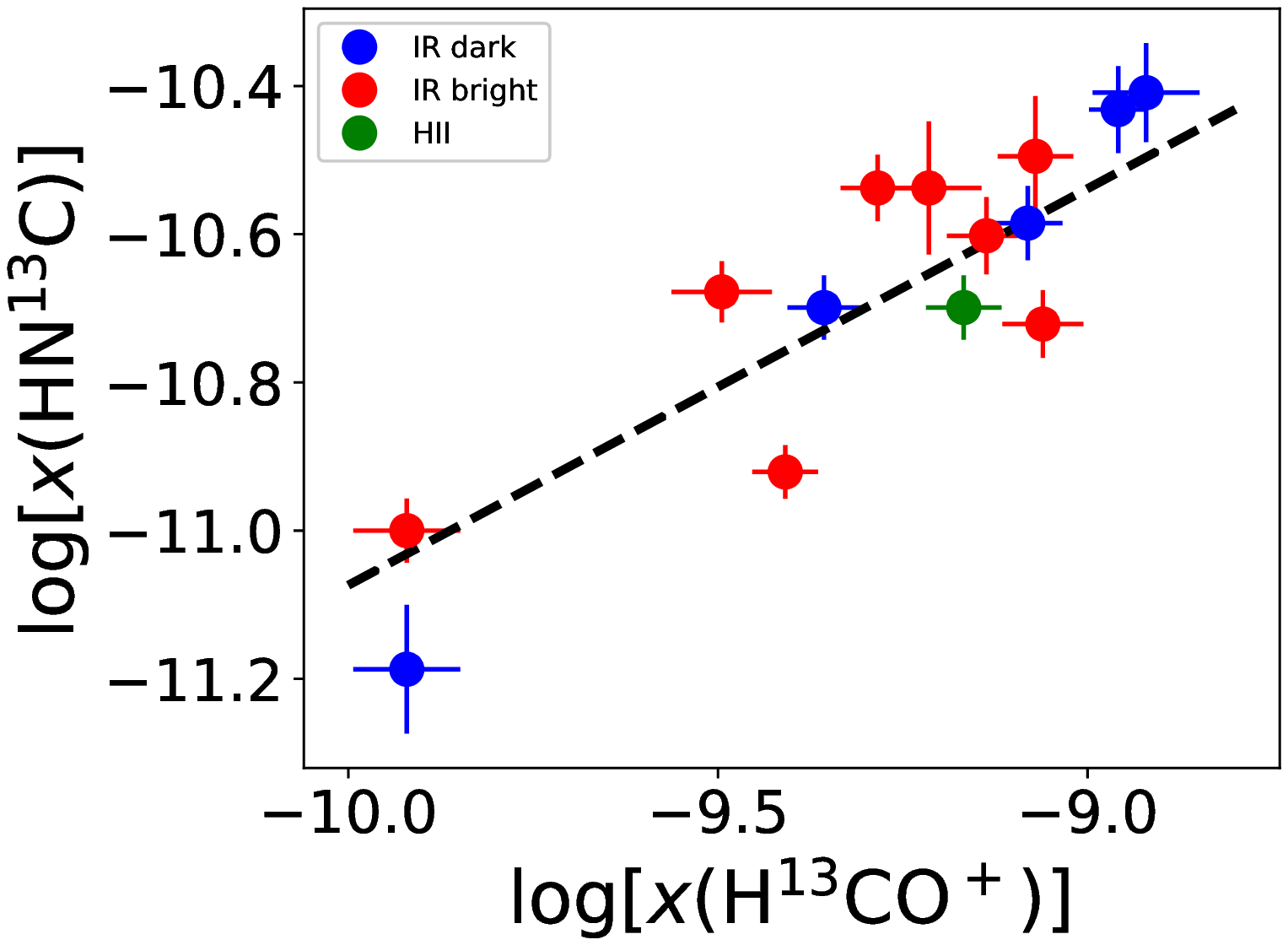}
\includegraphics[scale=0.33]{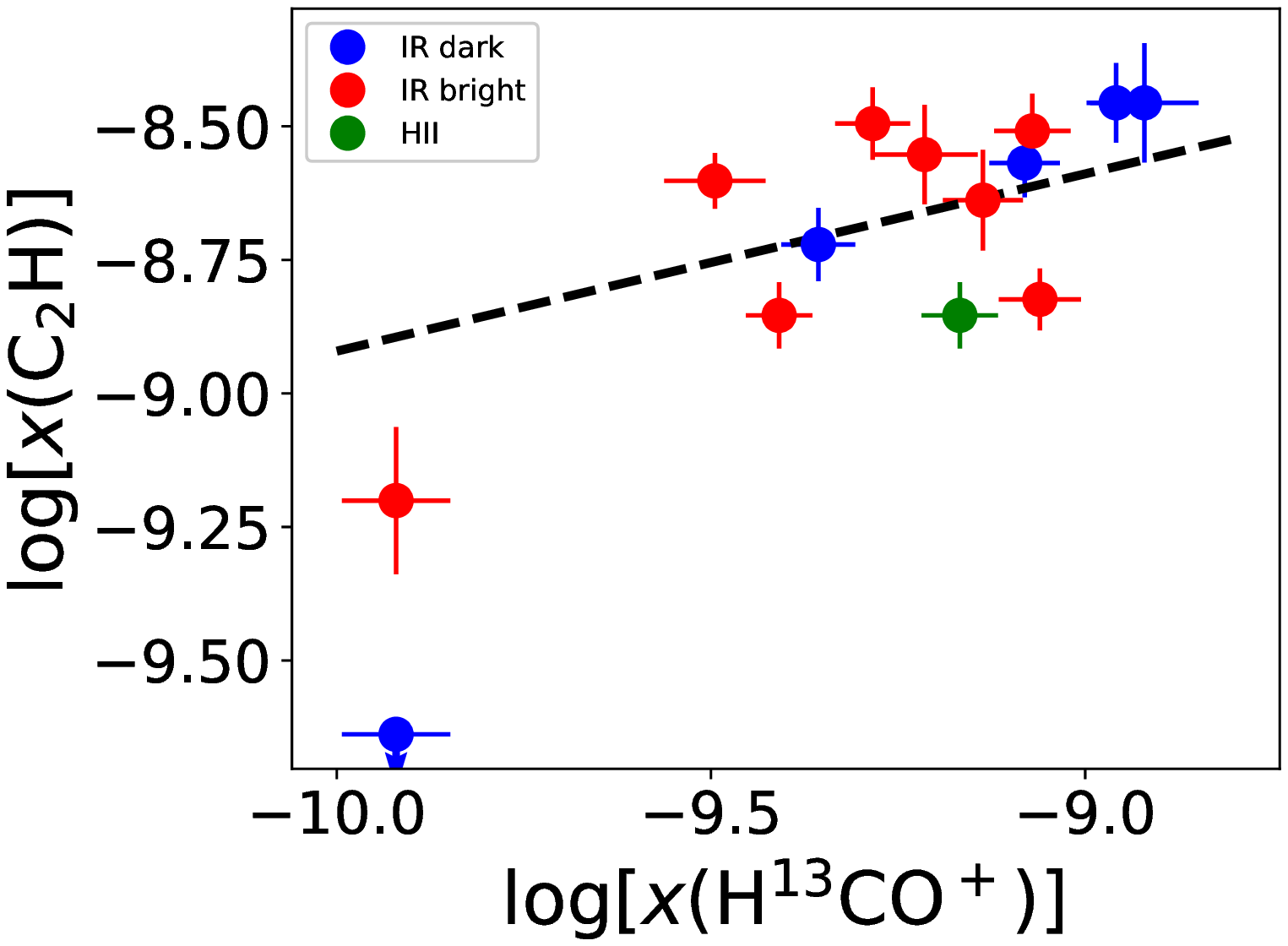}
\includegraphics[scale=0.33]{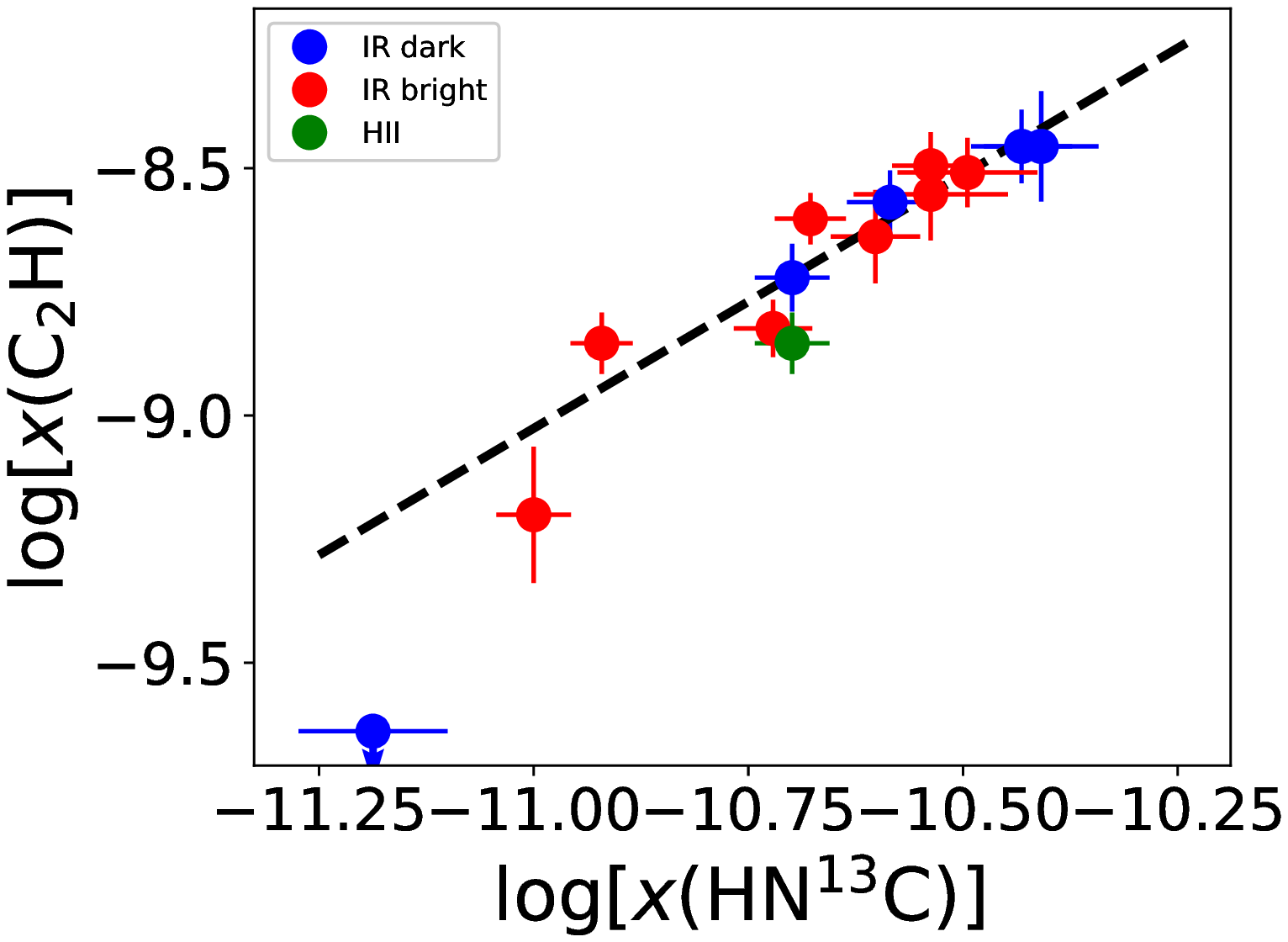}
\caption{Scatter plots between the variables that were found to be linearly correlated with a Pearson correlation coefficient of 
$r > 0.7$. The arrows pointing down indicate upper limits. The dashed lines show the linear regression fits to the uncensored data points (i.e. the upper limits were not taken into account).}
\label{figure:corr}
\end{center}
\end{figure*}

To search for correlations between the fractional molecular abundances and abundance ratios derived in the present study and the dust temperatures and H$_2$ number densities listed in Table~\ref{table:sources}, we first created a correlation matrix where each cell shows the linear or Pearson correlation coefficient, $r$, between two variables. We note that in this analysis we used the logarithms of the abundances and densities. The relationship between two variables is generally considered strong when $\left|r\right|>0.7$ (e.g. \cite{moore2018}). Hence, as a threshold for potential correlation, we adopted a value of $\left|r\right|>0.7$. Such correlations were found between the H$^{13}$CO$^+$ abundance and those of HN$^{13}$C and C$_2$H, and between the abundances of HN$^{13}$C and C$_2$H. The corresponding scatter plots are shown in Fig.~\ref{figure:corr}. 

In Fig.~\ref{figure:corr}, the IR dark, IR bright, and the \ion{H}{ii} sources are shown in different colours (blue, red, and green, respectively) to better illustrate how the corresponding data points populate the plotted parameter spaces. Linear regression fits were made to the aforementioned relationships, and the derived formulas together with the $r$ values are provided in Table~\ref{table:regression}. The $y$ error bars were taken into account in the linear fits, but the upper limits were omitted.

\section{Discussion}

\subsection{Detection rates and spectral line profiles}

The observed spectral lines of SO, H$^{13}$CO$^+$, and HN$^{13}$C were detected in all the target sources. The next highest detection rates 
were for C$_2$H (92.9\%) and H$^{13}$CN (85.7\%). The SiO line was detected in only one of the target sources, SMM~3, which means that 
the corresponding detection rate was only 7.1\%. The clump SMM~3 was also the only source in the Searhorse IRDC that was found to show a hint of SiO$(5-4)$ emission ($\sim2\sigma$) by Miettinen (2012).

Interestingly, the SO, H$^{13}$CN, H$^{13}$CO$^+$, HN$^{13}$C, and C$_2$H lines are seen in absorption towards SMM~5, SMM~6, and SMM~7. 
However, the H$^{13}$CN line towards SMM~6 and the H$^{13}$CN and C$_2$H lines towards SMM~7 were not found to fulfil our detection threshold of $\geq3\sigma$, but there is clearly a hint of line absorption, which is further supported by the positive detection of SO, H$^{13}$CO$^+$, and HN$^{13}$C absorption. 

\begin{figure}[!htb]
\centering
\resizebox{\hsize}{!}{\includegraphics{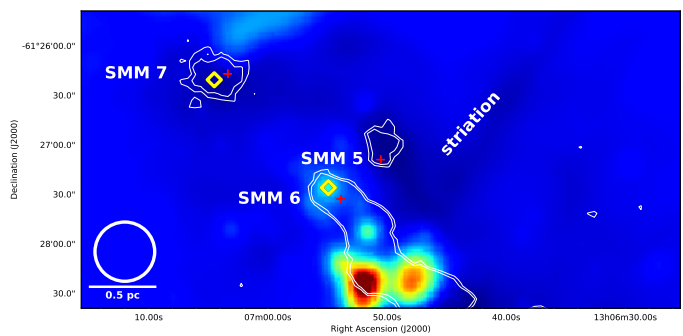}}
\caption{Zoom-in of Fig.~\ref{figure:map} towards SMM~5, 6, and 7 for which some of the spectral lines are seen in absorption. Only 
the $3\sigma$ and $6\sigma$ LABOCA contours are plotted for clarity. Besides the clumps SMM~5, 6, and 7, also the striation seen 
as a dark absorption feature is labelled (cf.~Fig.~\ref{figure:spire}, where the striation is seen in dust emission). The red plus signs indicate the target positions of the present spectral line observations, while the yellow diamond symbols show the positions of the \textit{WISE} sources seen towards SMM~6 and SMM~7. The latter one is likely to be an extragalactic object, that is a star-forming galaxy or an active galactic nucleus, and hence SMM~7 is considered an IR dark clump (\cite{miettinen2018}). A scale bar of 0.5~pc projected length, and the beam size at the tuning frequency of the present spectral line observations are shown in the bottom left corner.}
\label{figure:zoom}
\end{figure}

A zoom-in version of Fig.~\ref{figure:map} towards SMM~5, 6, and 7 is shown in Fig.~\ref{figure:zoom}. As indicated in the figure, SMM~6 is associated with an embedded young stellar object (YSO) seen in the mid-IR. Hence, the detected absorption lines might be caused by the outer layers of the clump absorbing the millimetre-wavelength continuum photons emitted by the warm or hot ($\sim100$~K; \cite{rathborne2011}) dust close to the central YSO (e.g. \cite{doty1997}). There is also an IR source detected with \textit{WISE} only at 22~$\mu$m towards SMM~7, but the source, J130704.50-612620.6, is likely a chance projection of a star-forming galaxy or an active galactic nucleus (\cite{miettinen2018}). This extragalactic object seen towards SMM~7 could act as the continuum source against which the absorption lines are generated by the clump medium. The clump SMM~5 appears to not have any IR point sources within the $\sim36\arcsec$ beam of the present line observations, and hence the detection of absorption lines towards it is somewhat surprising. As can be seen in Fig.~\ref{figure:map}, SMM~5 is associated with an elongated IR dark feature that could be associated with the striation on its western side. In principle, the system's geometry could be such that the dense gas associated with the striation acts as a foreground absorbing layer for the sight-line towards SMM~5, but this is highly speculative at present. Spectral line imaging of the Seahorse and its surroundings are needed to quantitatively study these issues. 

We note that the C$^{17}$O$(2-1)$ lines were detected in emission towards SMM~5, SMM~6, and SMM~7 by Miettinen (2012; Fig.~2 therein), but the beam size of those observations was 1.3 times smaller than that of the present observations, and the target positions towards the aforementioned clumps were offset from the present ones by $5\farcs7-25\farcs6$ because they were based on an earlier version of our LABOCA map and different method of clump extraction (\cite{miettinenharju2010}; \cite{miettinen2018}).  

Aside being visible in absorption, the H$^{13}$CO$^+$ and HN$^{13}$C lines towards SMM~6 exhibit a double-peaked profile with a stronger blueshifted peak compared to the redshifted peak. Such blue asymmetric profiles were also detected by Miettinen (2012) in the $^{13}$CO$(2-1)$ lines towards clumps in the southern half of the Seahorse IRDC, most notably towards (SMM~2; Fig.~3 therein), and they 
are indicative of large-scale infall gas motions. This supports the hypothesis that the Seahorse IRDC is still accreting mass 
from its surrounding gas reservoirs (Sect.~1). 

Finally, the spectral line profile of H$^{13}$CO$^+$ towards SMM~3 shows an 
absorption-like feature next to the actual spectral line, while the H$^{13}$CO$^+$ and HN$^{13}$C absorption lines towards SMM~5, 
H$^{13}$CN absorption line towards SMM~6 (below the detection threshold), and the C$_2$H absorption lines towards SMM~5 and SMM~6 exhibit emission-like features next to the actual lines. These are indications of the presence of H$^{13}$CN, H$^{13}$CO$^+$, HN$^{13}$C, and C$_2$H gas in the observations' off-position ($50\arcsec$ azimuthal offset from the target position). If there is molecular gas emission in the observations' off-position, then, owing to the wobbler switching technique, when the off-spectrum is subtracted from the on-source spectrum, one can see an absorption-like feature in the final spectrum. Similarly, an absoprtion line in the off-position can appear as an emission-like feature in the final spectrum.

\subsection{Relationships between the molecular abundances}

We found three potential cases of correlation between fractional abundances of two molecular species (Fig.~\ref{figure:corr}). Below, we discuss each of these in the order of decreasing significance (as quantified by the Pearson correlation coefficient). 

The strongest correlation was found between the abundances of HN$^{13}$C and C$_2$H ($r=0.93$). According to the gas-phase chemistry models, the most abundant isotopologue of HN$^{13}$C, HNC, is primarily formed through the dissociative recombination reaction 
${\rm HCNH^+} + {\rm e^-} \rightarrow {\rm HNC} + {\rm H}$ (e.g. \cite{herbst1978}). Other dissociative recombination reactions that can form HNC are ${\rm H_2CN^+} + {\rm e^-}$ and ${\rm H_2NC^+} + {\rm e^-}$ (\cite{pearson1974}; \cite{allen1980}). The neutral-neutral reaction ${\rm NH_2} + {\rm C} \rightarrow {\rm HNC} + {\rm H}$ can also act as the source of HNC (e.g. \cite{herbst2000}). In the PDRs, C$_2$H can form as a result of the photodissociation of C$_2$H$_2$ (${\rm C_2H_2} + h\nu \rightarrow {\rm C_2H} + {\rm H}$; e.g. \cite{fuente1993}). The C$_2$H molecules can also form via the dissociative recombination reaction ${\rm C_2H_3^+} + {\rm e^-} \rightarrow {\rm C_2H} + {\rm H_2}$ (e.g. \cite{turner2000}). Hence, if C$_2$H is predominantly formed via the aforementioned dissociative recombination reaction, the correlation we found between HN$^{13}$C and C$_2$H could reflect the gas-phase electron abundance (i.e. ionisation degree). If C$_2$H is mainly formed via the photodissociation of C$_2$H$_2$, the positive relationship with HN$^{13}$C could be linked to photoionisation that yields free electrons from which HN$^{13}$C can form. On the other hand, when ultraviolet (UV) photons destroy CO via photodissociation (e.g. \cite{visser2009}), the resulting C atoms can react with CH$_2$ to form C$_2$H (e.g. \cite{turner2000}). The aforementioned correlation could therefore also reflect the gas-phase abundance of C atoms (and hence the importance of UV photodissociation in the formation of both HNC (and HN$^{13}$C) and C$_2$H).

With a correlation coefficient of $r=0.88$, the correlation between the H$^{13}$CO$^+$ and HN$^{13}$C abundances was found 
to be the second strongest of those plotted in Fig.~\ref{figure:corr}. The HCO$^+$ molecules are predominantly formed in the 
reaction ${\rm H_3^+} + {\rm CO} \rightarrow {\rm HCO^+} + {\rm H_2}$ (e.g. \cite{herbst1973}), where the required CO can form in 
the reaction ${\rm C} + {\rm OH} \rightarrow {\rm CO} + {\rm H}$ (e.g. \cite{millar2015}). Ionised carbon can also react with 
water to form HCO$^+$ (${\rm C^+} + {\rm H_2O} \rightarrow {\rm HCO^+} + {\rm H}$; e.g. \cite{rawlings2004}). 
The $^{13}$C isotopologue of HCO$^+$ we observed, H$^{13}$CO$^+$, forms in a similar was as HCO$^+$ but from $^{13}$CO. The isotope 
exchange reaction ${\rm HCO^+} + {\rm ^{13}CO} \rightarrow {\rm H^{13}CO^+} + {\rm CO}$ provides another route for the formation of H$^{13}$CO$^+$ (e.g. \cite{langer1984}; \cite{mladenic2017}). The correlation we found between H$^{13}$CO$^+$ and HN$^{13}$C could therefore be a manifestation of their dependence on carbon. However, the chemical network is complicated by the fact that HCO$^+$ is mainly destroyed via the dissociative recombination ${\rm HCO^+} + {\rm e^-} \rightarrow {\rm CO} + {\rm H}$ (e.g. \cite{liszt2017}). As described above, dissociative recombination reactions can also lead to the formation of HNC, and because the reaction ${\rm HCO^+} + {\rm e^-}$ is the main source of CO from which HCO$^+$ can re-form, our finding is also likely to reflect the degree of ionisation in the gas.  

The positive correlation we found between the H$^{13}$CO$^+$ and C$_2$H abundances ($r=0.74$) is to be expected from the aforementioned two relationships. However, the slope of the linear least squares fit shown in the middle panel in Fig.~\ref{figure:corr} deviates from a flat trend by only $1.5\sigma$. On the other hand, the upper limit to the C$_2$H abundance at the lowest H$^{13}$CO$^+$ abundance, which was not used in the fit, supports the presence of a positive correlation.

\begin{table*}
\renewcommand{\footnoterule}{}
\caption{Equations of the linear regressions plotted in Fig.~\ref{figure:corr} and the corresponding correlation coefficients.}
{\normalsize
\begin{minipage}{2\columnwidth}
\centering
\label{table:regression}
\begin{tabular}{c c}
\hline\hline 
Equation & $r$\tablefootmark{a}  \\ 
       &  \\ 
\hline 
$\log [x({\rm C_2H})]=(1.02\pm0.16)\times \log [x({\rm HN^{13}C})] - (2.25\pm 1.73)$ & 0.93 \\[1ex]
$\log [x({\rm HN^{13}C})]=(0.54\pm0.11)\times \log [x({\rm H^{13}CO^+})] - (5.72\pm 1.05)$ & 0.88   \\[1ex]
$\log [x({\rm C_2H})]=(0.33\pm0.22)\times \log [x({\rm H^{13}CO^+})] - (5.60\pm 2.07)$ & 0.74   \\[1ex]
\hline
\end{tabular} 
\tablefoot{The relationships are listed in the order of decreasing correlation coefficient.\tablefoottext{a}{The Pearson correlation coefficient.}}
\end{minipage} }
\end{table*}

\subsection{Comparison of the molecular abundances in the Seahorse IRDC to those in other IRDCs}

\subsubsection{SO}

Observations of SO spectral lines towards IRDCs are quite rare. 
Sakai et al. (2010) used the Nobeyama Radio Observatory (NRO) 45~m telescope to observe the SO$(N_J=2_2-1_1)$ line at 86~GHz 
towards 20 clumps in different IRDCs. The detection rate was 70\%. The authors reported the abundances with respect to 
H$^{13}$CO$^+$ (their Table~5), so we decided to compare the molecular column densities instead of the fractional abundances with respect to H$_2$. The $N({\rm SO})$ distribution derived in the present work is compared with that from Sakai et al. (2010) in the top left panel in Fig.~\ref{figure:comparison}. All our values are lower than those derived by Sakai et al. (2010), and our mean (median) SO column density is lower by a factor of 53 (42). 

Sanhueza et al. (2012) used the Mopra 22~m telescope to observe the SO$(2_2-1_1)$ line towards a sample of 92 clumps in IRDCs. 
However, the line was detected in only $<8$ clumps, and hence the SO properties were not analysed further in their statistical study. 
To our knowledge, the present detections of SO$(N_J=4_4-3_3)$ line emission at 172~GHz towards an IRDC is the first of its kind. 

Through observations of SO$(N_J=6_5-5_4)$ towards 59 high-mass star-forming objects in total, Gerner et al. (2014) derived median SO abundances of $<1.3\times10^{-10}$, $9\times10^{-10}$, $9\times10^{-10}$, and $3.8\times10^{-9}$ for their IRDCs, high-mass protostellar objects (HMPOs), hot molecular cores (HMCs), and ultracompact (UC) \ion{H}{ii} regions, respectively (assuming excitation temperatures of 15~K, 50~K, 100~K, and 100~K, respectively; see their Tables~3--6). The authors assumed that the gas-to-dust mass ratio is 100, 870~$\mu$m dust opacity is 1.42~cm$^2$~g$^{-1}$, and that the mean molecular weight is 2.8. Hence, we scaled down their reported molecular abundances by a factor of 0.6942 to take these different assumptions into account. The aforementioned value for the IRDCs is comparable (although it is an upper limit) to the present median SO abundance of $1.5\times10^{-10}$ for IR dark clumps, but we found a factor of 8.2 lower median SO abundance for the IR bright clumps than in the HMPOs and HMCs in the Gerner et al. (2014) sample. Finally, the one \ion{H}{ii} region in the Seahorse IRDC exhibits a $x({\rm SO})$ value that is over 40 times lower than the UC \ion{H}{ii} regions' median $x({\rm SO})$ in the Gerner et al. (2014) sample.

\subsubsection{H$^{13}$CN}

The H$^{13}$CN$(1-0)$ transition was part of the Mopra 90~GHz molecular line survey of 15 IRDCs (37 clumps in total) by Vasyunina et al. (2011). The line was detected in only five sources (13.5\% detection rate; their Table~3). The corresponding abundances 
are compared with the present values in the top middle panel in Fig.~\ref{figure:comparison}. We note that Vasyunina et al. (2011) adopted a gas-to-dust mass ratio of 100, and assumed that the dust opacity is 1~cm$^2$~g$^{-1}$ at their observed dust emission wavelength of 1.2~mm. Moreover, a hydrogen molecule mass was used in the denominator of their $N({\rm H_2})$ formula, which effectively means that the mean molecular weight was 2 (see Eq.~(2) in \cite{vasyunina2009}). Hence, we scaled down their molecular abundances by a factor of 0.7735 for a more meaningful comparison with our values (the dust emissivity index, $\beta$, used to scale the dust opacity as $\kappa_{\nu}\propto \nu^{\beta}$ was assumed to be $\beta\simeq1.8$; see \cite{miettinen2018}, and references therein). The H$^{13}$CN abundances in the Seahorse IRDC are similar to those in the Vasyunina et al. (2011) sample. For example, our mean (median) value is only 1.1 (0.6) times their value. However, we note that Vasyunina et al. (2011) calculated the rotational partition functions by interpolating data from the CDMS catalogue values, which complicates our comparison as discussed in Sect.~3.3. Hence, part of the discrepancy can still arise from the different methods of calculation rather than being physical (or the true difference could be larger than quoted above).

As in the case of SO, Sanhueza et al. (2012) observed the H$^{13}$CN$(1-0)$ line but detected it in $<8$ sources, and hence no 
quantitative comparison could be made with their study. Our detection rate of H$^{13}$CN towards the Seahorse IRDC, 85.7\%, seems 
quite high compared to the studies by Vasyunina et al. (2011) and Sanhueza et al. (2012), although the mean abundance of 
the molecule in the Seahorse appears to be very similar to that derived for the Vasyunina et al. (2011) sources.

\subsubsection{H$^{13}$CO$^+$}

The H$^{13}$CO$^+$ species has commonly been included in the previous studies of chemistry in the IRDCs (e.g. \cite{sakai2010}; \cite{vasyunina2011}; \cite{sanhueza2012}; \cite{miettinen2014}; \cite{gerner2014}). The detection rate of this species has also been high in some of the previous surveys, most notably 100\% in the surveys carried out by Sakai et al. (2010) and Gerner et al. (2014), which is also the case in the present study. 

Sakai et al. (2010) reported H$^{13}$CO$^+$ column densities in the range of $(1.3-13.5)\times10^{12}$~cm$^{-3}$ with a mean (median) 
of $(6.2\pm0.8)\times10^{12}$~cm$^{-3}$ ($6.0\times10^{12}$~cm$^{-3}$). The H$^{13}$CO$^+$ column densities we derived, $(1.1-11)\times10^{12}$~cm$^{-3}$ with a mean (median) of $(6.3\pm0.8)\times10^{12}$~cm$^{-3}$ ($6.2\times10^{12}$~cm$^{-3}$) are remarkably similar to those from Sakai et al. (2010). 

In the top right panel in Fig.~\ref{figure:comparison}, we visualise the comparison of the H$^{13}$CO$^+$ abundances with those from previous studies. Again, the values from Vasyunina et al. (2011) were multiplied by 0.7735 because of the different assumptions used in the analysis (Sect.~4.3.2). Sanhueza et al. (2012) made the same assumptions about the dust properties (opacity and amount with respect to gas) as Vasyunina et al. (2011). However, no information about the mean molecular weight used in the calculation of $N({\rm H_2})$ was given. Nevertheless, from the reported 1.2~mm surface brightnesses and the corresponding $N({\rm H_2})$ values we derived a mean molecular weight similar to that assumed in the present study. Hence, the fractional abundances reported by Sanhueza et al. (2012; Table~10 therein) were scaled down by a factor of 0.549 for a proper comparison with our results. Moreover, we note that Sanhueza et al. (2012) reported the HCO$^+$ abundancies (rather than H$^{13}$CO$^+$), which were divided by 50 to obtain the H$^{13}$CO$^+$ abundances (the authors assumed an [HCO$^+$]/[H$^{13}$CO$^+$] abundance ratio of 50 for all sources). 
Miettinen (2014) assumed a same dust opacity value as we did, but the assumed dust-to-gas mass ratio and mean molecular 
weight were different (1/100 and 2.8, respectively). Hence, the abundances from Miettinen (2014) were multipled by 0.714 to bring 
them to level where a direct comparison is meaningful. The H$^{13}$CO$^+$ abundances in the Seahorse IRDC are 
in the same ballpark with those in other IRDCs. For example, the mean (median) abundances for the Vasyunina et al. (2011) and 
Sanhueza et al. (2012) samples are $(2.7\pm0.3)\times10^{-10}$ ($2.3\times10^{-10}$) and $(4.3\pm0.5)\times10^{-10}$ ($2.8\times10^{-10}$), while our corresponding values are $(6.2\pm1.0)\times10^{-10}$ ($6.1\times10^{-10}$). 

Gerner et al. (2014) used H$^{13}$CO$^+(J=1-0)$ observations, and derived the median abundances of $3.1\times10^{-11}$, 
$3.1\times10^{-10}$, $5.7\times10^{-10}$, and $6.9\times10^{-10}$ for IRDCs, HMPOs, HMCs, and UC \ion{H}{ii} regions, respectively. Again, 
their abundances were scaled down for a better comparison with the present results (Sect.~4.3.1). We derived about 27 times higher median H$^{13}$CO$^+$ abundance for our IR dark clumps than in the aforementioned IRDCs, but our median H$^{13}$CO$^+$ abundance for IR bright clumps
($5.7\times10^{-10}$) is very similar to those in the HMPOs and HMCs observed by Gerner et al. (2004). Also, the H$^{13}$CO$^+$ abundance in our \ion{H}{ii} region, $(6.8\pm0.8)\times10^{-10}$, is in excellent agreement with their median abundance in UC \ion{H}{ii} regions.

Saral et al. (2018) studied 30 IRDC clumps that were drawn from the 870~$\mu$m APEX Telescope Large Area Survey
(ATLASGAL; \cite{schuller2009}) and that appear dark at 8~$\mu$m or 24~$\mu$m wavelengths. The mean (median) H$^{13}$CO$^+$ 
column density reported by the authors, $(1.8\pm0.1)\times10^{12}$~cm$^{-3}$ ($1.4\times10^{12}$~cm$^{-3}$), is 3.5 (4.4) times 
lower than the present value, which could simply reflect the aforementioned selection effect, namely that the Saral et al. (2018) 
sample is solely composed of mid-IR dark sources. On the other hand, our IR dark clumps have a mean H$^{13}$CO$^+$ 
column density of $(6.7\pm1.7)\times10^{12}$~cm$^{-3}$, which is very similar to our full sample mean of 
$(6.3\pm0.8)\times10^{12}$~cm$^{-3}$. 

\subsubsection{SiO}

We detected SiO in only one of the target clumps, SMM~3. This source was also the only one of the five target sources 
in the Seahorse where SiO$(J=5-4)$ was tentatively detected ($\sim2\sigma$) by Miettinen (2012). In the bottom left panel in Fig.~\ref{figure:comparison}, we plot the SiO abundance distributions from previous studies and 
compare them with the SiO abundance derived for SMM~3. The abundances from Vasyunina et al. (2011), Sanhueza et al. (2012), 
and Miettinen (2014) were scaled as described above. Sanhueza et al. (2013) employed the same assumptions in their analysis as 
Sanhueza et al. (2012), and hence their SiO abundances were scaled down by a factor of 0.549.
Clearly the SiO abundance in SMM~3 appears to be very low compared to the typical values in clumps in other IRDCs. For example, 
the mean $x({\rm SiO})$ values in the aforementioned reference studies are 4--9.7 times higher than a value of $x({\rm SiO})=7.4\times10^{-11}$ in SMM~3.

Gerner et al. (2014) observed the $J=2-1$ transition of SiO, and derived the median abundances of $5.8\times10^{-11}$, 
$2.8\times10^{-10}$, $9.0\times10^{-10}$, and $2.6\times10^{-10}$ for IRDCs, HMPOs, HMCs, and UC \ion{H}{ii} regions, respectively (where the quoted values were scaled to the present assumptions). The SiO abundance we derived for the IR bright clump SMM~3 is as low as observed by Gerner et al. (2014) in IRDCs, but about 4--12 times lower than the median abundance in their HMPOs and HMCs.

Li et al. (2019) carried out an SiO$(J=5-4)$ survey of a sample of 44 massive clumps in IRDCs, where the detection rate was 57\%. 
The mean (median) SiO abundance reported by the authors is $3.7\times10^{-11}$ ($2.6\times10^{-11}$), which was scaled down by a factor of 0.64 to take into account the different assumptions in the analysis (Li et al. (2019) used the values $R_{\rm dg}=1/100$, $\kappa_{\rm 870\,,\mu m}=1.54$~cm$^2$~g$^{-1}$, and $\mu_{\rm H_2}=2.8$). This is within a factor of 2 (2.8) of the SiO abundance in SMM~3. 

Both Sakai et al. (2010) and Saral et al. (2018) determined SiO column densities in IRDC clumps. In the former study, 
the mean (median) SiO column density was $(1.5\pm0.2)\times10^{13}$~cm$^{-3}$ ($1.3\times10^{13}$~cm$^{-3}$), while in 
the latter one it was $(2.0\pm0.1)\times10^{12}$~cm$^{-3}$ ($1.4\times10^{12}$~cm$^{-3}$). The derived value of $N({\rm SiO})$ 
towards SMM~3, $(1.0\pm0.1)\times10^{12}$~cm$^{-3}$, is more similar to the mean value from Saral et al. (2018), and 15 times lower than the mean for the Sakai et al. (2010) sample.

Because we have observed a fairly high-$J$ transition of SiO, $J=4-3$, whose upper-state energy is 20.84~K
the low detection rate could be explained by the transition not being excited at the physical conditions of the Seahorse IRDC. 
This hypothesis could be tested through observations of low-$J$ transitions of SiO (preferably $J=1-0$; $E_{\rm u}/k_{\rm B}=2.08$~K). In particular, mapping observations of low-$J$ SiO line emission would be useful to explore the possibility that the Seahorse filament 
is formed through converging, supersonic turbulent flows. In some other filamentary IRDCs, detection of widespread ($\gtrsim 2$~pc) SiO emission is suggested to be a potential manifestation of the large-scale shock associated with the cloud formation process via colliding turbulent flows (\cite{jimenezserra2010}; \cite{cosentino2018}). For example, Jim{\'e}nez-Serra et al. (2010) mapped the filamentary IRDC G035.39-00.33 in the $J=2-1$ transition of SiO ($E_{\rm u}/k_{\rm B}=6.25$~K), and single-pointing observations were also done in the 
$J=3-2$ transition of SiO ($E_{\rm u}/k_{\rm B}=12.50$~K). The SiO abundance derived for the narrow-line ($\sim0.8$~km~s$^{-1}$)
component in G035.39-00.33 (only detected in SiO$(2-1)$) was found to be $\sim10^{-11}$, and, as one potential explanation, the authors suggested that such SiO emission could have its origin in the processing of dust grains by a shock associated with the formation of the IRDC.
As pointed out by the authors, the narrow-line SiO emission could also be caused by outflows from YSOs like the broader lines 
($\sim4-7$~km~s$^{-1}$) they detected, but possibly caught in a decelerated or recently activated stage.

The SiO$(4-3)$ line we detected towards SMM~3 is broad, $8.30\pm0.50$~km~s$^{-1}$ in FWHM, and is very likely to arise from a YSO outflow-driven shock. However, the SiO abundance we derived for SMM~3, $(7.4 \pm 0.9)\times10^{-11}$, appears lower than those derived by Jim{\'e}nez-Serra et al. (2010) towards YSOs in G035.39-00.33 ($\sim10^{-10}$ to $\geq10^{-8}$). Also, the upper limits to the SiO abundance we estimated along the Seahorse IRDC, ranging from $<5\times10^{-12}$ to $<2.2\times10^{-11}$, are even lower than expected from a large-scale shock if the filament was formed in the same way as G035.39-00.33 and is of comparable age. However, Jim{\'e}nez-Serra et al. (2010) calculated the SiO column densities using the large velocity gradient modelling, and derived the abundances with respect to $N({\rm H_2})$ estimated from CO isotopologues, which makes a direct comparison with their results difficult. Again, observations of $J_{\rm up}<4$ transitions of SiO towards the Seahorse are needed to shed light on these issues.

\subsubsection{HN$^{13}$C}

In the bottom middle panel in Fig.~\ref{figure:comparison}, the derived HN$^{13}$C abundances 
are compared with those derived by Sanhueza et al. (2012) and Miettinen (2014), where the 
aforementioned scaling factors were again applied. Sanhueza et al. (2012) detected HN$^{13}$C 
in 34\% of their target clumps, and to obtain their HN$^{13}$C abundances we divided the reported HNC 
abundances by 50 (the authors assumed an [HNC]/[HN$^{13}$C] abundance ratio of 50 for all their sources). 
The mean (median) value of $x(\rm HN^{13}C)$ for the Sanhueza et al. (2012) sample is $(5.3\pm0.4)\times10^{-10}$ ($4.1\times10^{-10}$),
while for the present sample it is $(2.4\pm0.3)\times10^{-11}$ ($2.5\times10^{-11}$), that is over 20 (16) times lower 
than found by Sanhueza et al. (2012). 

Gerner et al. (2014) observed the HN$^{13}$C$(J=1-0)$ line towards their sample, and their scaled-down median HN$^{13}$C abundances 
for IRDCs, HMPOs, HMCs, and UC \ion{H}{ii} regions are $4.0\times10^{-11}$, $2.6\times10^{-10}$, $3.4\times10^{-10}$, and $5.3\times10^{-10}$, respectively. We derived a median HN$^{13}$C abundance of a few times $10^{-11}$ in every source types in our sample (IR dark and bright clumps, and an \ion{H}{ii} region), which is comparable to that found by Gerner et al. (2014) in their IRDCs, but over an order of magnitude lower for clumps clearly associated with YSOs.

Column densities of HN$^{13}$C towards IRDCs were determined by Sakai et al. (2010) and Saral et al. (2018). 
The mean (median) values for their samples are $(7.8\pm0.8)\times10^{12}$~cm$^{-3}$ ($7.0\times10^{12}$~cm$^{-3}$) and 
$(3.0\pm0.1)\times10^{12}$~cm$^{-3}$ ($1.6\times10^{12}$~cm$^{-3}$). Compared to the present value, $(2.4\pm0.3)\times10^{11}$~cm$^{-3}$ ($2.1\times10^{11}$~cm$^{-3}$), those values are over an order of magnitude higher.

\subsubsection{C$_2$H}

\begin{figure*}
\begin{center}
\includegraphics[scale=0.32]{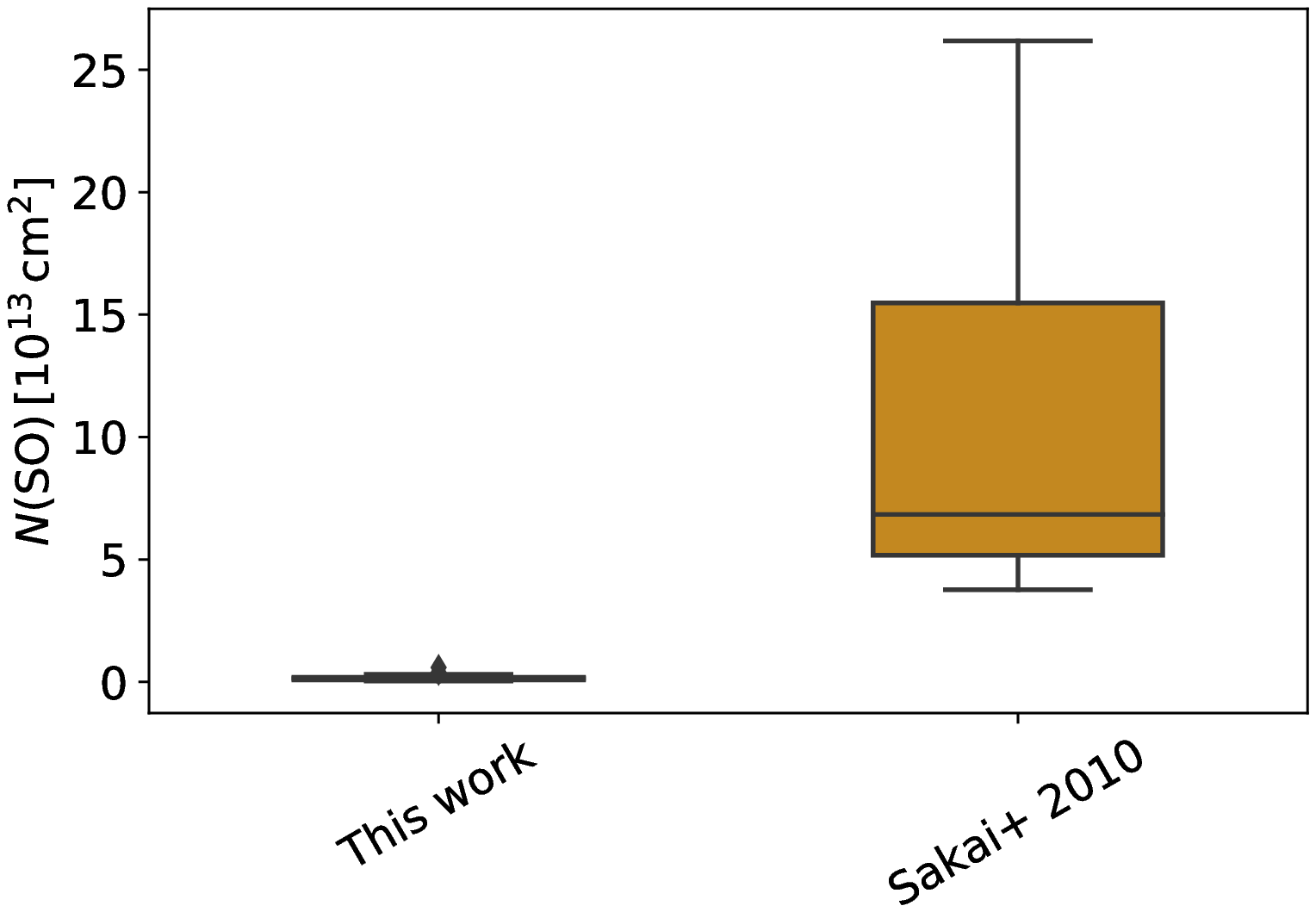}
\includegraphics[scale=0.32]{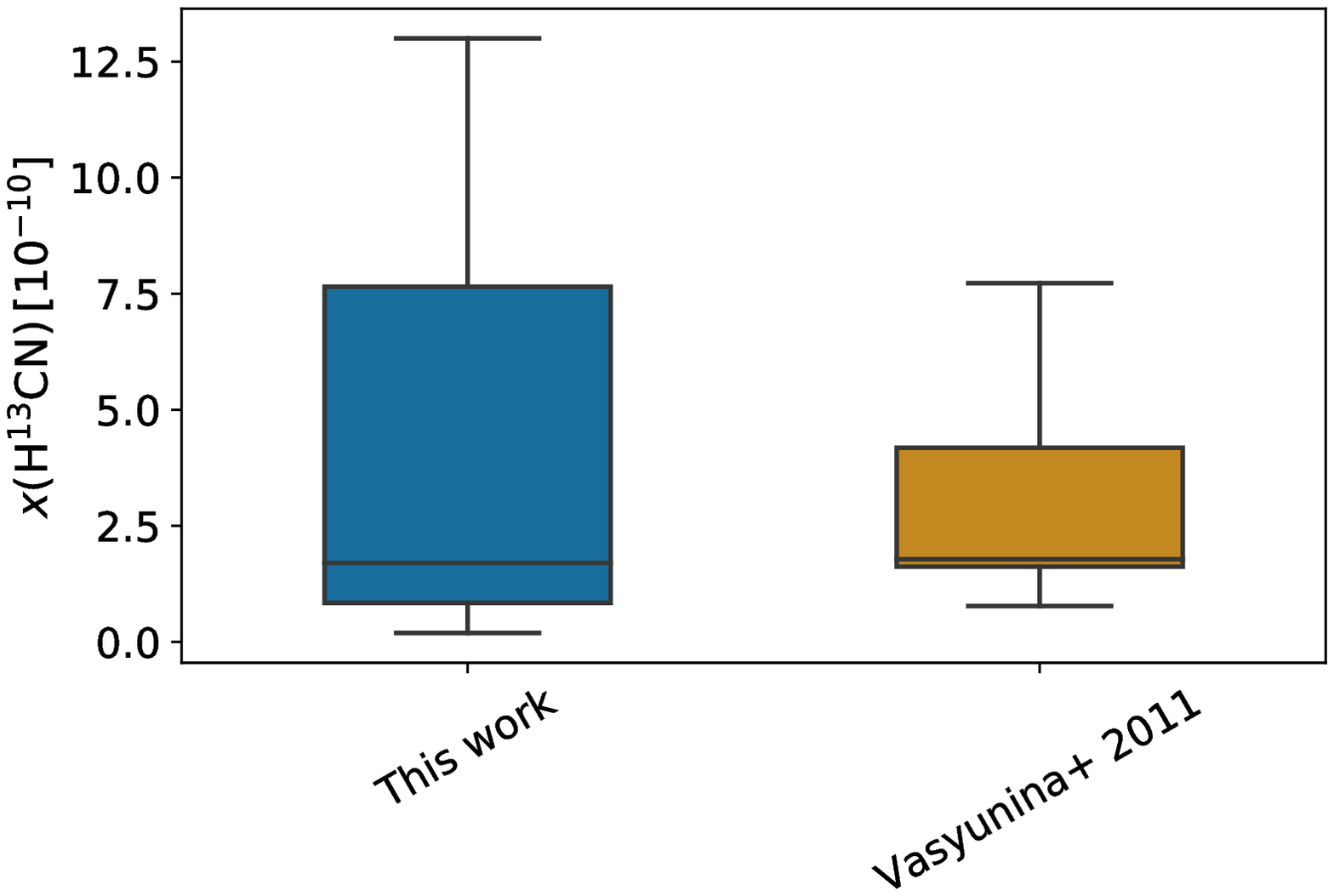}
\includegraphics[scale=0.32]{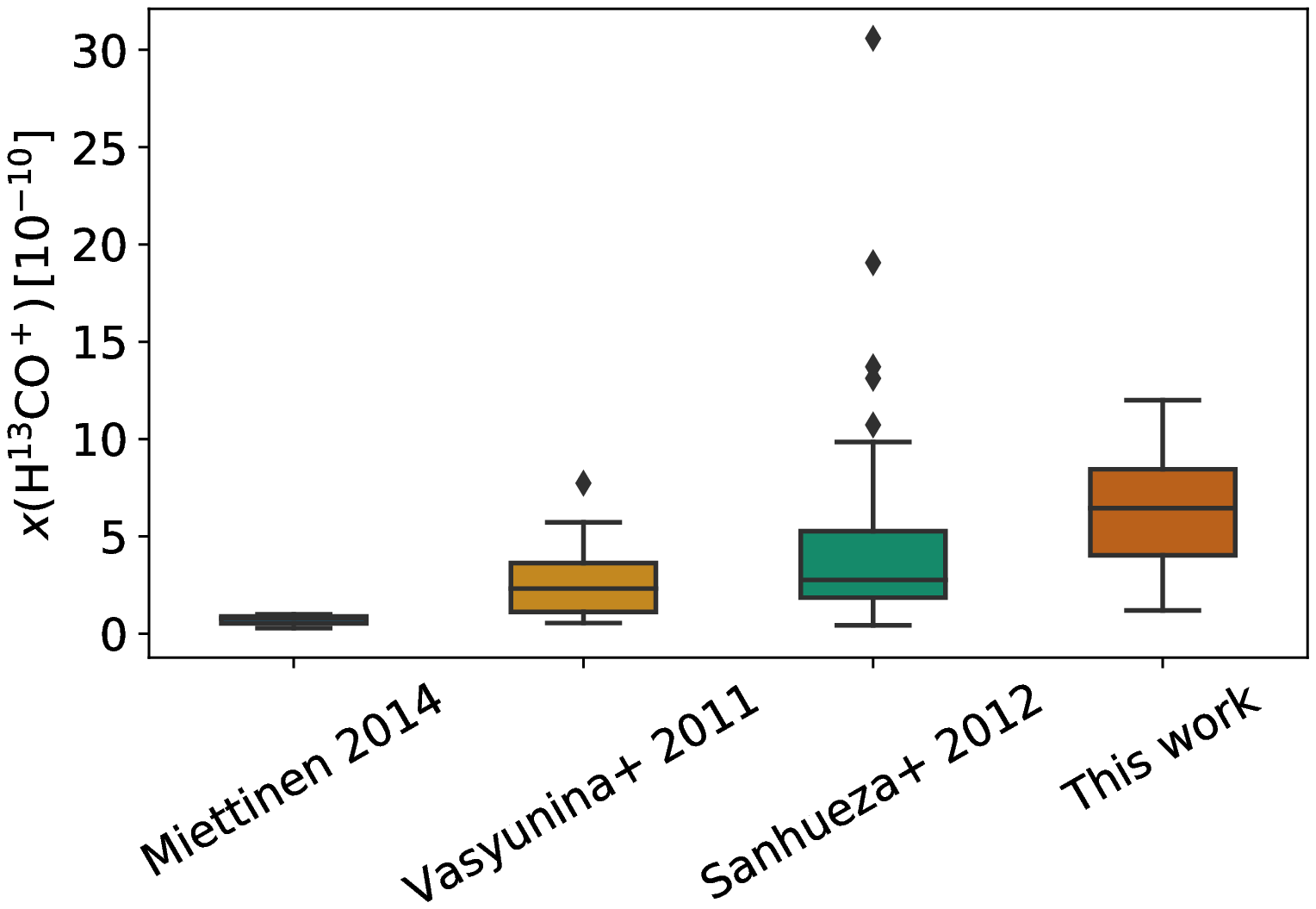}
\includegraphics[scale=0.32]{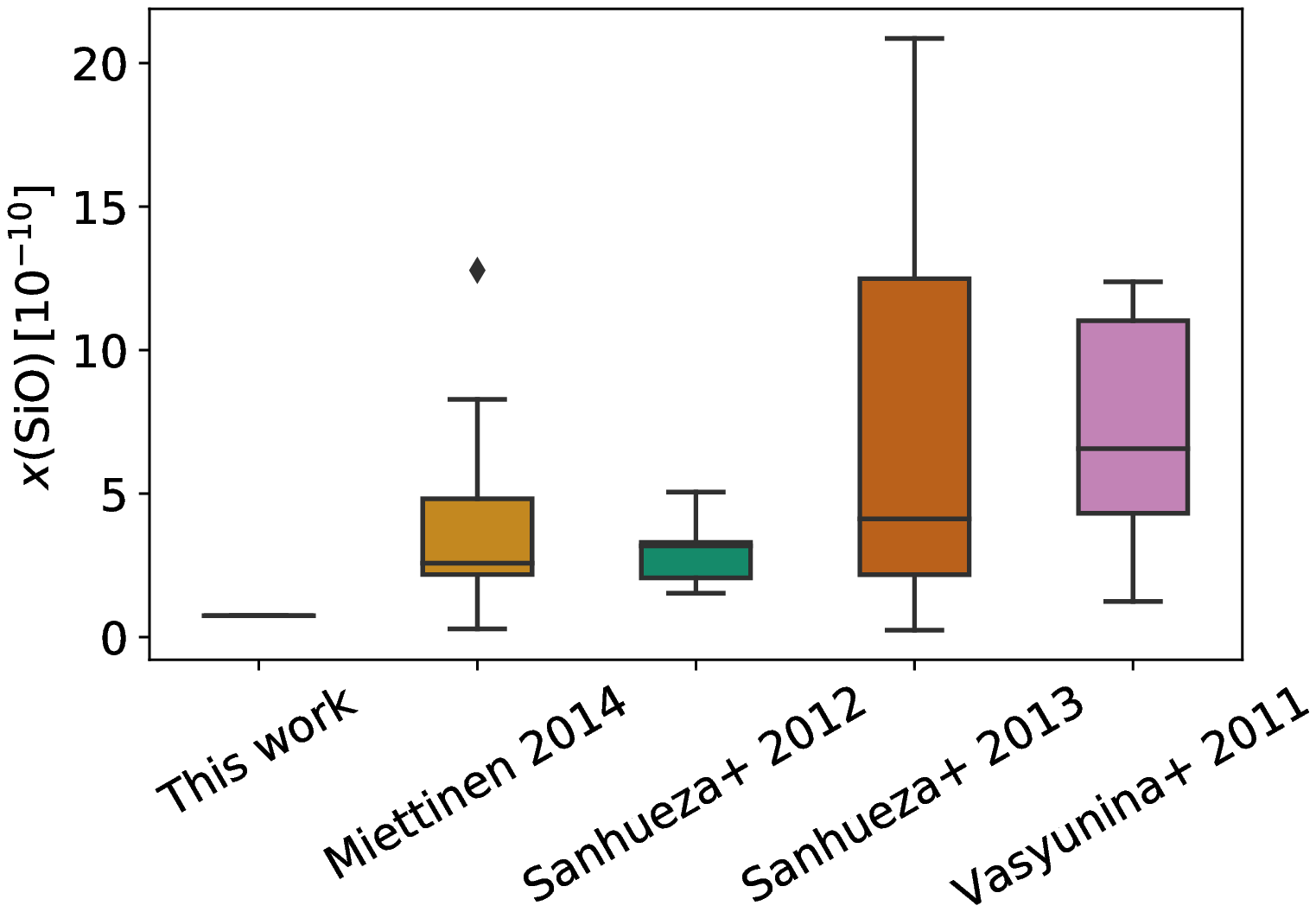}
\includegraphics[scale=0.32]{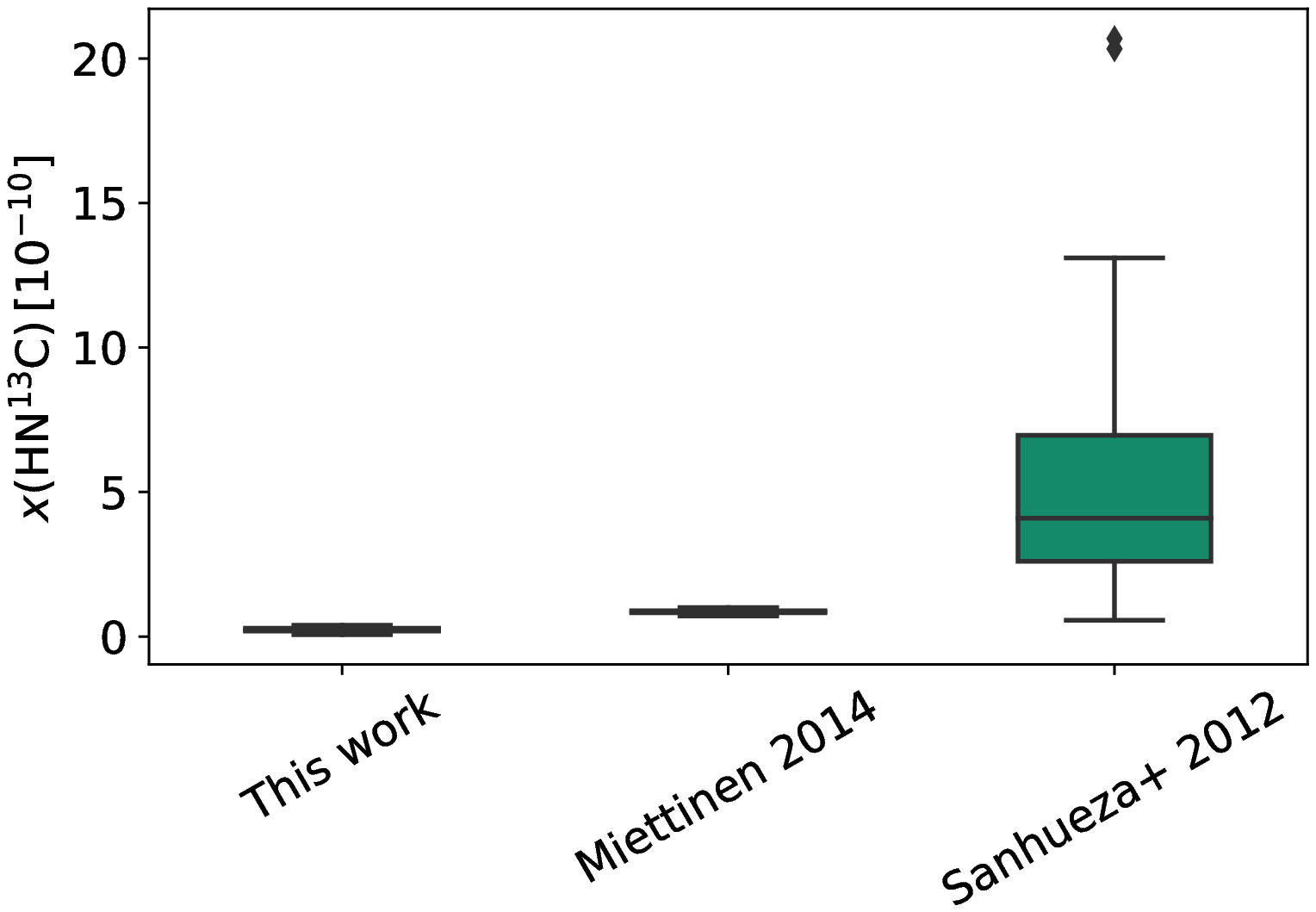}
\includegraphics[scale=0.32]{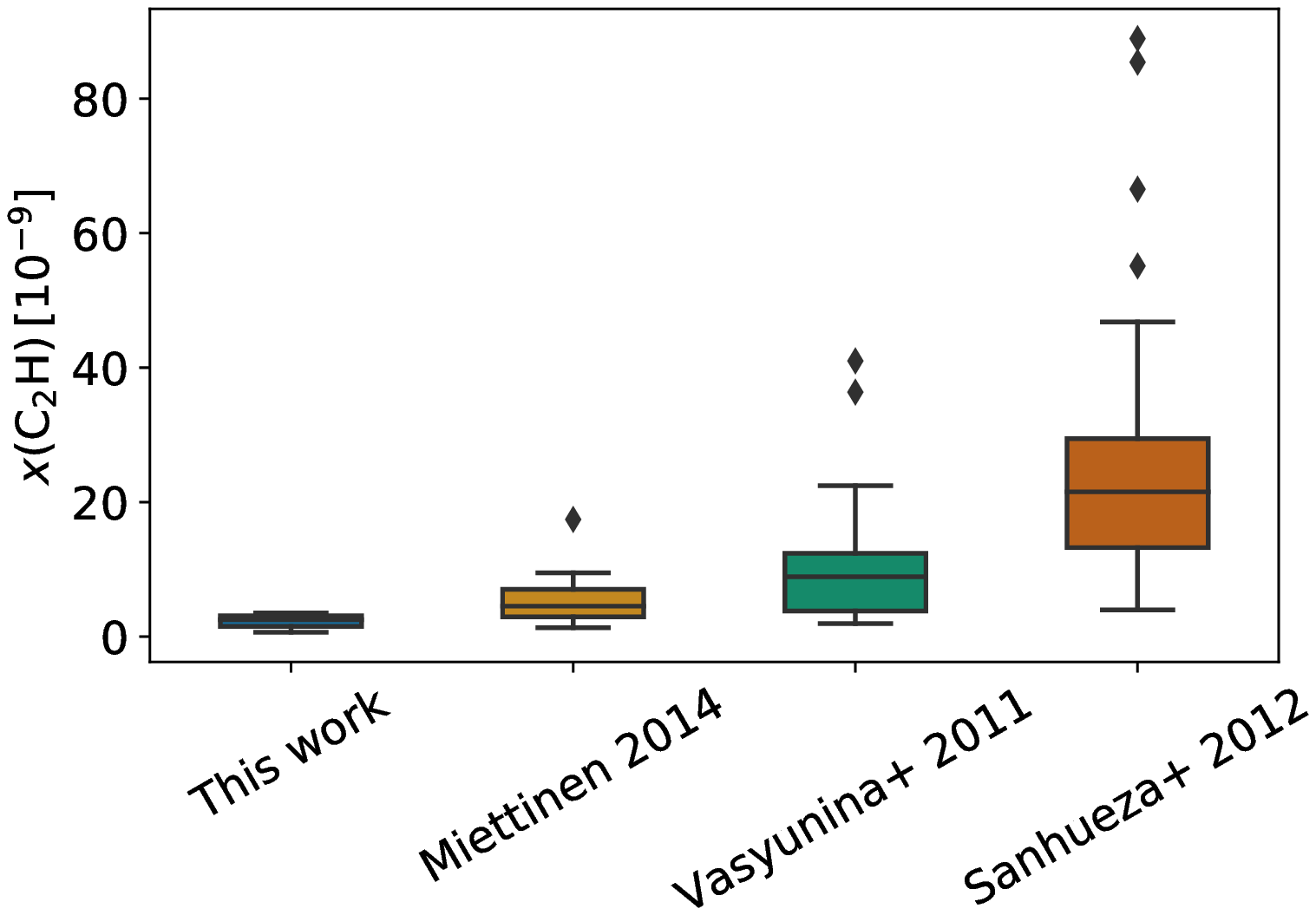}
\caption{Comparison of the molecular abundance distributions (column density for SO) in the Seahorse IRDC with those in other IRDC sources presented in the literature. The box plots are sorted by increasing median abundance. The data points above some of the boxes 
represent outlier values. Upper limits are neglected from the distributions.}
\label{figure:comparison}
\end{center}
\end{figure*}

As illustrated in Fig.~\ref{figure:boxall}, C$_2$H was found to be the most abundant molecule in the present study, but 
we note that together with H$^{13}$CN it was also the only case where the optical thickness and hence excitation temperature could be derived via hyperfine structure fitting. An abundance comparison with previous studies (\cite{vasyunina2011}; \cite{sanhueza2012}; \cite{miettinen2014}) is visualised in the bottom right panel in Fig.~\ref{figure:comparison}. The reference abundances were scaled down as discussed above. 

The C$_2$H abundances in the Seahorse appear to be lower than in many other IRDCs. The mean (median) abundances for the 
Vasyunina et al. (2011), Sanhueza et al. (2012), and Miettinen (2014) samples are, respectively, $(1.1\pm0.2)\times10^{-8}$ ($8.9\times10^{-9}$), $(2.6\pm0.3)\times10^{-8}$ ($2.2\times10^{-8}$), and $(5.8\pm1.2)\times10^{-9}$ ($4.5\times10^{-9}$). 
The mean (median) C$_2$H abundance in the Seahorse, $(2.3\pm0.3)\times10^{-9}$ ($2.5\times10^{-9}$), is 4.8 (3.6), 11.3 (8.8), 
and 2.5 (1.8) times lower than the aforementioned values, respectively. 

The median C$_2$H abundances in the IRDCs, HMPOs, HMCs, and UC \ion{H}{ii} regions studied by Gerner et al. (2014) were derived 
to be $8.3\times10^{-9}$, $9.7\times10^{-8}$, $2.2\times10^{-7}$, and $2.7\times10^{-7}$, respectively. These are all larger than those we derived for the Seahorse clumps, by a factor of 3.1 for the IR dark clumps, factors of $\sim40-92$ for the clumps associated with YSOs, and by a factor of over 190 when comparing the \ion{H}{ii} region values.

The mean (median) C$_2$H column density we derived, $(1.3\pm0.2)\times10^{15}$~cm$^{-3}$ ($1.1\times10^{15}$~cm$^{-3}$), is almost an order of magnitude higher than those derived by Sakai et al. (2010; $(1.5\pm0.2)\times10^{14}$~cm$^{-3}$ ($1.4\times10^{14}$~cm$^{-3}$)).

\subsubsection{Abundance ratios}

In the present study, we also derived the $[{\rm HN^{13}C}]/[{\rm H^{13}CN}]$ and $[{\rm HN^{13}C}]/[{\rm H^{13}CO^+}]$ 
abundance ratios (see Table~\ref{table:ratios}). The full sample means are $0.30\pm 0.07$ and $0.04\pm0.005$, respectively. 
If the $^{13}$C fractionation is equally (un-)important or (in-)efficient in the molecules of the aforementioned ratios, then the ratios are expected to be equal to $[{\rm HNC}]/[{\rm HCN}]$ and $[{\rm HNC}]/[{\rm HCO^+}]$ (e.g. \cite{roberts2012}).

The $[{\rm HNC}]/[{\rm HCN}]$ ratio in IRDCs is found to be about unity on average (e.g. \cite{vasyunina2011}; \cite{liu2013}; \cite{miettinen2014}; \cite{jin2015}), which is consistent with theoretical expectations for dark interstellar clouds (e.g. \cite{sarrasin2010}; \cite{loison2014}). For example, the mean $[{\rm HNC}]/[{\rm HCN}]$ ratio and its standard error for the IRDC clump sample of Miettinen (2014) is $1.26 \pm 0.46$, which is $4.2\pm1.8$ times higher than the present mean $[{\rm HN^{13}C}]/[{\rm H^{13}CN}]$ ratio. Hence, 
the $[{\rm HN^{13}C}]/[{\rm H^{13}CN}]$ ratio in the Seahorse IRDC appears to be lower than expected for the $^{12}$C isotopologues of the observed species. 

On the other hand, the $[{\rm HNC}]/[{\rm HCN}]$ ratio is subject to temperature variations. 
For example, the $[{\rm HNC}]/[{\rm HCN}]$ ratio in Orion was found to increase from 0.003--0.005 in the hot plateau and hot core regions ($\sim150-200$~K) to 0.017--0.17 in the colder ($\sim30-60$~K) quiescent parts of the cloud (\cite{goldsmith1986}; see also \cite{schilke1992}; \cite{hacar2020}). Hirota et al. (1998) found that the $[{\rm HNC}]/[{\rm HCN}]$ ratio drops rapidly when the gas kinetic temperature becomes $T_{\rm kin}>24$~K, that is at temperatures higher than typically found in dark clouds ($\sim10$~K). The authors suggested that the reason why the $[{\rm HNC}]/[{\rm HCN}]$ ratio drops is that HNC is being destroyed by neutral-neutral reactions some of which can form HCN (${\rm HNC}+{\rm H}\rightarrow {\rm HCN}+{\rm H}$ and ${\rm HNC}+{\rm O}\rightarrow {\rm NH}+{\rm CO}$). We note that the neutral-neutral reaction ${\rm CH_2} + {\rm N} \rightarrow {\rm HCN} + {\rm H}$ could also play a role in lowering the $[{\rm HNC}]/[{\rm HCN}]$ ratio (e.g. \cite{herbst2000}). However, Hirota et al. (1998) did not found any significant differences in $[{\rm HNC}]/[{\rm HCN}]$ between star-forming cores and starless cores, which is in reasonable agreement with our comparable mean $[{\rm HN^{13}C}]/[{\rm H^{13}CN}]$ ratios derived for the IR dark and IR bright clumps ($0.38\pm0.11$ and $0.27\pm0.10$, respectively, whose ratio is $1.4\pm0.7$). However, the median $[{\rm HN^{13}C}]/[{\rm H^{13}CN}]$ ratios derived for our IR dark and IR bright clumps show a more significant difference (being 2.8 times higher in the IR dark clumps). In the \ion{H}{ii} region clump IRAS~13039-6108, the $[{\rm HN^{13}C}]/[{\rm H^{13}CN}]$ ratio was derived to be $0.05\pm0.03$, which is lower than the averages in the IR dark and IR bright sources by factors of $7.6\pm5.1$ and $5.4\pm3.8$, respectively. The potential evolutionary trend in the $[{\rm HN^{13}C}]/[{\rm H^{13}CN}]$ ratio in the Seahorse is further discussed in Sect.~4.4. 

Gerner et al. (2014) modelled the chemical evolution of different stages of high-mass star formation using a 
one-dimensional physical model of density and temperature evolution coupled with a time-dependent gas-grain chemical model. Their best-fit 
model followed the aforementioned trend observed in Orion, and the $[{\rm HNC}]/[{\rm HCN}]$ ratio reached a minimum of $\sim0.02$ at 
the beginning of the hot core stage (see their Fig.~6). However, their observed $[{\rm HNC}]/[{\rm HCN}]$ ratio did not exhibit such a clear trend, but instead the ratio was found to be comparable for all the four types of sources they studies, namely IRDCs, HMPOs, HMCs, and UC \ion{H}{ii} regions (all values were in the range $\sim1.7-3.3$).

Jin et al. (2015) found that the mean $[{\rm HNC}]/[{\rm HCN}]$ ratio drops from about unity in quiescent IRDC cores (neither 4.5~$\mu$m nor 24~$\mu$m emission) to $\sim0.4$ in active IRDC cores (associated with both 4.5~$\mu$m and 24~$\mu$m emission) to 
$\sim0.2$ in HMPOs to $\sim0.1$ in UC \ion{H}{ii} regions. This behaviour 
can be understood in terms of gas temperature increase with source evolution as discussed above. On the other hand, 
we did not find any correlation between $[{\rm HN^{13}C}]/[{\rm H^{13}CN}]$ and $T_{\rm dust}$. 
The dust temperatures in the Seahorse clumps might be too similar to each other for the temperature-dependent 
$[{\rm HNC}]/[{\rm HCN}]$ evolutionary indicator to be manifested (the dust temperatures range from 11.5~K to 22.2~K; Table~\ref{table:sources}). 

Colzi et al. (2018) observed the $J=1-0$ transition of different isotopologues of HCN and HNC towards 
27 high-mass star-forming objects including high-mass starless cores (HMSCs), HMPOs, and UC \ion{H}{ii} regions. The authors used a Galactocentric dependent $[{\rm ^{12}C}]/[{\rm ^{13}C}]$ ratio to convert the column densities of H$^{13}$CN and HN$^{13}$C to those of the main isotopologues. and found that the mean $[{\rm HNC}]/[{\rm HCN}]$ ratios for HMSCs, HMPOs, and UC \ion{H}{ii} regions are $1.3\pm0.3$, $0.4\pm0.03$, and $0.5\pm0.1$, respectively. For the first evolutionary stages, the results obtained by Colzi et al. (2018) are similar to those from Jin et al. (2015), but the mean $[{\rm HNC}]/[{\rm HCN}]$ ratio derived by Colzi et al. (2018) for their UC \ion{H}{ii} regions does not appear to drop compared to HMPOs as found by Jin et al. (2015). Finally, we note that Saral et a. (2018) derived a high mean $[{\rm HNC}]/[{\rm HCN}]$ ratio of $8.9\pm1.0$ for their sample (median was 7.1), which could be caused by their sample being, by construction, biased towards clumps in very early stages of evolution (see Sect.~4.3.3).

The average $[{\rm HNC}]/[{\rm HCO^+}]$ ratios in IRDCs are found to be $\sim2$ (e.g. $1.7\pm0.2$ for the Liu et al. (2013) sample, 
and $2.3\pm0.1$ for the Saral et al. (2018) sample). This is in fair agreement with the gas-phase chemical models of cold dark
clouds, which suggest comparable abundances for these species (e.g. \cite{roberts2012} and references therein). 
However, Miettinen (2014) derived a higher mean value of $4.3 \pm 0.7$ (median was 2.8). Our mean $[{\rm HN^{13}C}]/[{\rm H^{13}CO^+}]$ ratio is $\sim50-108$ times lower than the main isotopologue ratio in the aforementioned studies. At least partly, this could be assigned to $^{13}$C fractionation effects being different for the two molecules. Miettinen (2014) derived comparable 
mean abundances for HN$^{13}$C and H$^{13}$CO$^+$, their ratio being 1.2, which is 30 times higher than our mean ratio. On the other hand, Saral et a. (2018) suggested that the $[{\rm HNC}]/[{\rm HCO^+}]$ ratio could be subject to environmental effects (density, turbulence; \cite{godard2010}), and hence an unreliable source evolutionary indicator. Observations of the main isotopologues HNC and HCO$^+$ towards the Seahorse are required to examine how the $[{\rm HNC}]/[{\rm HCO^+}]$ ratio compares with those in other Galactic IRDCs, and to quantify the $^{13}$C fractionation differences in the species. 

\subsection{Potential evolutionary indicators}

\begin{figure*}
\begin{center}
\includegraphics[scale=0.46]{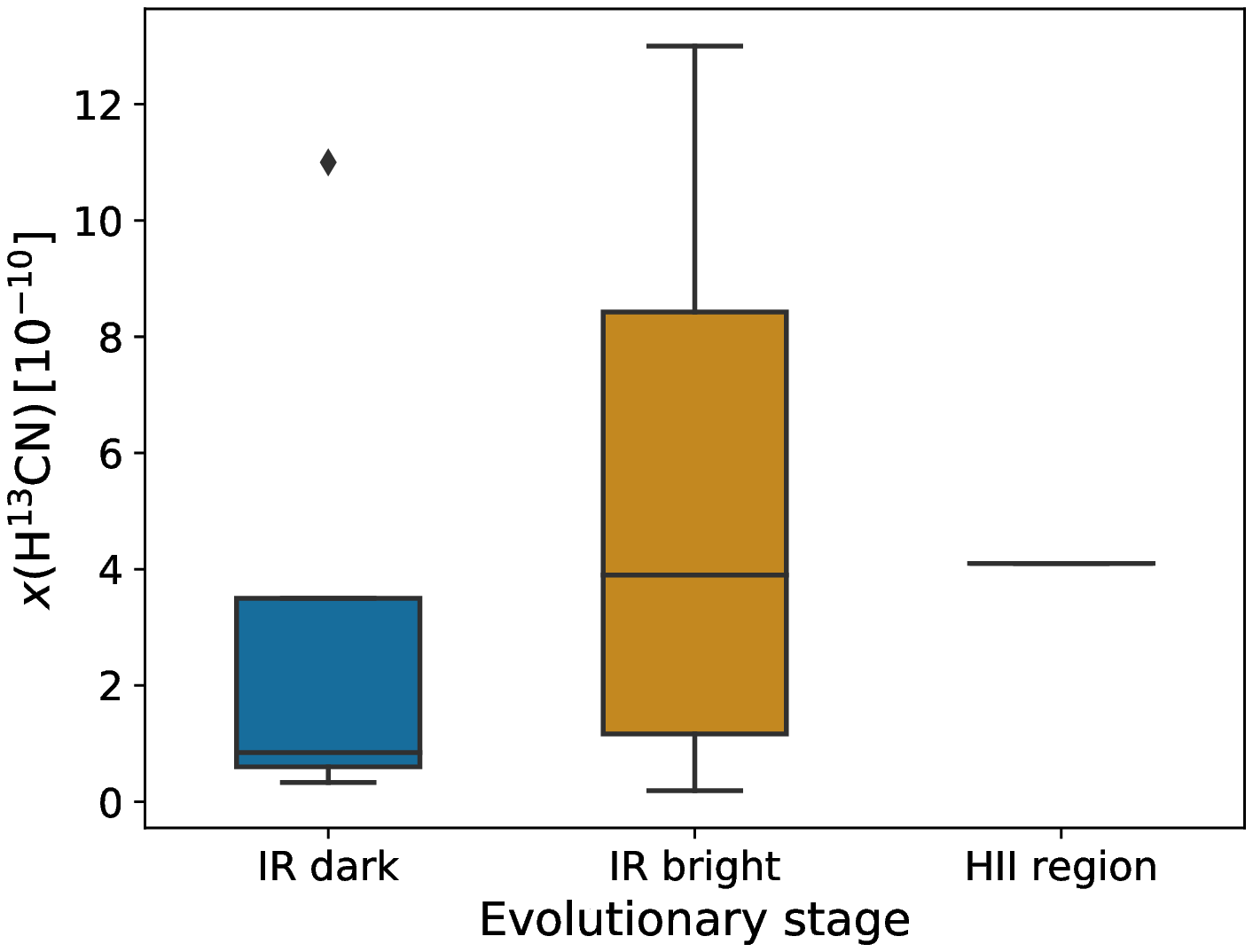}
\includegraphics[scale=0.46]{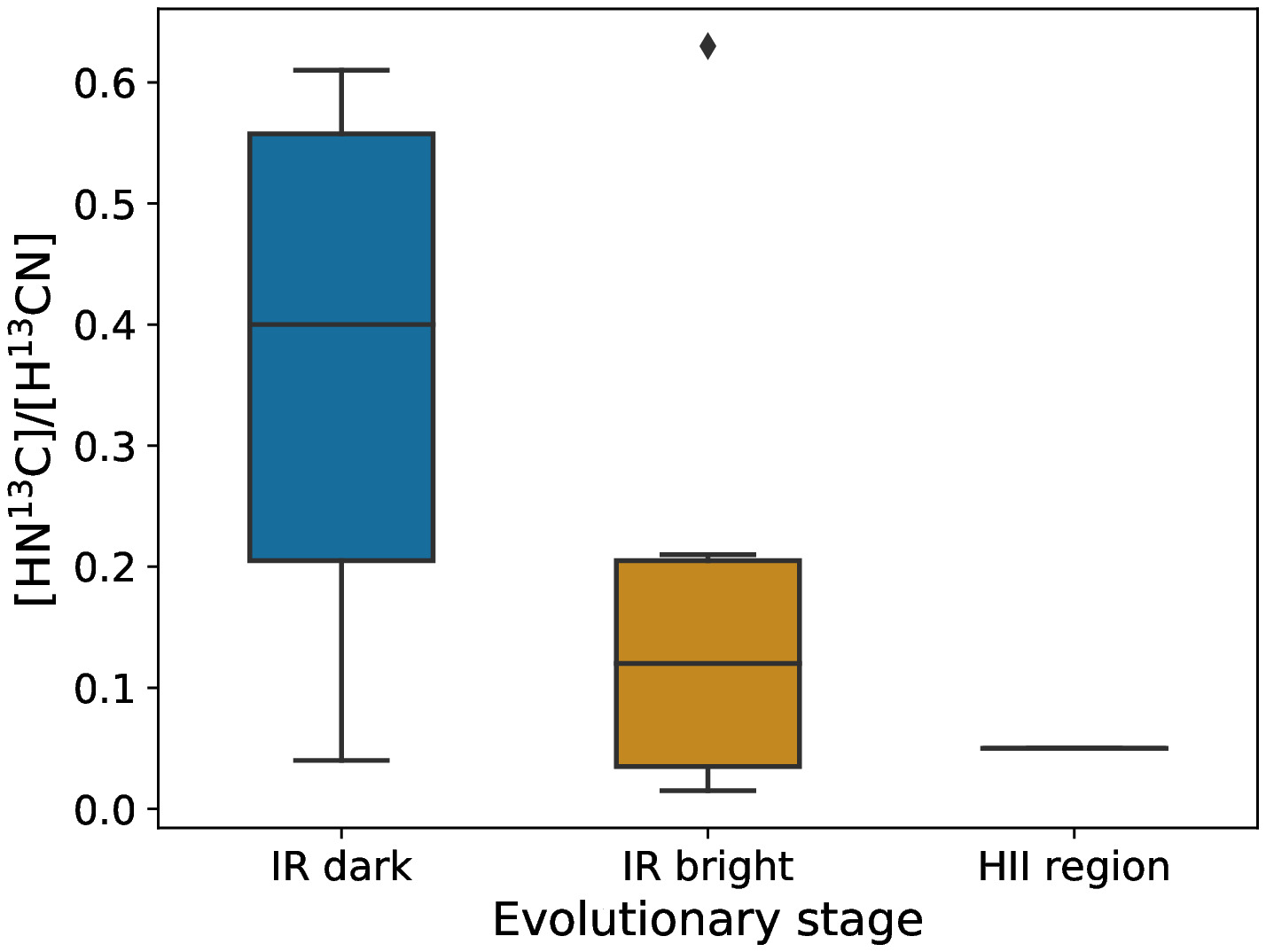}
\includegraphics[scale=0.46]{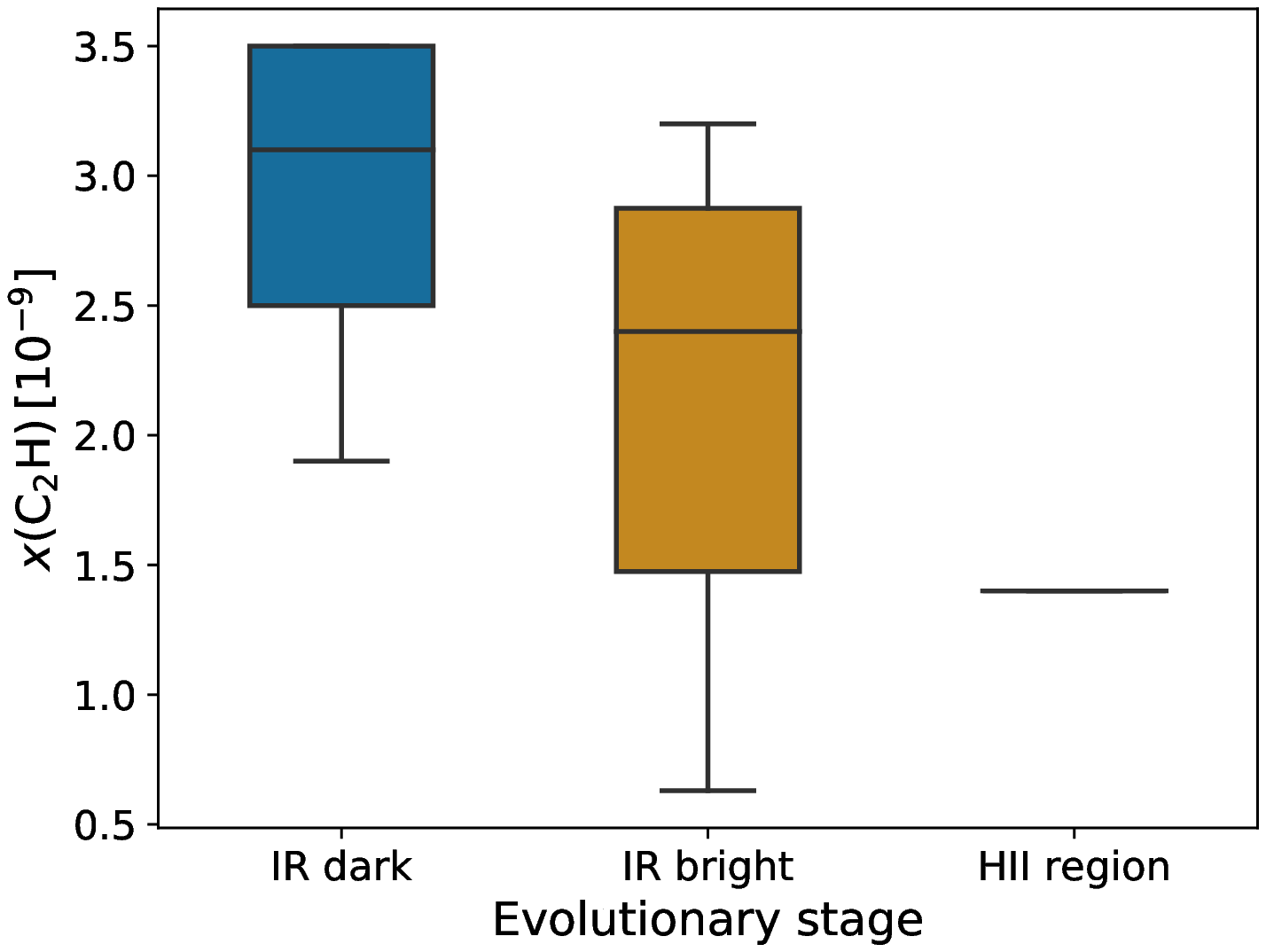}
\includegraphics[scale=0.46]{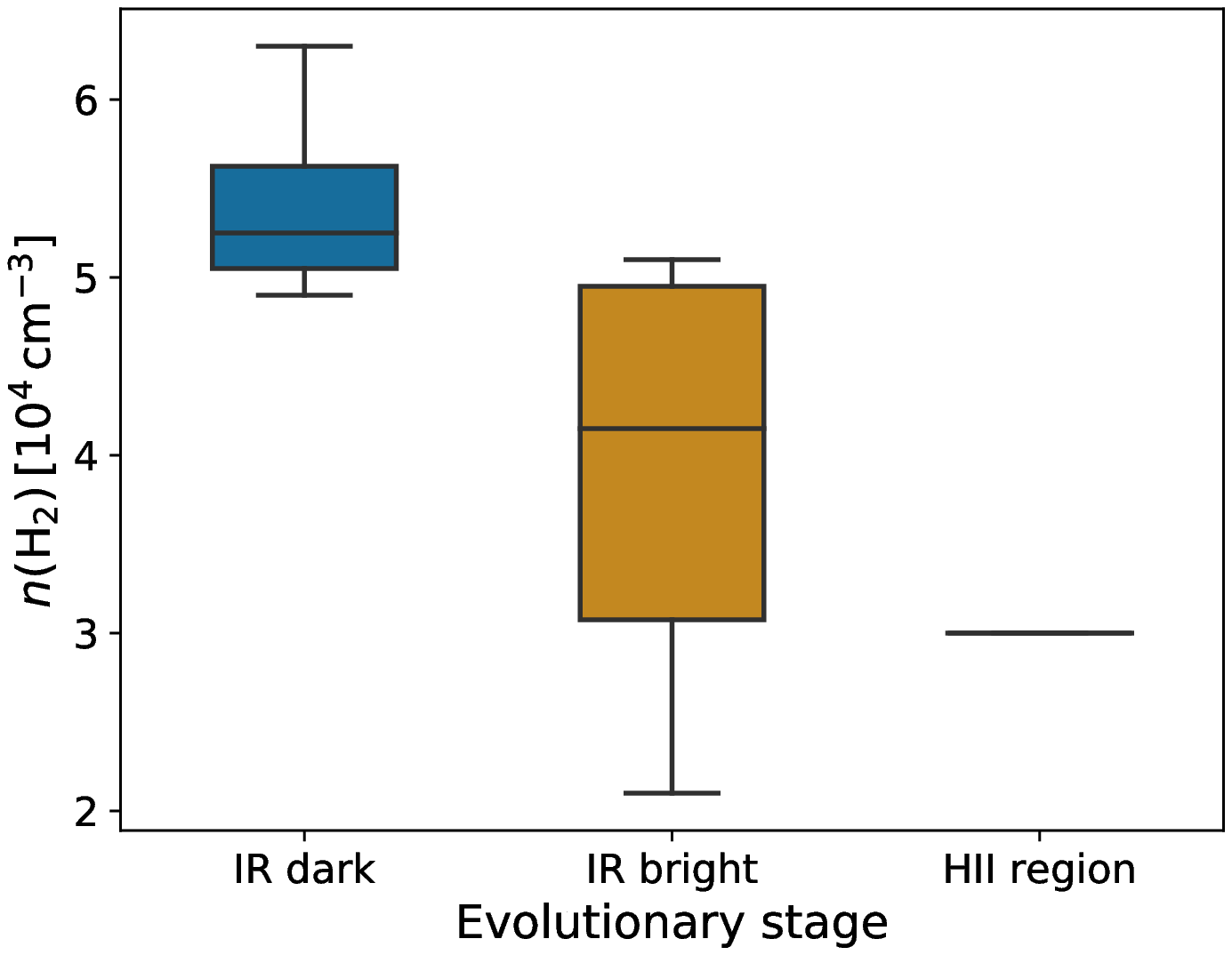}
\caption{Box plots showing the clump properties that were found to be different on average for the different types of clumps in the present sample, namely clumps that are either dark or bright in the mid-IR and clumps associated with an \ion{H}{ii} region (only one source, IRAS~13039-6108). Censored data are not included in the plot. Some of the trends could be manifestations of clump evolution (Sect.~4.4).}
\label{figure:evolution}
\end{center}
\end{figure*}

Because our sample is composed of different types of clumps, namely five IR dark clumps, eight IR bright clumps, and 
one clump associated with an \ion{H}{ii} region, we searched for potential evolutionary trends among the different 
physical and chemical parameters discussed in the present paper. The four possible trends that we found are shown in 
Fig.~\ref{figure:evolution}. We note that although our sample is small, and hence subject to small number statistics, our results benefit from the fact that all the sources belong to the same parent cloud, and hence lie at the same distance and are expected to share similar initial chemical and physical conditions. Thanks to a common distance between all the sources, distance-dependent beam-dilution effects are expected to play no role in the derived chemical abundances (or rather the effect is similar for all sources).

As shown in the top left panel in Fig.~\ref{figure:evolution}, we found a hint that the H$^{13}$CN abundance increases as the clump evolves from the IR dark stage to IR bright and \ion{H}{ii} region stages. The mean (median) abundances derived for these three types of sources are $(2.7\pm2.2)\times10^{-10}$ ($0.7\times10^{-10}$), $(4.0\pm1.8)\times10^{-10}$ ($1.0\times10^{-10}$), and $(4.1\pm2.4)\times10^{-10}$ (only one source). These are not statistically significant differences and, as noted above, we are dealing with a 
fairly small sample, especially a sample of one for the \ion{H}{ii} regions. Nevertheless, our result is consistent with the increasing trend in the HCN abundance from an IRDC stage to the ignition of a HMC found by Gerner et al. (2014) in their observational data and through chemical modelling (see their Fig.~5). Also, Jin et al. (2015) found that the peak intensity of the sub-sample averaged H$^{13}$CN spectra gets stronger from quiescent IRDC cores to UC \ion{H}{ii} regions. As discussed in Sect.~4.3.7, an increasing evolutionary trend in $x({\rm H^{13}CN})$ can be understood to arise from the neutral-neutral reactions ${\rm HNC}+{\rm H}\rightarrow {\rm HCN}+{\rm H}$ and ${\rm CH_2} + {\rm N} \rightarrow {\rm HCN} + {\rm H}$ being favoured by the increasing abundances of hydrogen and nitrogen atoms as the gas and dust temperature of the source rises.

In the top right panel in Fig.~\ref{figure:evolution}, a decreasing trend in the $[{\rm HN^{13}C}]/[{\rm H^{13}CN}]$ ratio is seen as the clump evolves. The mean (median) ratio in the three types of clumps was found to decrease from $0.38\pm0.11$ (0.54) in the IR dark clumps to $0.27\pm0.10$ (0.19) in the IR bright clumps, and further to $0.05\pm0.03$ in the \ion{H}{ii} region stage. This behaviour is consistent with the positive evolutionary trend of $x({\rm H^{13}CN})$ shown in the top left panel in Fig.~\ref{figure:evolution}, and 
was already addressed in Sect.~4.3.7.

The decreasing evolutionary trend in the C$_2$H abundance is shown in the bottom left panel in Fig.~\ref{figure:evolution}. 
The mean (median) $x({\rm C_2H})$ drops from $(2.7\pm0.4)\times10^{-9}$ ($2.7\times10^{-9}$) to $(2.2\pm0.3)\times10^{-9}$ ($2.4\times10^{-9}$) and to $(1.4\pm0.2)\times10^{-9}$ for our three types of clumps. Sakai et al. (2010) concluded that the C$_2$H abundance (with respect to H$^{13}$CO$^+$) does not enhance significantly when star formation in the clump kicks in. Sanhueza et al. (2012) found that $N({\rm C_2H})$ increases with clump evolution, but such trend is not visible in their $x({\rm C_2H})$ values (Table~8 therein). On the other hand, 
Miettinen (2014) found that both the mean and median values of $x({\rm C_2H})$ are higher in IR dark clumps than in IR bright clumps, 
which is similar to our result. 

Gerner et al. (2014) observed an increasing trend in the C$_2$H abundance when the source evolves from an IRDC stage to a HMPO, and further to a HMC and then to an UC \ion{H}{ii} region (see their Fig.~5). However, their modelled values 
showed a different behaviour so that $x({\rm C_2H})$ first increased from IRDCs to HMPOs, but then dropped for HMCs and UC \ion{H}{ii} regions. The behaviour at the beginning is opposite to our finding, but the later decreasing evolutionary trend resembles our  observational results. The latter is consistent with a suggestion that the C$_2$H abundance decreases 
in the hot core stage of evolution (\cite{beuther2008}). For example, C$_2$H molecules in hot cores react rapidly with O atoms 
to form CO (${\rm C_2H}+{\rm O}\rightarrow {\rm CO}+{\rm CH}$; e.g. \cite{watt1988}).
Furthermore, there is observational evidence that C$_2$H is destroyed in \ion{H}{ii} regions (e.g. \cite{walsh2010}); the far-UV 
photons can dissociate C$_2$H into H atoms and C$_2$ molecules (e.g. \cite{nagy2015}). This could also explain the low value of 
$x({\rm C_2H})$ observed in IRAS~13039-6108.

Finally, we found that the volume-averaged H$_2$ number density appears to decrease as the clump evolves (Fig.~\ref{figure:evolution}, bottom right panel). The mean (median) density drops from $(5.2\pm0.4)\times10^4$~cm$^{-3}$ ($5.1\times10^4$~cm$^{-3}$) to 
$(4.0\pm0.4)\times10^4$~cm$^{-3}$ ($4.9\times10^4$~cm$^{-3}$) and further to $(3.0\pm1.6)\times10^4$~cm$^{-3}$ for the three stages of clump evolution. This is a counter-intuitive result because one would expect the density to increase as the clump contracts under the influence of gravity. Previous studies of high-mass clumps have indeed demonstrated that the average densities of star-forming clumps are higher than those that show no signs of star formation (e.g. \cite{chambers2009}; \cite{giannetti2013}). Although one can think of multiple sources of uncertainty in the H$_2$ number density calculation, such as assuming a spherical geometry for clumps embedded in a filamentary parent cloud and using the same dust opacity for all types of clumps in calculating their mass and density ($n\propto M\propto \kappa_{\nu}^{-1}$), it is still puzzling to see a decreasing trend shown in Fig.~\ref{figure:evolution}. One could speculate that the H$_2$ content of the Seahorse is complicated by its potential accretion of gas from the surrounding medium and stellar feedback effects (H$_2$ destruction by UV radiation in star-forming clumps; e.g. \cite{mckee2010}), but more detailed studies are needed to test such speculations.

\section{Summary and conclusions}

We used the APEX telescope to carry out a molecular line survey at 170~GHz towards the clumps in the 
filamentary IRDC G304.74+01.32, nicknamed the Seahorse Nebula. The new spectral line data were combined
with our previous LABOCA 870~$\mu$m dust continuum data to calculate the fractional abundances of the detected molecules.
Our main results are summarised as follows:

\begin{enumerate}
\item Altogether six different molecular species were detected, namely SO, H$^{13}$CN, H$^{13}$CO$^+$, SiO, HN$^{13}$C, and C$_2$H. 
The species SO, H$^{13}$CO$^+$, and HN$^{13}$C were detected in all the 14 target sources. The next highest detection rates 
were for C$_2$H (92.9\%) and H$^{13}$CN (85.7\%). Only one source (SMM~3) was found to be an SiO emitter at $\sim174$~GHz (7.1\% detection rate).
\item The SO, H$^{13}$CN, H$^{13}$CO$^+$, HN$^{13}$C, and C$_2$H lines were seen in absorption towards three target positions
(SMM~5, SMM~6, and SMM~7). In SMM~6, the absorption line spectra are likely seen against the continuum emission arising from the 
central YSO, while towards SMM~7 an extragalactic background source could act as the source of continuum emission against which the line absorption in the clump medium is caused. However, the origin of the absorption lines in the IR dark clump SMM~5 remains unclear.
\item The C$_2$H molecules were found to be the most abundant ones of the detected species ($\sim2\times10^{-9}$ on average), while HN$^{13}$C is the least abundant species ($\sim2\times10^{-11}$ on average).
\item We found three potential positive correlations between the derived fractional abundances of the molecules. With a correlation coefficient of $r\simeq0.9$, the most significant of these are between C$_2$H and HN$^{13}$C, and between HN$^{13}$C and H$^{13}$CO$^+$. These relationships can be understood as manifestations of the gas-phase electron (ionisation degree) and atomic carbon abundances. 
\item The fractional abundances of the detected molecules in the Seahorse IRDC are typically low compared to other IRDC clumps. The 
only exceptions are H$^{13}$CN and H$^{13}$CO$^+$, which appear to be comparably abundant (or even a few times more abundant) in the Seahorse than in other Galactic IRDCs .
\item The statistically most signifcant evolutionary trends we found are the drop in the C$_2$H abundance and in the $[{\rm HN^{13}C}]/[{\rm H^{13}CN}]$ ratio as the clump evolves from an IR dark stage to IR bright stage, and further to an \ion{H}{ii} region stage. These are in agreement with previous studies, and the former trend can be explained by the conversion of C$_2$H to species like CO when the clump temperature rises, especially owing to the emergence of hot cores in high-mass star-forming clumps. The latter trend is likely to be a manifestation of the fairly well established negative temperature dependence of the $[{\rm HNC}]/[{\rm HCN}]$ ratio, which in turn is likely predominantly driven by neutral-neutral reactions ${\rm HNC}+{\rm H}$ and ${\rm HNC}+{\rm O}$ that destroy HNC (and form HCN in the first case) in the heated gas phase.
\end{enumerate}

The Seahorse IRDC lies about $1\fdg3$ or $\sim60$~pc above the Galactic plane and, hence, it was not part of the Galactic regions mapped with the \textit{Spitzer} IR satellite. In this sense, the Seahorse is different from the more mainstream IRDCs that have been studied in more detail and which are typically selected on the basis of their appearance in the \textit{Spitzer} mid-IR images.
The Seahorse's environment and its hypothetical evolutionary stage of still being in the process of accreting mass along the dusty streams, as seen in \textit{Herschel} images, might have an effect on the physical and chemical properties of the cloud. Molecular line imaging of the cloud would be useful to get the picture of how the emission from different molecules is spatially distributed and to understand whether the striations projected towards the Seahorse are physically related to the main filament. In particular, mapping the filament in a low rotational transition of SiO could help to understand if the cloud is associated with a large-scale shock front driven by its precursor colliding gas flows. On the other hand, high-resolution interferometric spectral line imaging is required to search for HMCs within the clumps of the Seahorse IRDC and, hence, to see whether high-mass star formation is taking place elsewhere in the cloud besides the \ion{H}{ii} region associated with IRAS~13039-6108.

\begin{acknowledgements}

I would like to thank the anonymous referee for providing constructive, scientific comments that led to improvements in this paper.
I am grateful to the staff at the APEX telescope for performing the service mode SEPIA 
observations presented in this paper. This research has made use of NASA's Astrophysics Data System Bibliographic Services. 
This research made use of {\tt Astropy}\footnote{\url{http://www.astropy.org}}, a community-developed core {\tt Python} package for Astronomy 
(\cite{astropy2013}, \cite{astropy2018}). I also acknowledge the community effort devoted to the development of the following open-source 
{\tt Python} packages and libraries that were used in this work: {\tt matplotlib} (\cite{hunter2007}), {\tt pandas} (\cite{mckinney2010}), 
{\tt numpy} (\cite{vanderwawlt2011}), {\tt seaborn} (\cite{waskom2017}), and {\tt scipy} (\cite{virtanen2020}).
The title of this paper is inspired by the paper by Ward-Thompson et al. (2006), "SCUBA observations of the Horsehead nebula - what did the horse swallow?".

\end{acknowledgements}

\appendix

\section{Spectra and line parameters}

\begin{figure*}[!htb]
\begin{center}
\includegraphics[width=0.33\textwidth]{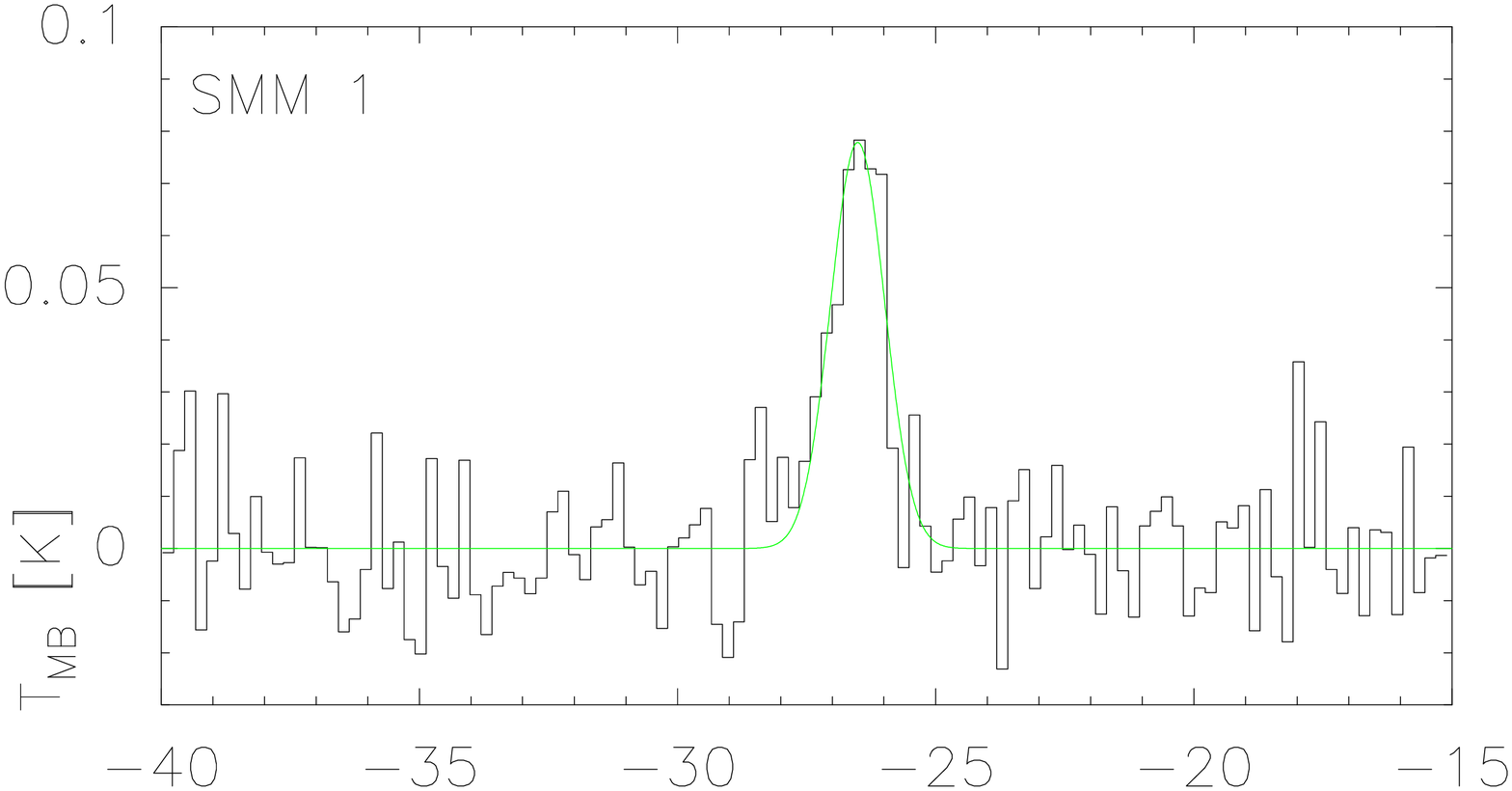}
\includegraphics[width=0.33\textwidth]{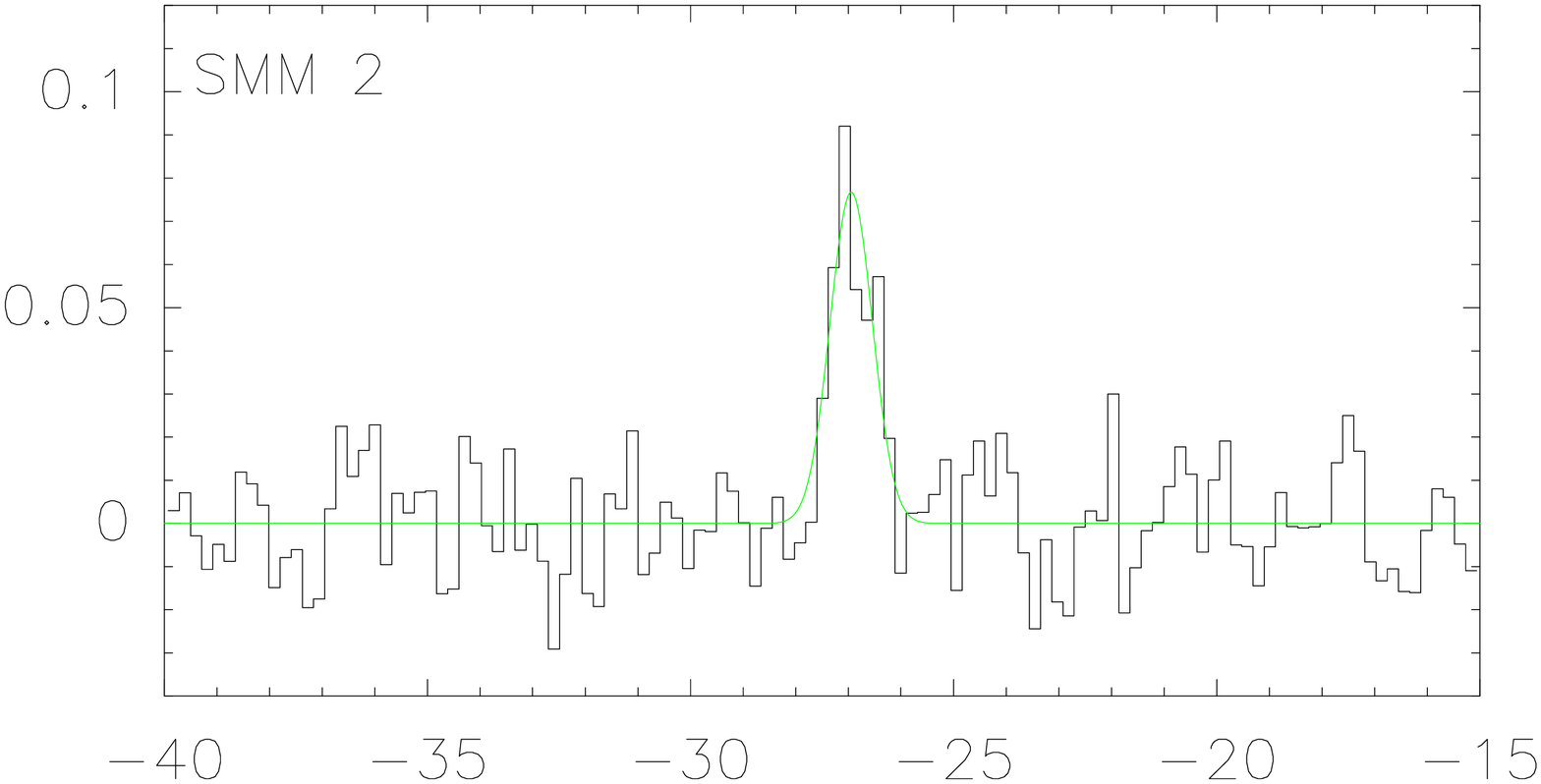}
\includegraphics[width=0.33\textwidth]{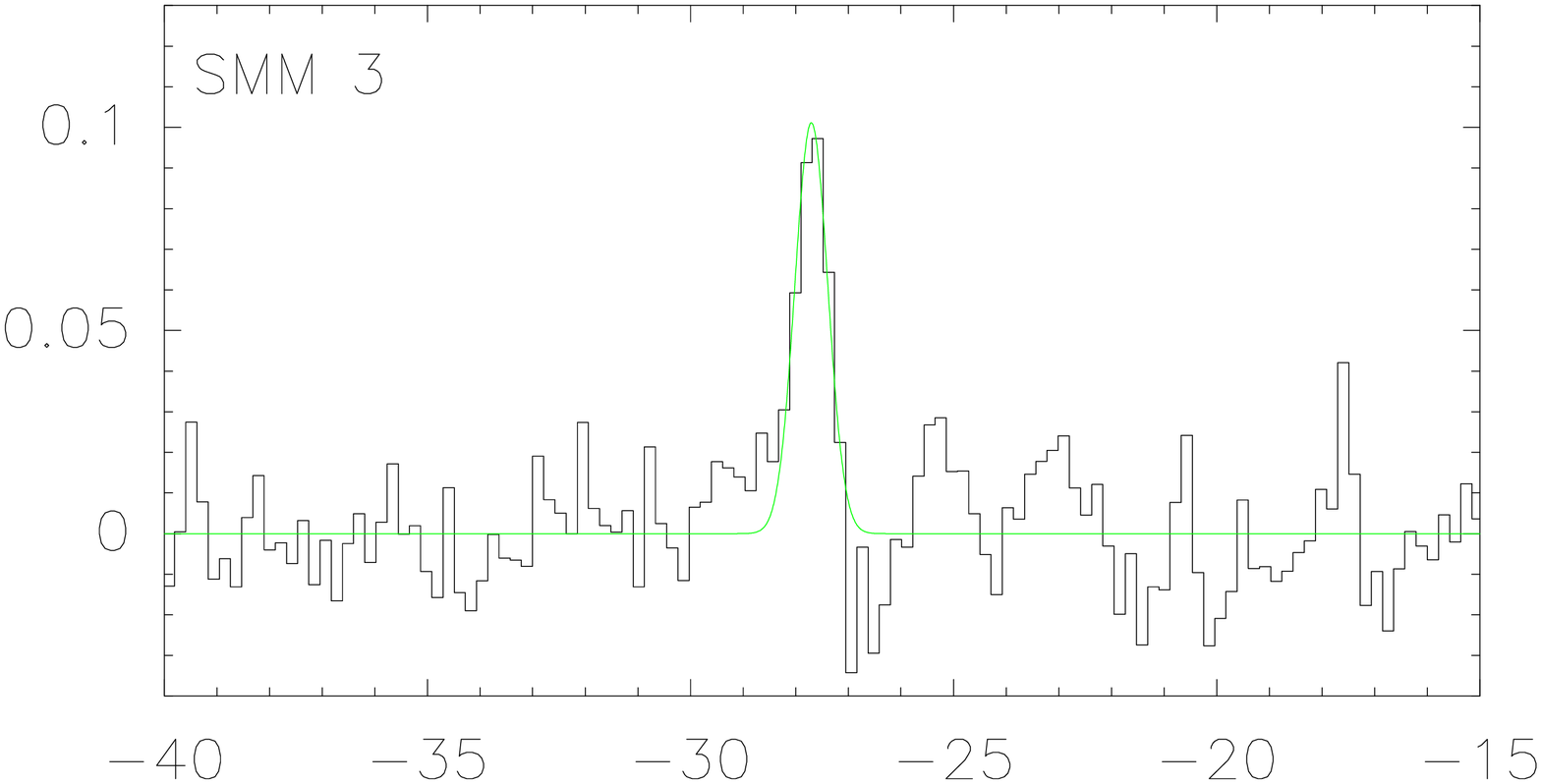}
\includegraphics[width=0.33\textwidth]{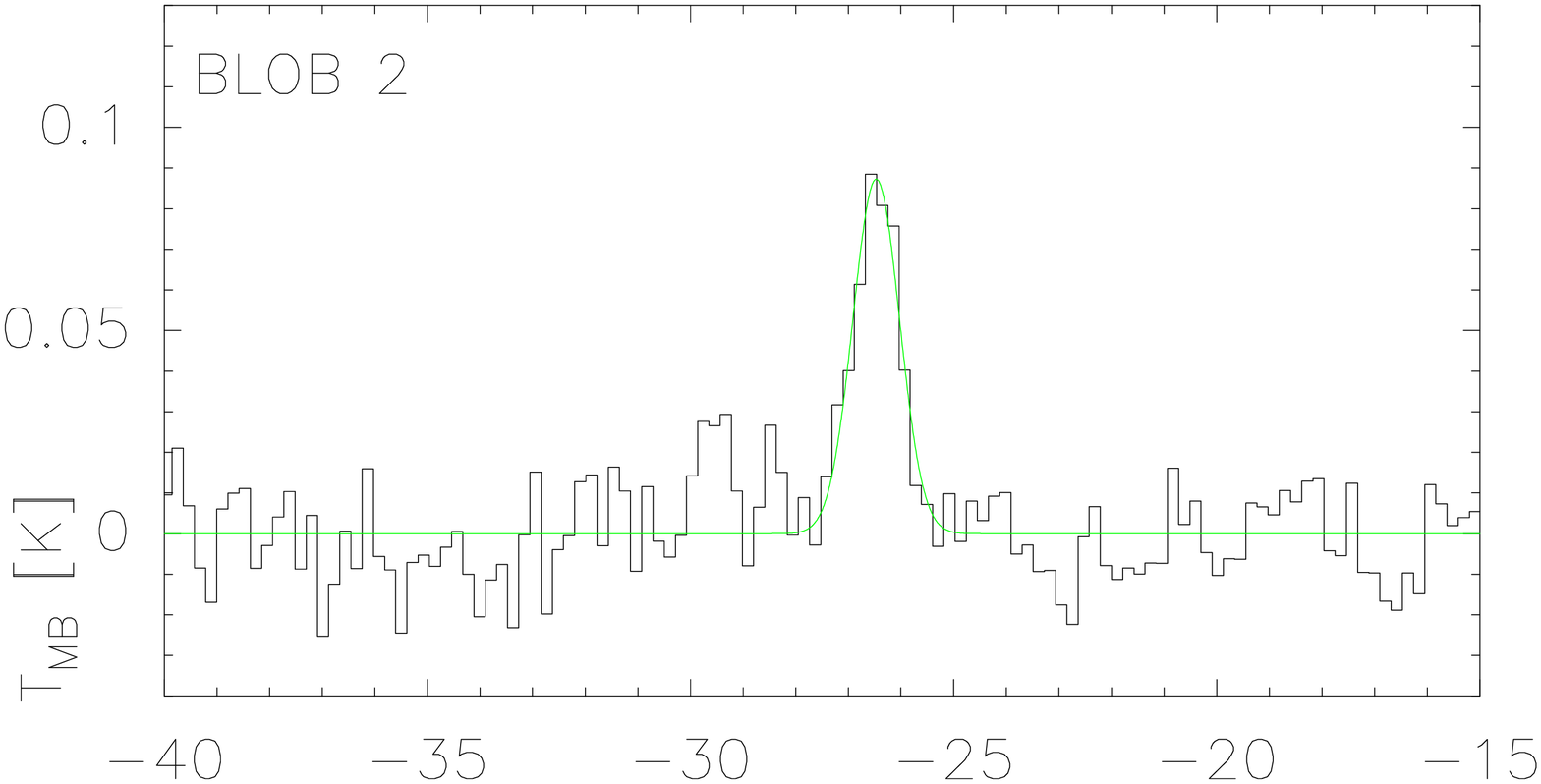}
\includegraphics[width=0.33\textwidth]{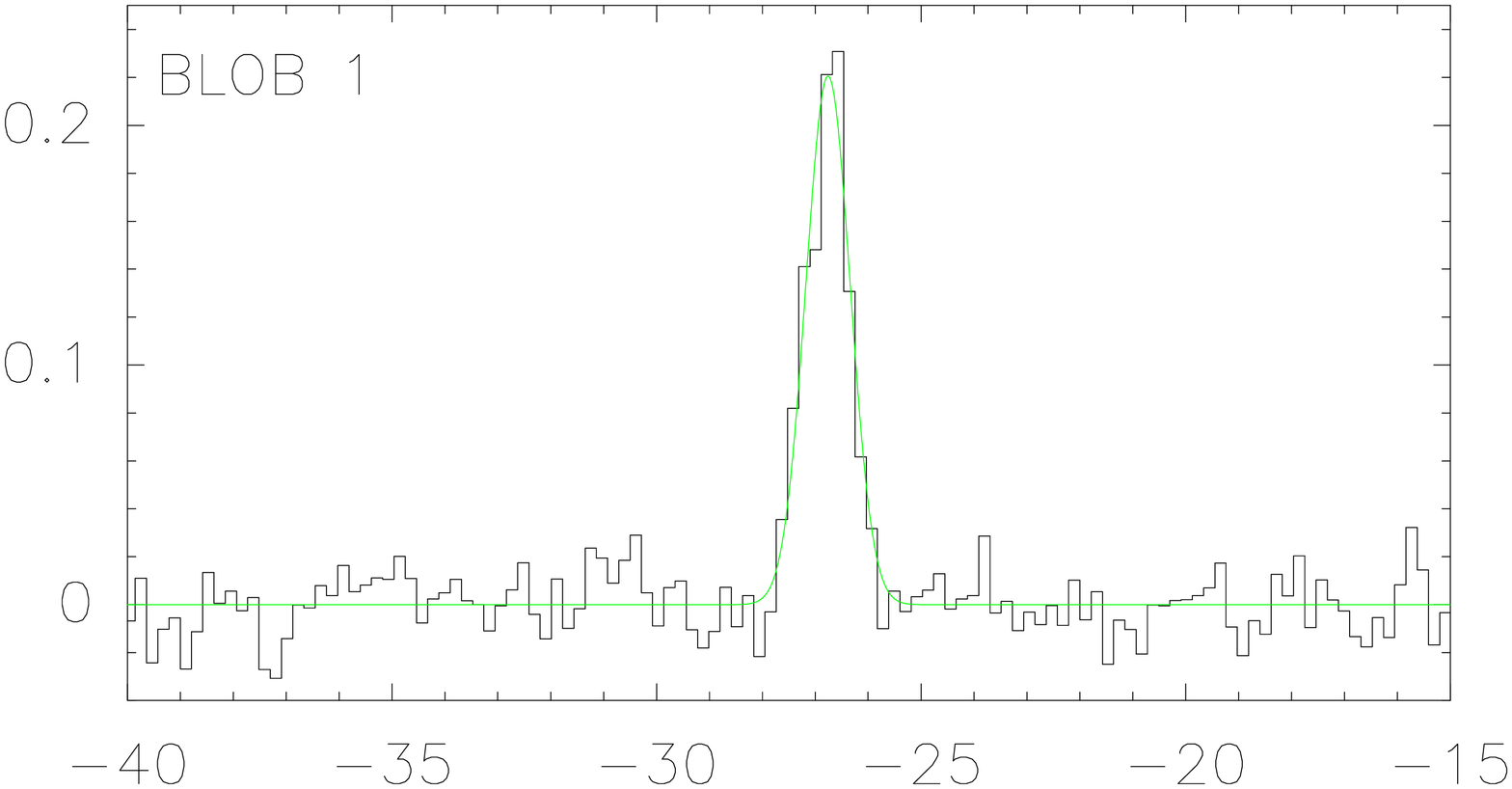}
\includegraphics[width=0.33\textwidth]{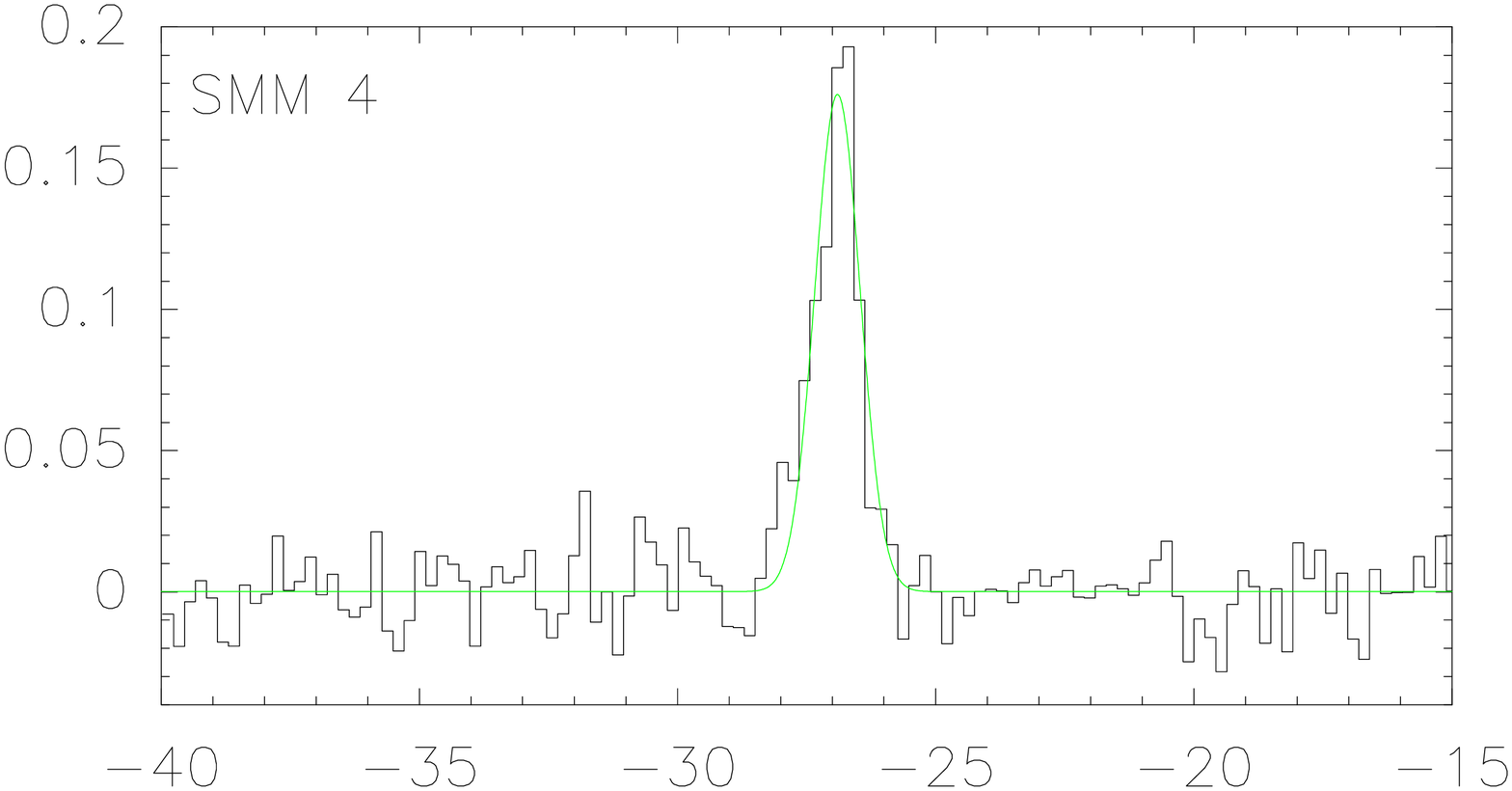}
\includegraphics[width=0.33\textwidth]{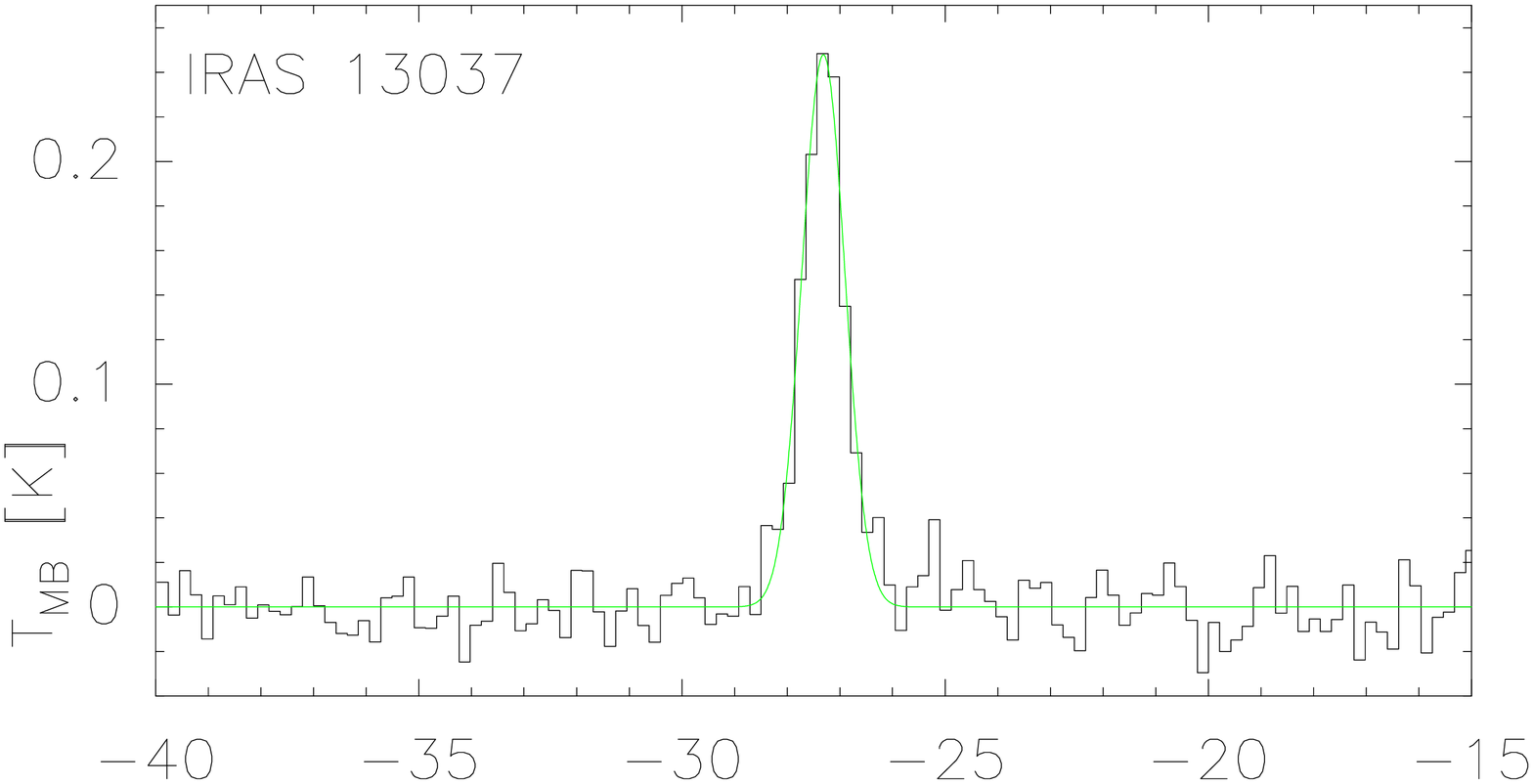}
\includegraphics[width=0.33\textwidth]{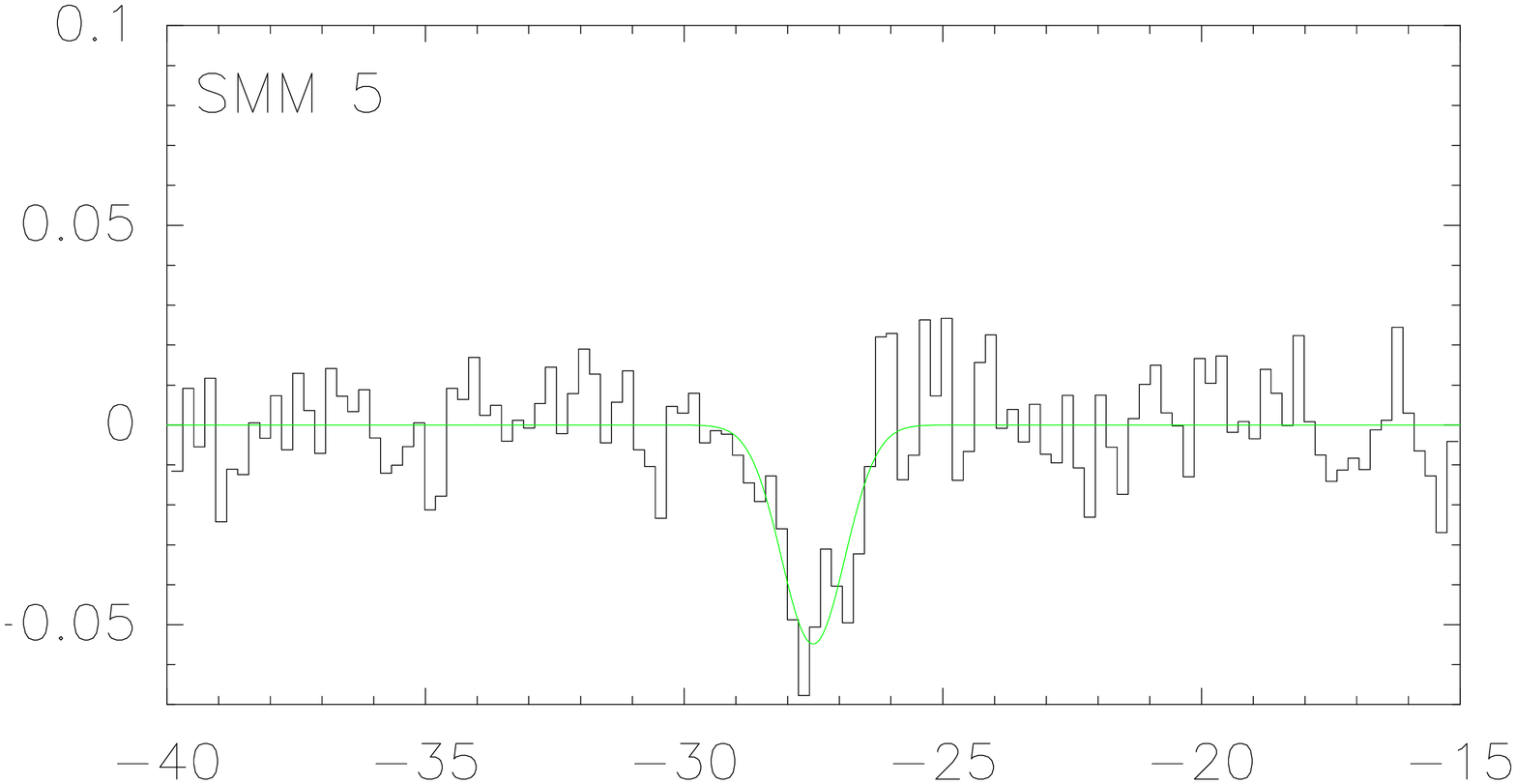}
\includegraphics[width=0.33\textwidth]{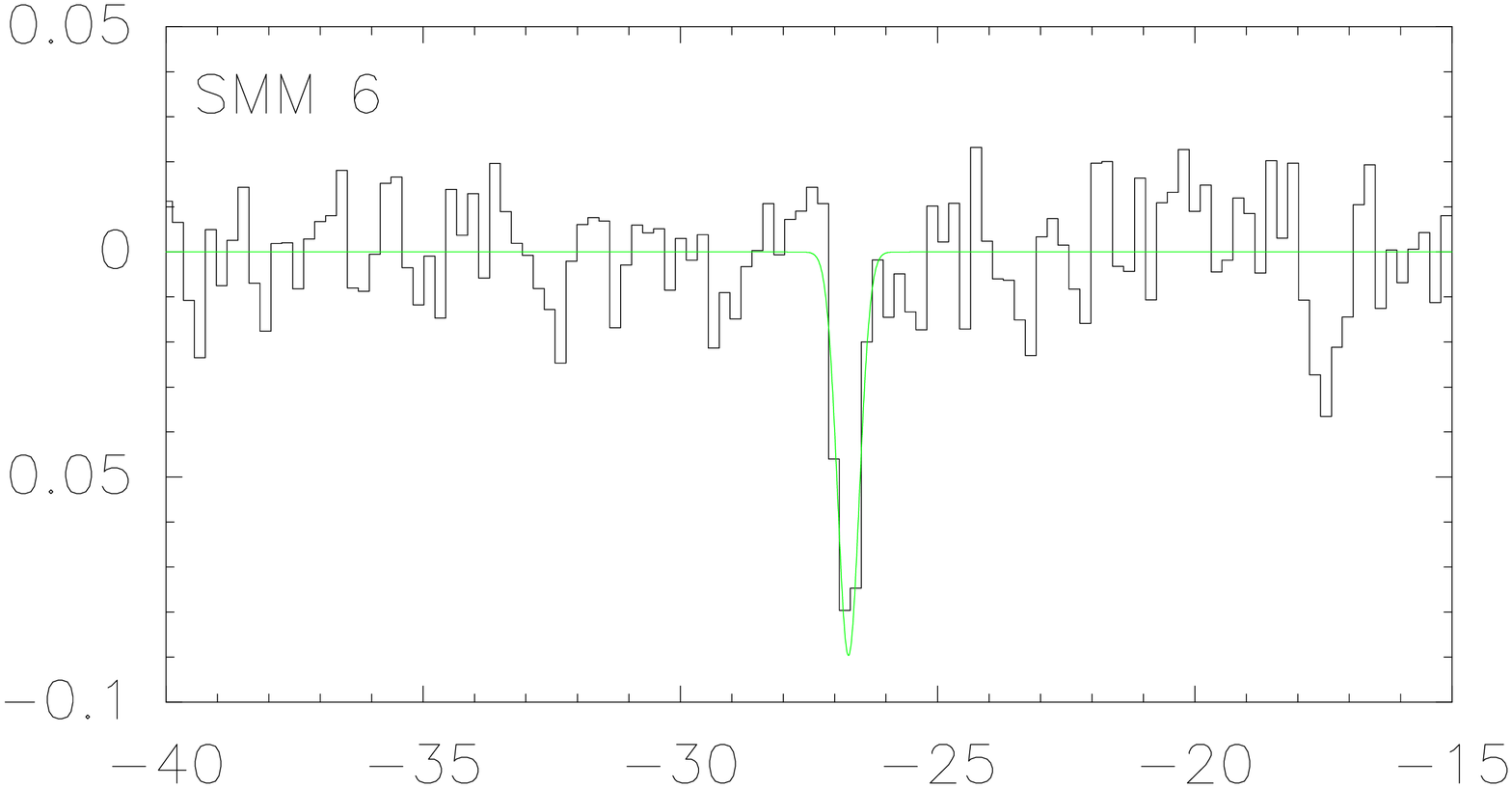}
\includegraphics[width=0.33\textwidth]{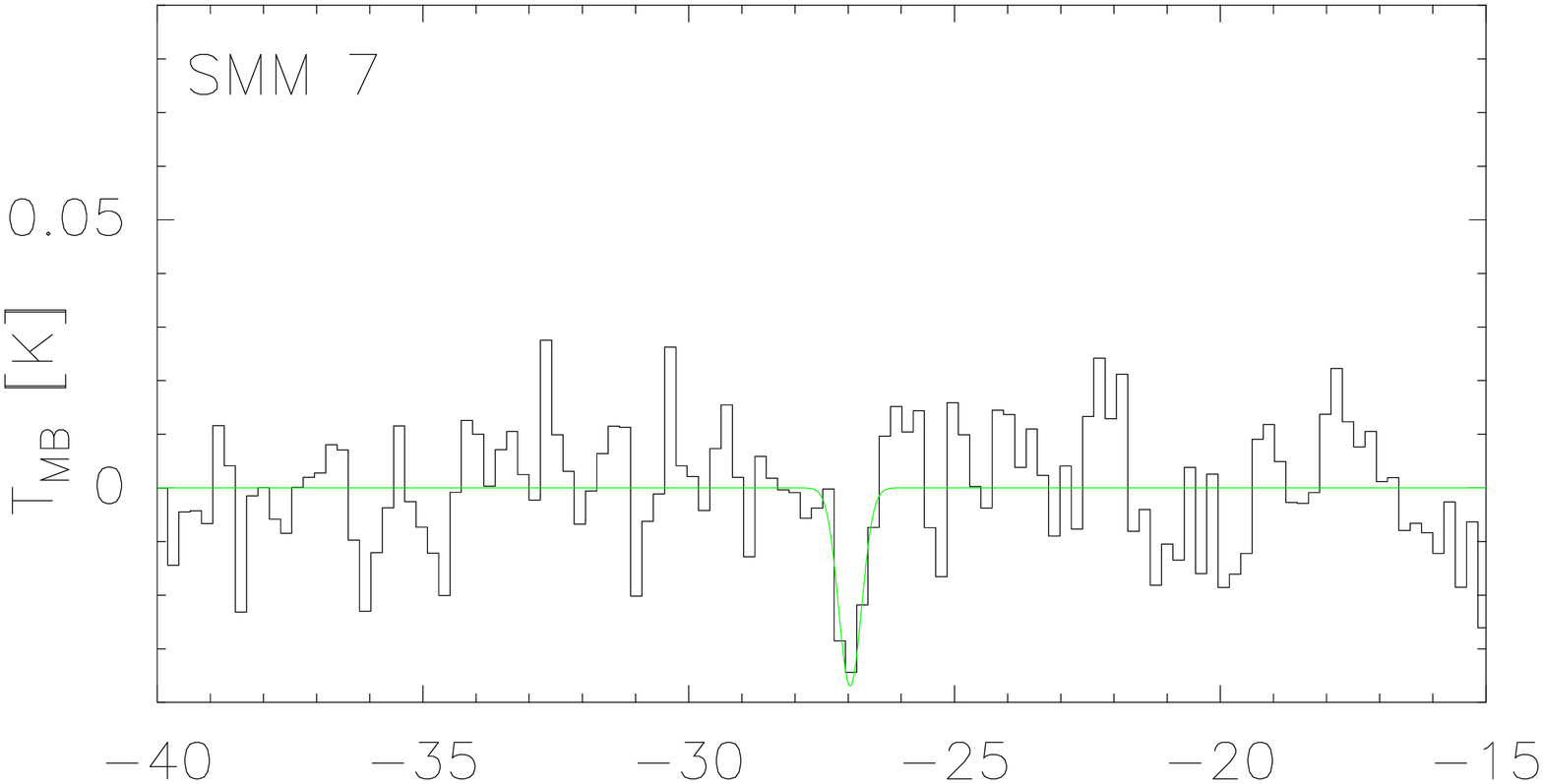}
\includegraphics[width=0.33\textwidth]{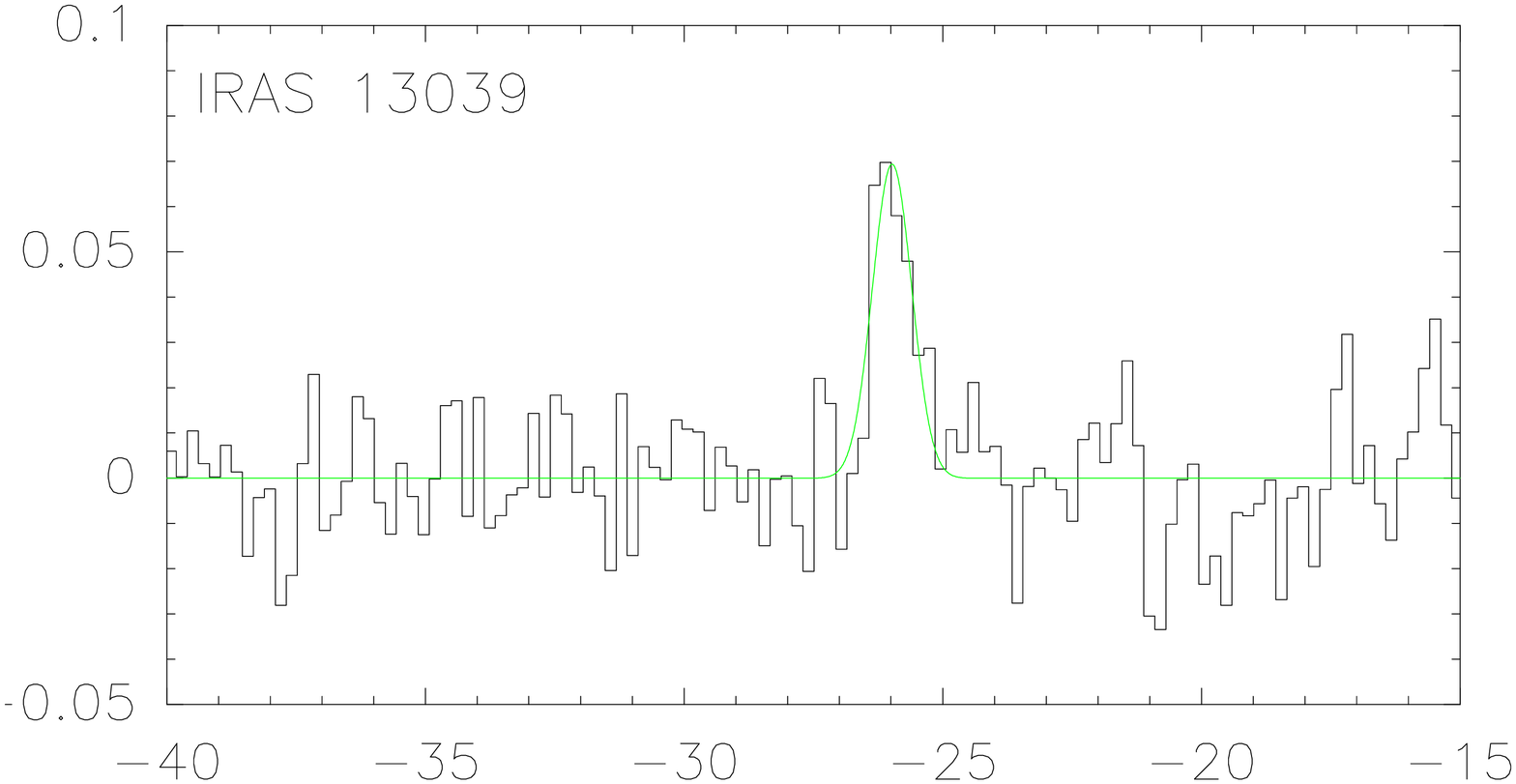}
\includegraphics[width=0.33\textwidth]{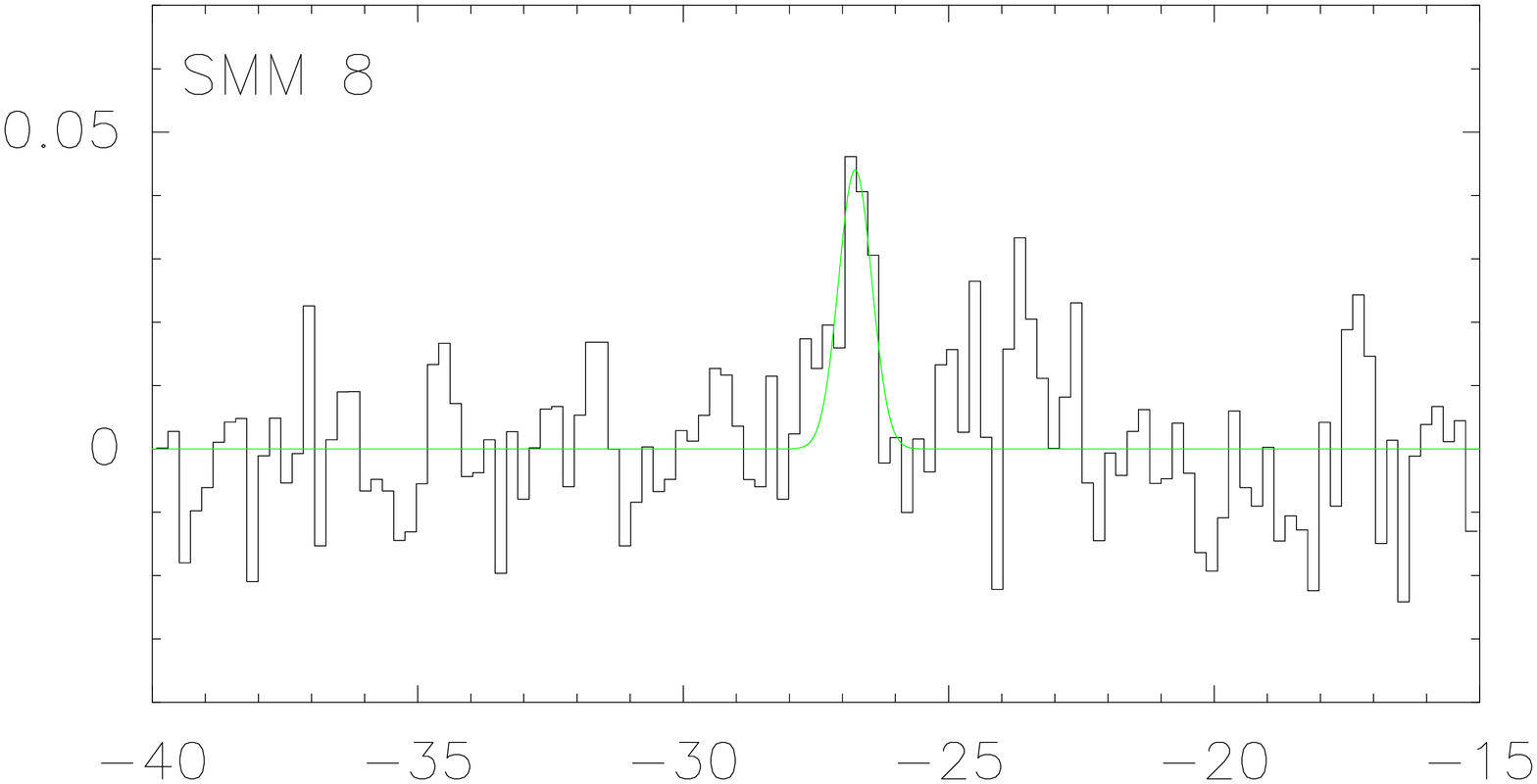}
\includegraphics[width=0.33\textwidth]{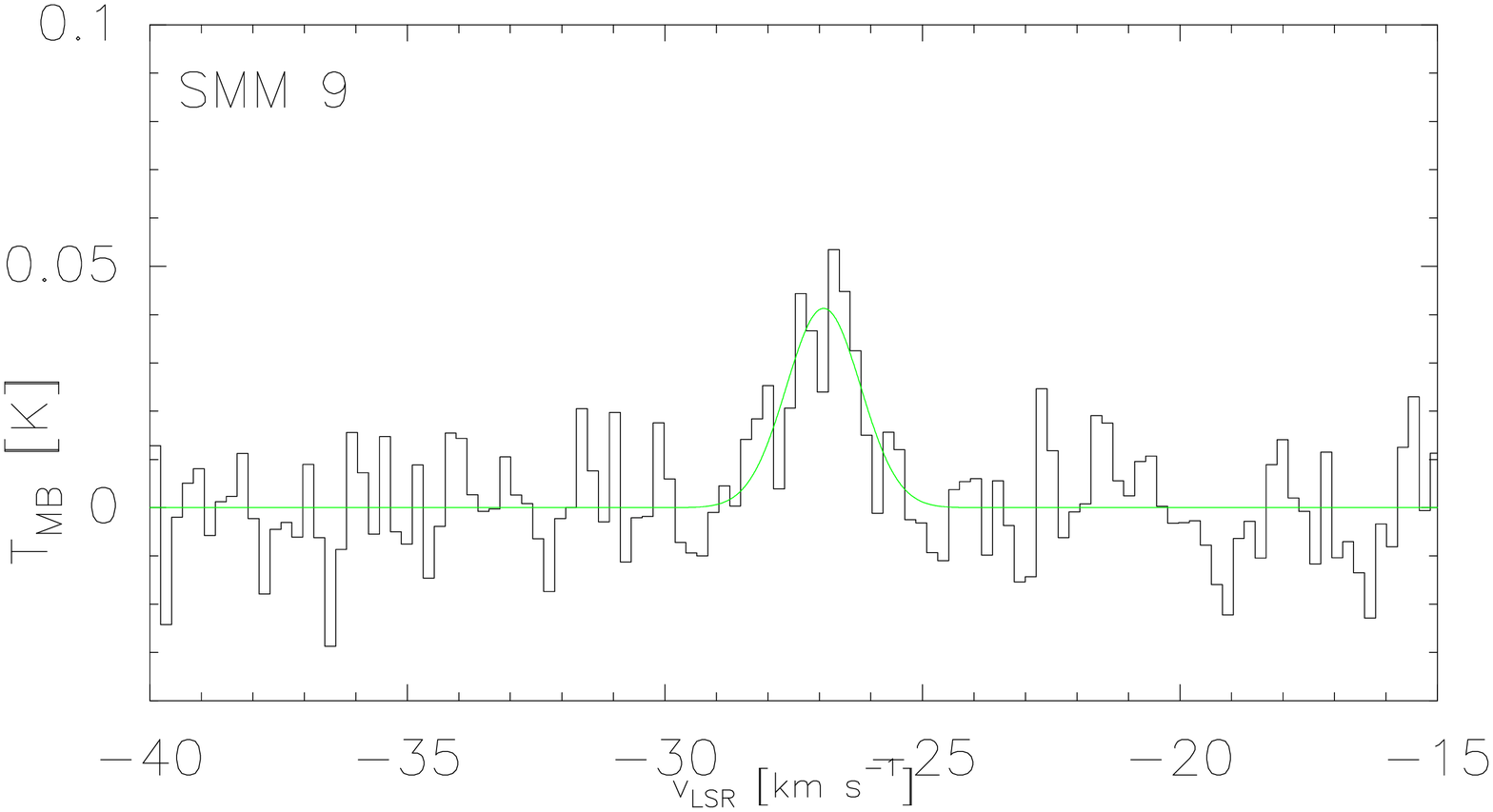}
\includegraphics[width=0.33\textwidth]{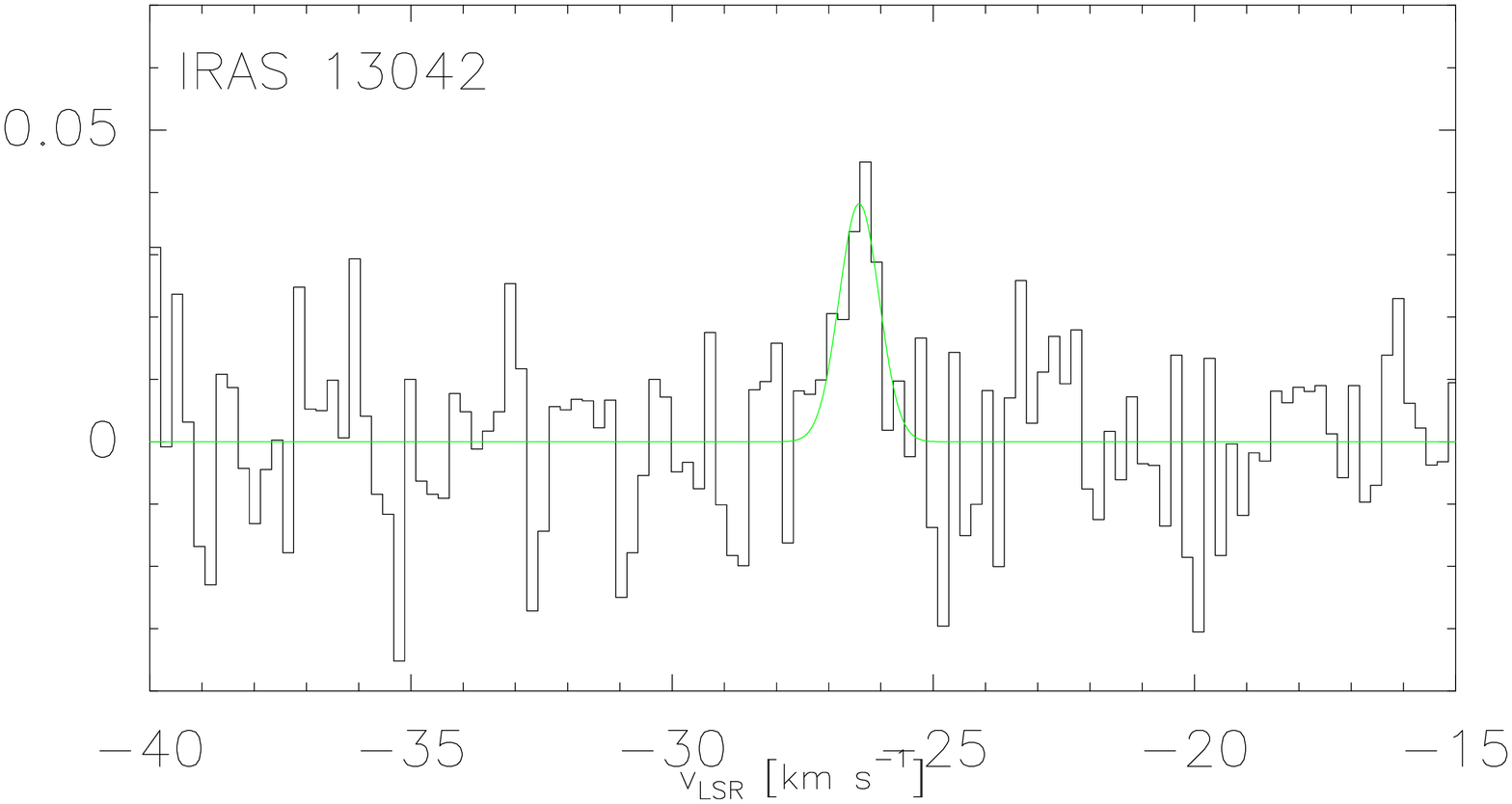}
\caption{SO$(4_4-3_3)$ spectra towards the Seahorse IRDC clumps. Gaussian fits to the lines are overlaid in green. While the velocity range shown in each panel is the same, the intensity range is different to better show the line profiles. The SO line is seen in absoprtion towards SMM~5--7.}
\label{figure:so}
\end{center}
\end{figure*}

\begin{figure*}[!htb]
\begin{center}
\includegraphics[width=0.33\textwidth]{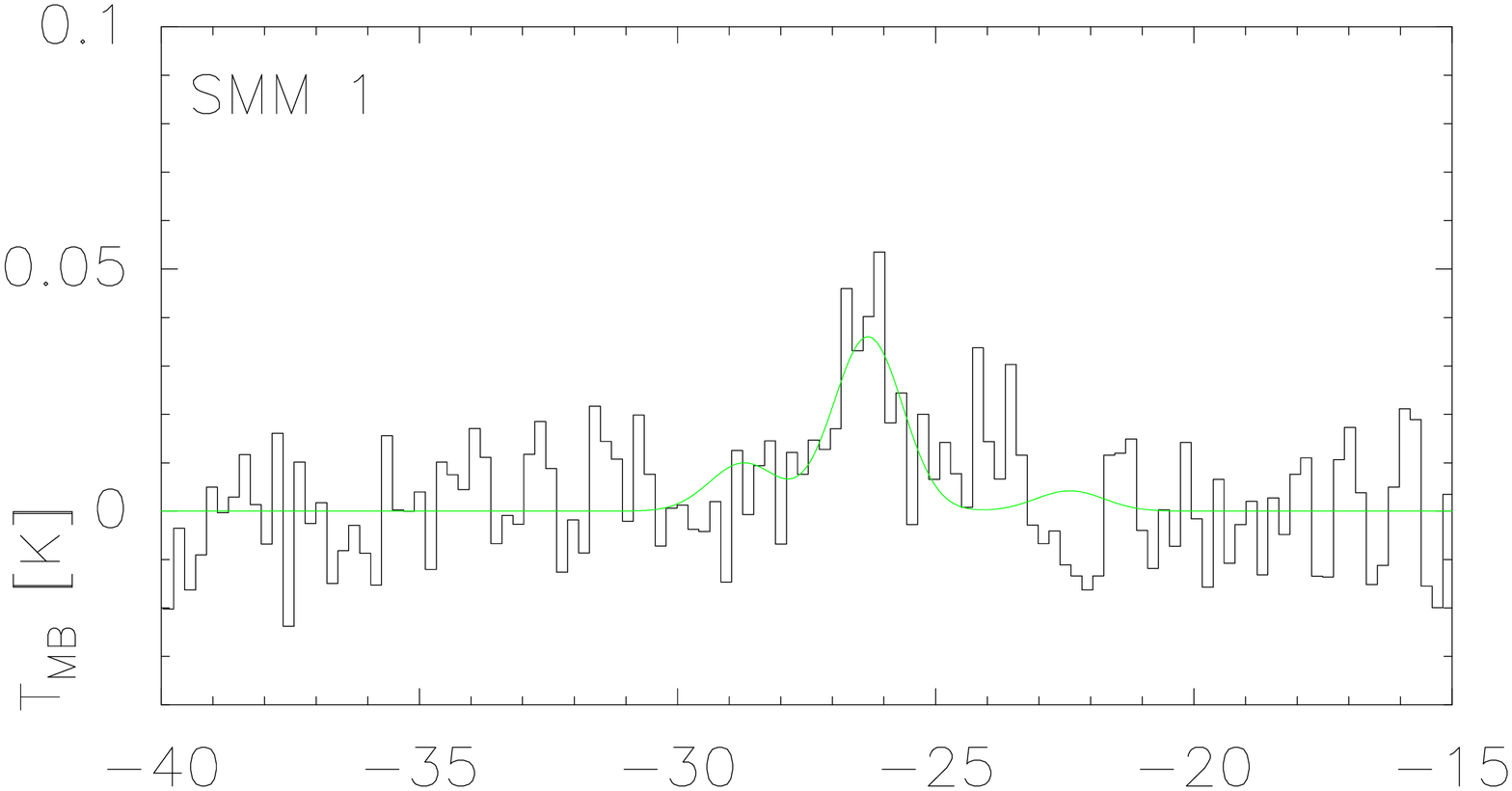}
\includegraphics[width=0.33\textwidth]{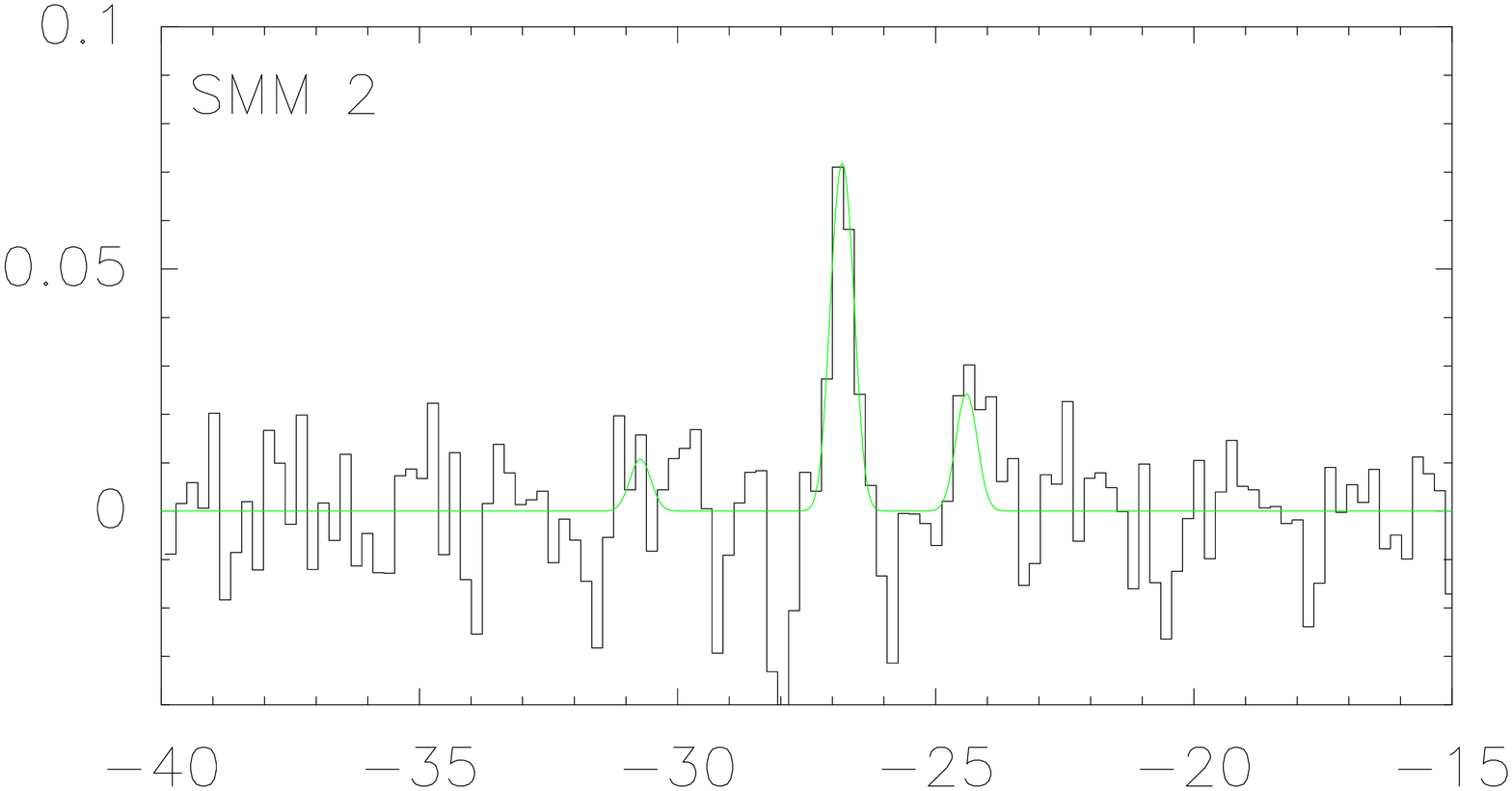}
\includegraphics[width=0.33\textwidth]{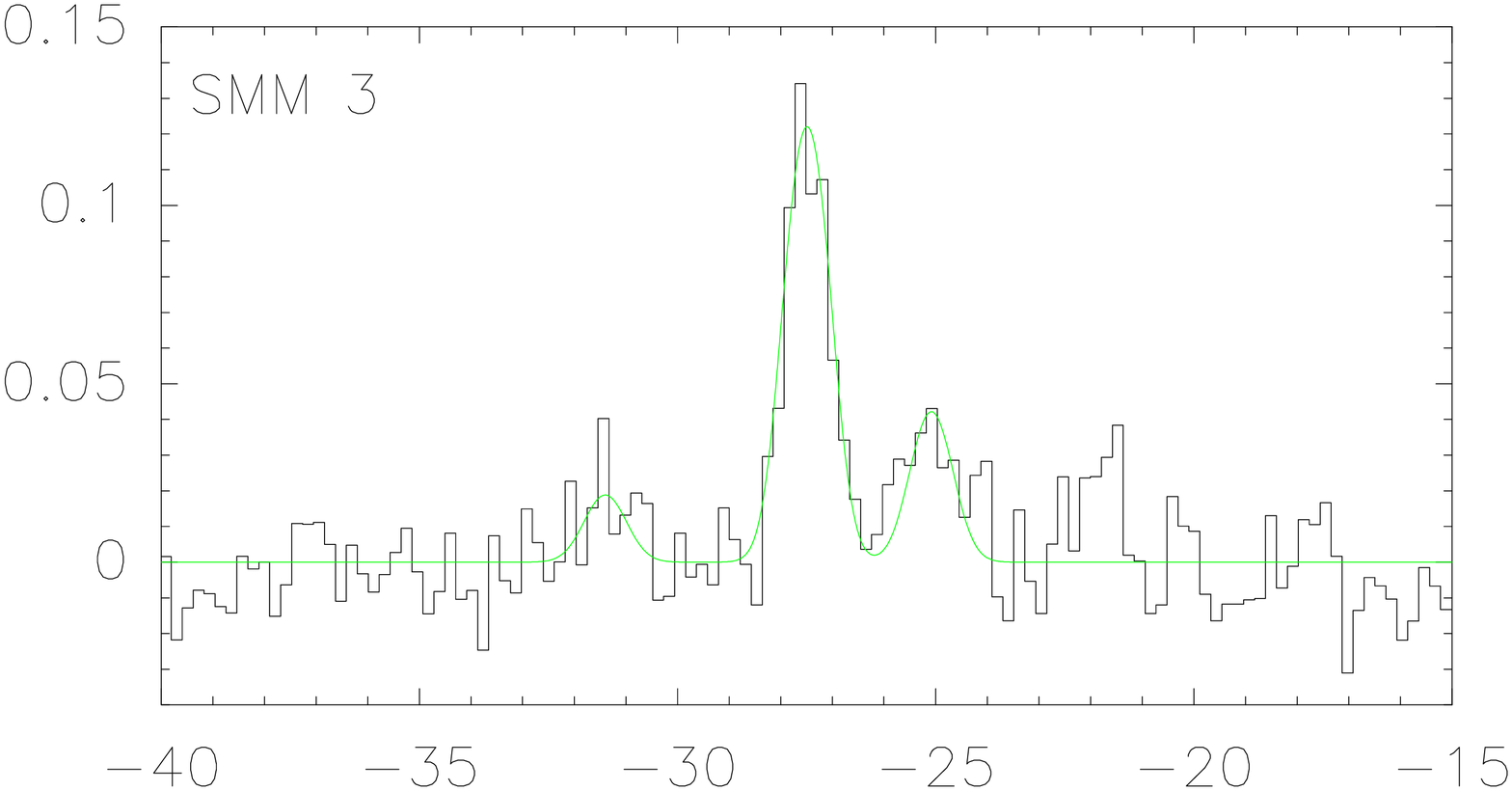}
\includegraphics[width=0.33\textwidth]{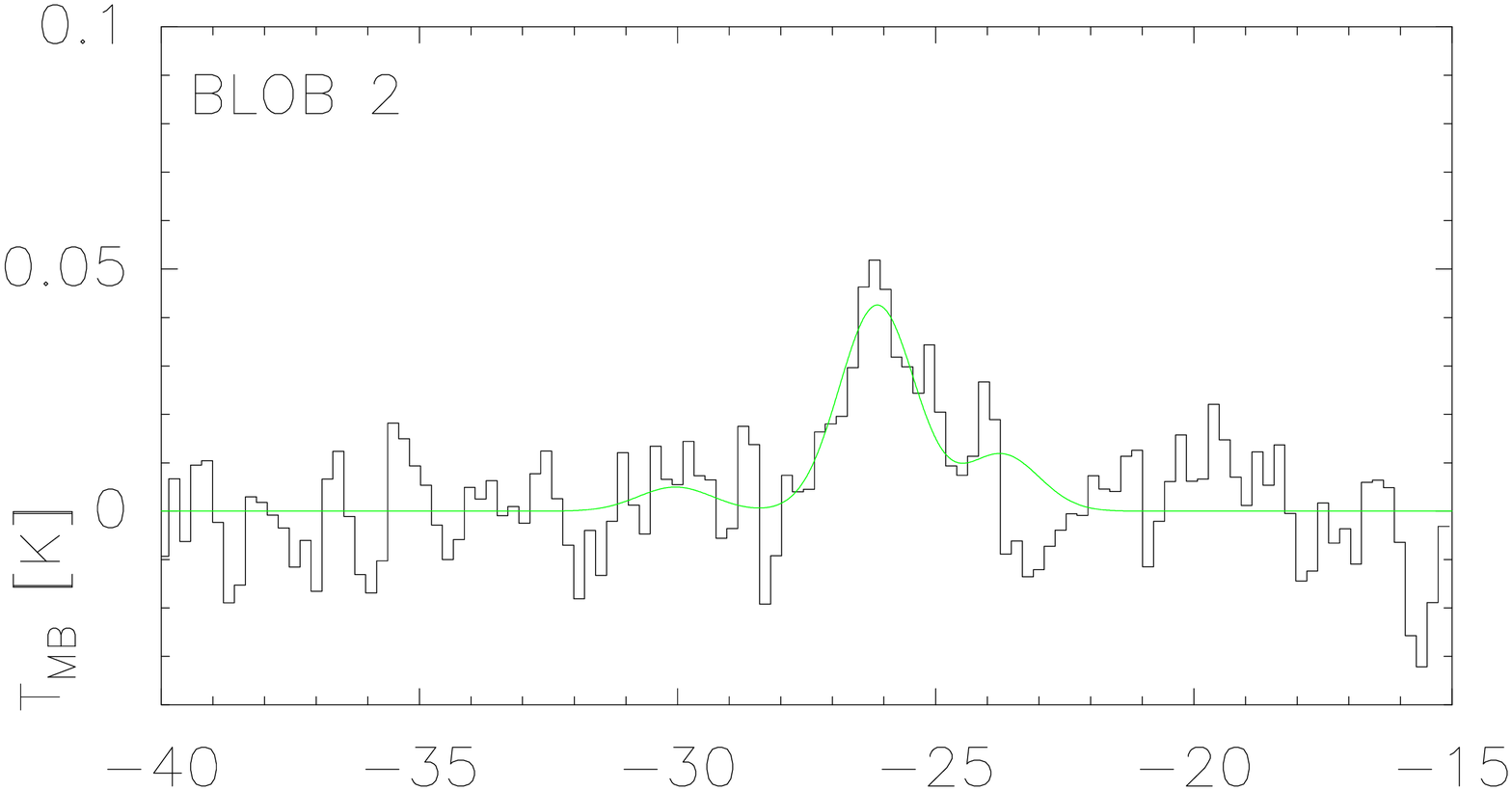}
\includegraphics[width=0.33\textwidth]{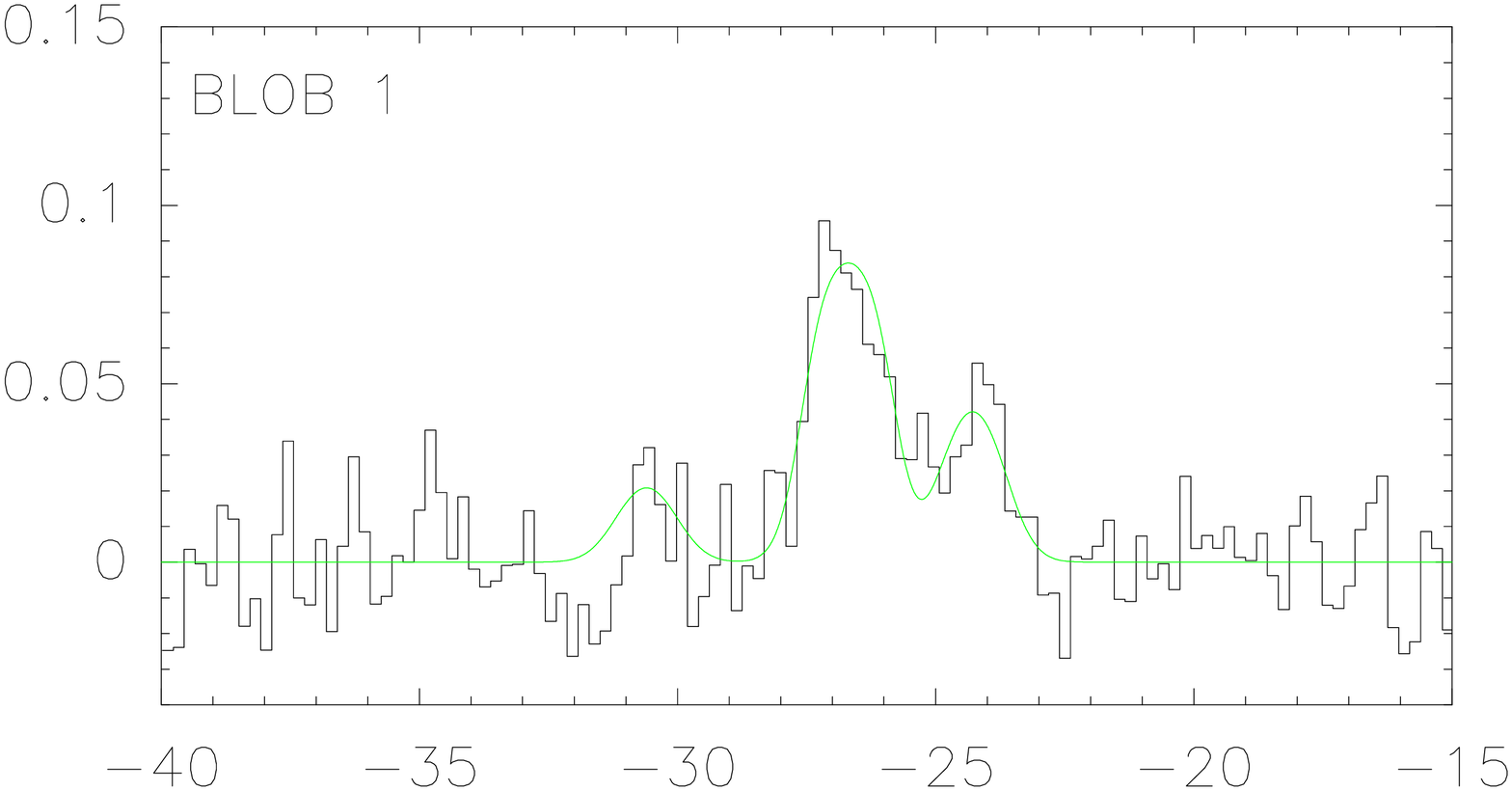}
\includegraphics[width=0.33\textwidth]{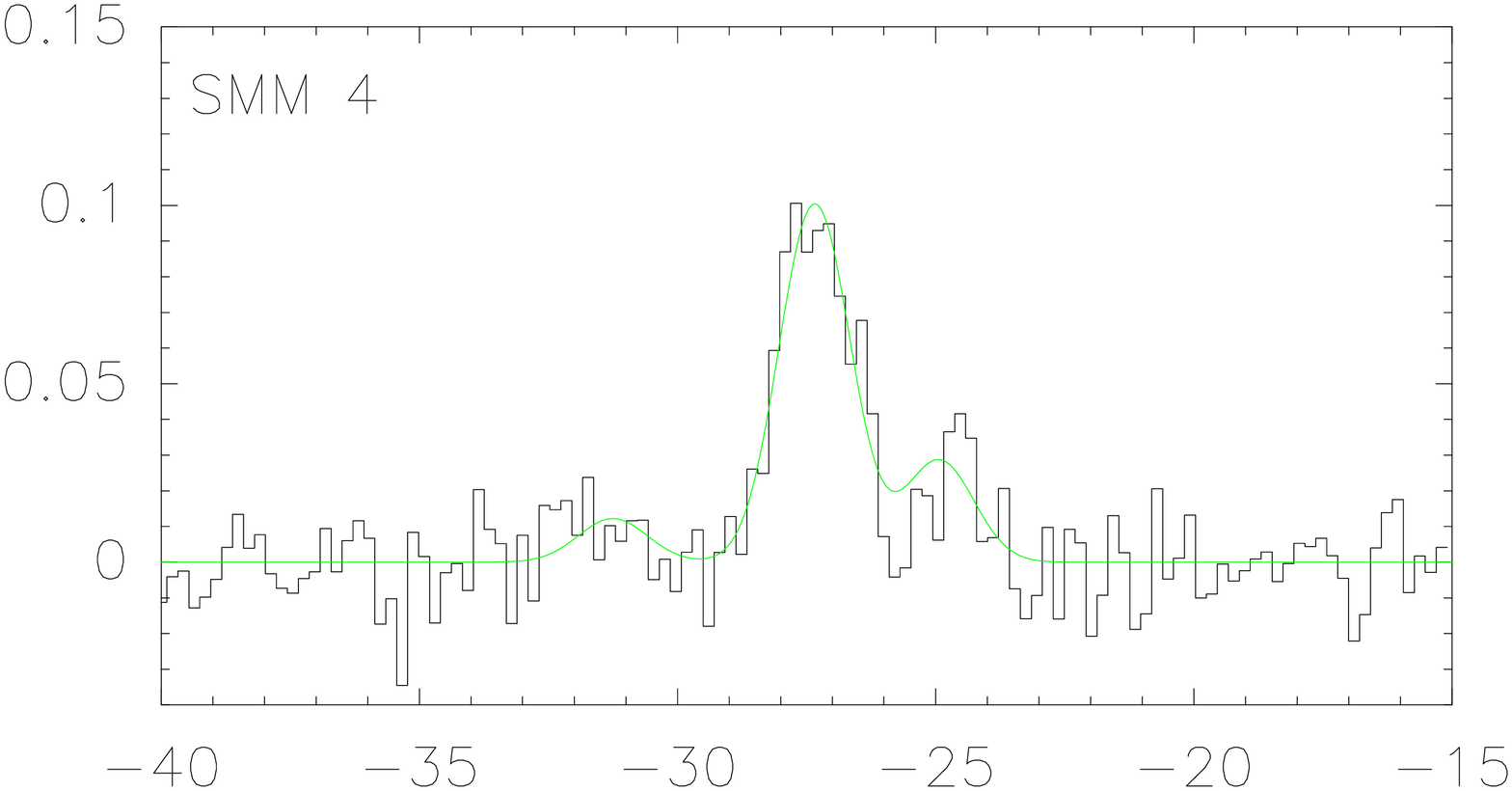}
\includegraphics[width=0.33\textwidth]{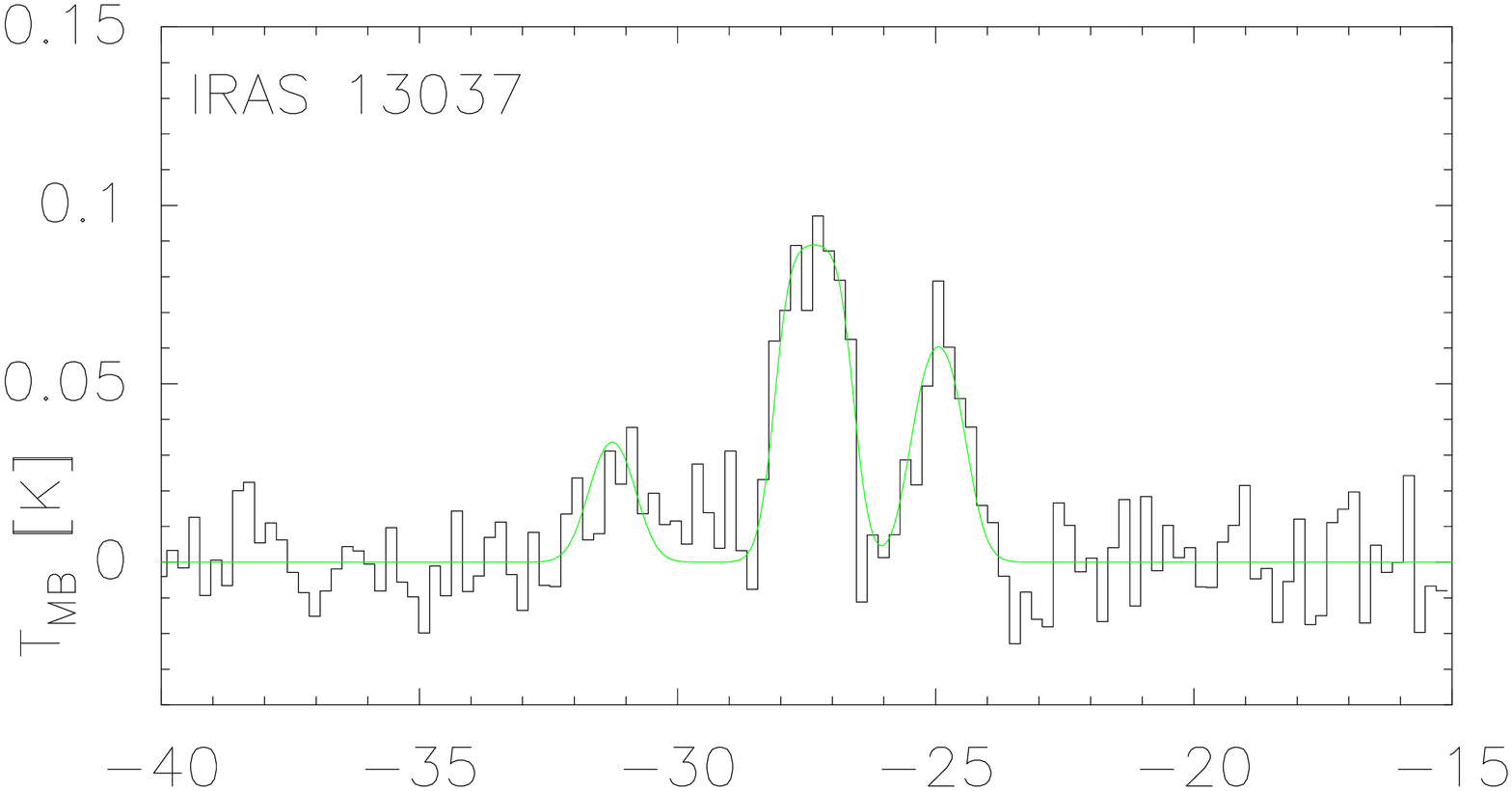}
\includegraphics[width=0.33\textwidth]{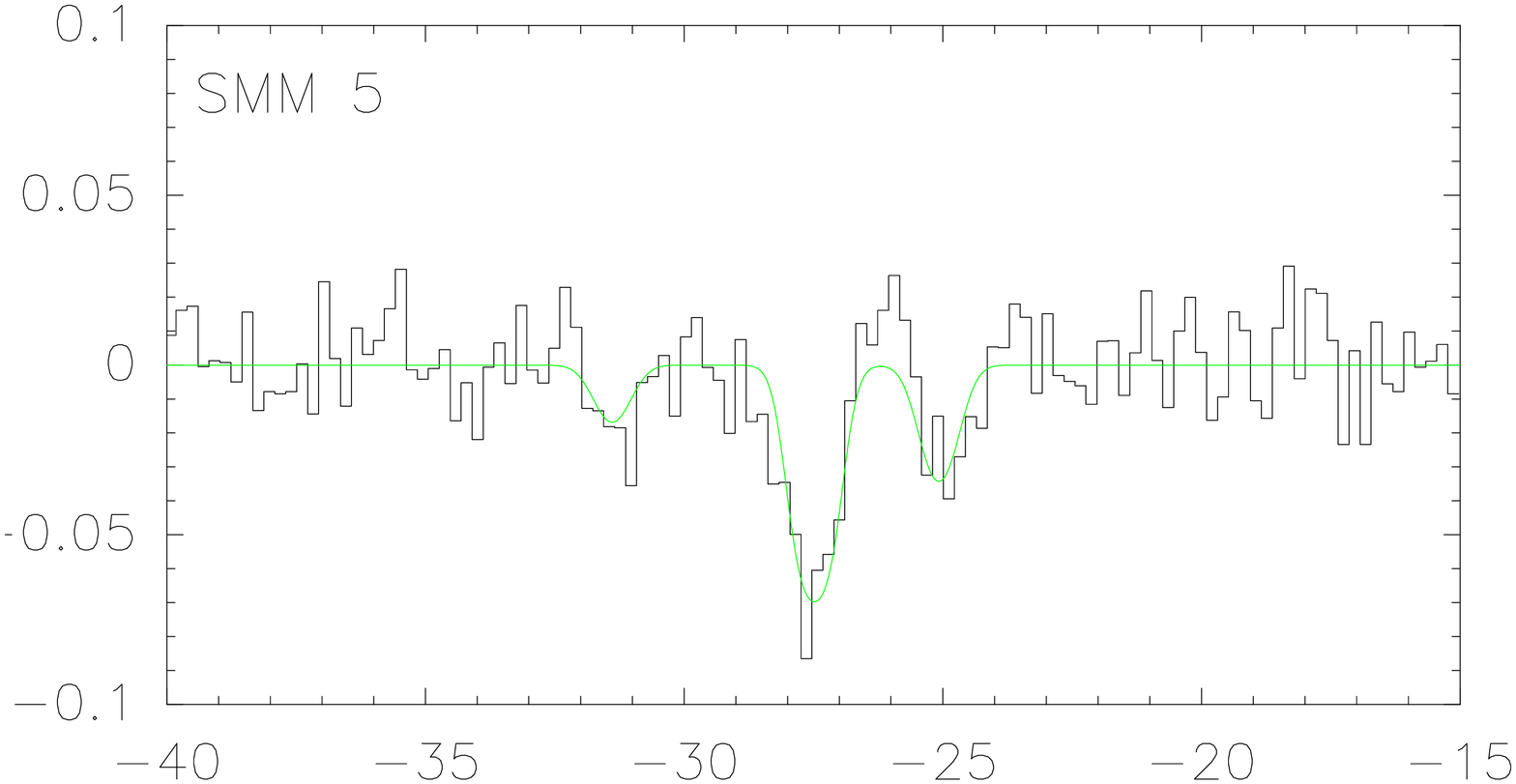}
\includegraphics[width=0.33\textwidth]{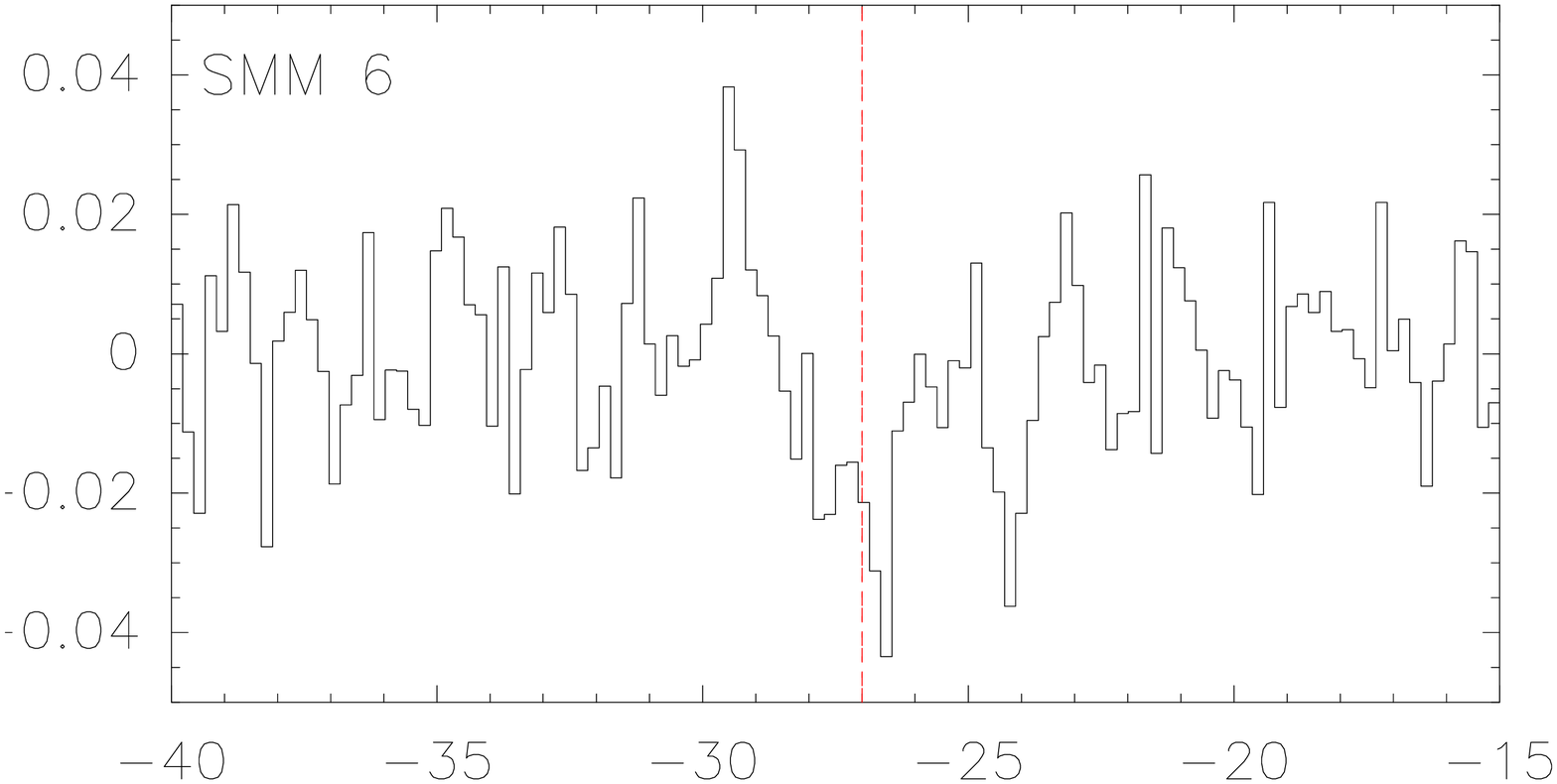}
\includegraphics[width=0.33\textwidth]{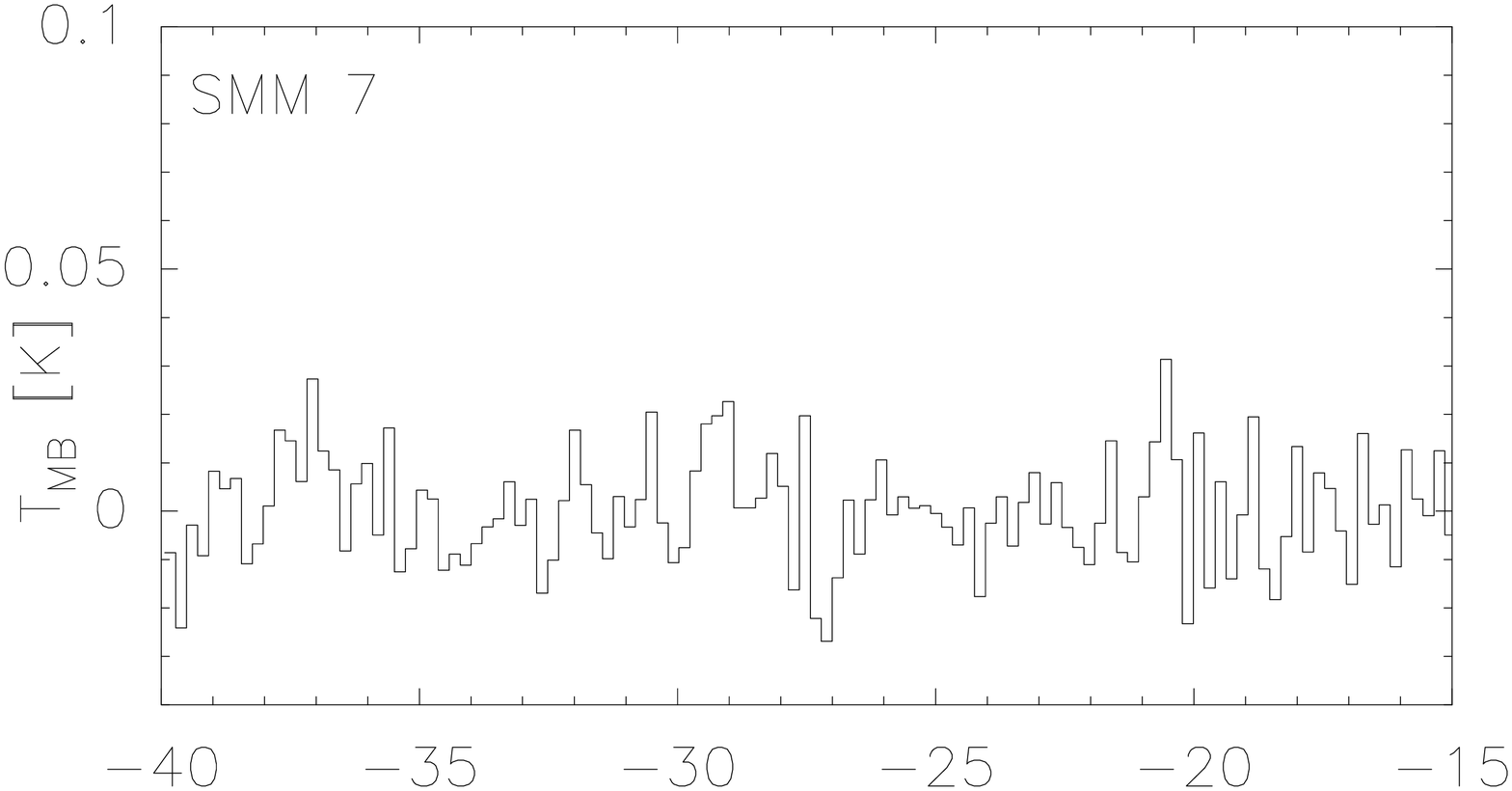}
\includegraphics[width=0.33\textwidth]{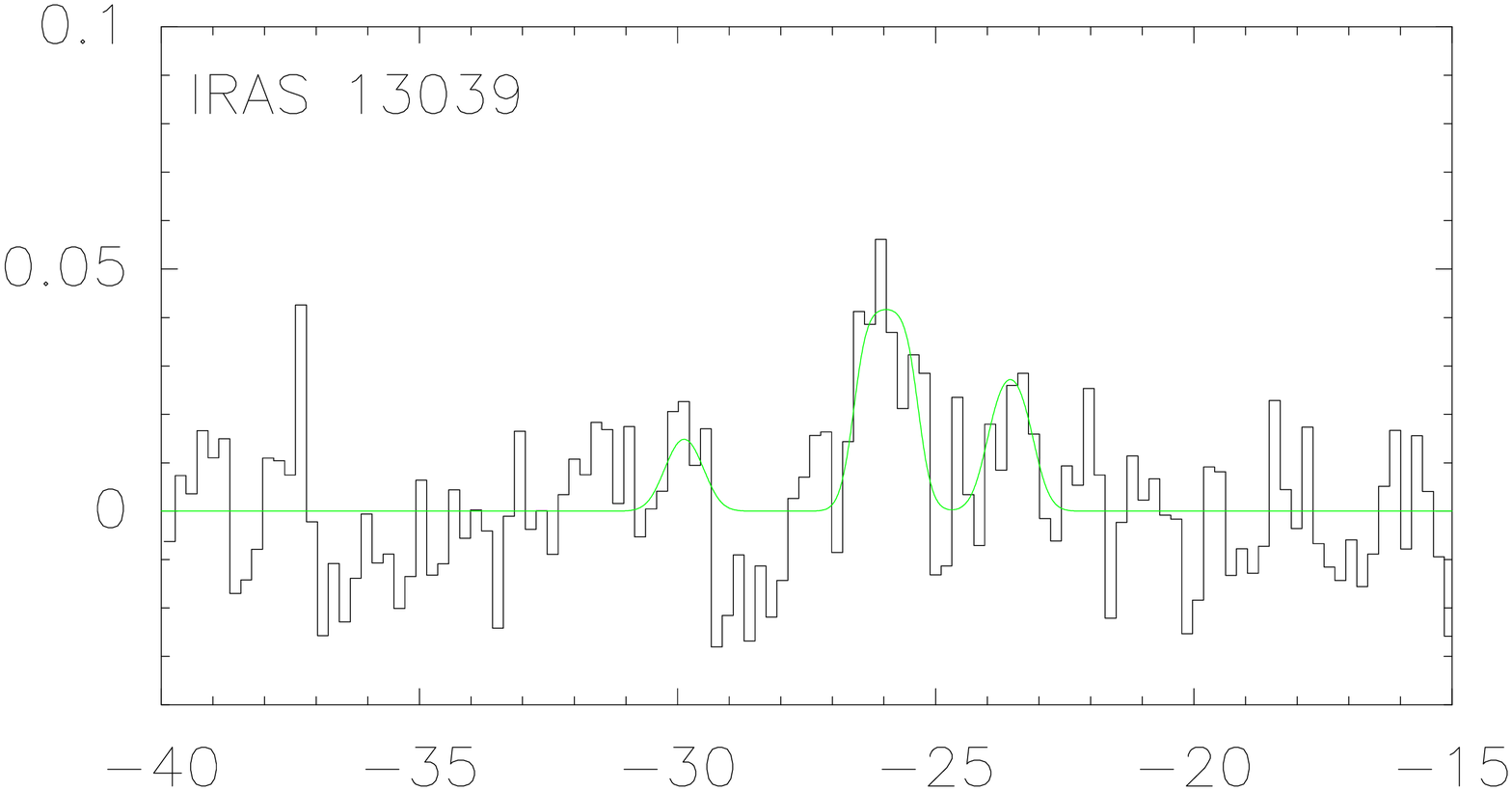}
\includegraphics[width=0.33\textwidth]{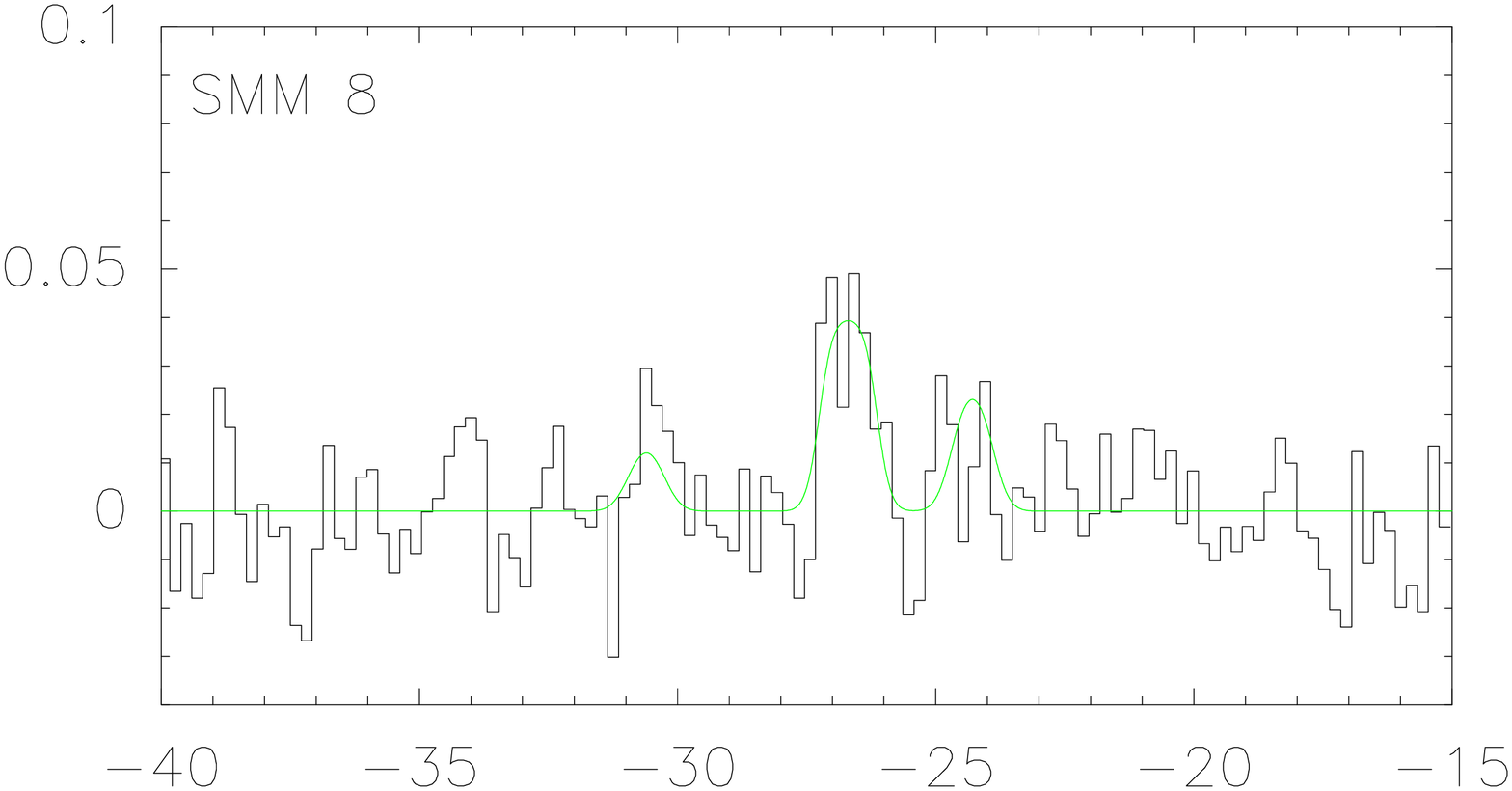}
\includegraphics[width=0.33\textwidth]{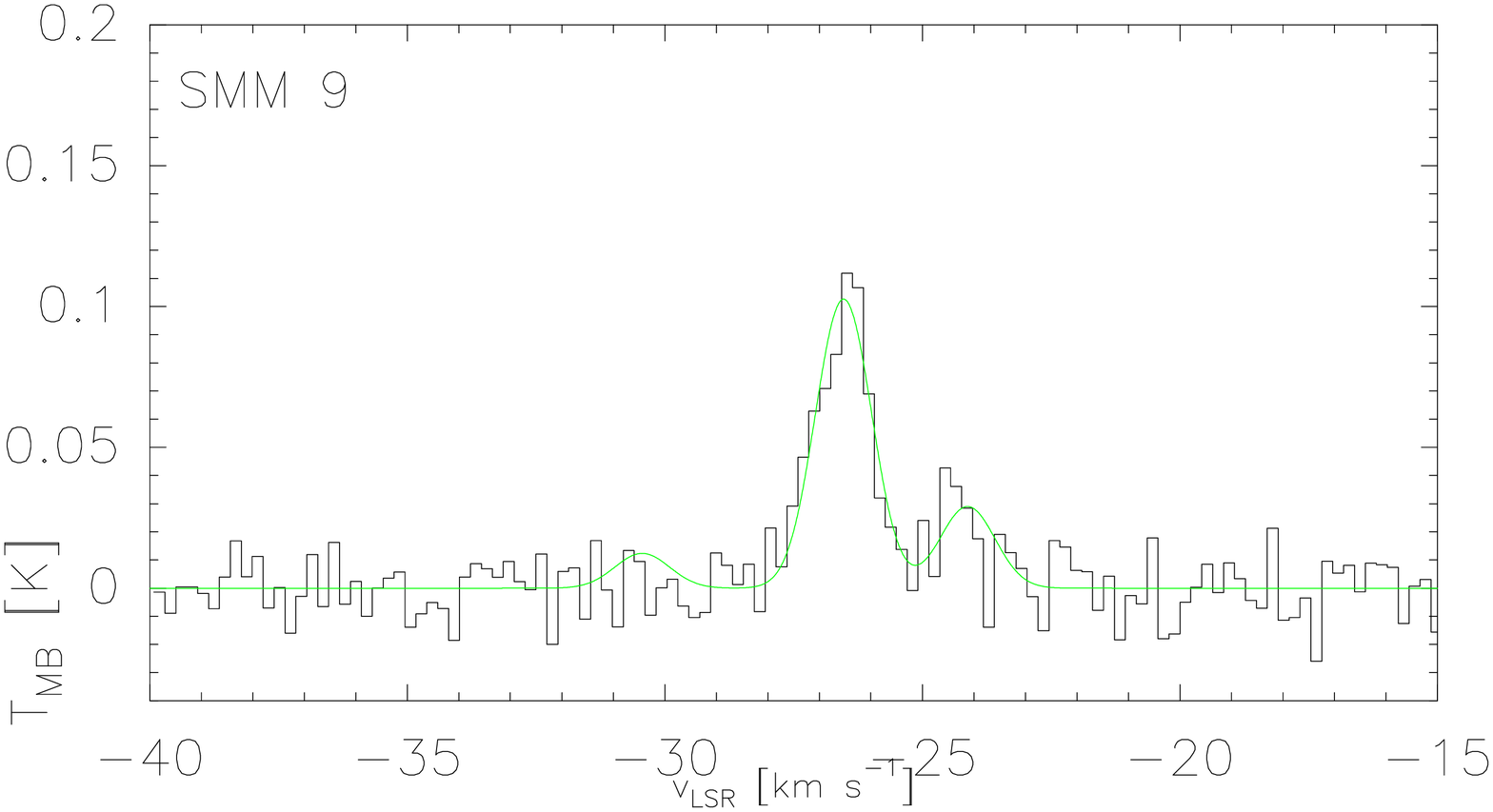}
\includegraphics[width=0.33\textwidth]{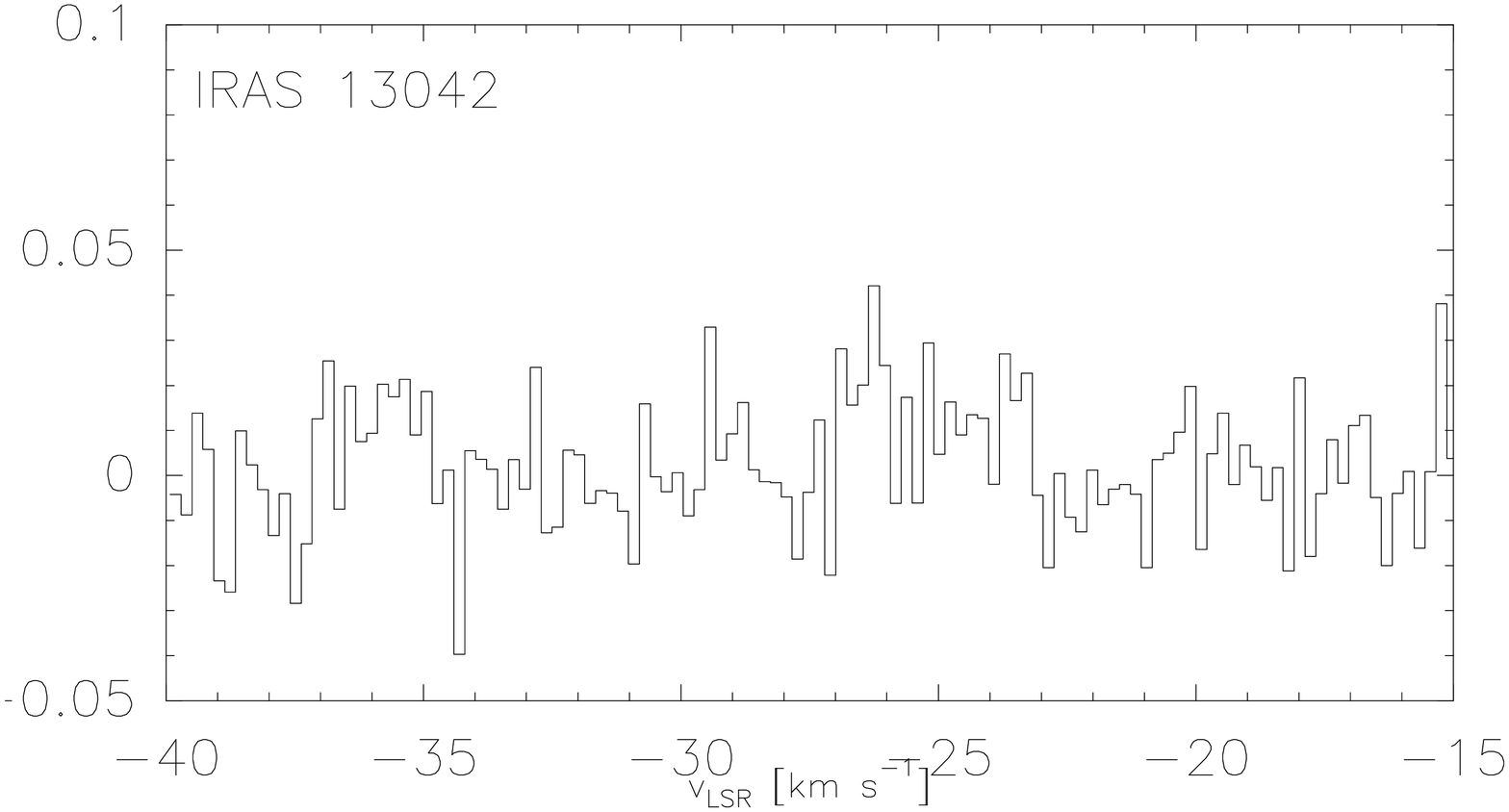}
\caption{H$^{13}$CN$(2-1)$ spectra towards the Seahorse IRDC clumps. Hyperfine structure fits to the lines are overlaid in green. While the velocity range shown in each panel is the same, the intensity range is different to better show the line profiles. The H$^{13}$CN line is seen in absoprtion towards SMM~5 (and a hint of it in SMM~6). The red, vertical dashed line in the SMM~6 panel shows 
the systemic velocity derived from C$^{17}$O$(J=2-1)$ by Miettinen (2012).}
\label{figure:h13cn}
\end{center}
\end{figure*}

\begin{figure*}[!htb]
\begin{center}
\includegraphics[width=0.33\textwidth]{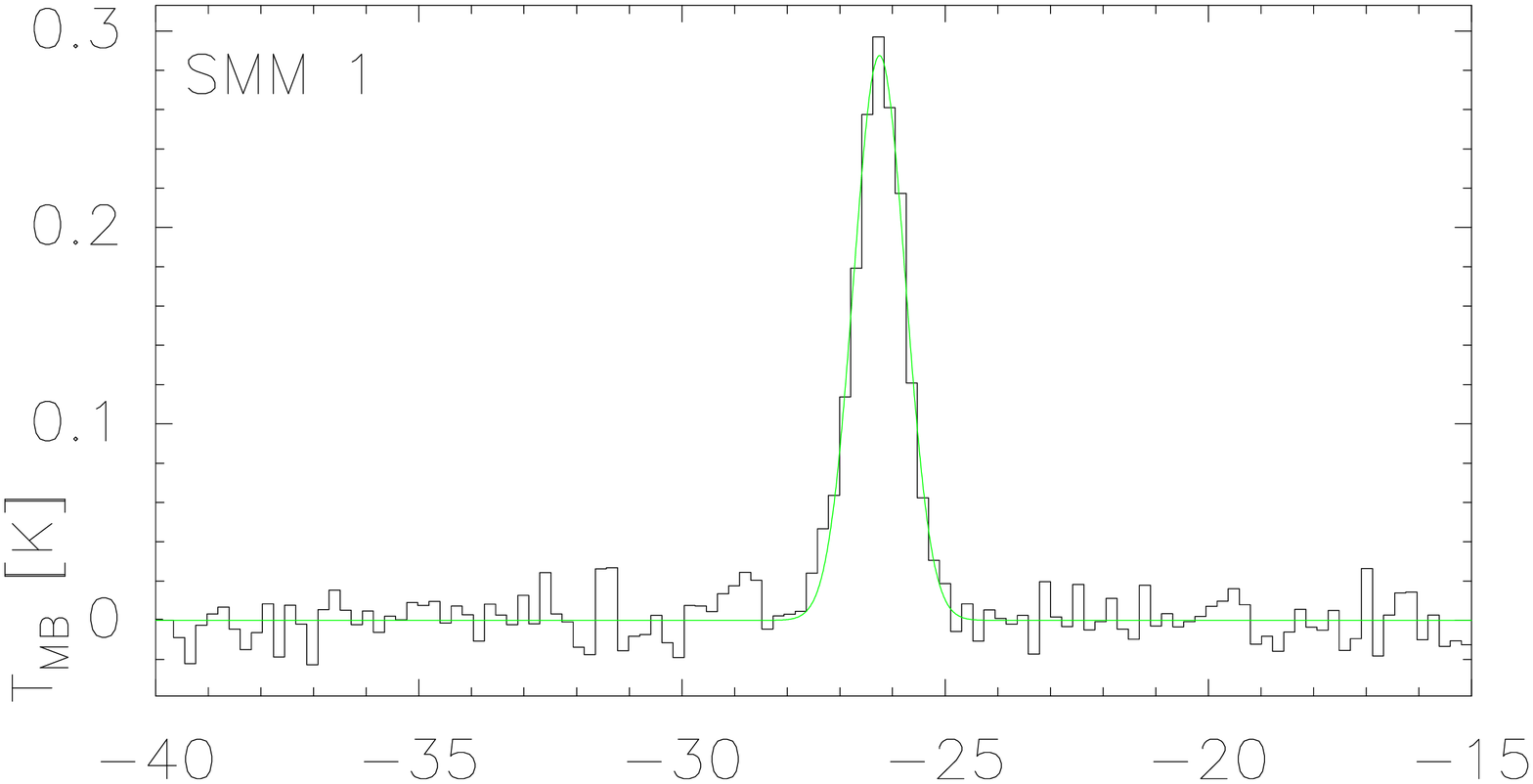}
\includegraphics[width=0.33\textwidth]{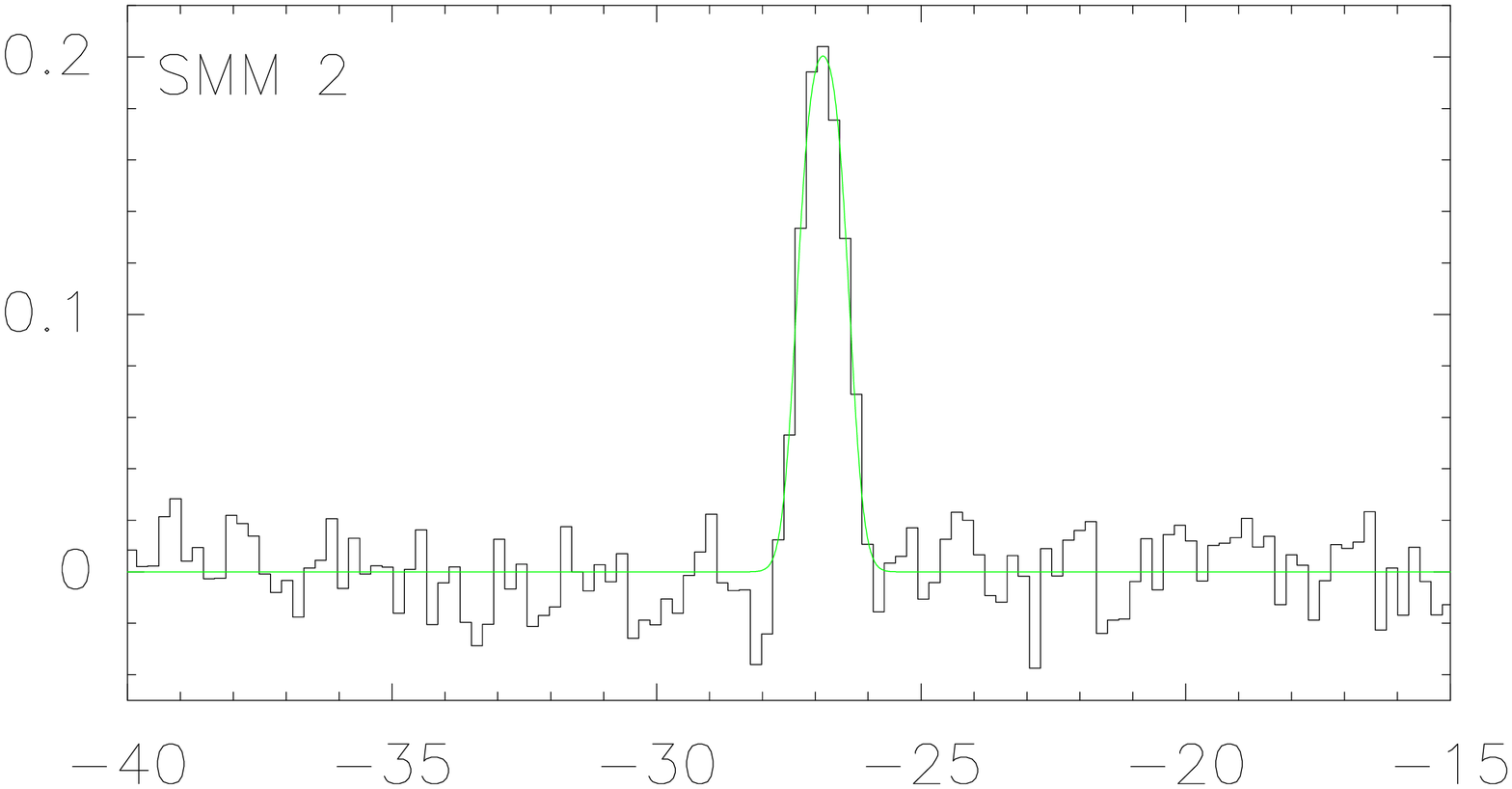}
\includegraphics[width=0.33\textwidth]{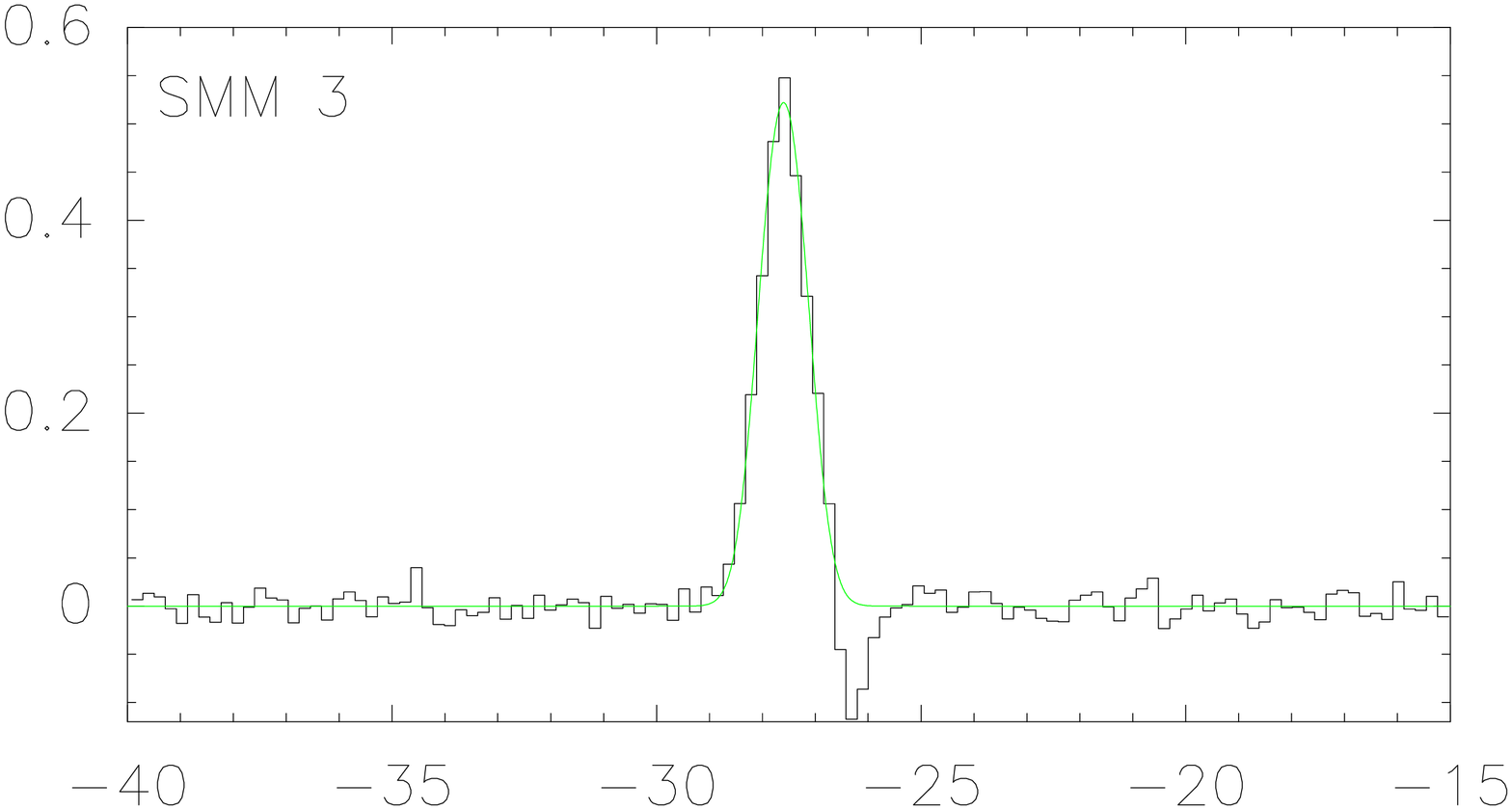}
\includegraphics[width=0.33\textwidth]{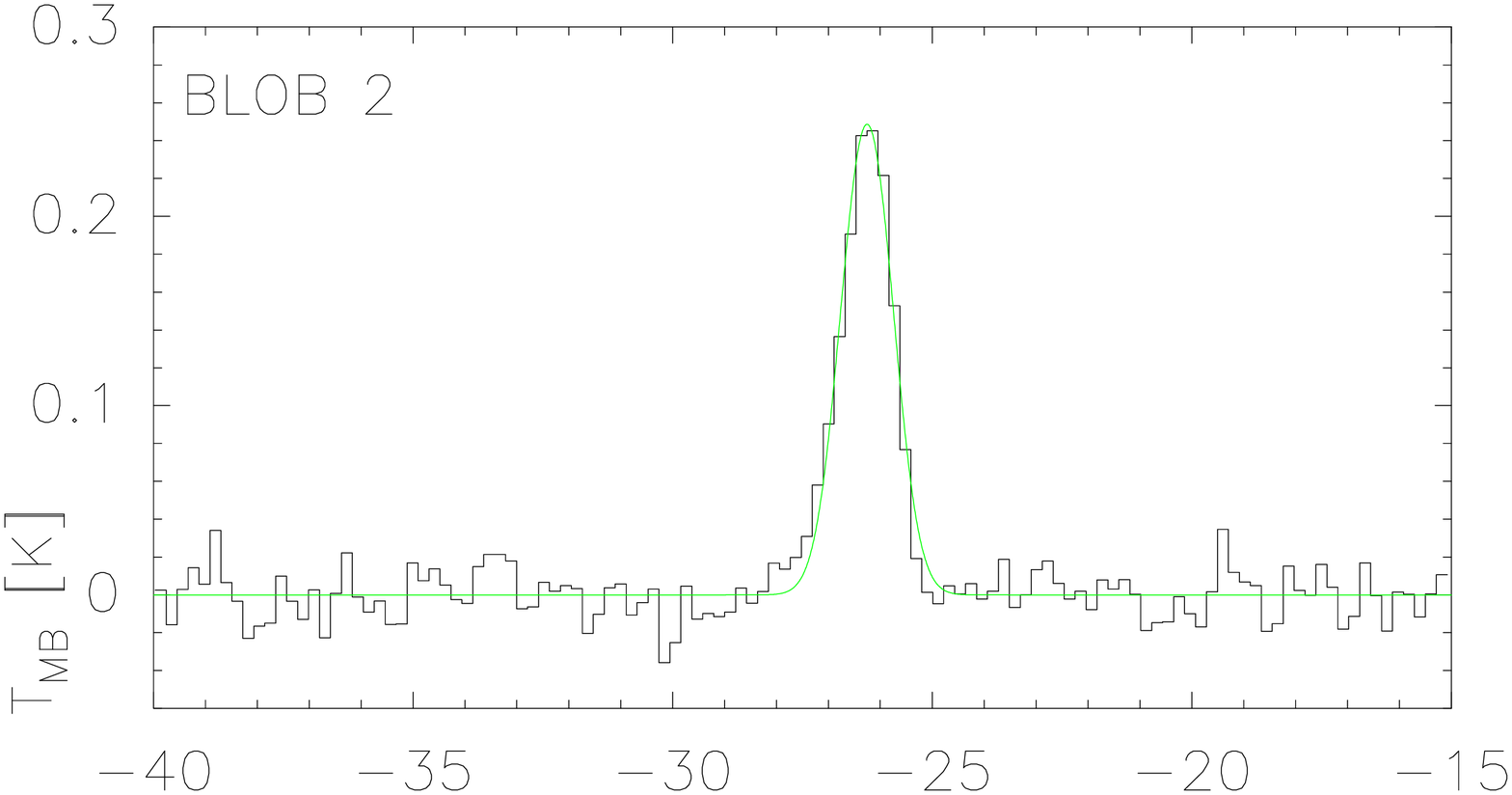}
\includegraphics[width=0.33\textwidth]{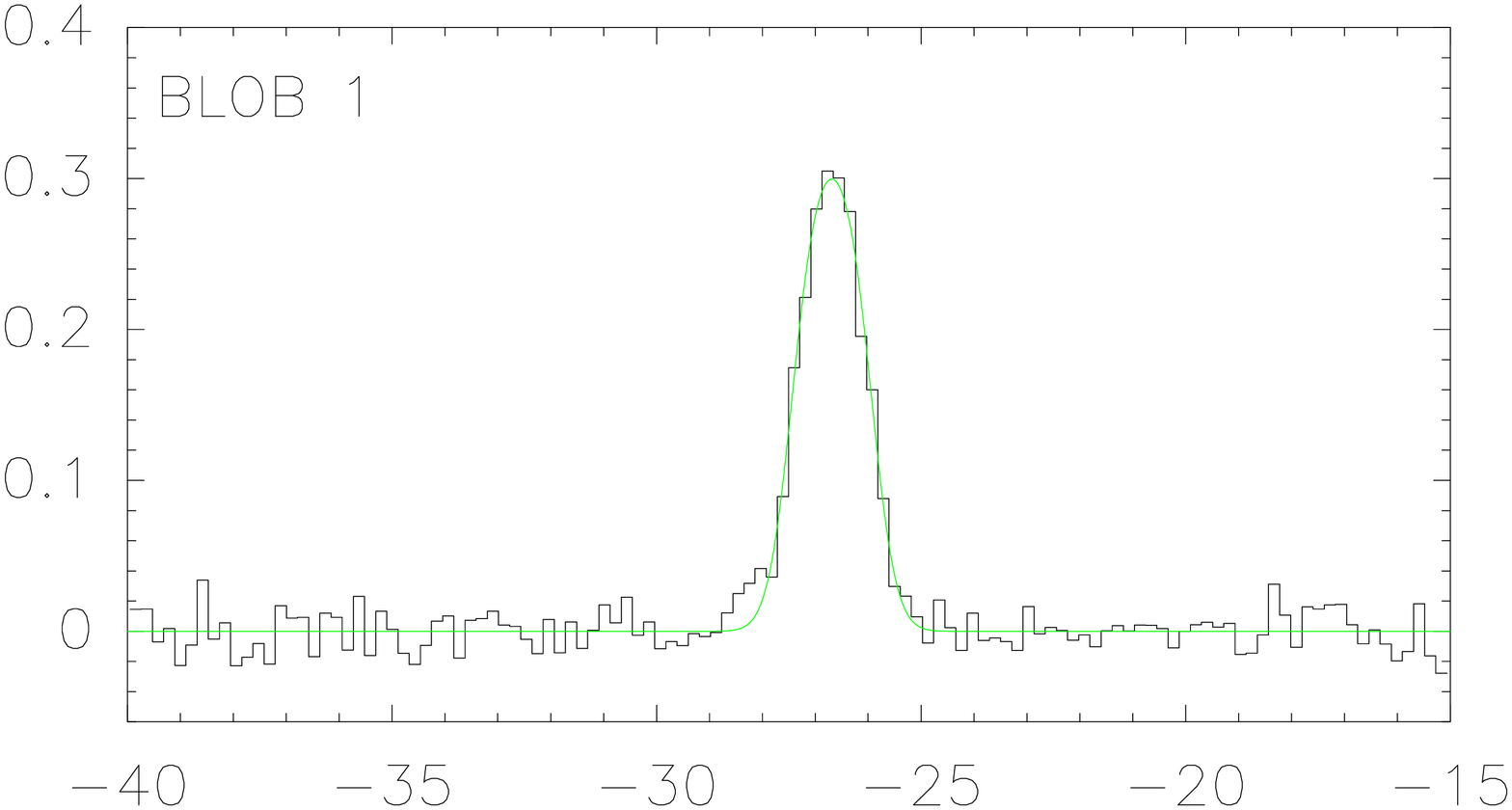}
\includegraphics[width=0.33\textwidth]{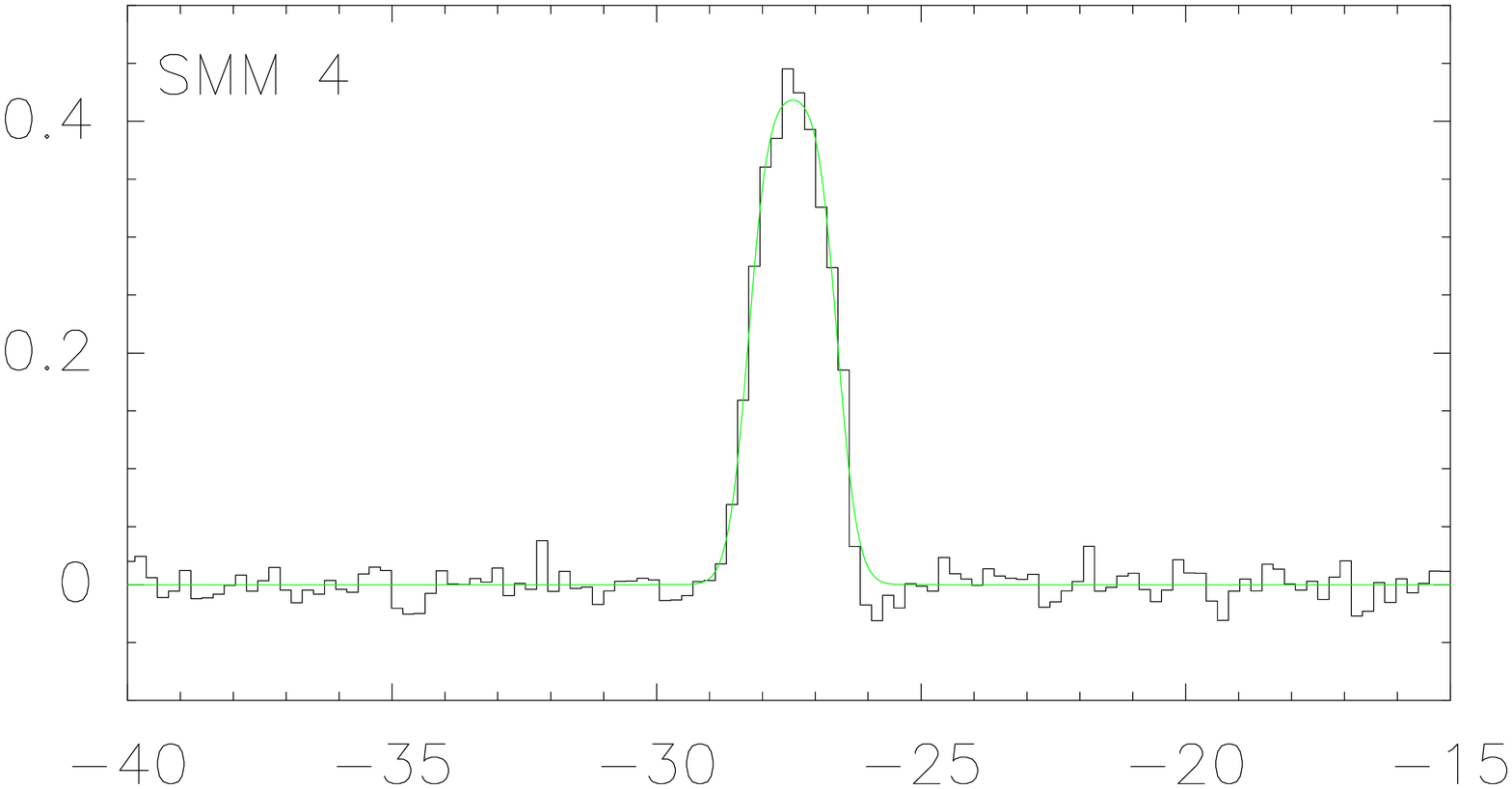}
\includegraphics[width=0.33\textwidth]{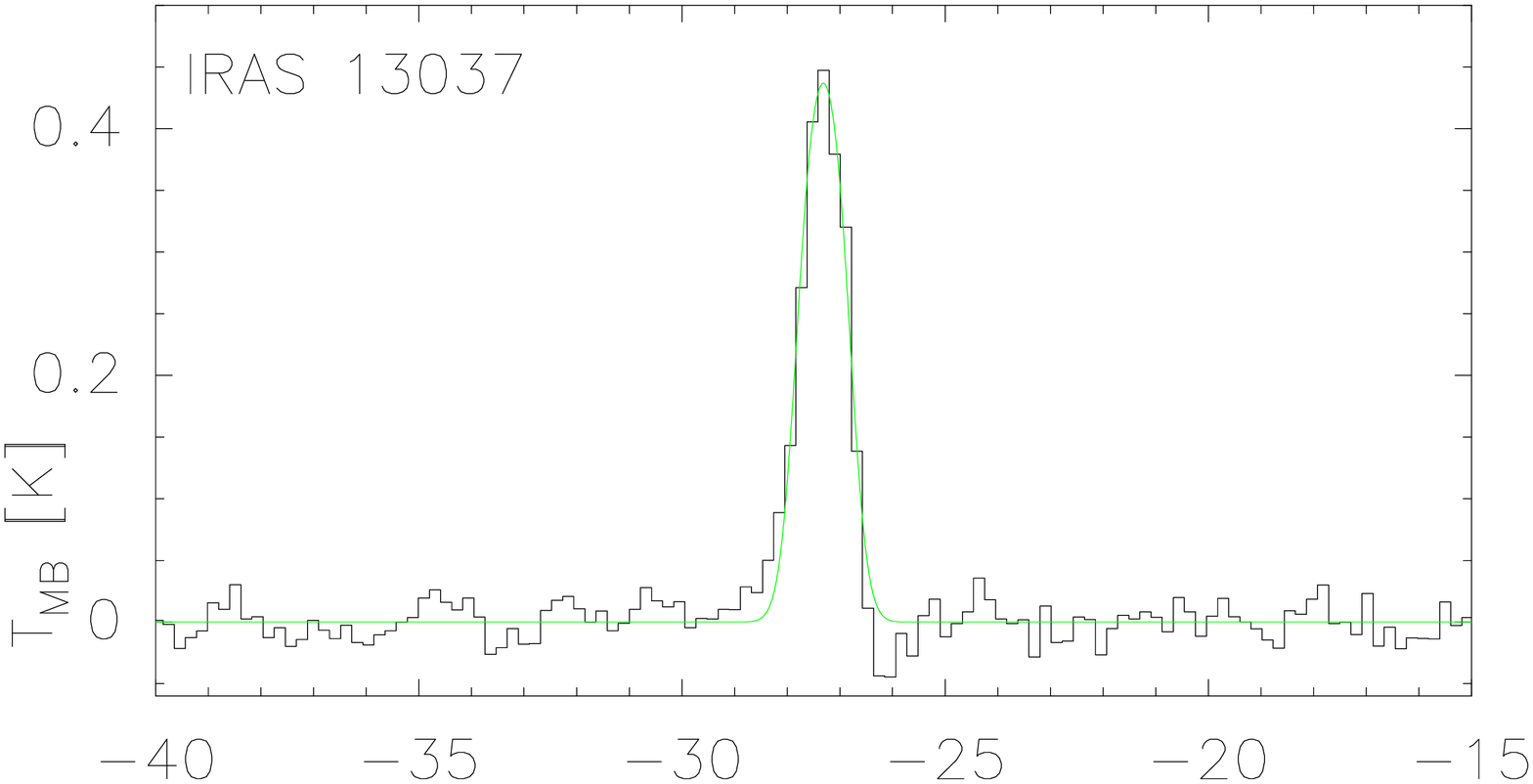}
\includegraphics[width=0.33\textwidth]{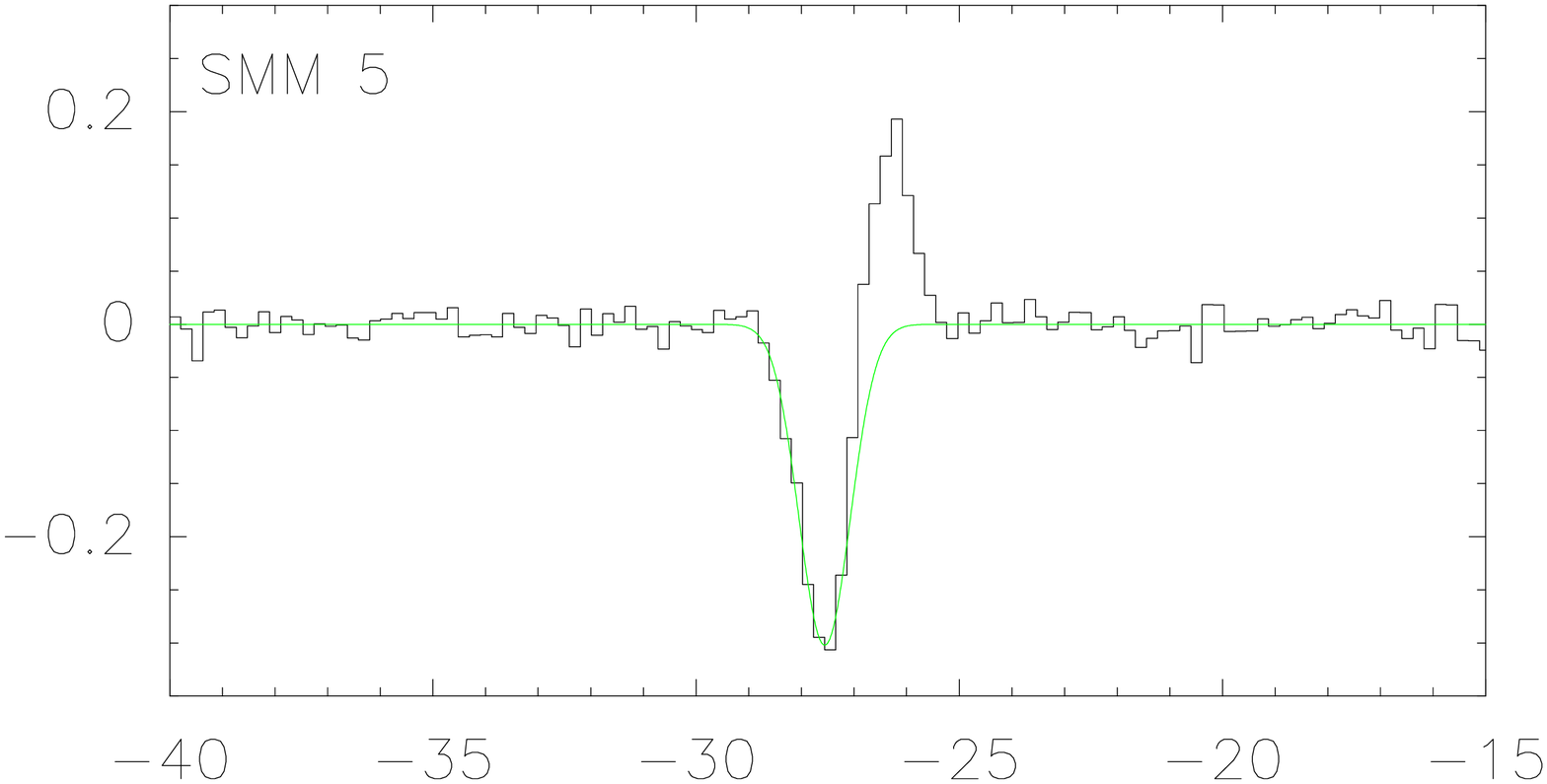}
\includegraphics[width=0.33\textwidth]{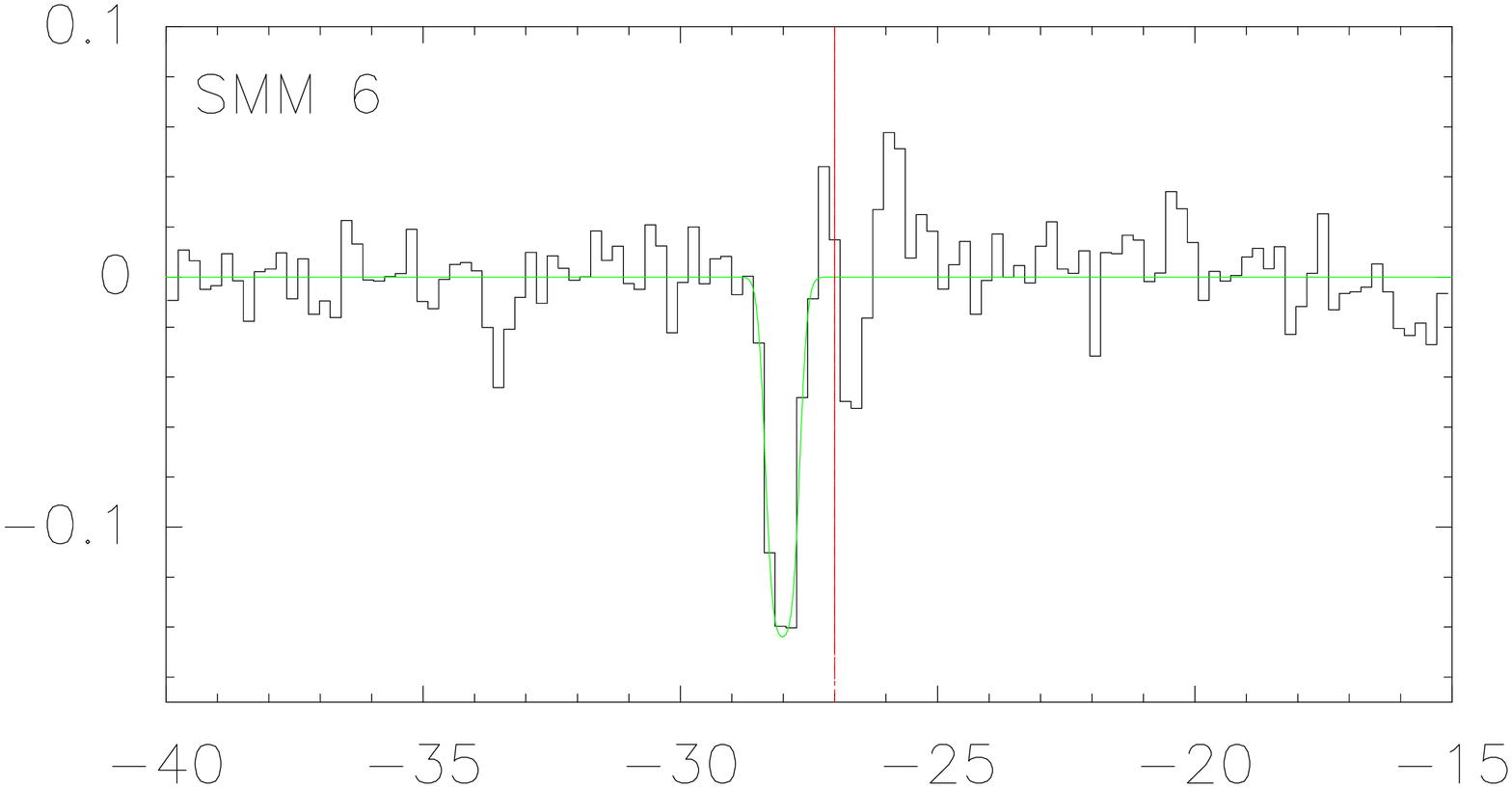}
\includegraphics[width=0.33\textwidth]{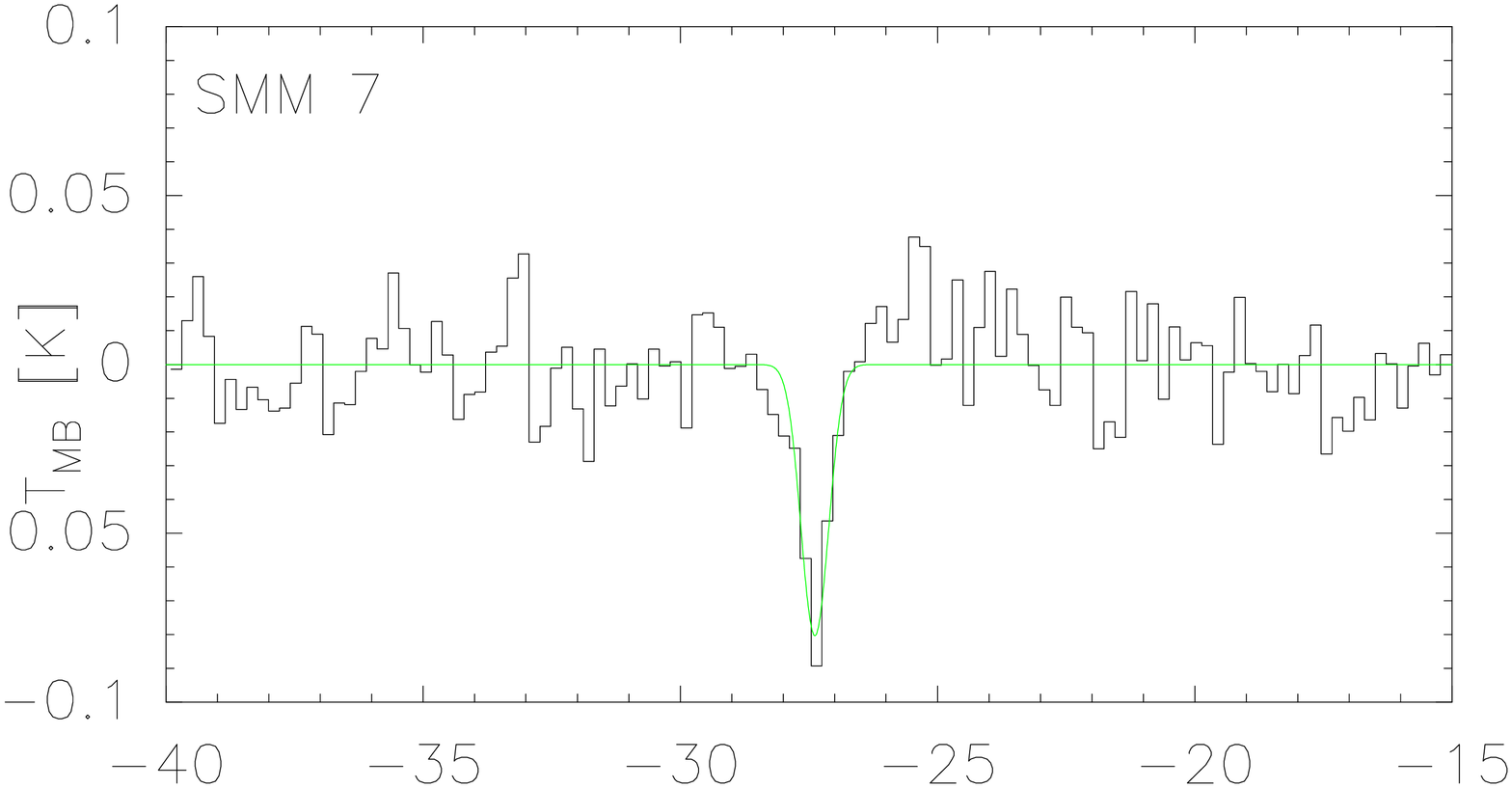}
\includegraphics[width=0.33\textwidth]{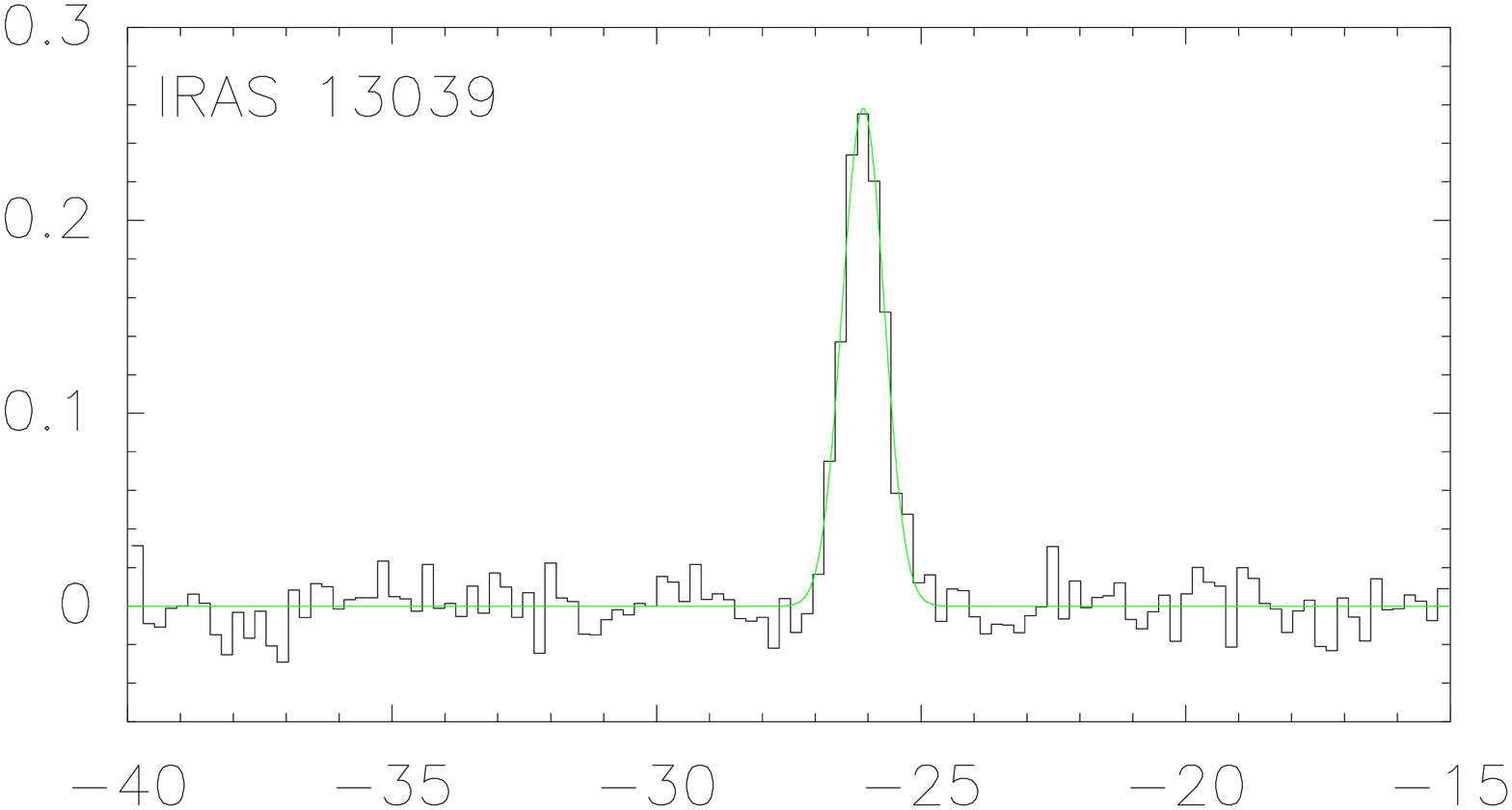}
\includegraphics[width=0.33\textwidth]{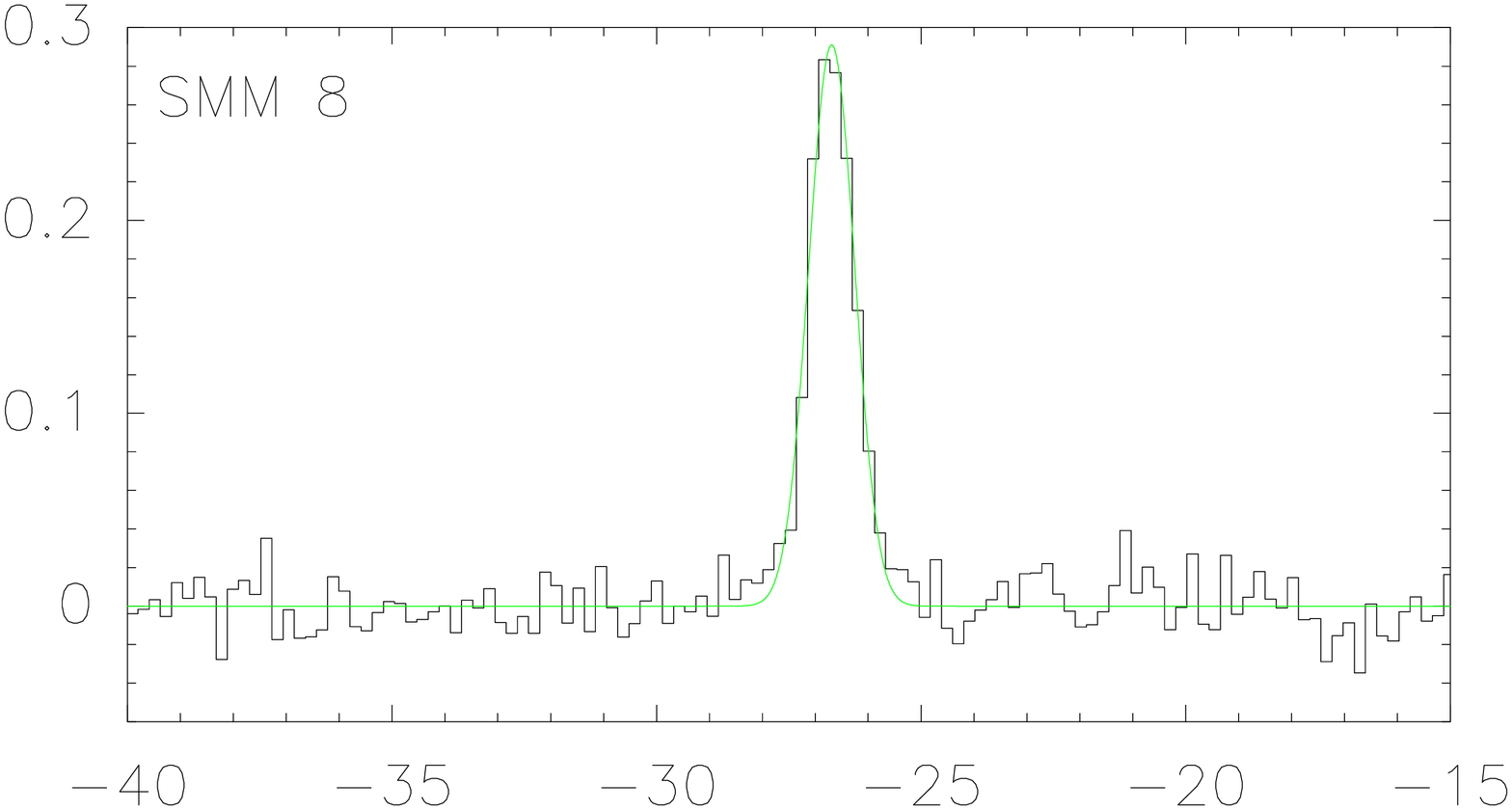}
\includegraphics[width=0.33\textwidth]{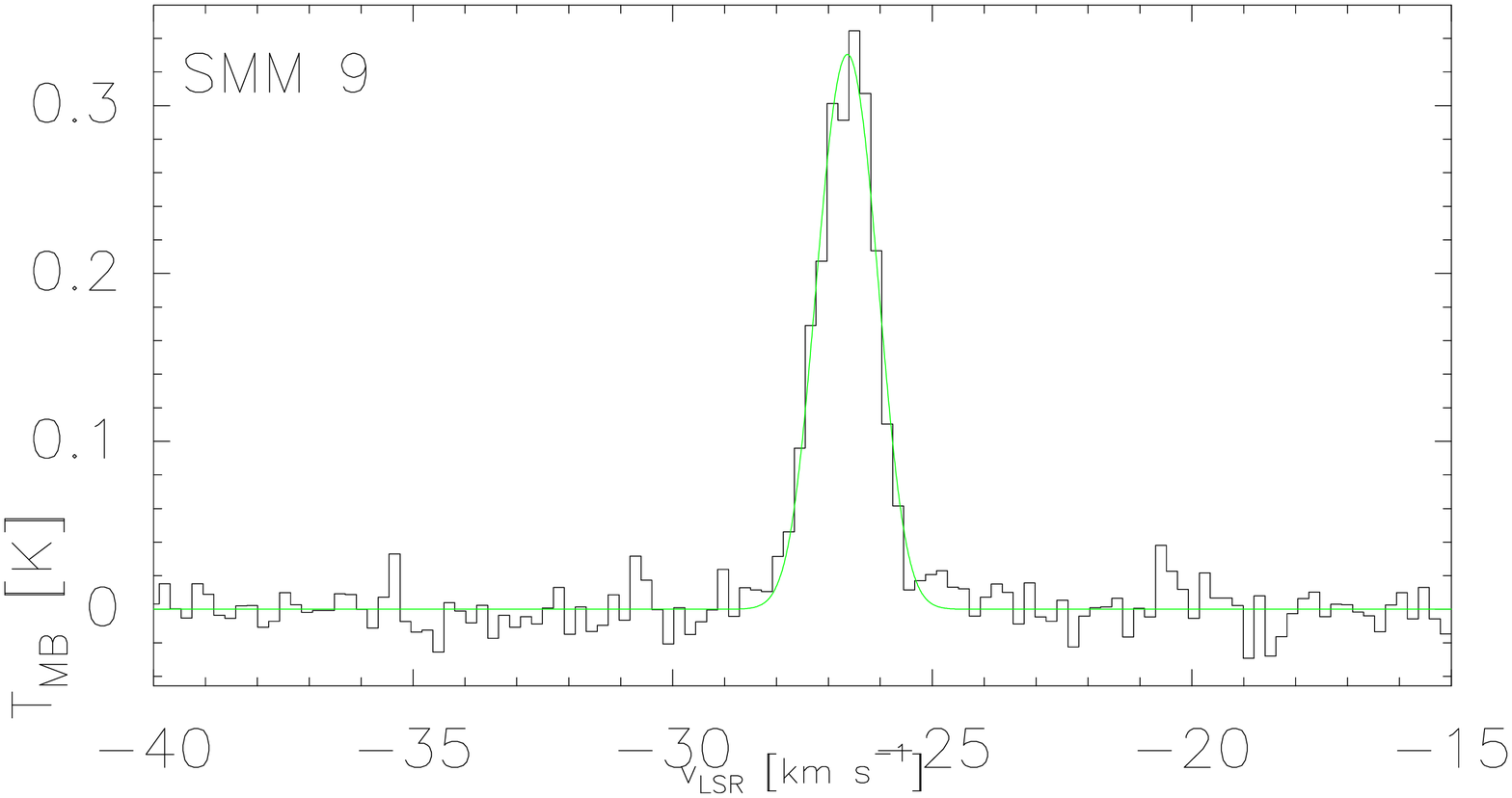}
\includegraphics[width=0.33\textwidth]{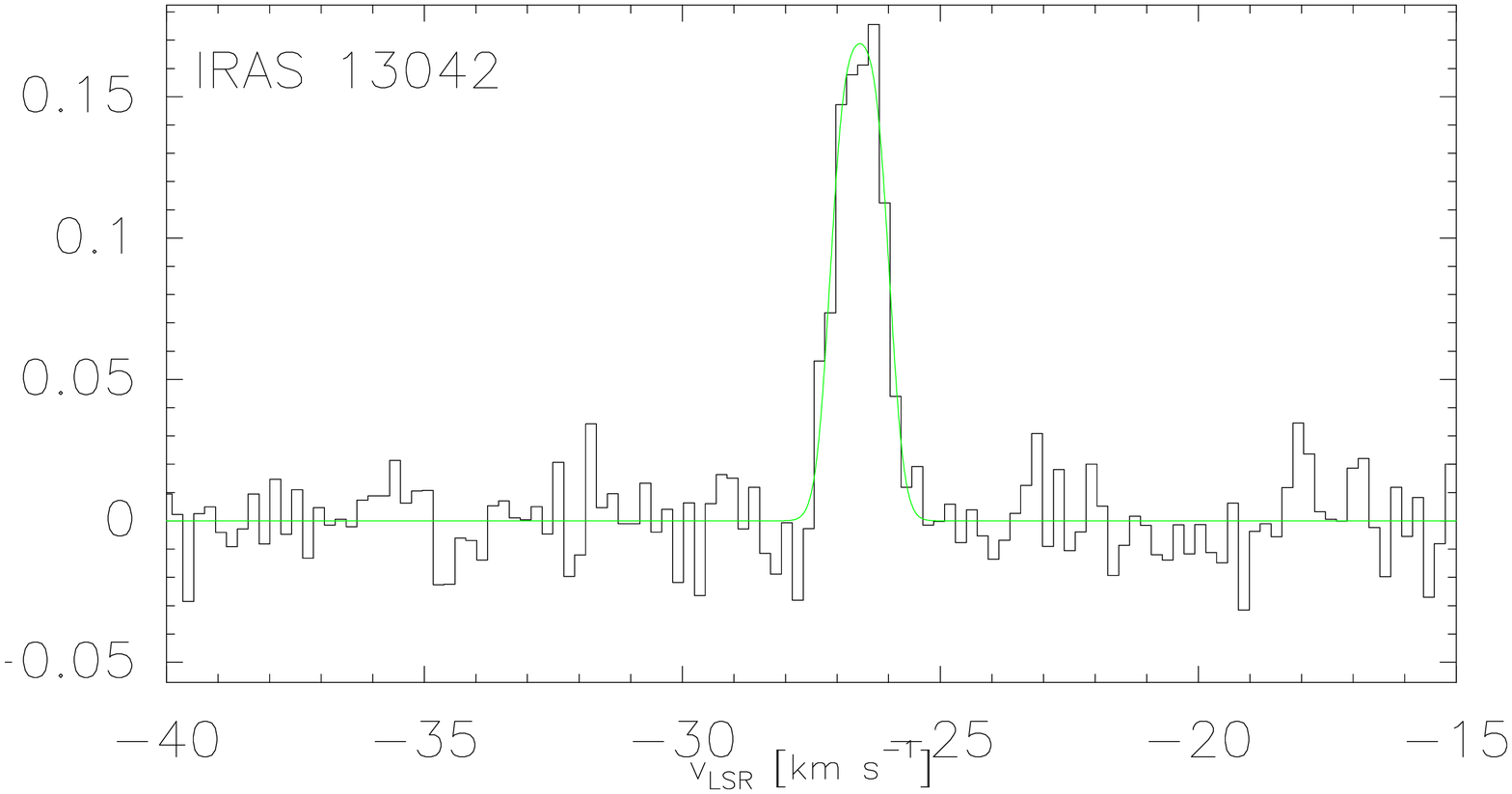}
\caption{H$^{13}$CO$^{+}(2-1)$ spectra towards the Seahorse IRDC clumps. Hyperfine structure fits to the lines are overlaid in green. While the velocity range shown in each panel is the same, the intensity range is different to better show the line profiles. The H$^{13}$CO$^+$ line is seen in absoprtion towards SMM~5--7. The red, vertical dashed line in the SMM~6 panel shows the systemic velocity derived from C$^{17}$O$(J=2-1)$ by Miettinen (2012).}
\label{figure:h13coplus}
\end{center}
\end{figure*}

\begin{figure*}[!htb]
\begin{center}
\includegraphics[width=0.33\textwidth]{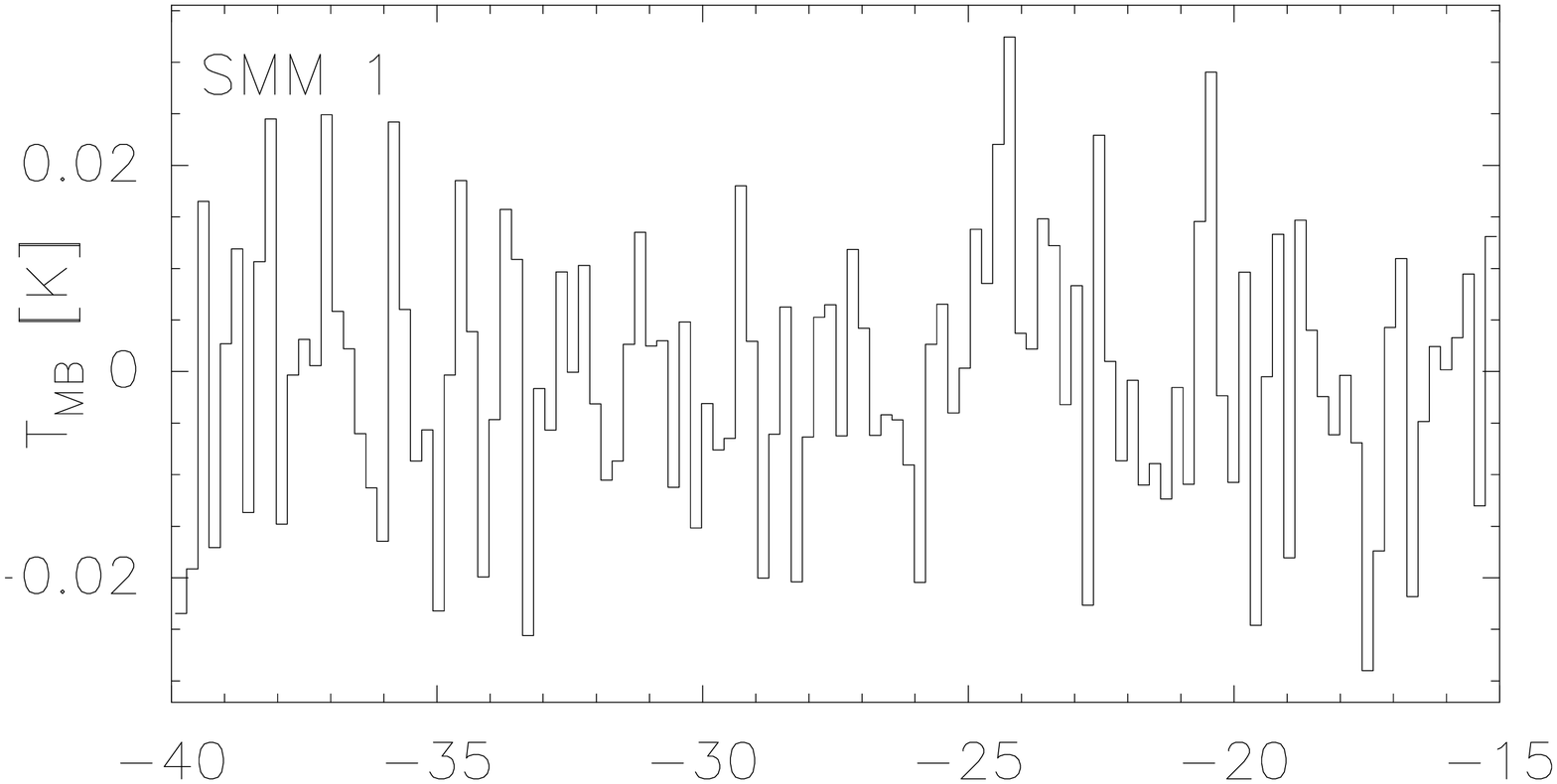}
\includegraphics[width=0.33\textwidth]{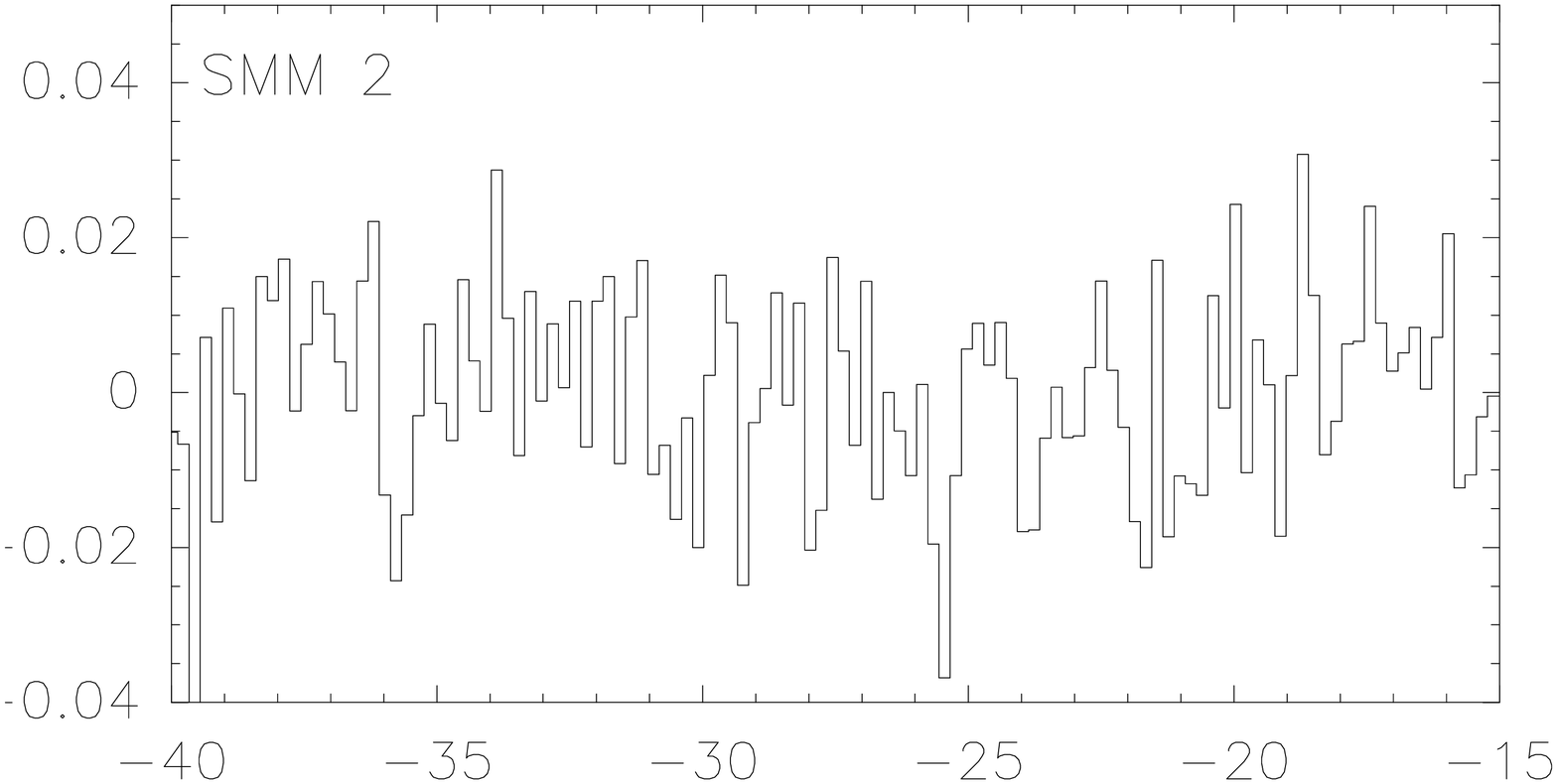}
\includegraphics[width=0.33\textwidth]{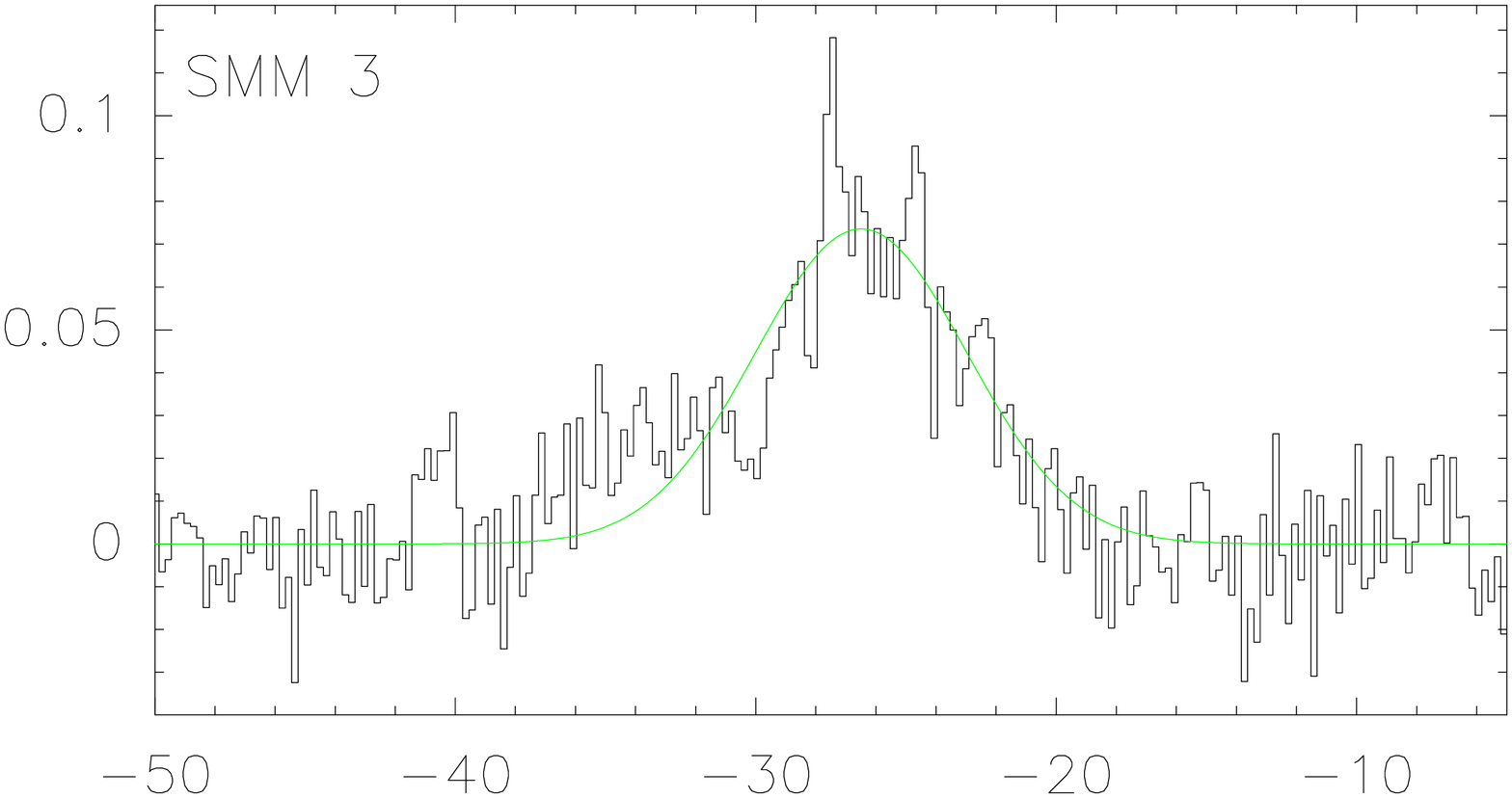}
\includegraphics[width=0.33\textwidth]{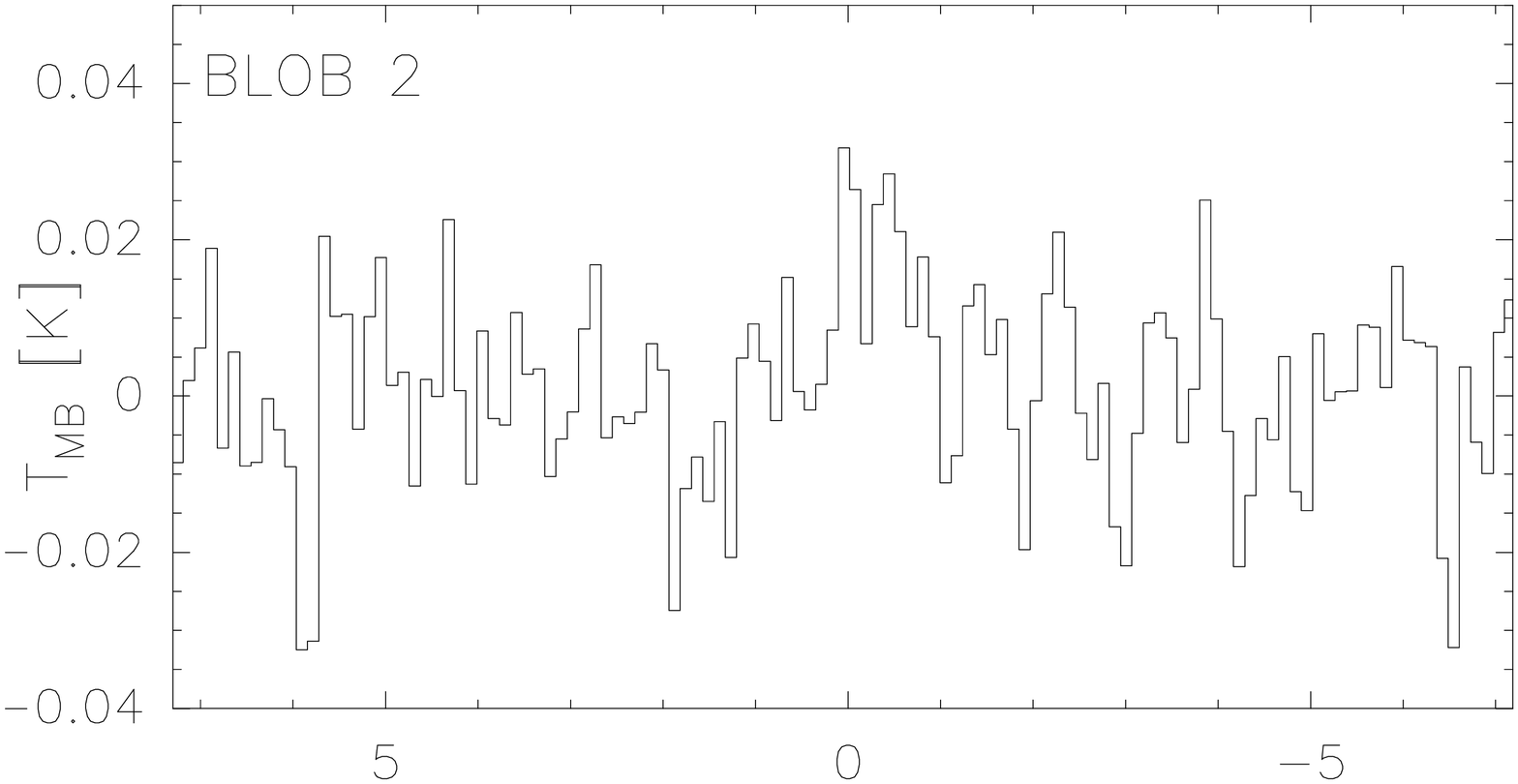}
\includegraphics[width=0.33\textwidth]{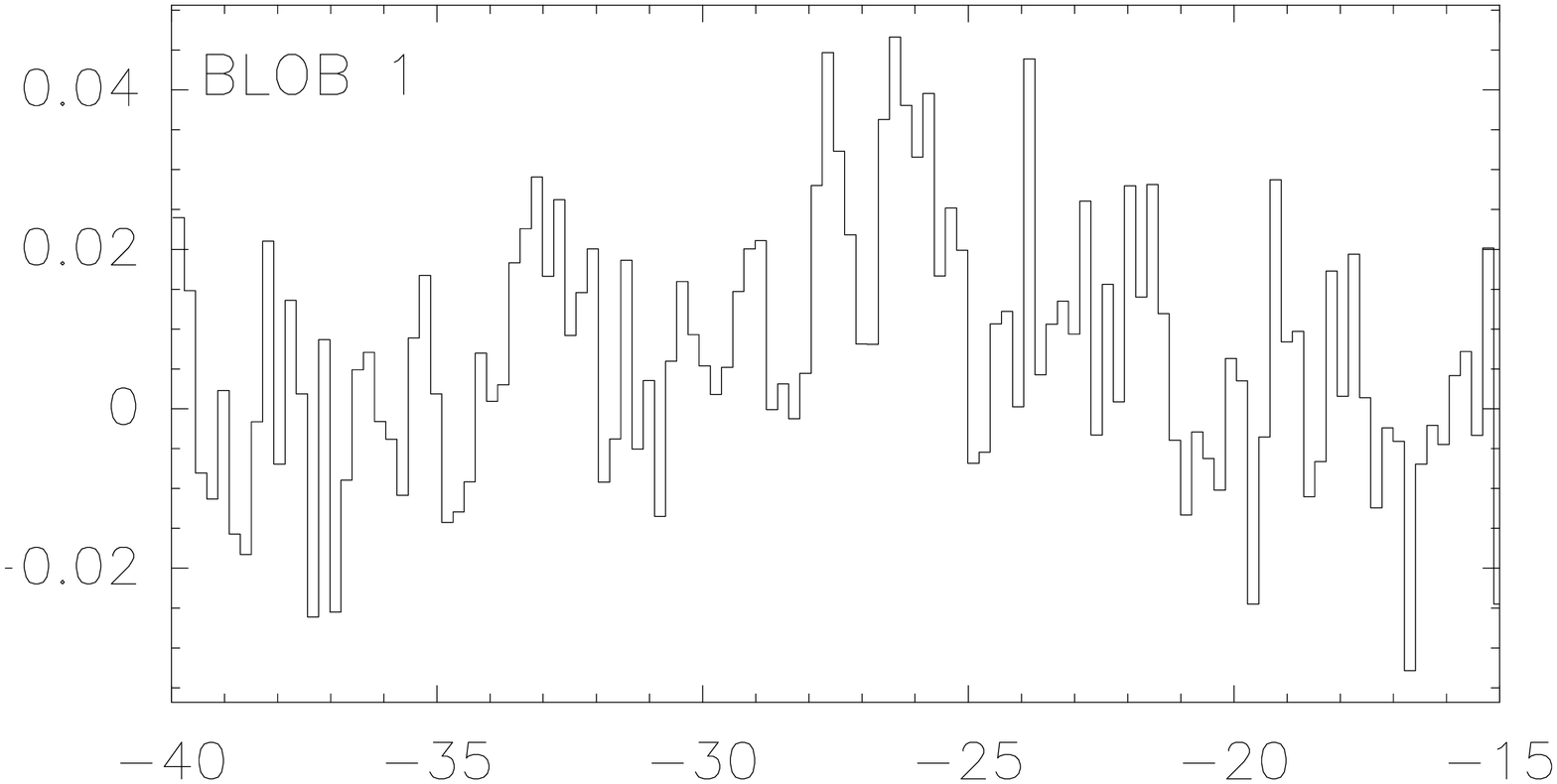}
\includegraphics[width=0.33\textwidth]{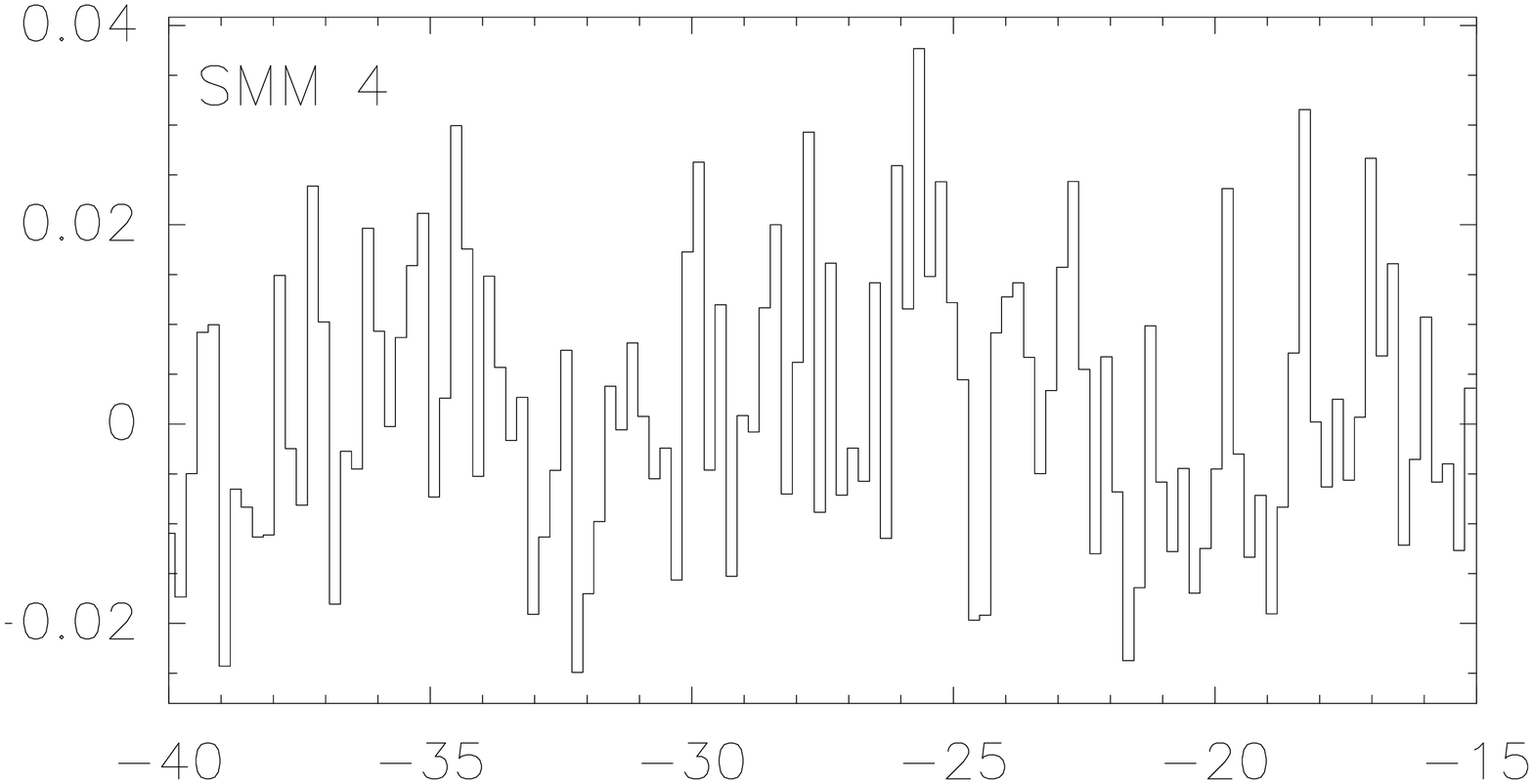}
\includegraphics[width=0.33\textwidth]{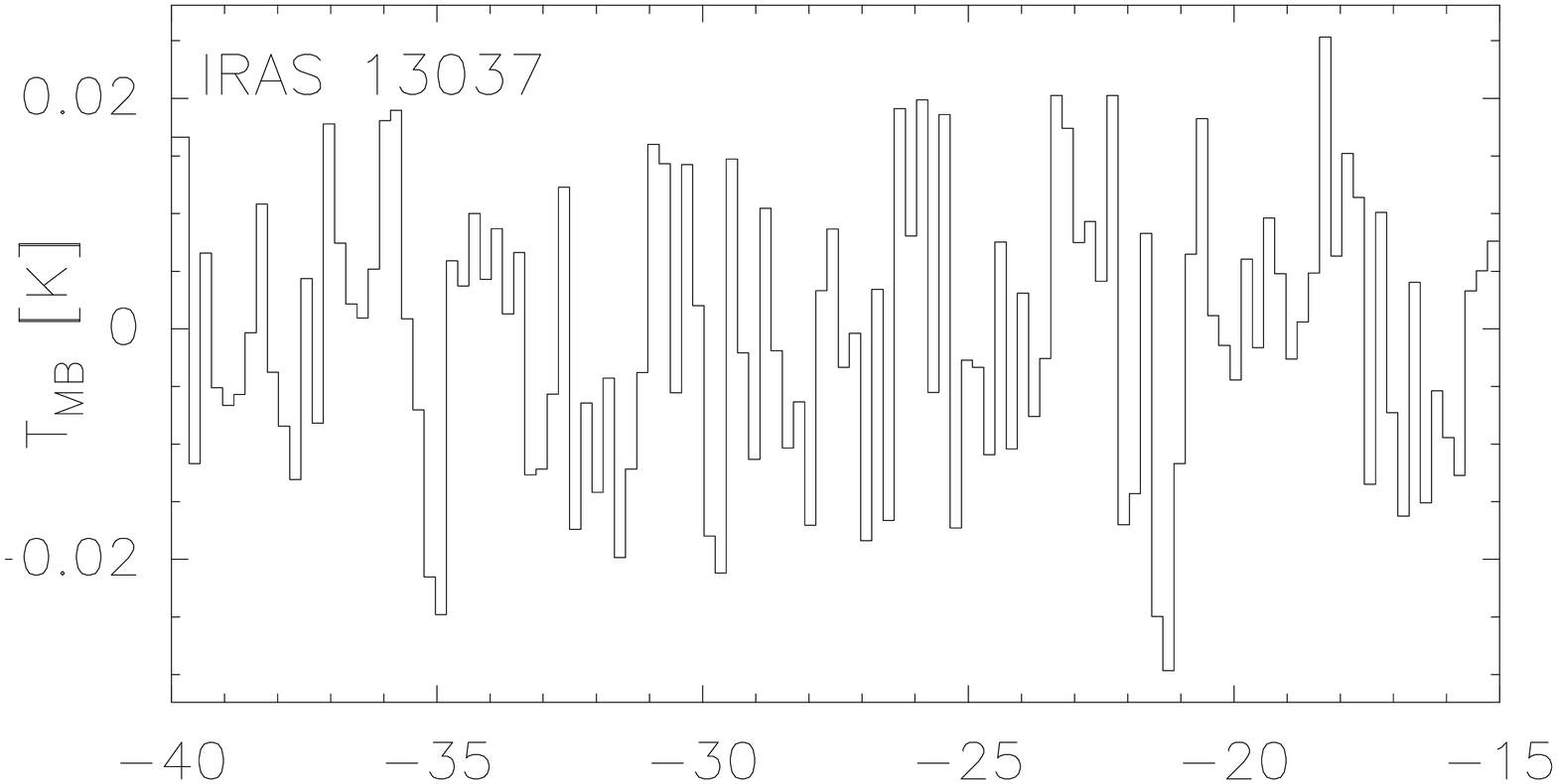}
\includegraphics[width=0.33\textwidth]{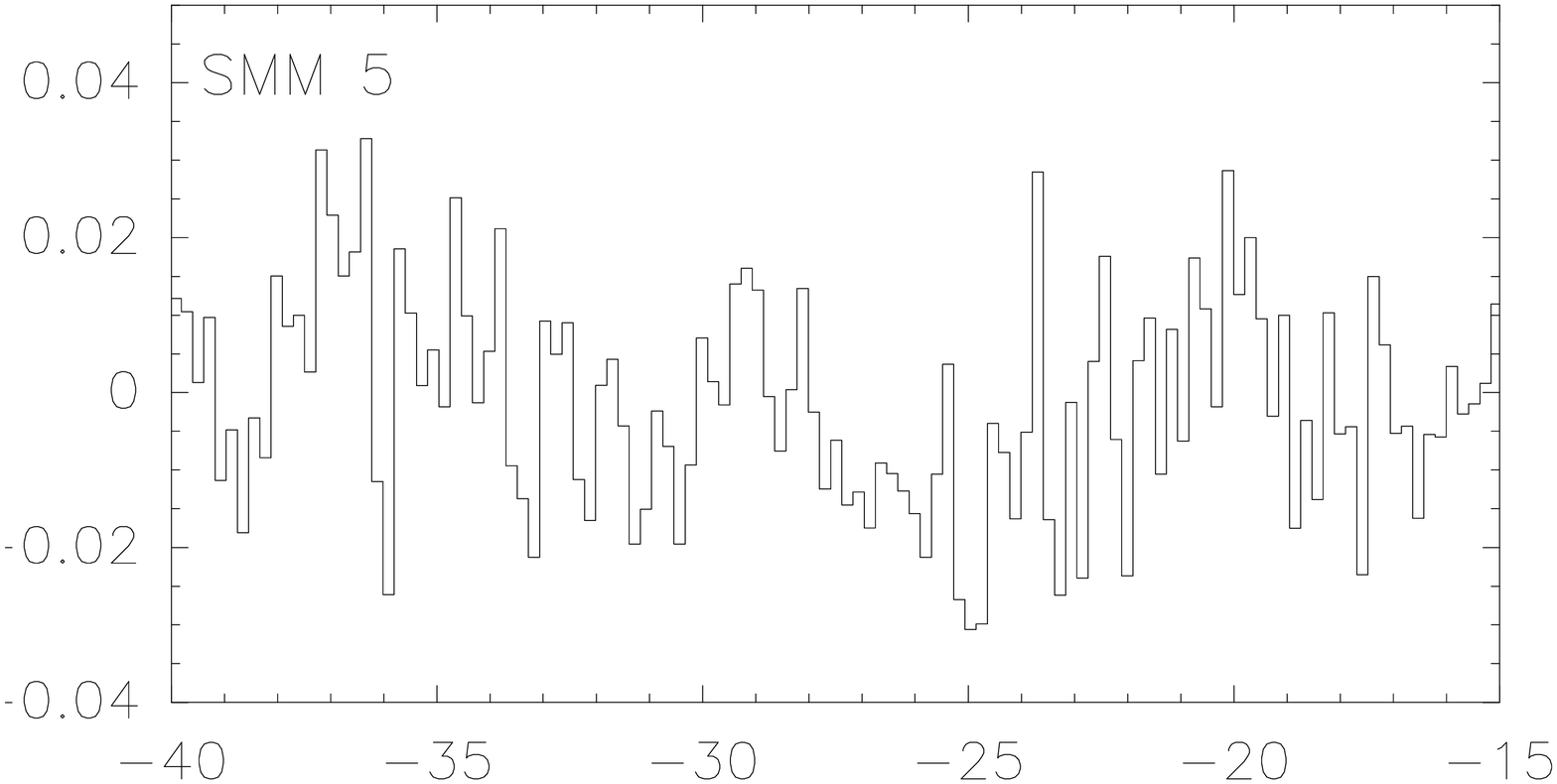}
\includegraphics[width=0.33\textwidth]{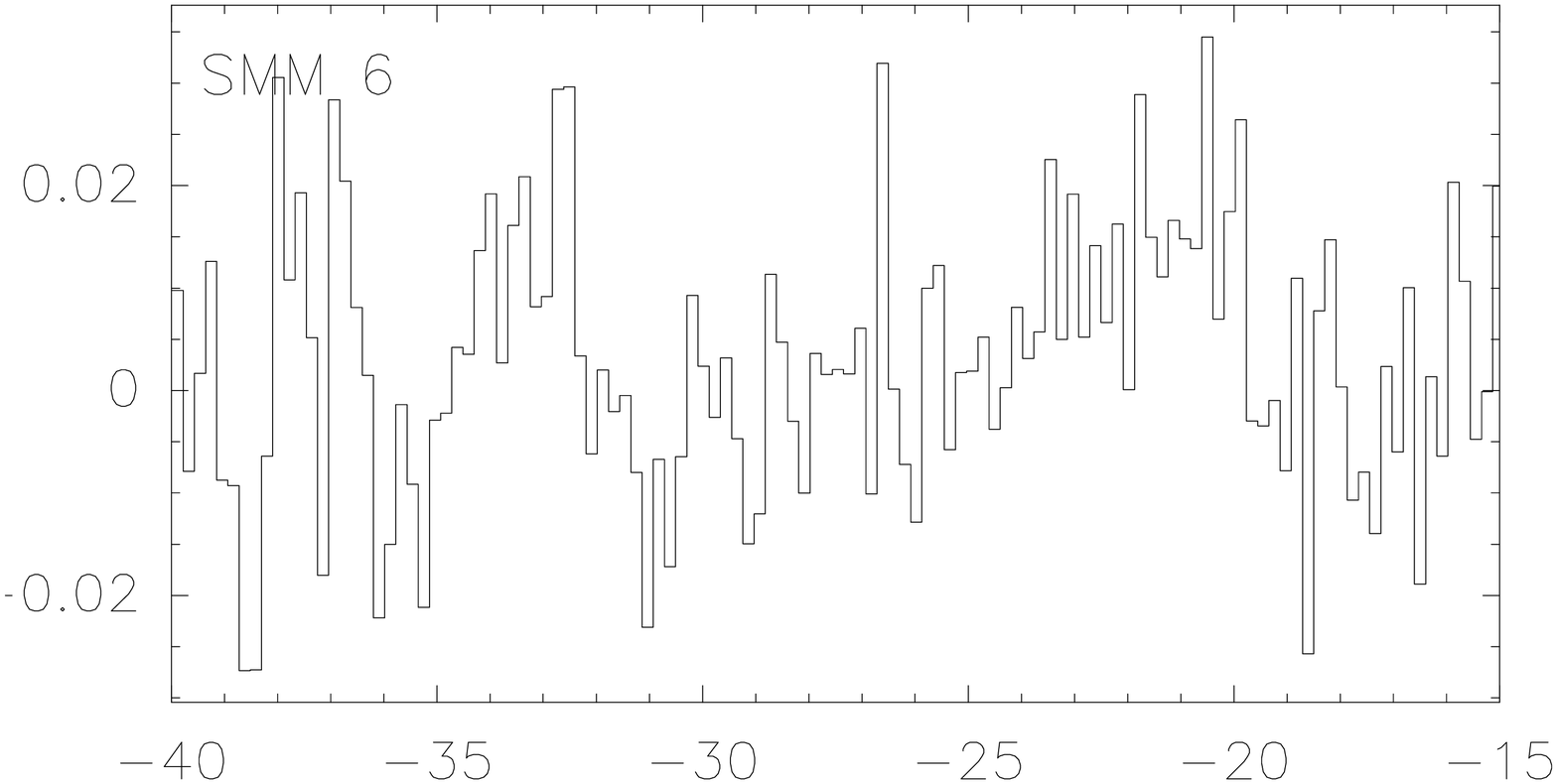}
\includegraphics[width=0.33\textwidth]{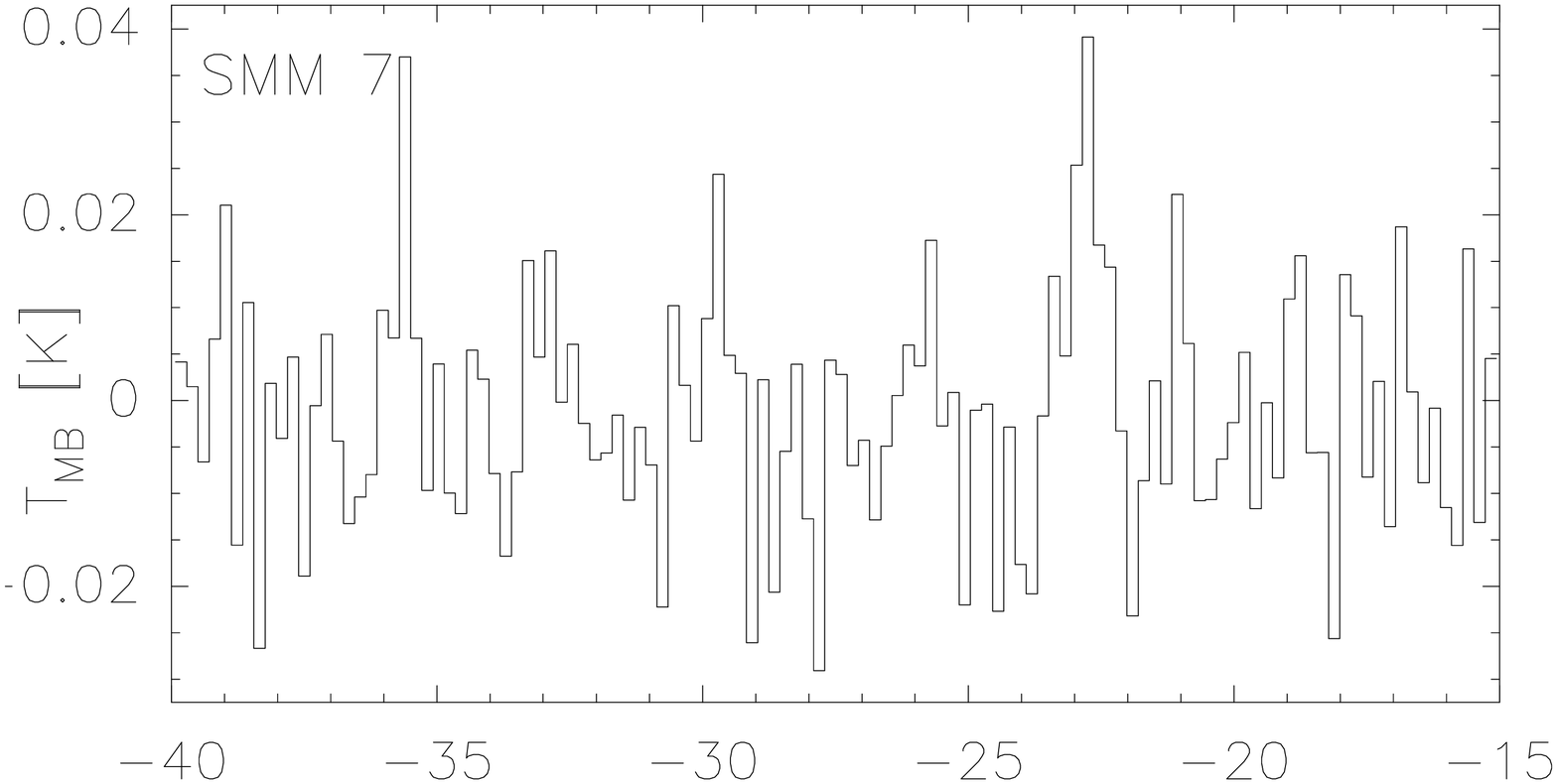}
\includegraphics[width=0.33\textwidth]{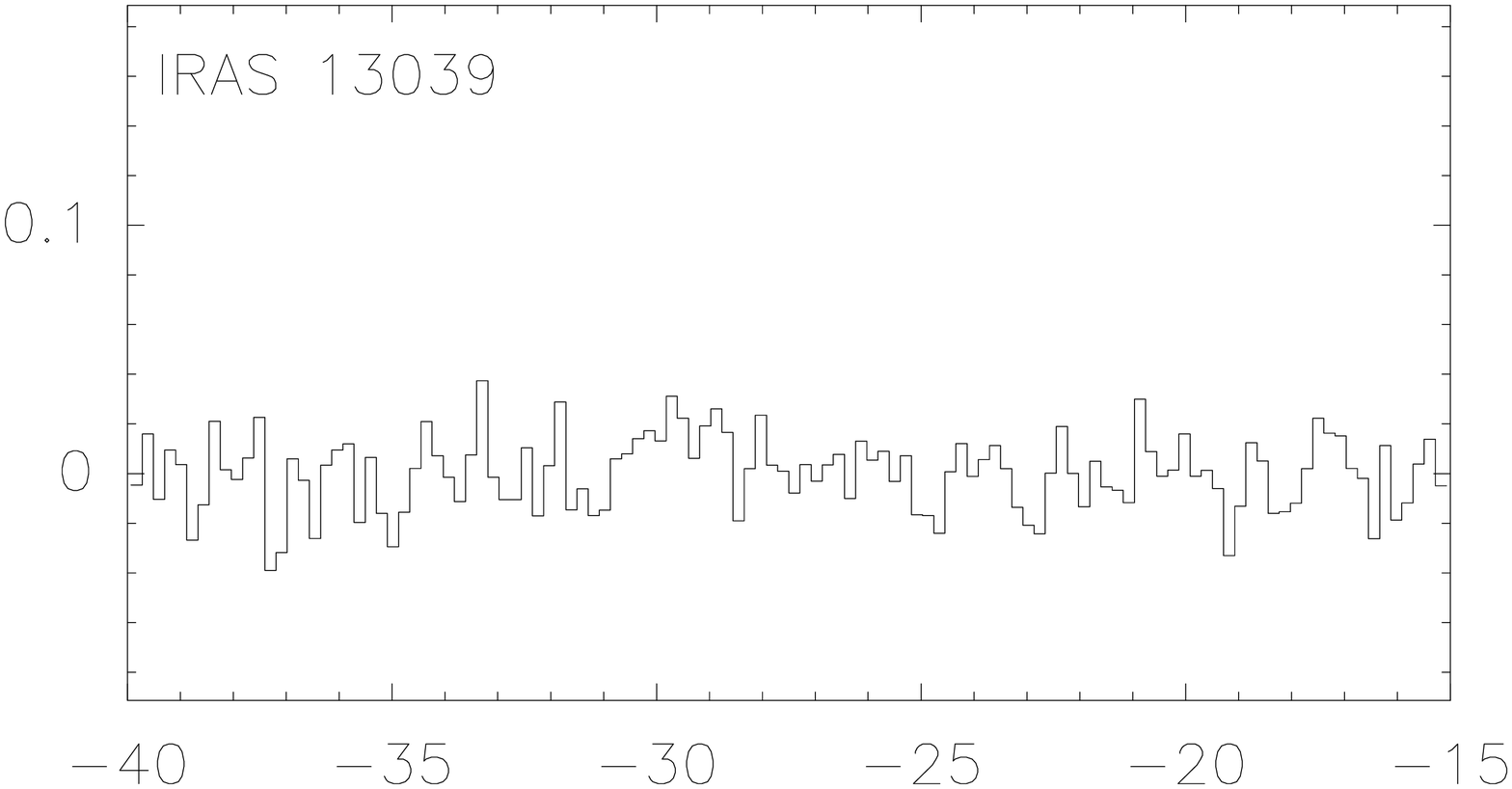}
\includegraphics[width=0.33\textwidth]{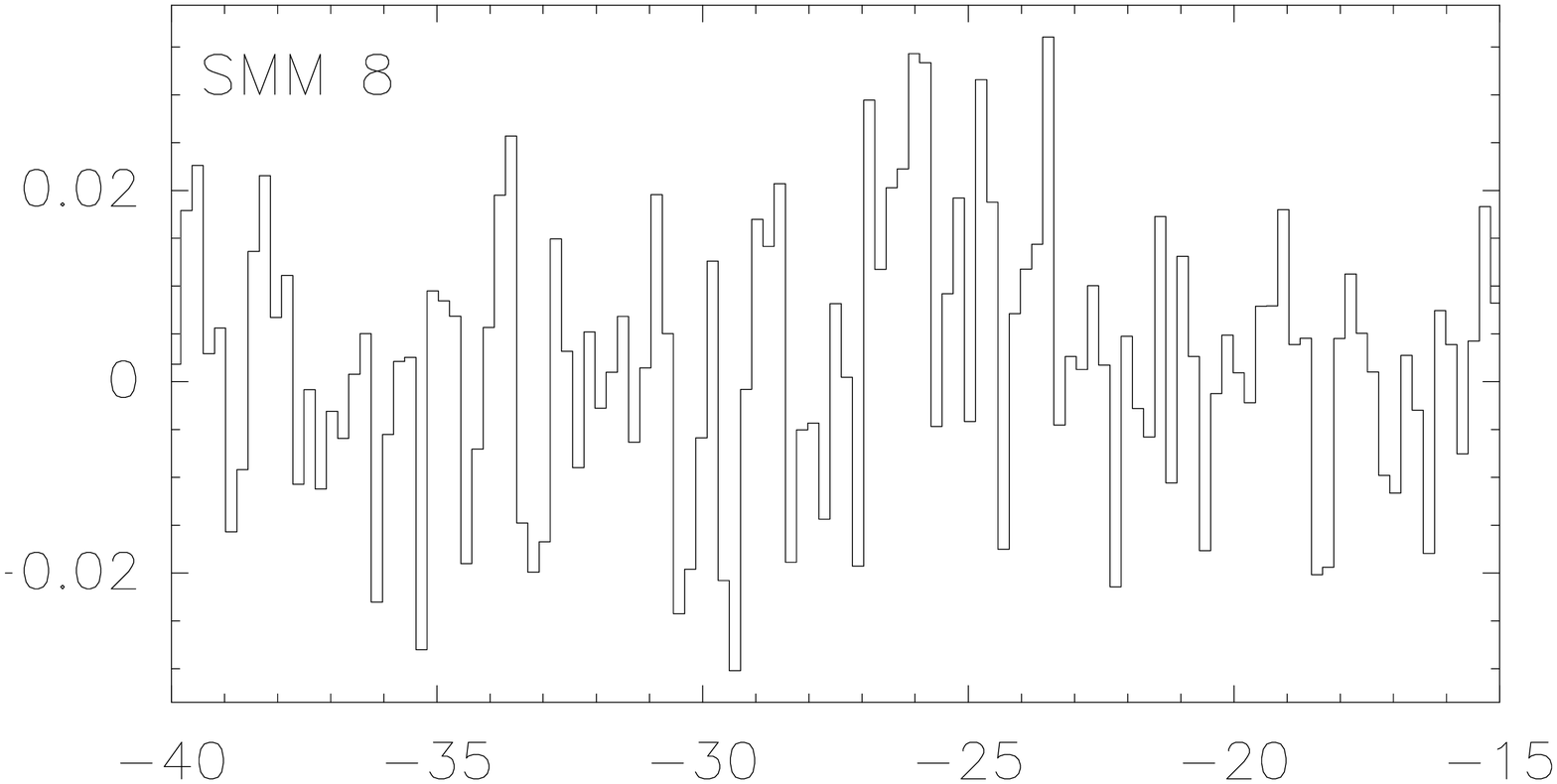}
\includegraphics[width=0.33\textwidth]{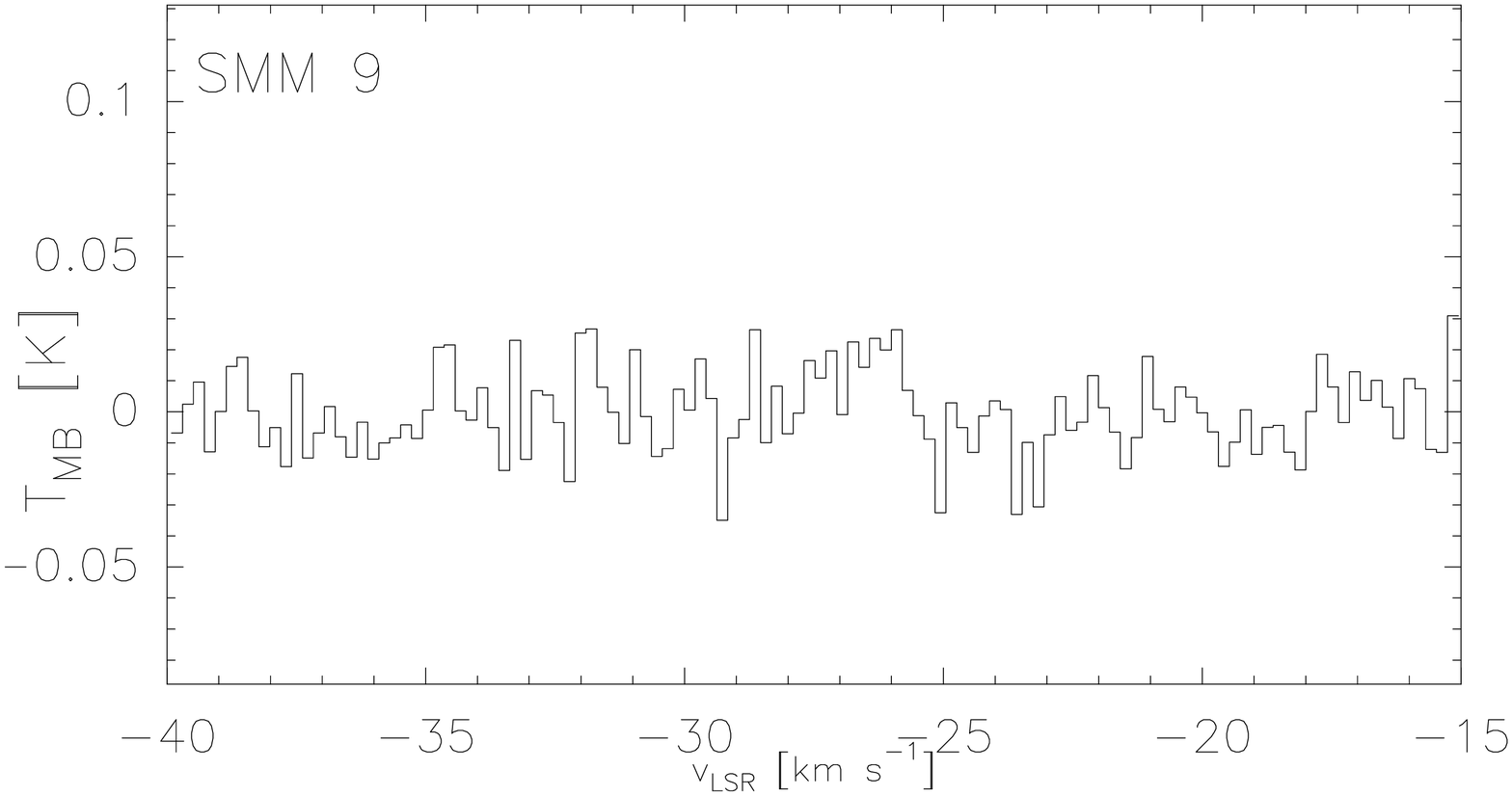}
\includegraphics[width=0.33\textwidth]{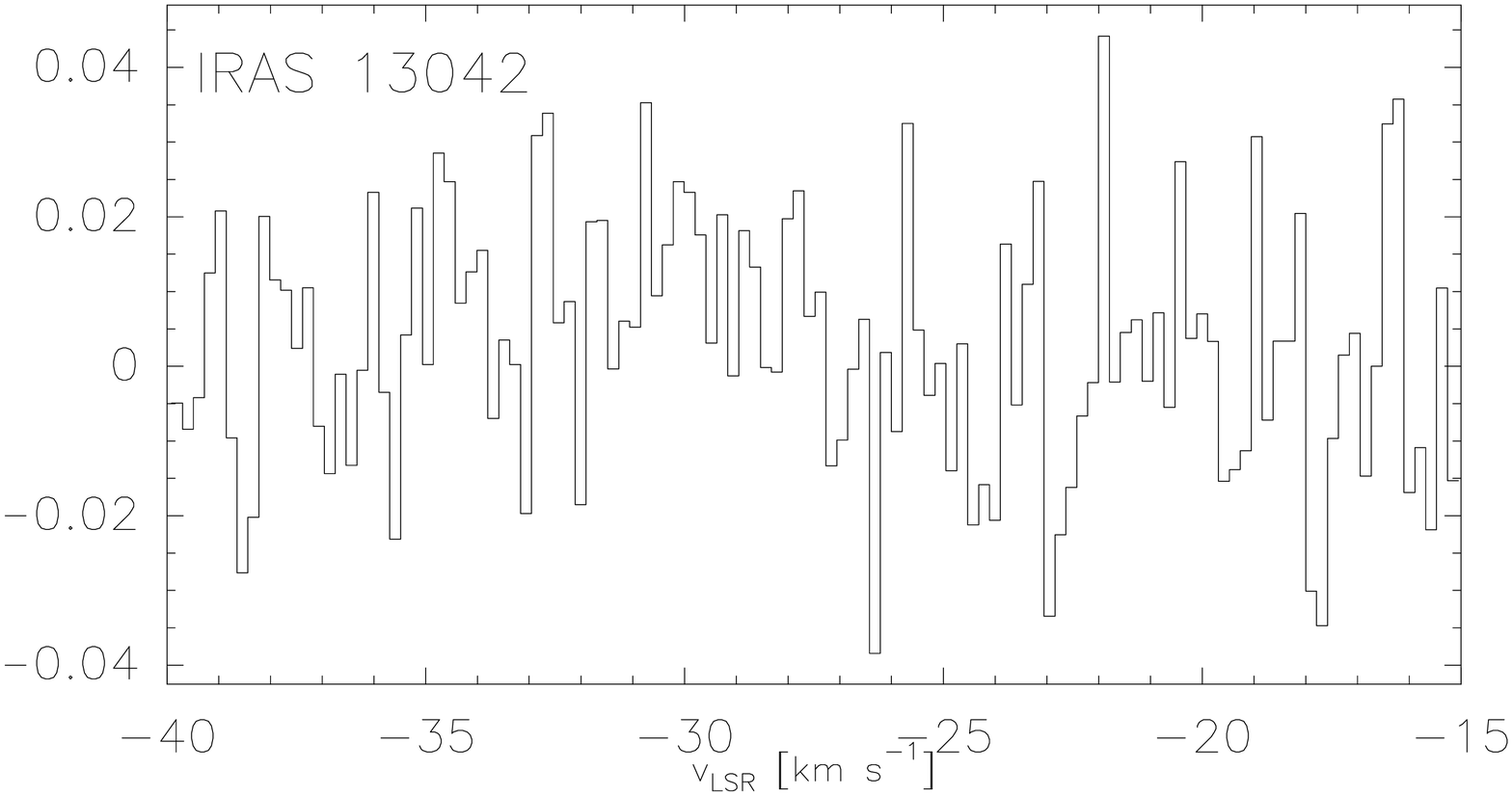}
\caption{SiO$(4-3)$ spectra towards the Seahorse IRDC clumps. Gaussian fits to the detected lines are overlaid in green (only one source, SMM~3). The velocity range in the SMM~3 panel is 1.8 times wider than in the other panels to better see the broad line profile. The intensity range in the panels is different to better show the line profiles.}
\label{figure:sio}
\end{center}
\end{figure*}

\begin{figure*}[!htb]
\begin{center}
\includegraphics[width=0.33\textwidth]{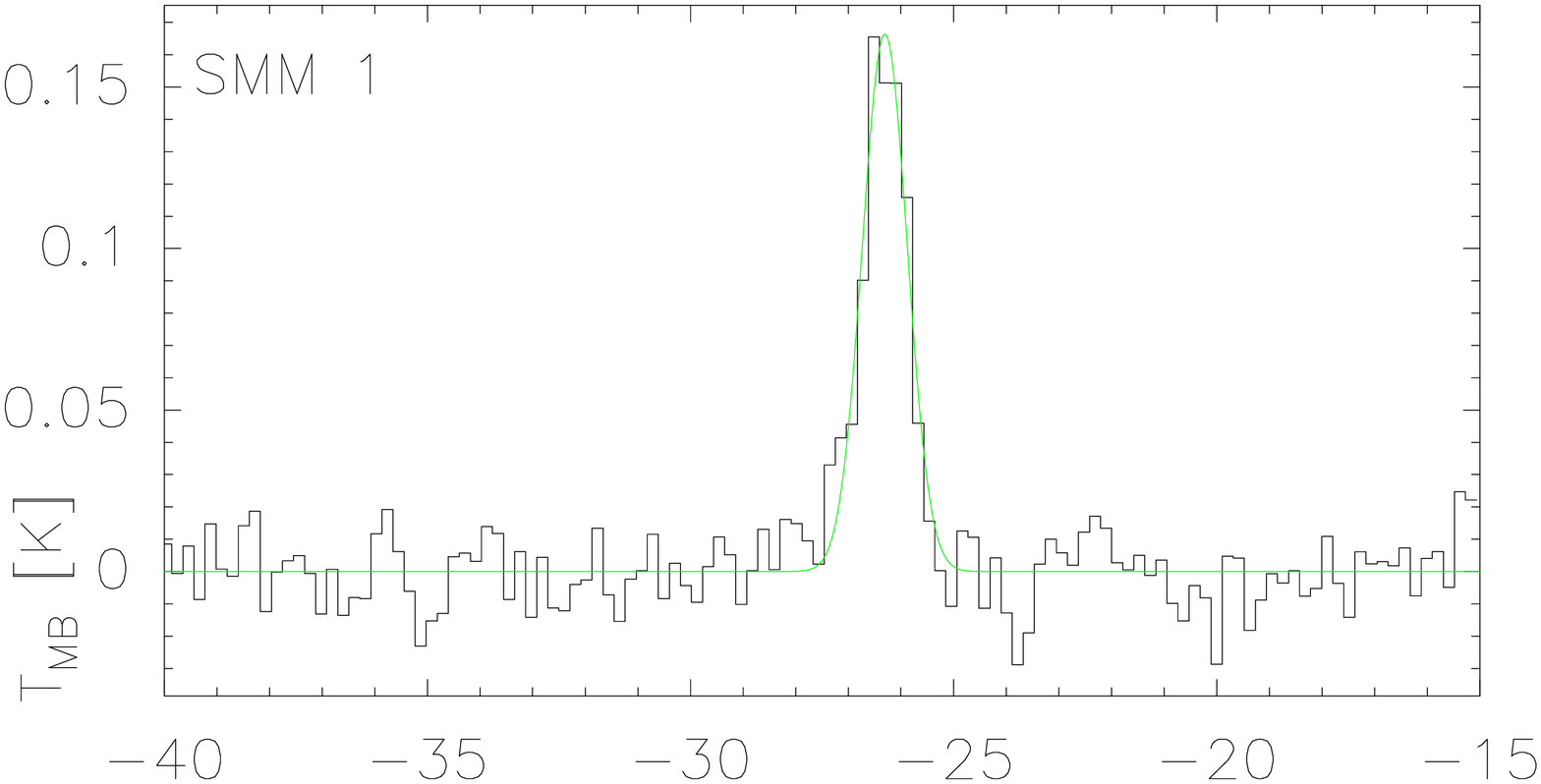}
\includegraphics[width=0.33\textwidth]{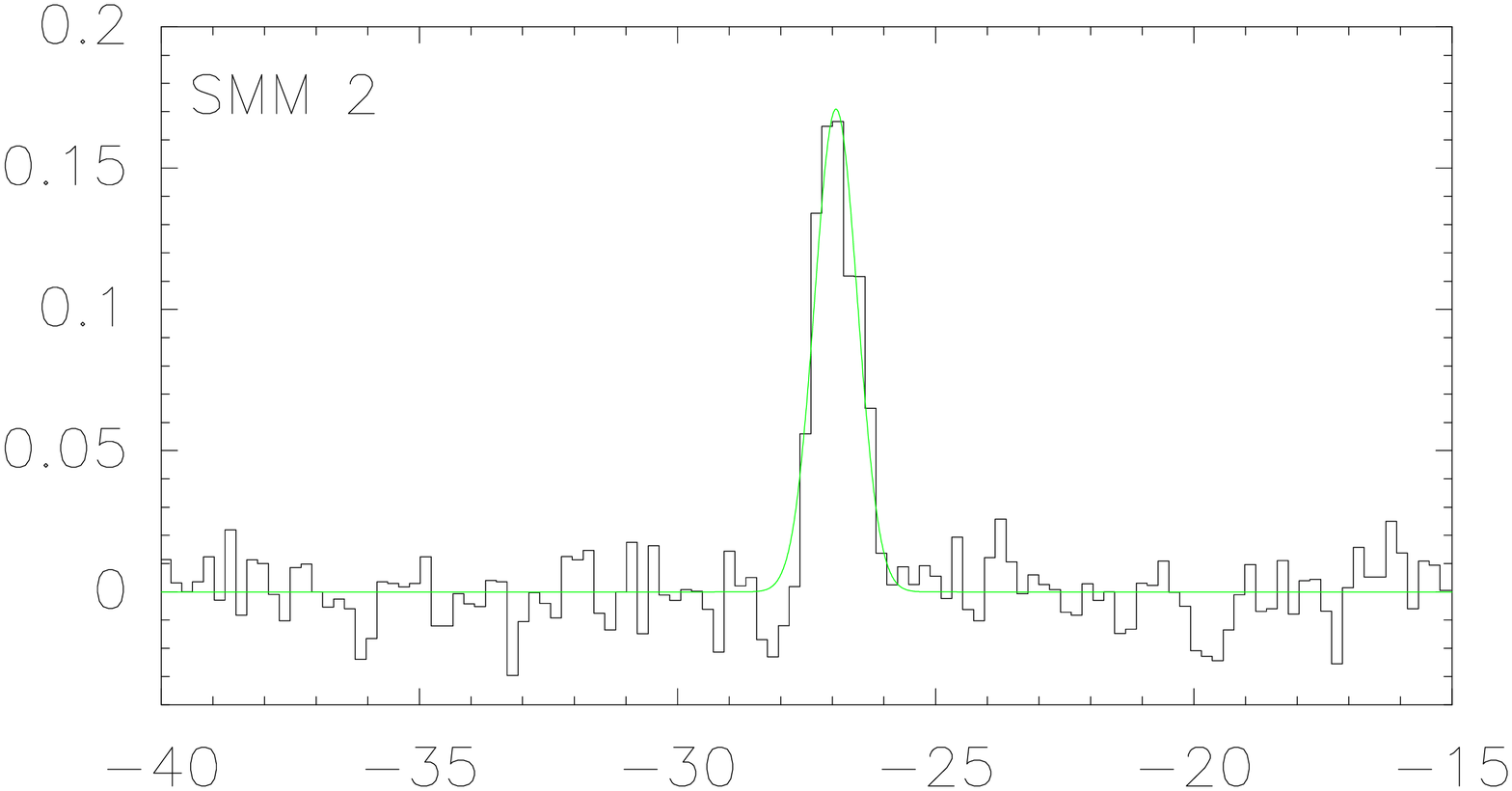}
\includegraphics[width=0.33\textwidth]{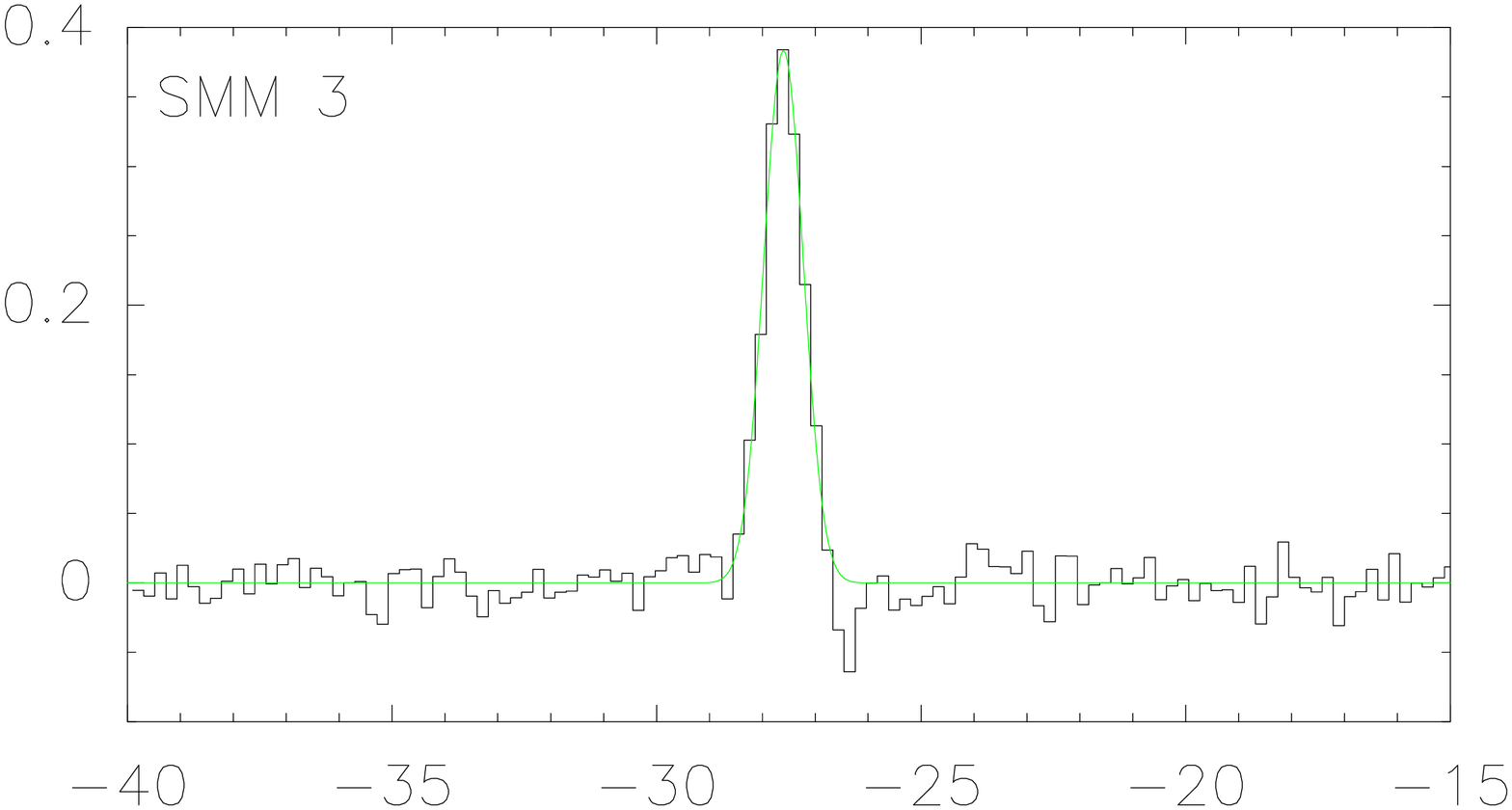}
\includegraphics[width=0.33\textwidth]{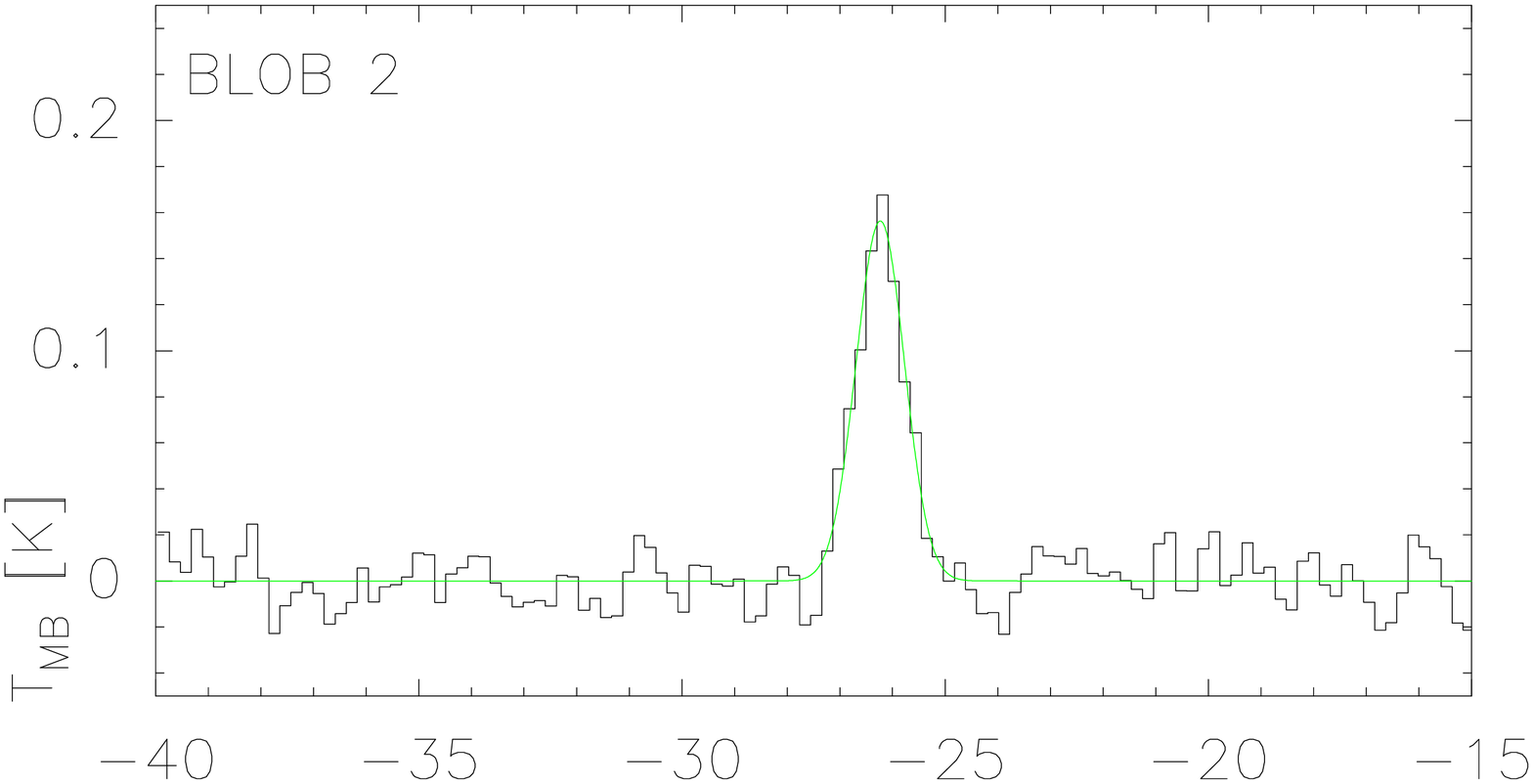}
\includegraphics[width=0.33\textwidth]{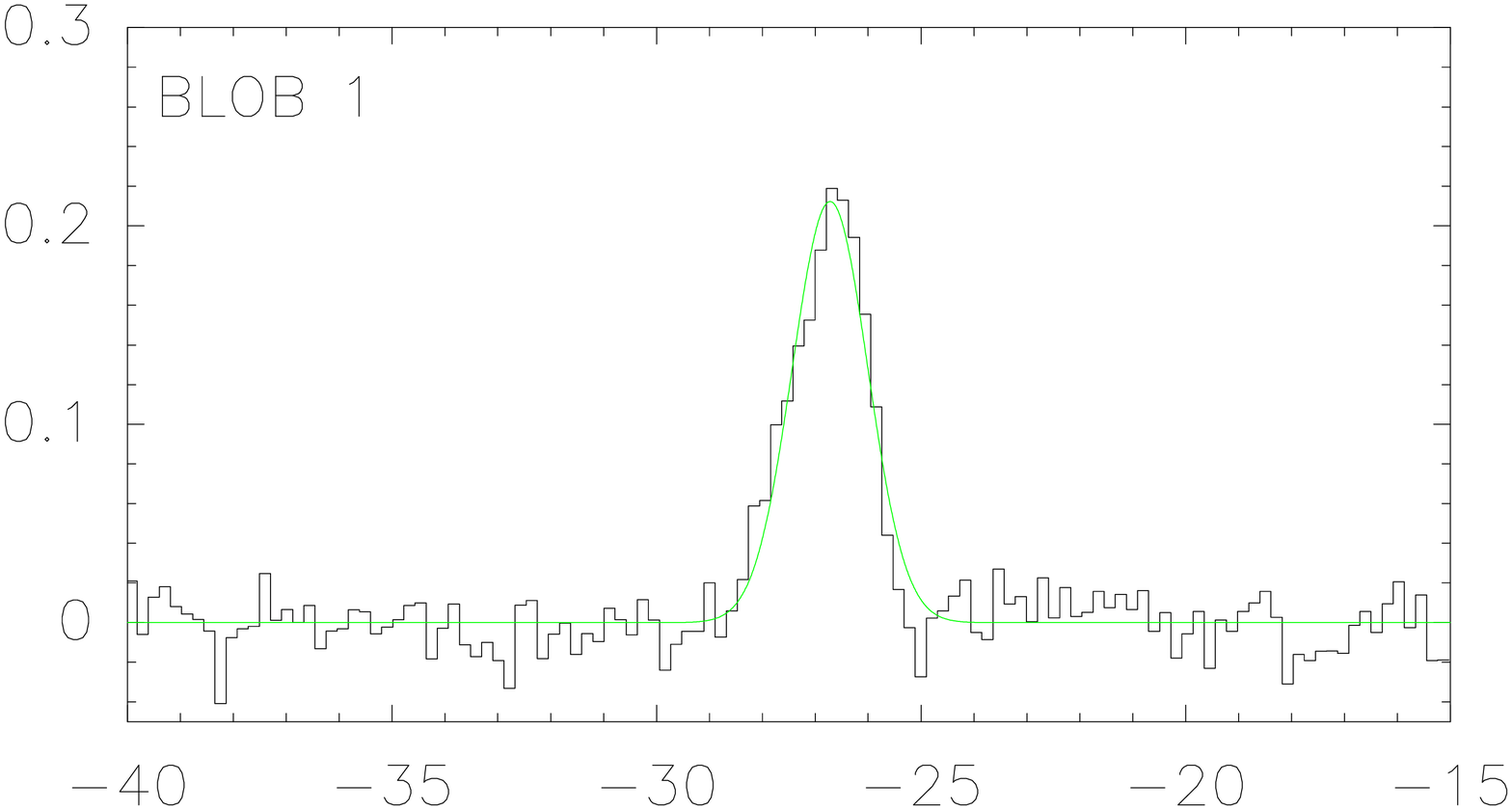}
\includegraphics[width=0.33\textwidth]{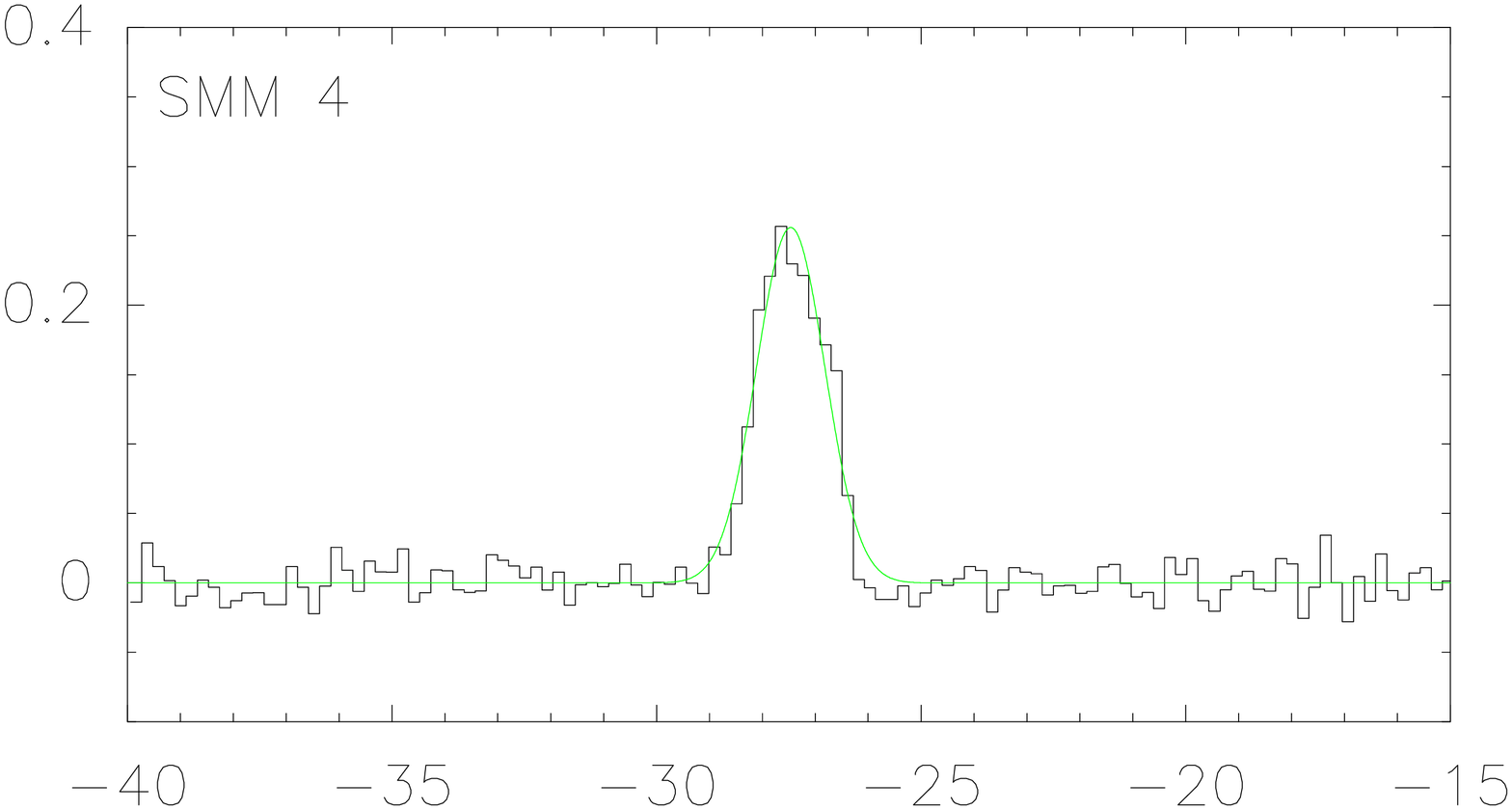}
\includegraphics[width=0.33\textwidth]{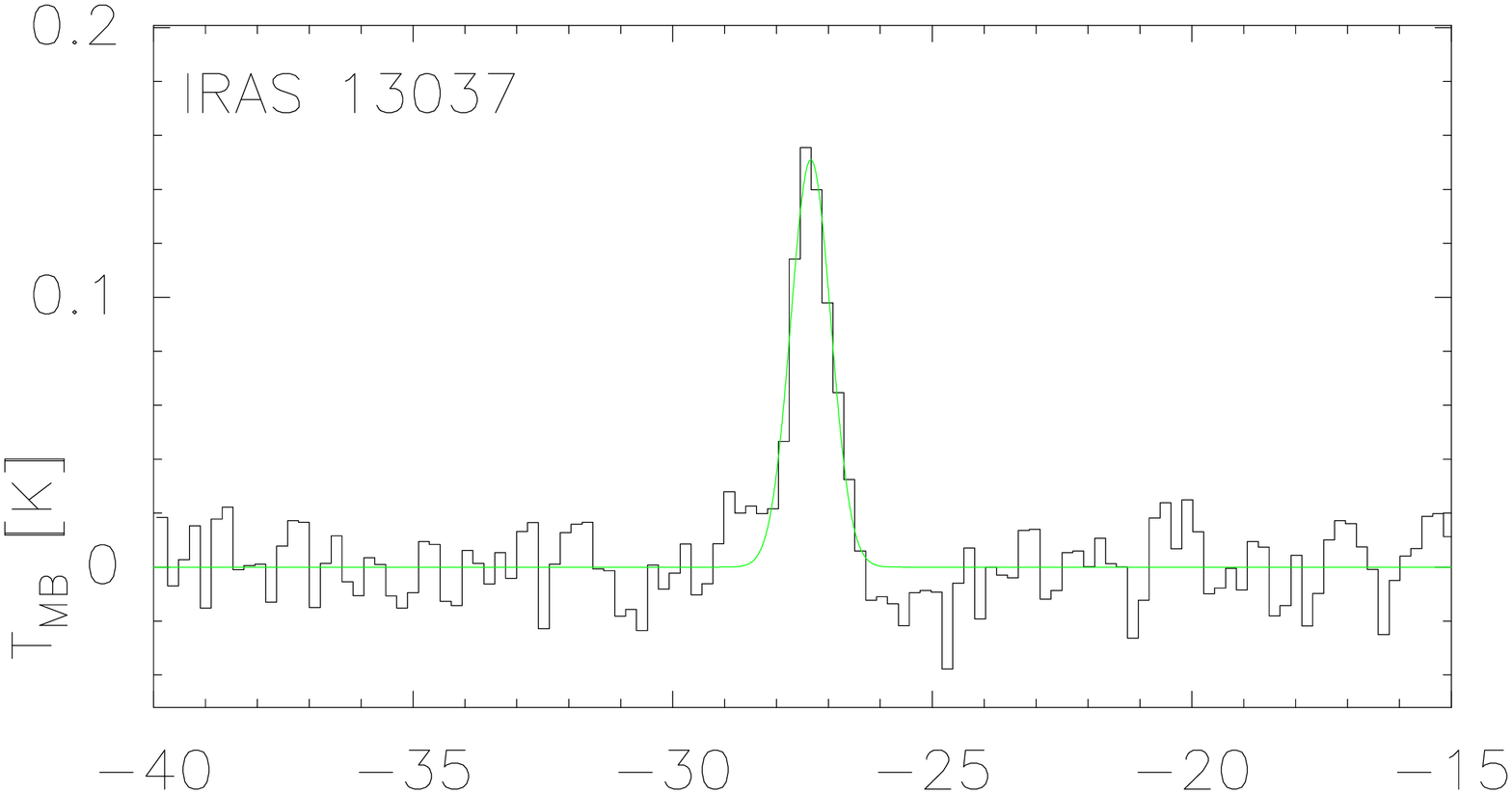}
\includegraphics[width=0.33\textwidth]{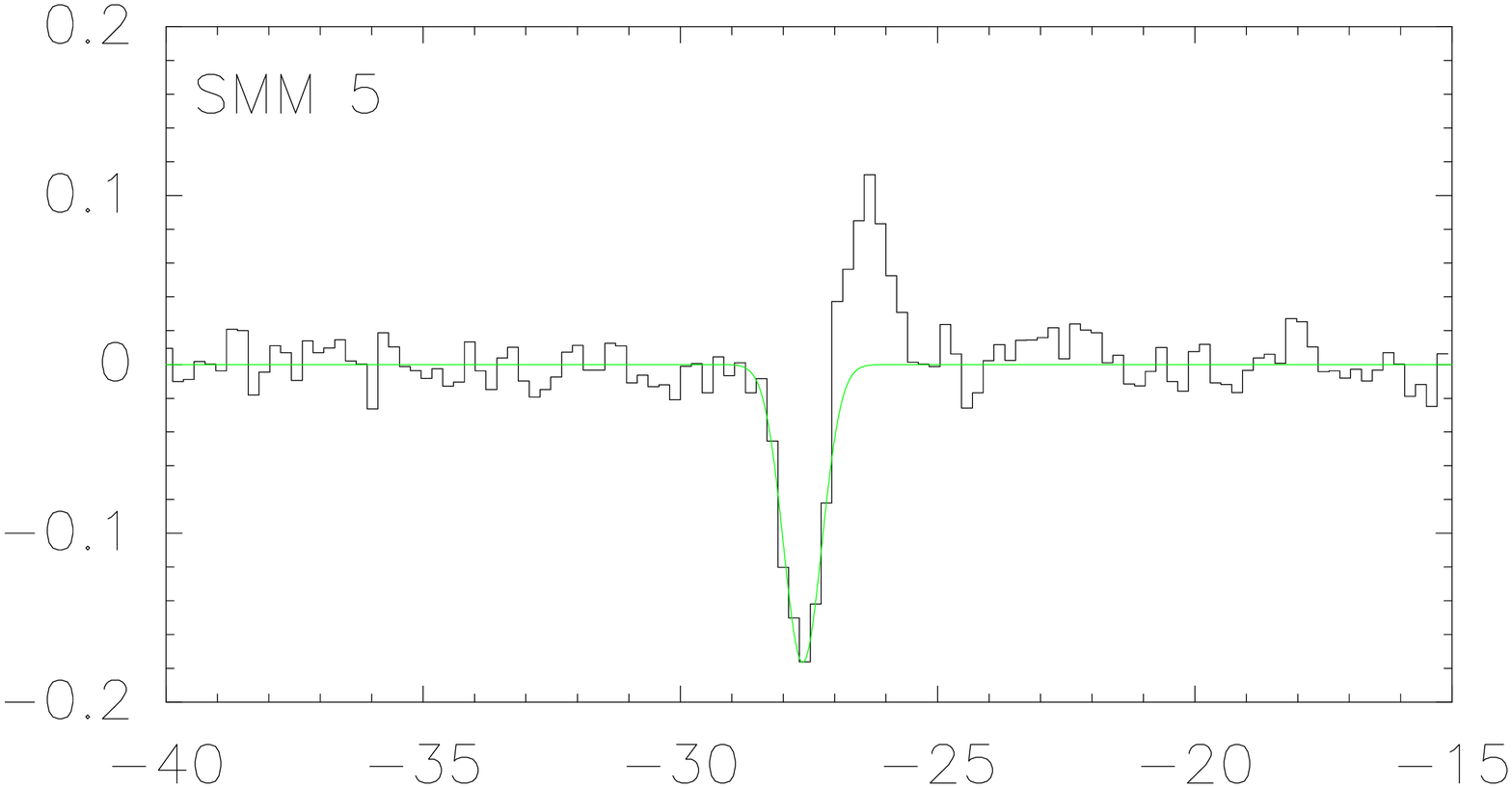}
\includegraphics[width=0.33\textwidth]{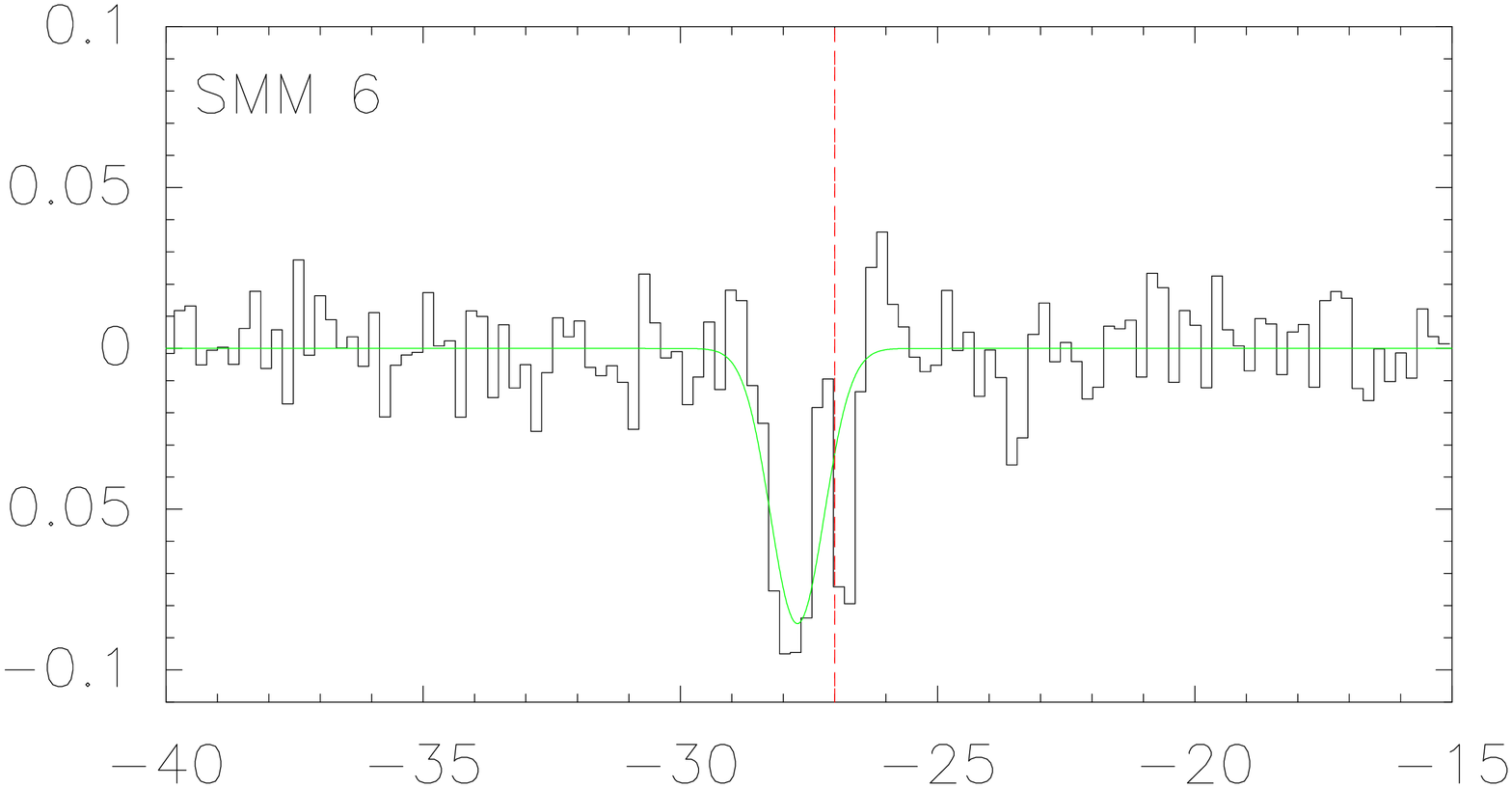}
\includegraphics[width=0.33\textwidth]{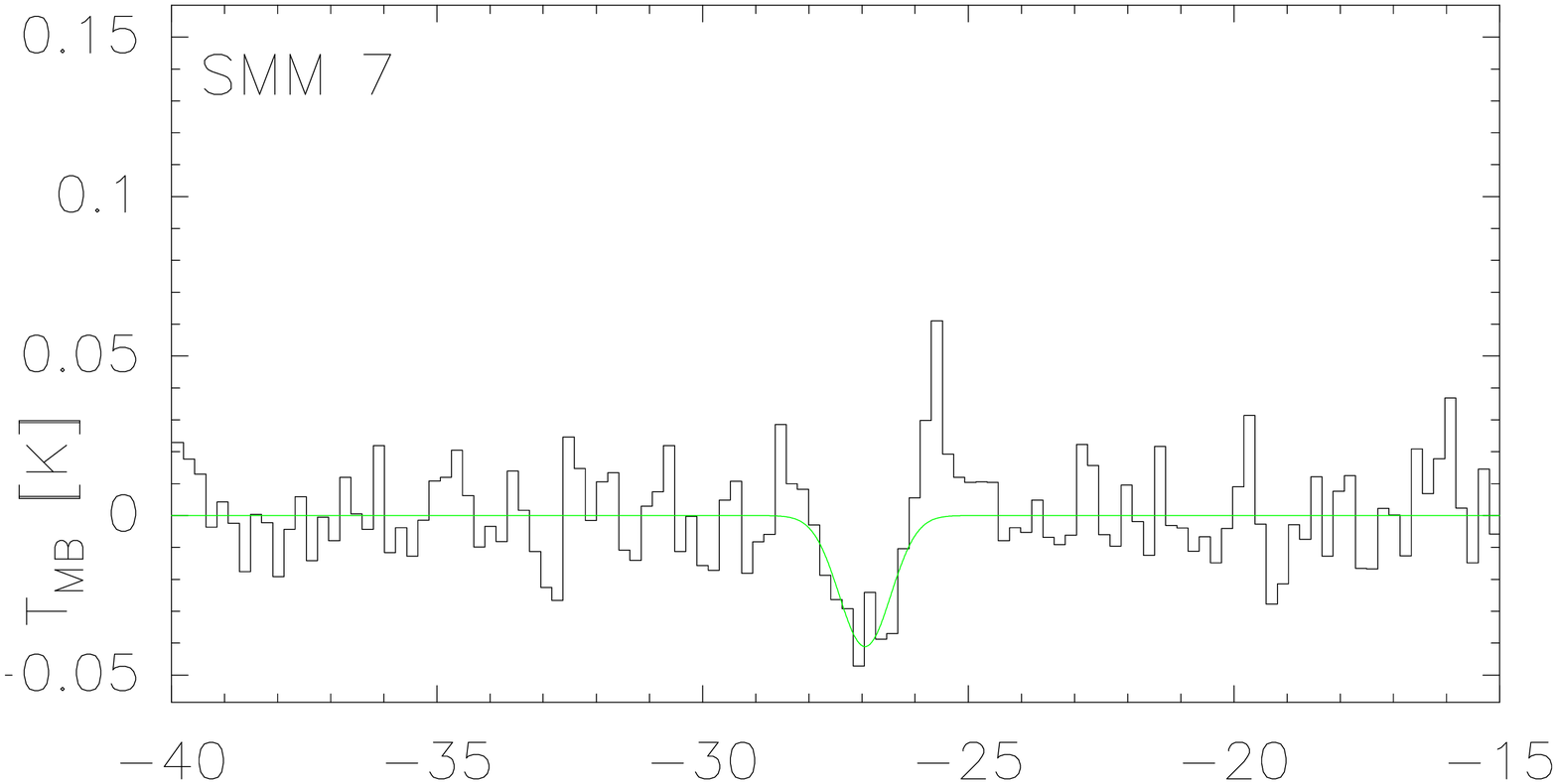}
\includegraphics[width=0.33\textwidth]{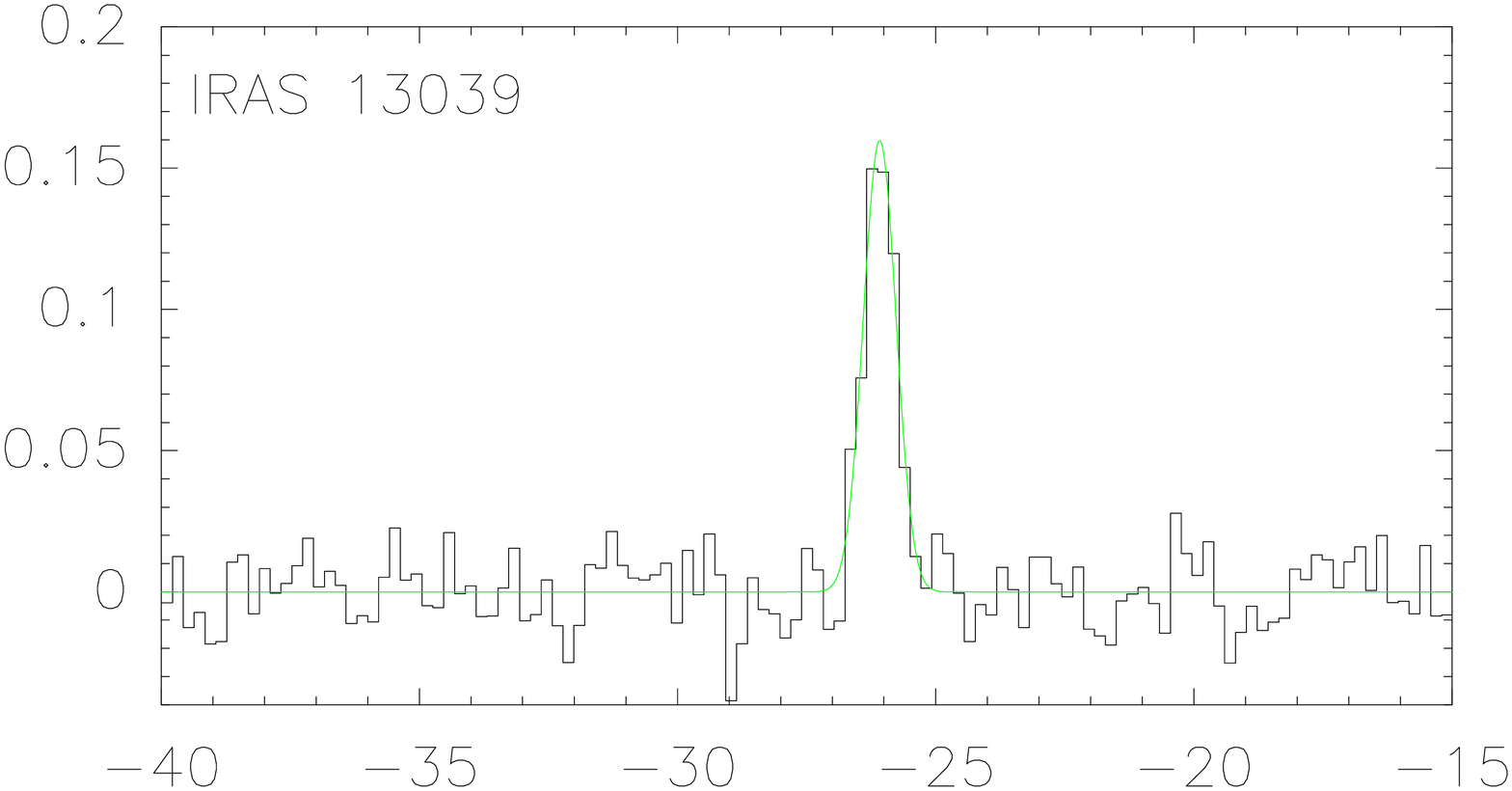}
\includegraphics[width=0.33\textwidth]{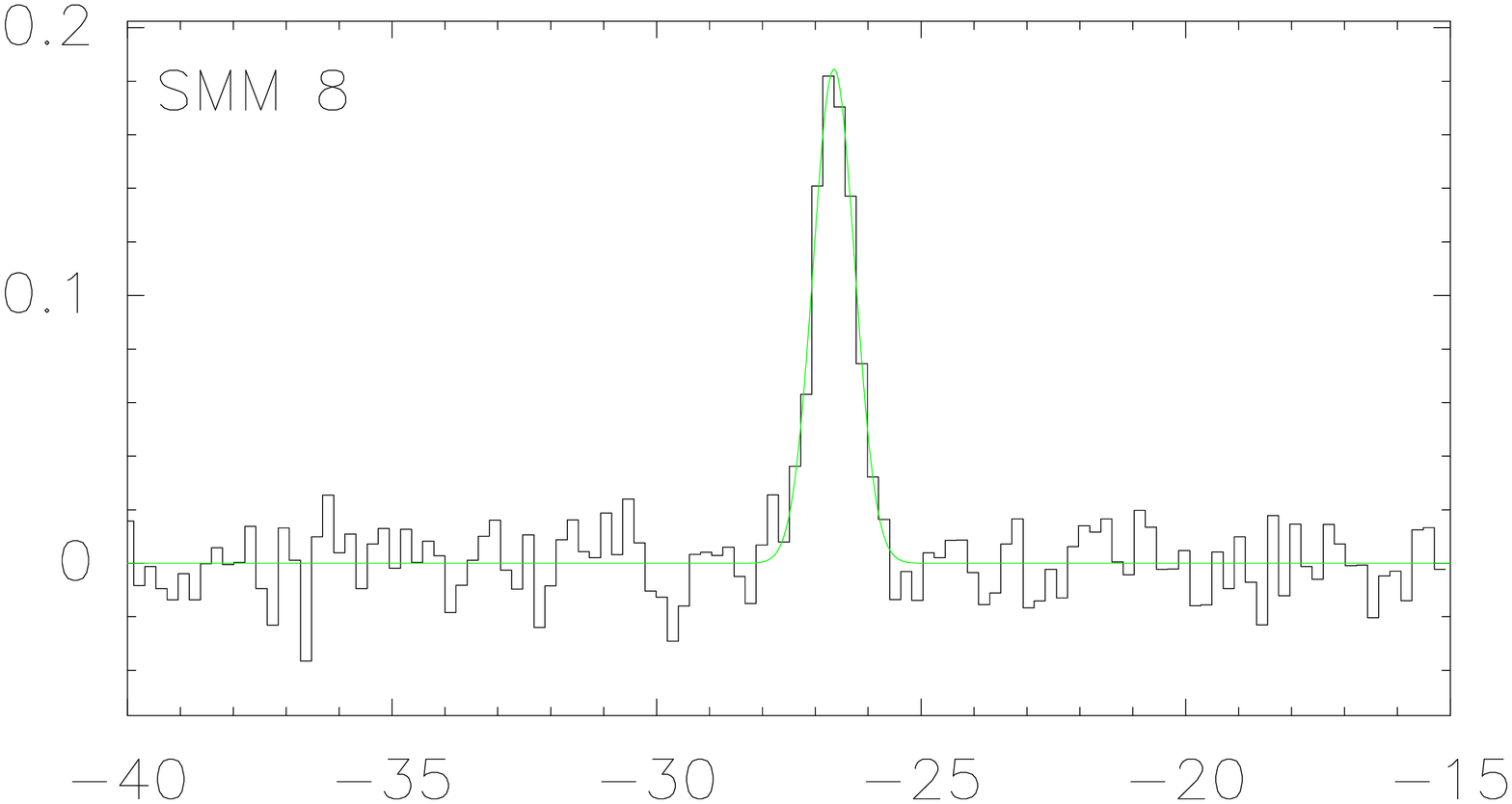}
\includegraphics[width=0.33\textwidth]{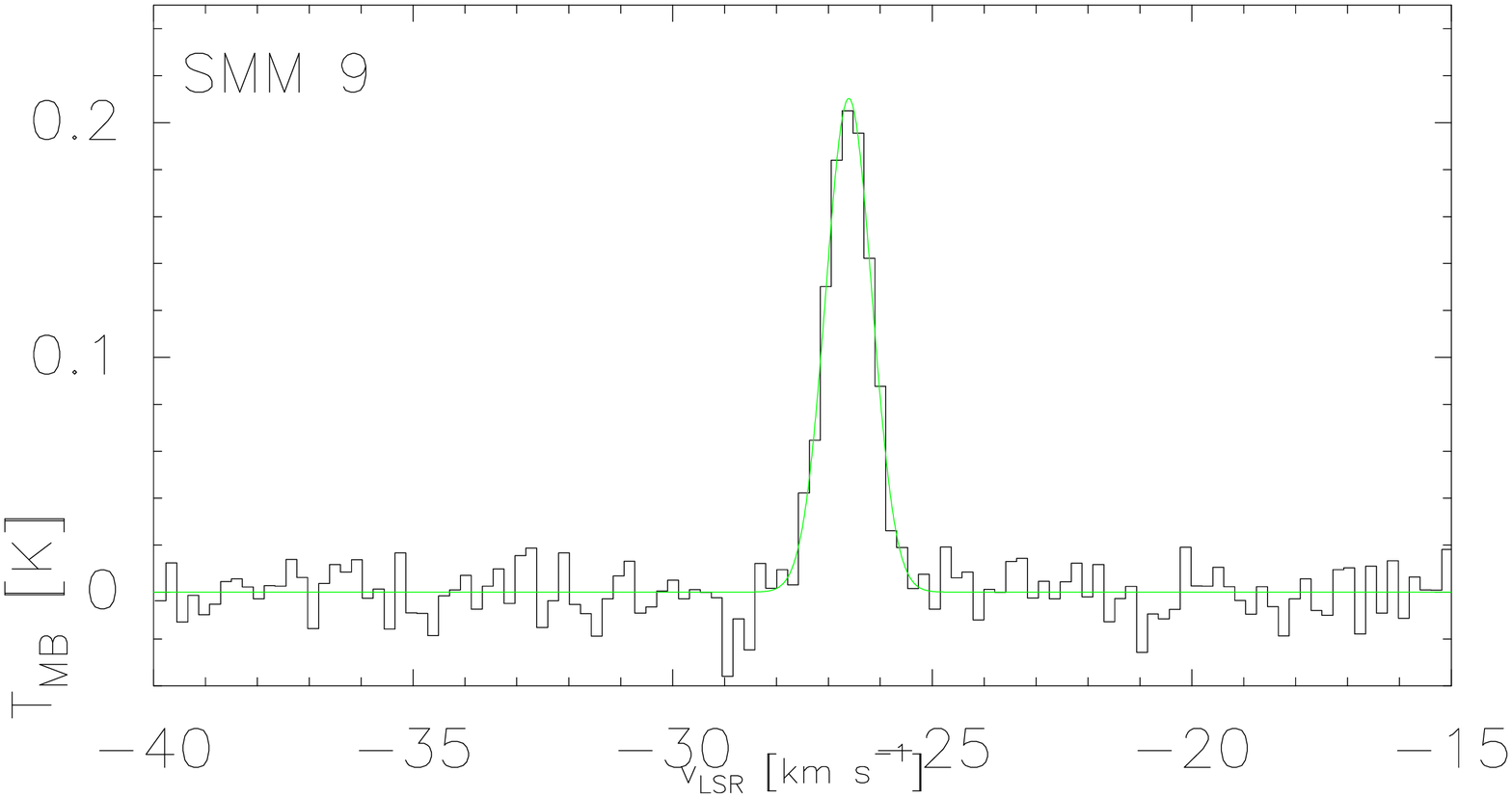}
\includegraphics[width=0.33\textwidth]{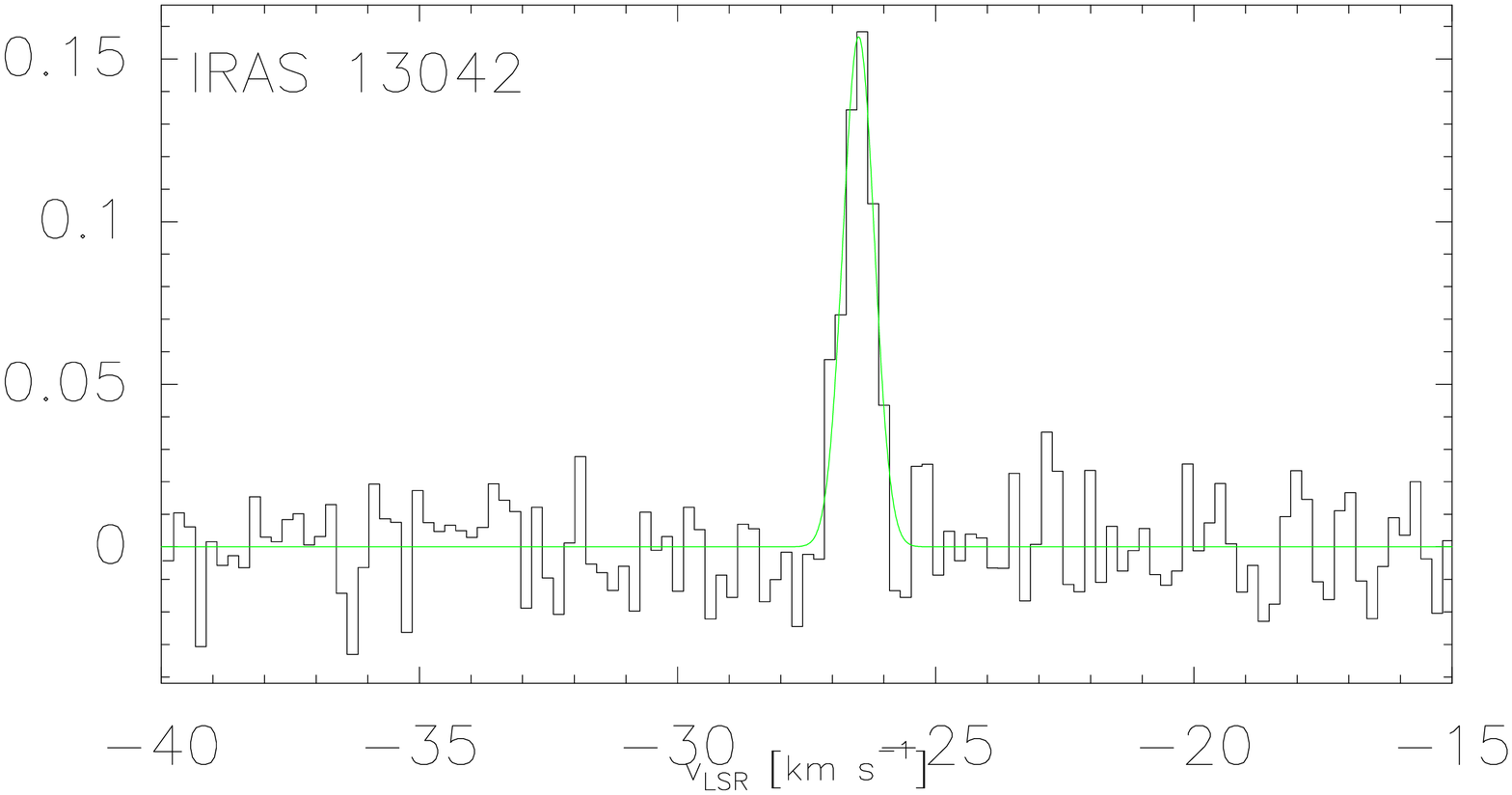}
\caption{HN$^{13}$C$(2-1)$ spectra towards the Seahorse IRDC clumps. Gaussian fits to the lines are overlaid in green. While the velocity range shown in each panel is the same, the intensity range is different to better show the line profiles. The HN$^{13}$C line is seen in absoprtion towards SMM~5--7. The red, vertical dashed line in the SMM~6 panel shows the systemic velocity derived from C$^{17}$O$(J=2-1)$ by Miettinen (2012).}
\label{figure:hn13c}
\end{center}
\end{figure*}

\begin{figure*}[!htb]
\begin{center}
\includegraphics[width=0.33\textwidth]{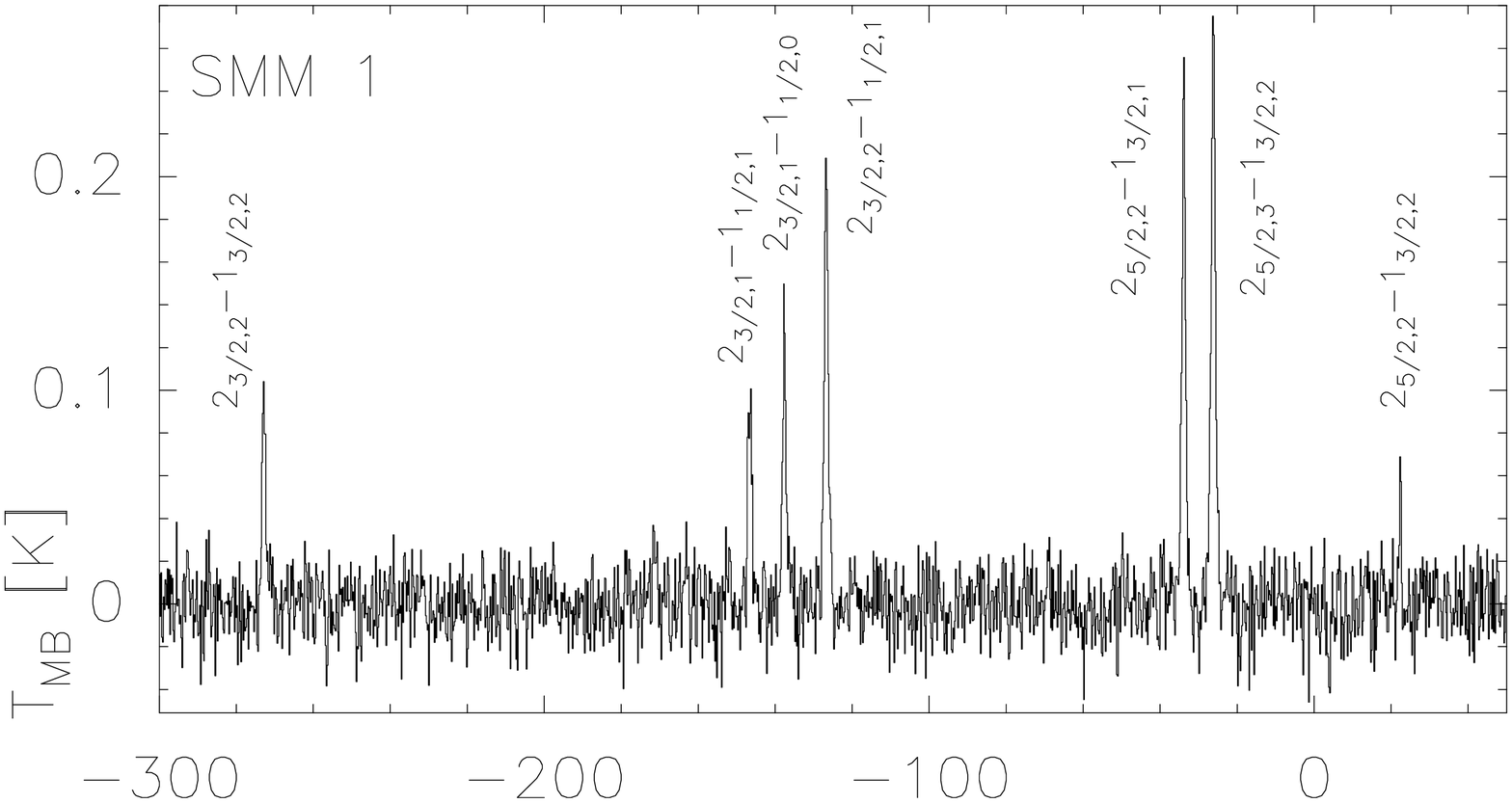}
\includegraphics[width=0.33\textwidth]{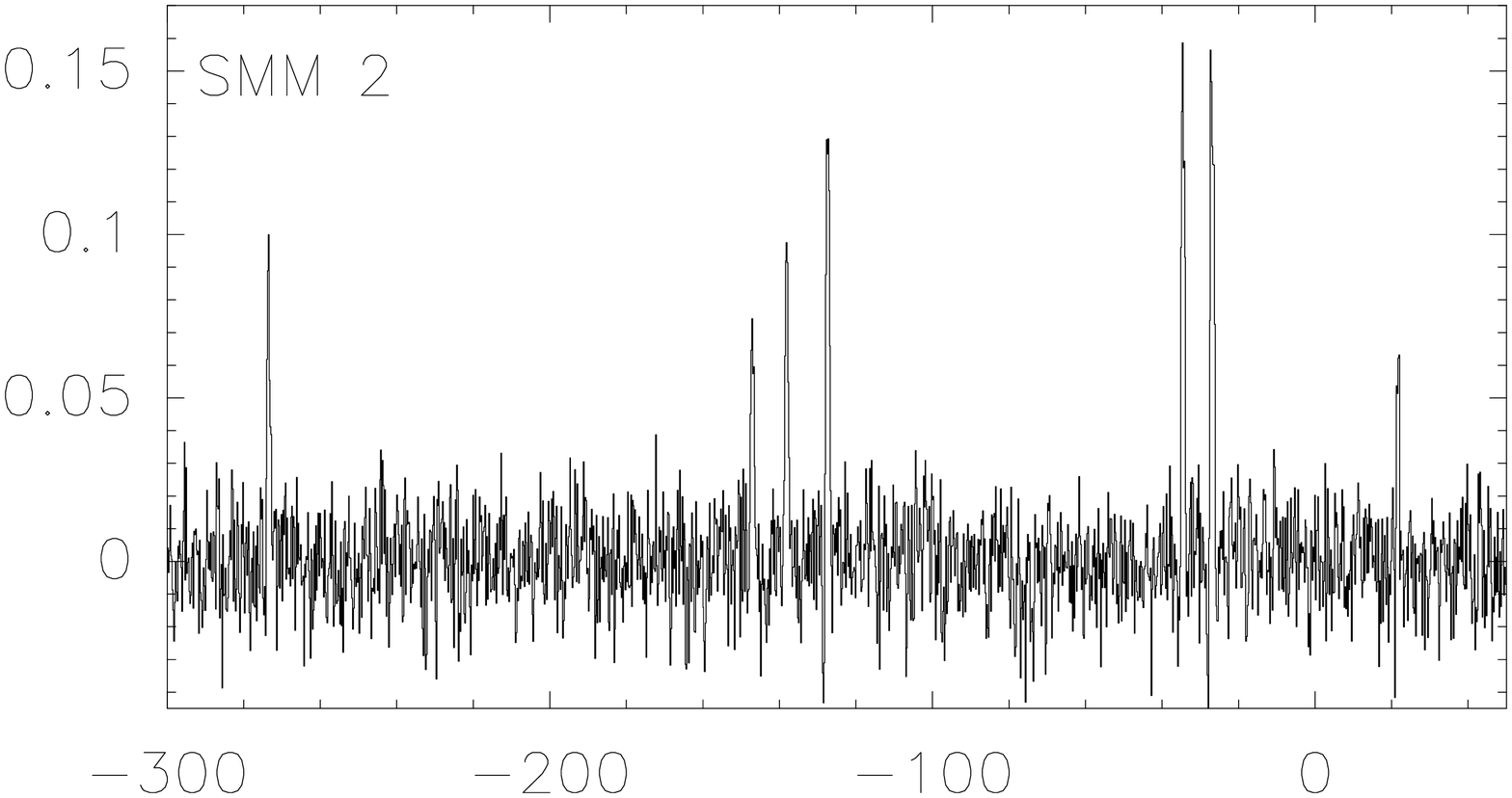}
\includegraphics[width=0.33\textwidth]{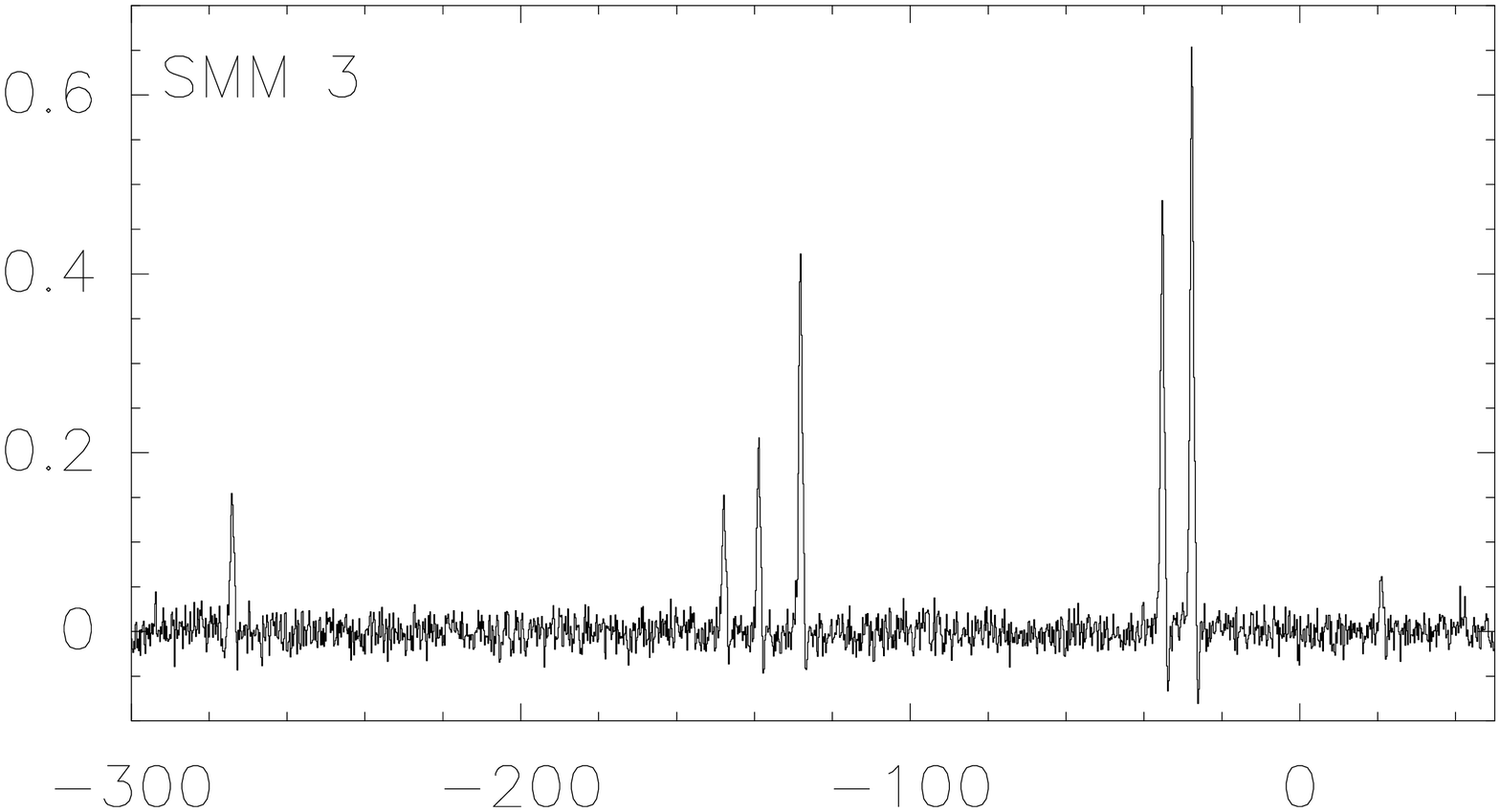}
\includegraphics[width=0.33\textwidth]{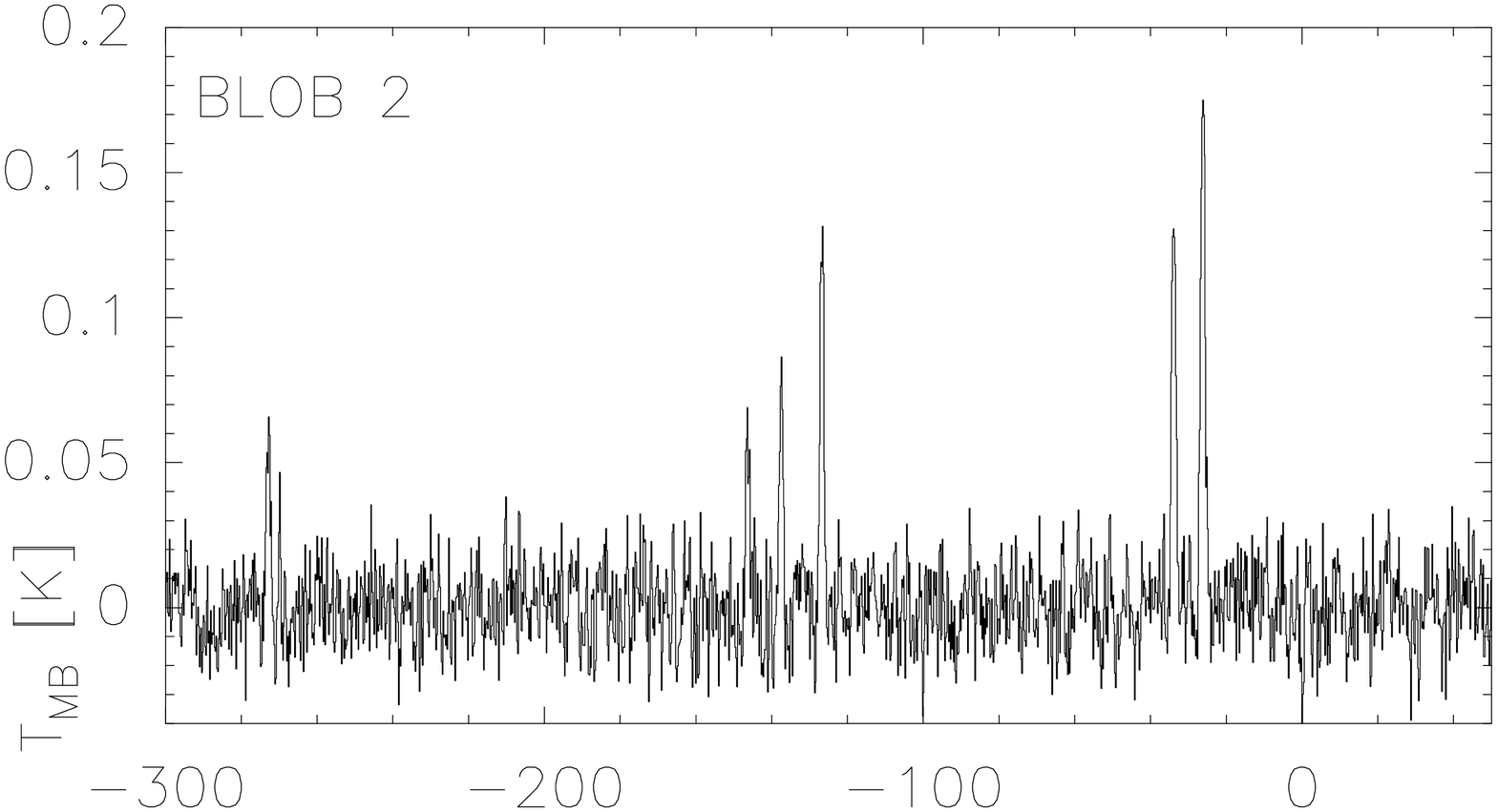}
\includegraphics[width=0.33\textwidth]{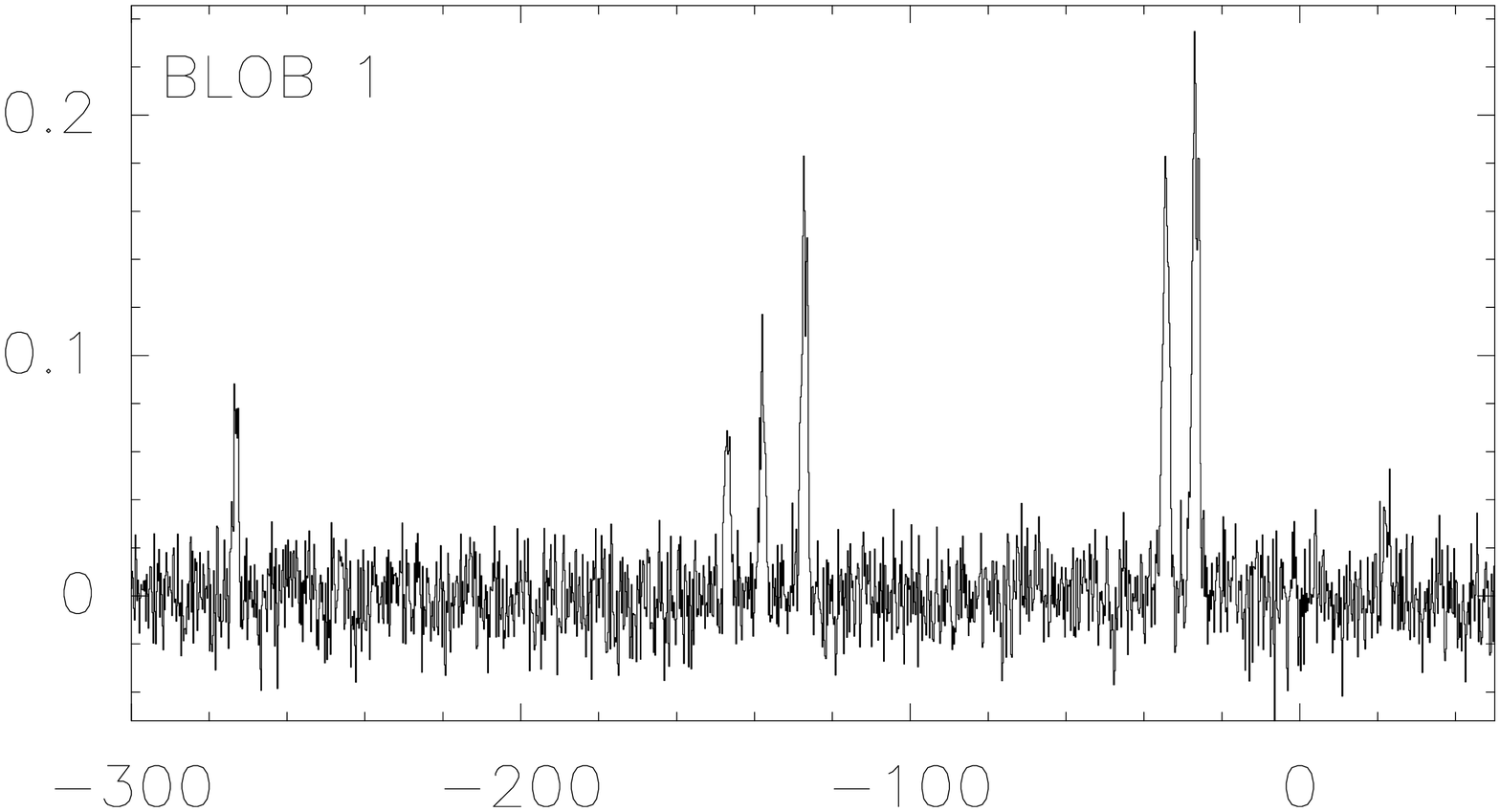}
\includegraphics[width=0.33\textwidth]{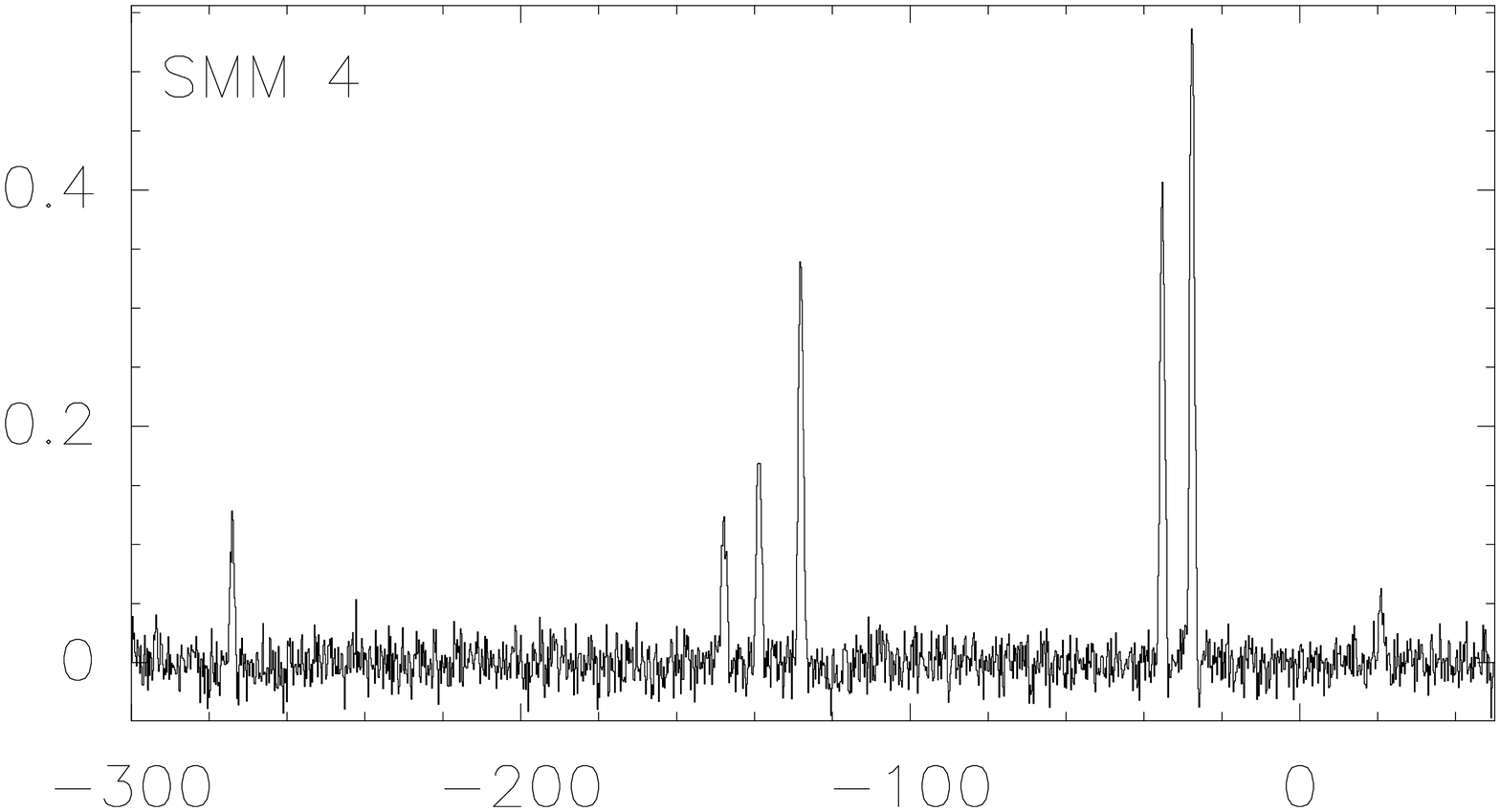}
\includegraphics[width=0.33\textwidth]{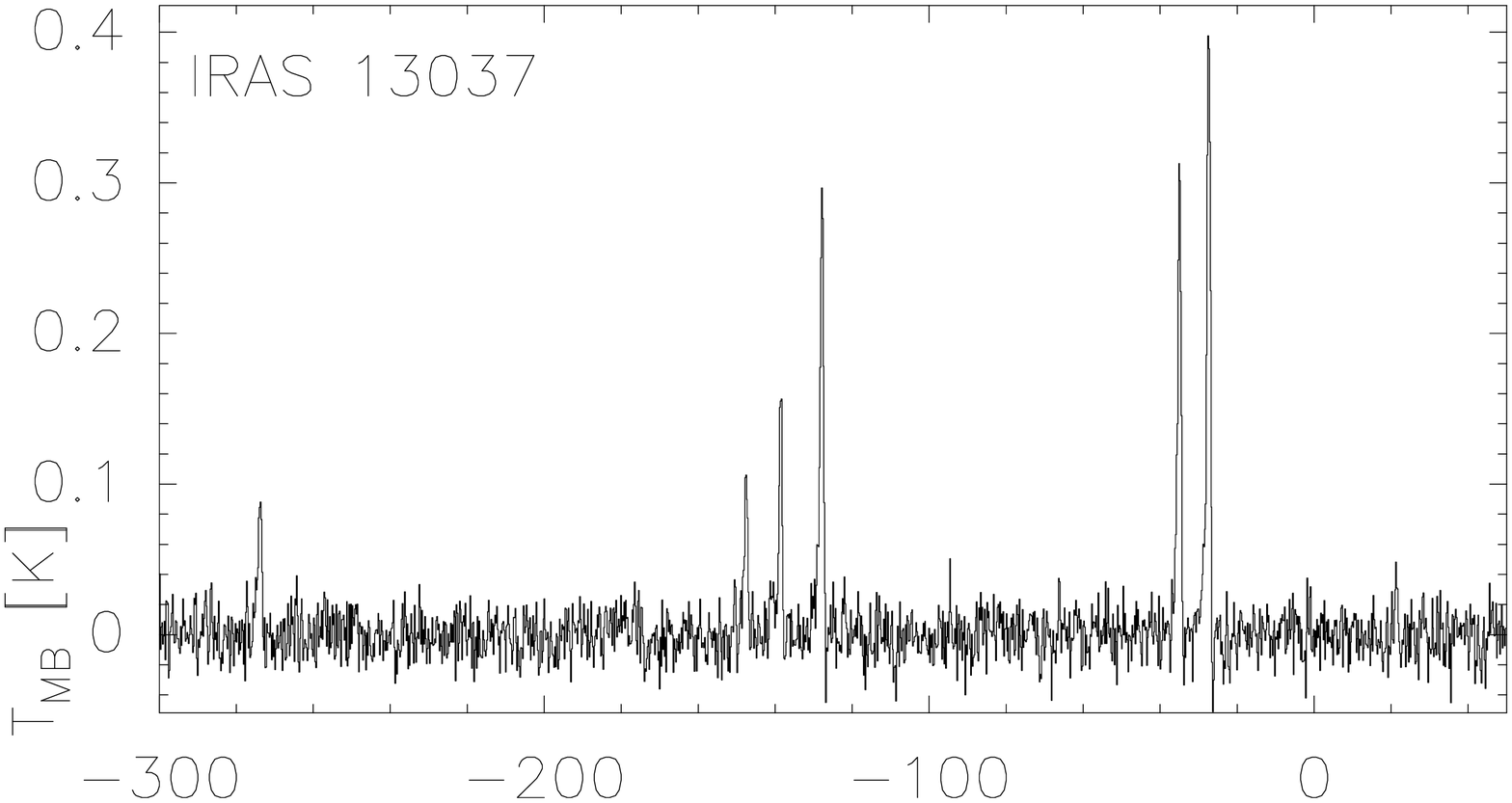}
\includegraphics[width=0.33\textwidth]{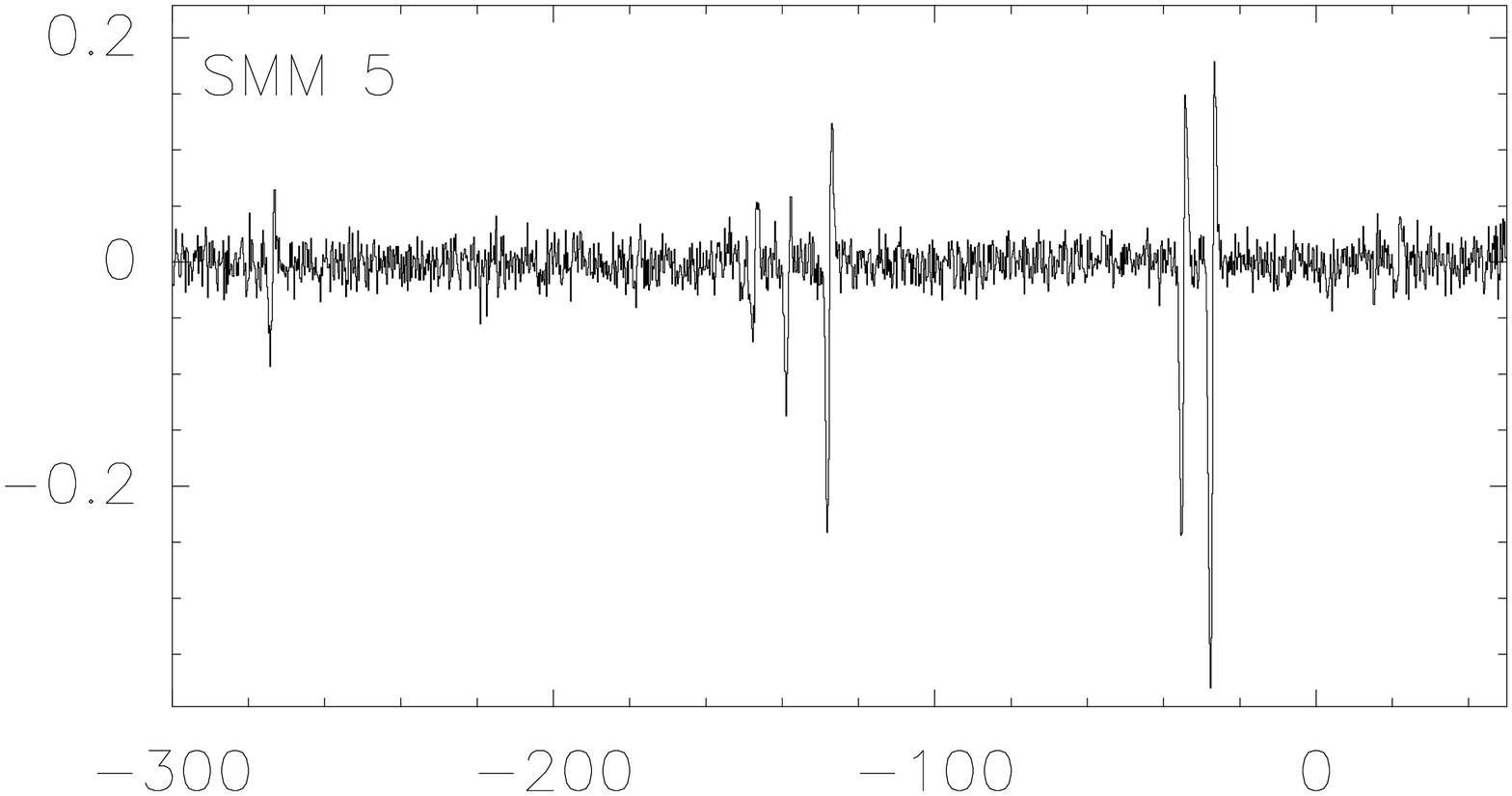}
\includegraphics[width=0.33\textwidth]{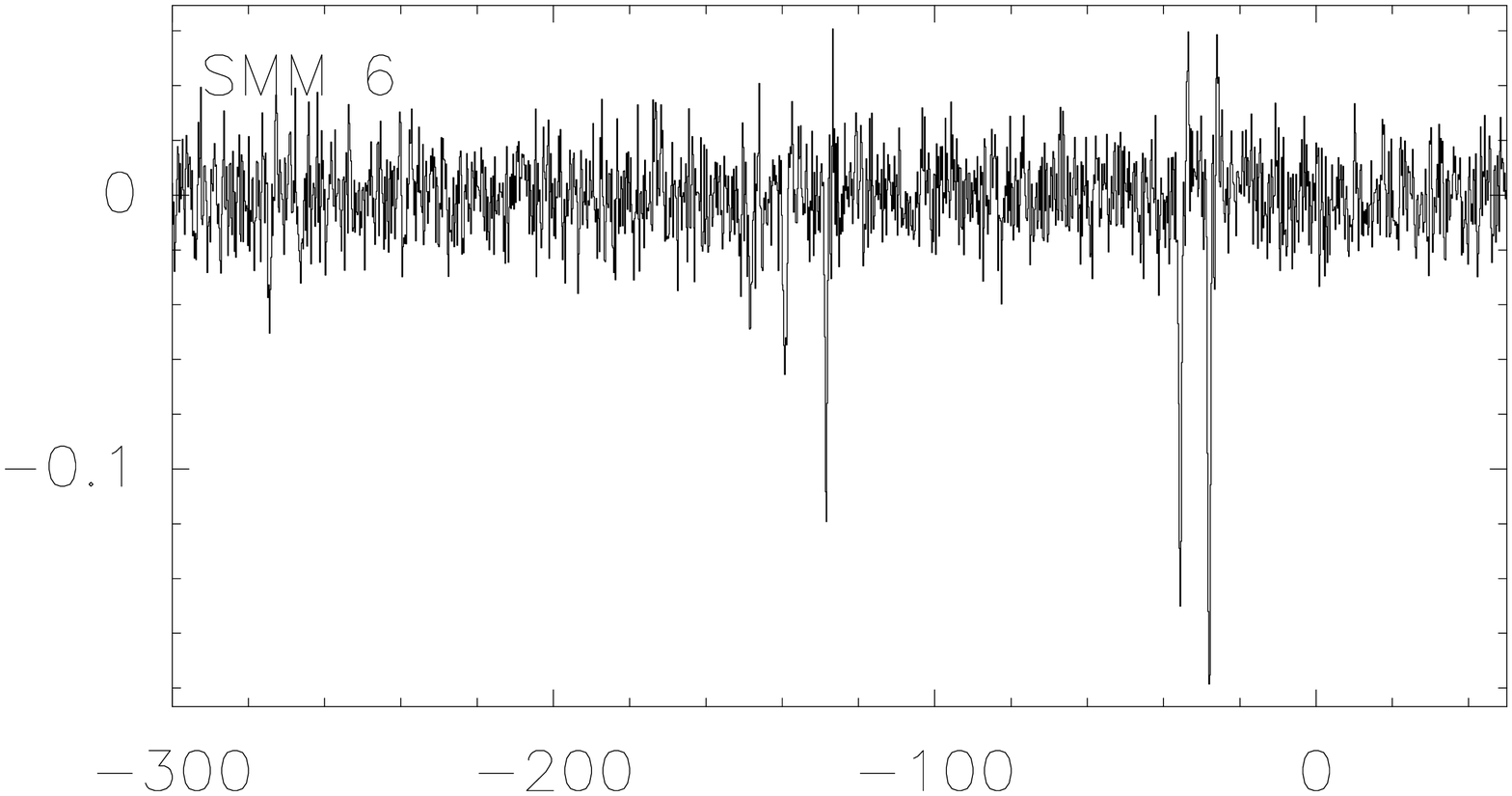}
\includegraphics[width=0.33\textwidth]{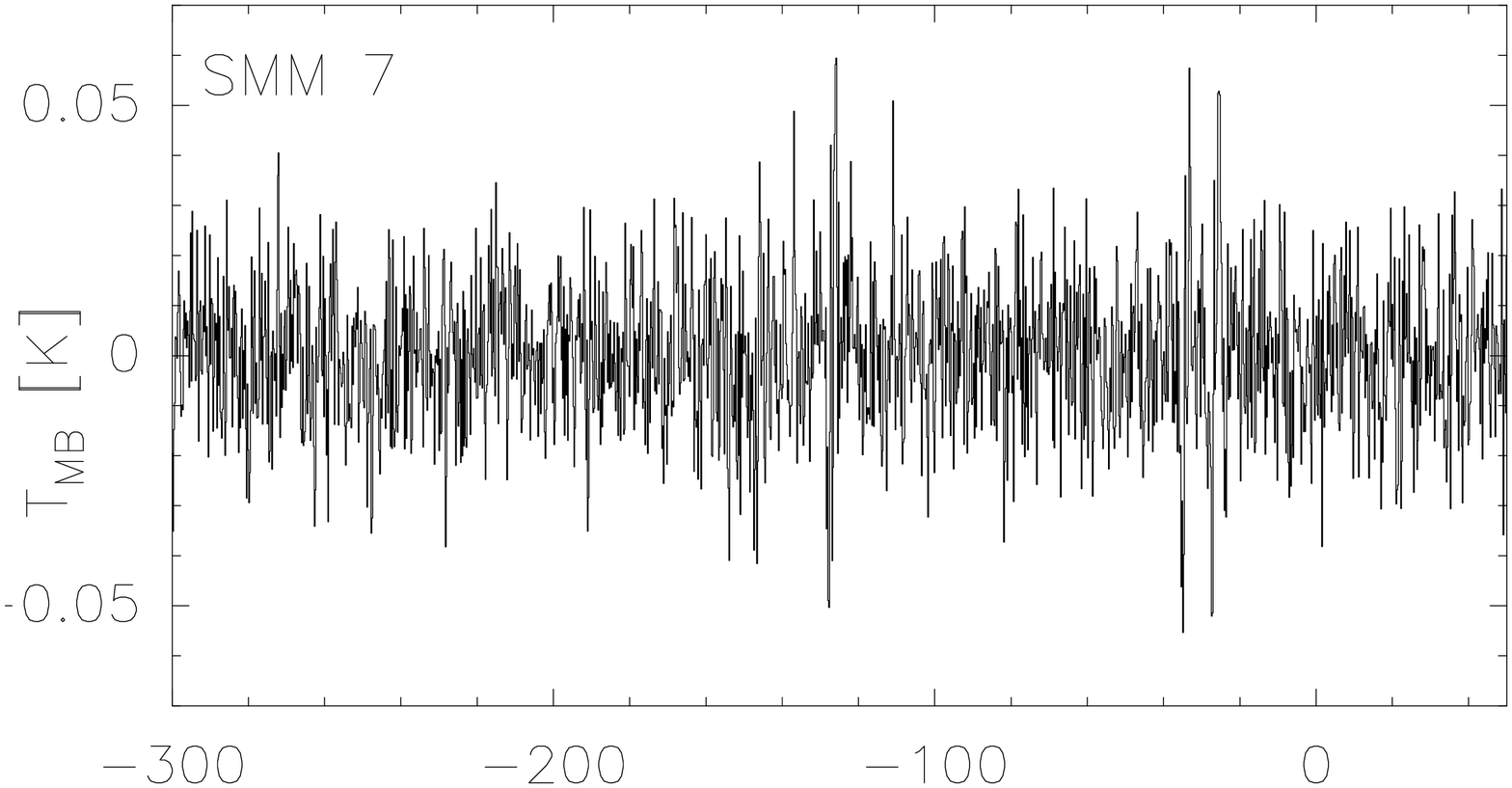}
\includegraphics[width=0.33\textwidth]{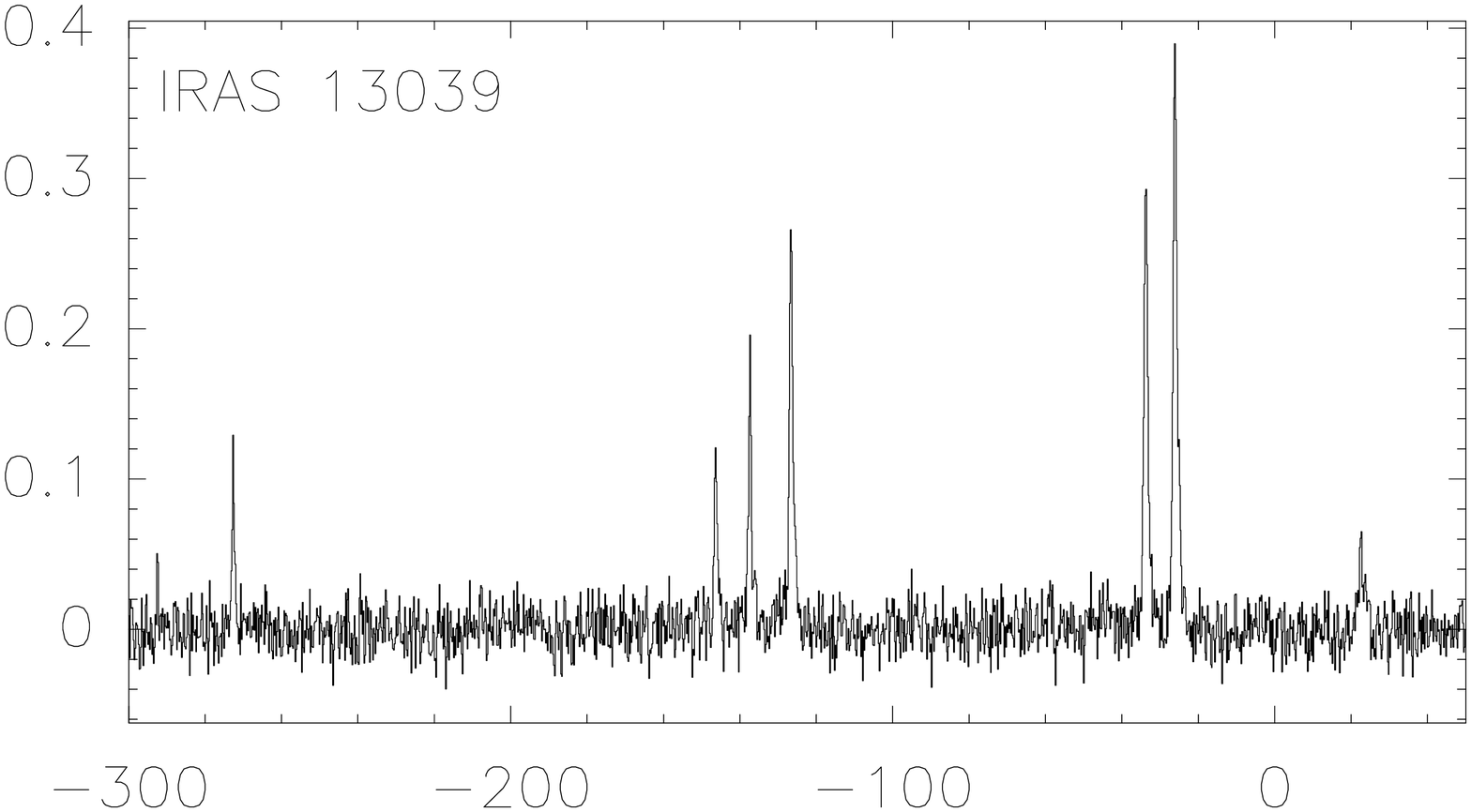}
\includegraphics[width=0.33\textwidth]{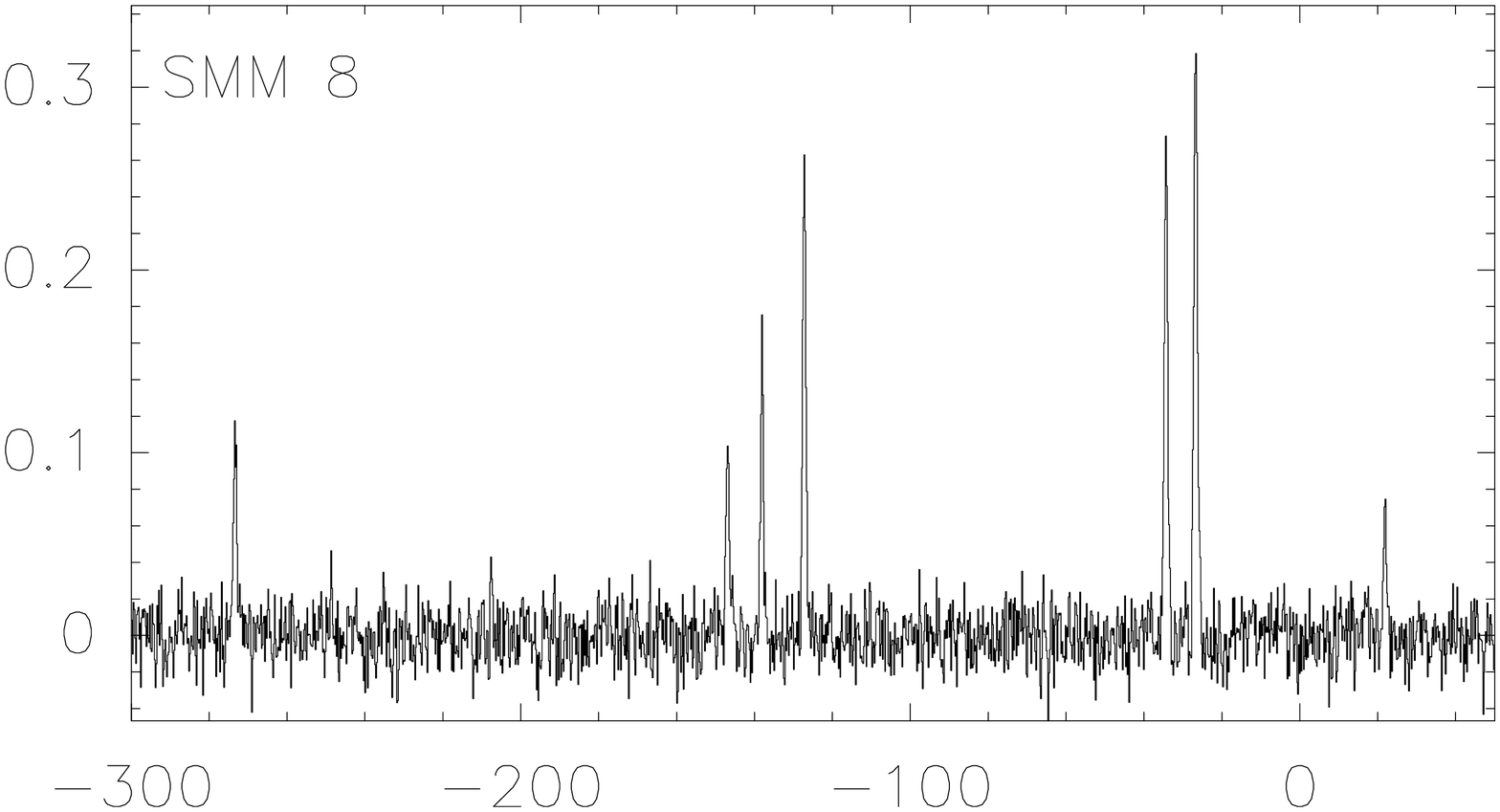}
\includegraphics[width=0.33\textwidth]{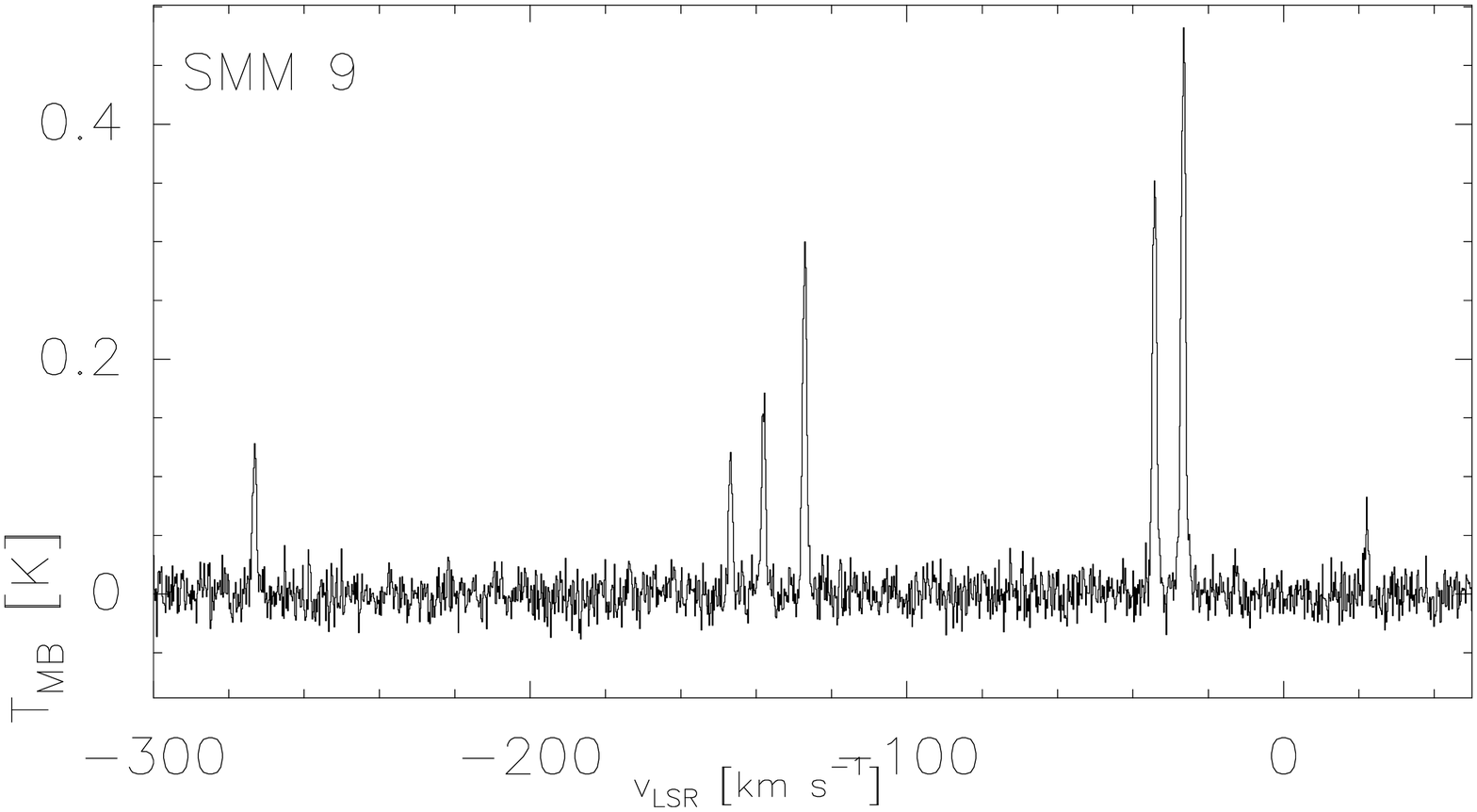}
\includegraphics[width=0.33\textwidth]{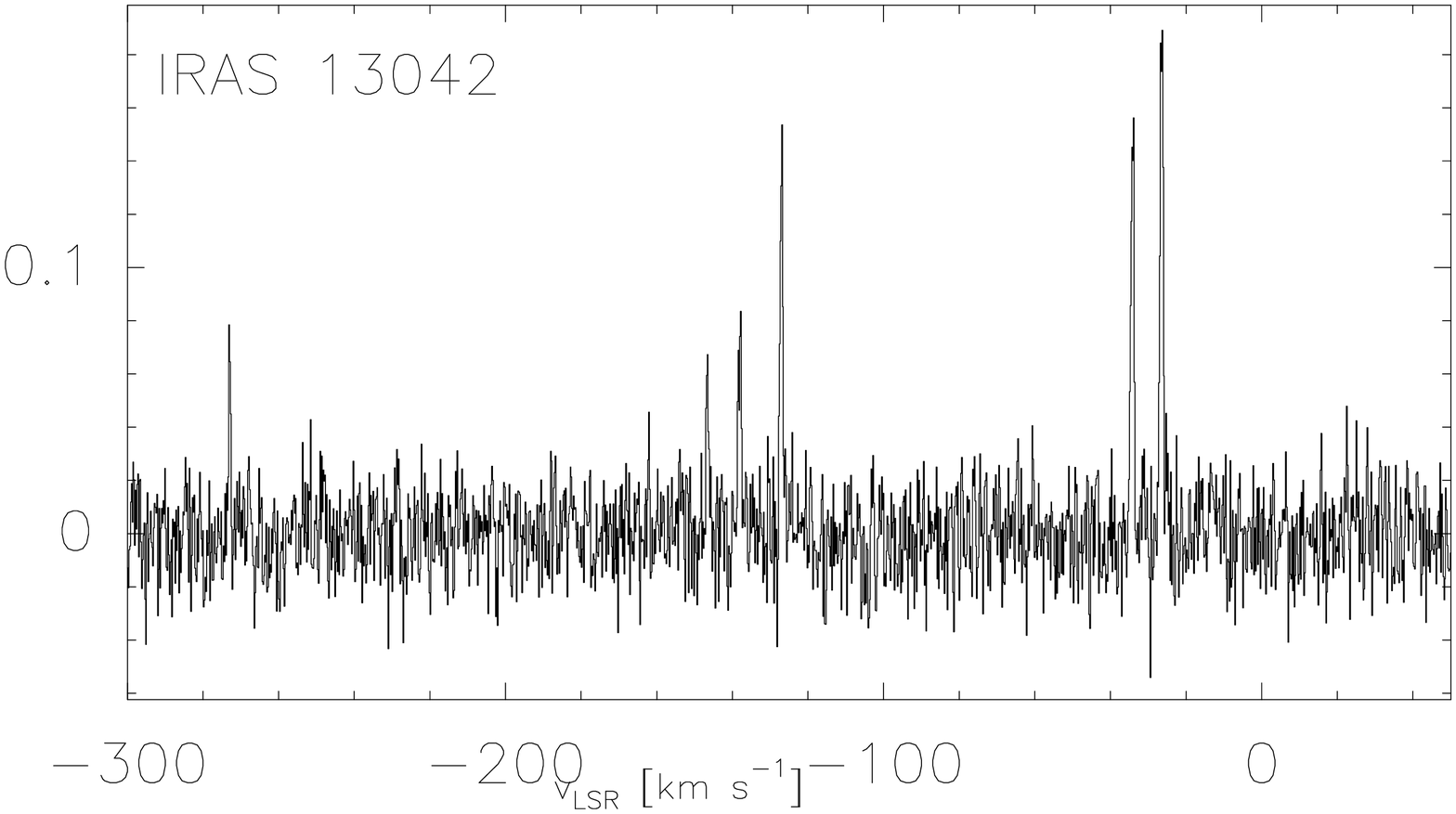}
\caption{C$_2$H$(2-1)$ spectra towards the Seahorse IRDC clumps. The seven detected hyperfine components are labelled in the SMM~1 panel. The hyperfine structure fits to the lines are not shown to better see the detected lines. While the velocity range shown in each panel is the same (we note that it is much wider than in Figs.~\ref{figure:so}--\ref{figure:hn13c}), the intensity range is different to better show the detected lines. The C$_2$H lines are seen in absoprtion towards SMM~5 and 6 (also towards SMM~7 but not at our detection limit of $\geq 3\sigma$).}
\label{figure:c2h}
\end{center}
\end{figure*}

The SO$(N_J=4_4-3_3)$, H$^{13}$CN$(J=2-1)$, H$^{13}$CO$^+(J=2-1)$, SiO$(J=4-3)$, HN$^{13}$C$(J=2-1)$, and C$_2$H$(N=2-1)$ spectra 
towards all the 14 target sources are shown in Figs.~\ref{figure:so}--\ref{figure:c2h}. The derived line parameters are given 
in Table~\ref{table:parameters}. The last two columns of Table~\ref{table:parameters} also list the derived molecular column densities 
and fractional abundances with respect to H$_2$.

\begin{table*}
\caption{Spectral line parameters, molecular column densities, and fractional abundances.}
{\scriptsize
\begin{minipage}{2\columnwidth}
\centering
\renewcommand{\footnoterule}{}
\label{table:parameters}
\begin{tabular}{c c c c c c c c c c}
\hline\hline 
Source & Transition & $v_{\rm LSR}$ & $\Delta v_{\rm LSR}$ & $T_{\rm MB}$\tablefootmark{a} & $\int T_{\rm MB} {\rm d}v$\tablefootmark{b} & $\tau$\tablefootmark{c} & $T_{\rm ex}$\tablefootmark{d} & $N$ & $x$\\
       &            & [km s$^{-1}$] & [km s$^{-1}$] & [K] & [K km s$^{-1}$] & & [K] & [cm$^{-2}$] & \\
\hline
SMM~1 & SO$(4_4-3_3)$ & $-26.51\pm0.06$ & $1.23\pm0.15$ & $0.08\pm0.01$ & $0.10\pm0.01$ & $0.008\pm0.001$ & 14.0 & $1.9\pm0.2(12)$ & $1.0\pm0.2(-10)$  \\ [1ex]
      & H$^{13}$CN$(2-1)$ & $-26.30\pm0.02$ & $1.03\pm0.05$ & $0.17\pm0.02$ & $0.18\pm0.02$ & $0.10\pm0.03$ & $7.5\pm1.3$ & $3.4\pm1.0(11)$ & $1.9\pm0.6(-11)$ \\ [1ex]
      & H$^{13}$CO$^+(2-1)$ & $-26.30\pm0.01$ & $1.13\pm0.03$ & $0.29\pm0.03$ & $0.35\pm0.04$ & $0.086\pm0.001$ & 14.0 & $7.1\pm0.2(12)$ &  $3.9\pm0.4(-10)$  \\ [1ex]
      & SiO$(4-3)$ & \ldots & \ldots & $<0.04$ & $<0.06$ & $<0.004$ & 14.0 & $<9.4(10)$ & $<5.3(-12)$ \\ [1ex]
      & HN$^{13}$C$(2-1)$ & $-26.31\pm0.02$ & $1.01\pm0.05$ & $0.17\pm0.02$ & $0.18\pm0.02$ & $0.017\pm0.001$ & 14.0 & $2.1\pm0.1(11)$ & $1.2\pm0.1(-11)$  \\ [1ex]
      & C$_2$H$(2-1)$ & $-26.20\pm0.01$ & $1.01\pm0.03$ & $0.26\pm0.03$ & $0.37\pm0.04$ & $7.31\pm0.70$ & $3.2\pm0.1$ & $2.0\pm0.2(15)$ & $1.4\pm0.2(-9)$\\ [1ex]
SMM~2 & SO$(4_4-3_3)$ & $-26.95\pm0.05$ & $0.95\pm0.10$ & $0.08\pm0.02$ & $0.08\pm0.01$ & $0.010\pm0.001$ & 12.5 & $1.7\pm0.2(12)$ & $1.7\pm0.3(-10)$ \\ [1ex]
      & H$^{13}$CN$(2-1)$ & $-26.80\pm0.01$ & $0.47\pm0.08$ & $0.07\pm0.01$ & $0.04\pm0.01$ & $0.91\pm0.35$ & $3.1\pm0.1$ & $1.0\pm0.4(12)$ & $9.9\pm4.3(-11)$ \\ [1ex]
      & H$^{13}$CO$^+(2-1)$ & $-26.90\pm0.02$ & $0.74\pm0.07$ & $0.21\pm0.03$ & $0.21\pm0.02$ & $0.072\pm0.001$ & 12.5 & $3.3\pm0.3(12)$ & $3.2\pm0.5(-10)$ \\ [1ex]
      & SiO$(4-3)$ & \ldots & \ldots & $<0.04$ & $<0.04$ & $<0.005$ & 12.5 & $<7.7(10)$ & $<7.8(-12)$ \\ [1ex]
      & HN$^{13}$C$(2-1)$ & $-26.93\pm0.03$ & $0.98\pm0.07$ & $0.17\pm0.02$ & $0.18\pm0.02$ & $0.021\pm0.001$ & 12.5 & $2.1\pm0.1(11)$ & $2.1\pm0.2(-11)$\\ [1ex]
      & C$_2$H$(2-1)$ & $-26.80\pm0.01$ & $0.77\pm0.02$ & $0.16\pm0.02$ & $0.17\pm0.02$ & $11.40\pm0.43$ & $3.0\pm0.1$ & $2.3\pm0.1(15)$ & $2.5\pm0.3(-9)$\\ [1ex]
SMM~3 & SO$(4_4-3_3)$ & $-27.70\pm0.04$ & $0.73\pm0.09$ & $0.10\pm0.02$ & $0.08\pm0.01$ & $0.011\pm0.001$ & 13.7 & $1.4\pm0.2(12)$ & $1.1\pm0.2(-10)$ \\ [1ex]
      & H$^{13}$CN$(2-1)$ & $-27.50\pm0.03$ & $0.95\pm0.09$ & $0.13\pm0.01$ & $0.13\pm0.02$ & $1.03\pm0.81$ & $3.3\pm0.3$ & $2.3\pm1.8(12)$ & $1.7\pm1.4(-10)$ \\ [1ex]
      & H$^{13}$CO$^+(2-1)$ & $-27.60\pm0.01$ & $0.98\pm0.05$ & $0.53\pm0.05$ & $0.60\pm0.06$ & $0.163\pm0.001$ & 13.7 & $1.1\pm0.1(13)$ & $8.5\pm1.0(-10)$ \\ [1ex]
      & SiO$(4-3)$ & $-26.49\pm0.18$ & $8.30\pm0.50$ & $0.07\pm0.01$ & $0.65\pm0.07$ & $0.007\pm0.001$ & 13.7 & $1.0\pm0.1(12)$ & $7.4\pm0.9(-11)$\\ [1ex]
      & HN$^{13}$C$(2-1)$ & $-27.60\pm0.01$ & $0.88\pm0.03$ & $0.38\pm0.04$ & $0.36\pm0.04$ & $0.041\pm0.001$ & 13.7 & $4.3\pm0.6(11)$ & $3.2\pm0.6(-11)$ \\ [1ex]
      & C$_2$H$(2-1)$ & $-27.70\pm0.05$ & $1.03\pm0.01$ & $0.64\pm0.07$ & $0.75\pm0.08$ & $2.43\pm0.25$ & $4.5\pm0.2$ & $6.9\pm0.7(14)$ & $3.1\pm0.5(-9)$ \\ [1ex]
BLOB~2 & SO$(4_4-3_3)$ & $-26.47\pm0.02$ & $1.04\pm0.10$ & $0.09\pm0.01$ & $0.10\pm0.01$ & $0.008\pm0.001$ & 14.9 & $1.6\pm0.2(12)$ & $2.8\pm0.4(-10)$  \\ [1ex]
      & H$^{13}$CN$(2-1)$ & $-26.10\pm0.11$ & $1.69\pm0.22$ & $0.04\pm0.01$ & $0.09\pm0.01$ & $0.10\pm0.44$\tablefootmark{e} & $4.1\pm0.3$ & $4.0\pm0.5(11)$ & $6.9\pm1.2(-11)$ \\ [1ex]
      & H$^{13}$CO$^+(2-1)$ & $-26.30\pm0.02$ & $1.14\pm0.08$ & $0.25\pm0.03$ & $0.31\pm0.03$ & $0.067\pm0.001$ & 14.9 & $6.2\pm0.4(12)$ & $1.1\pm0.1(-9)$ \\ [1ex]
      & SiO$(4-3)$ & \ldots & \ldots & $<0.04$ & $<0.08$ & $<0.004$ & 14.9 & $<1.3(11)$ & $<2.2(-11)$ \\ [1ex]
      & HN$^{13}$C$(2-1)$ & $-26.23\pm0.02$ & $1.08\pm0.05$ & $0.16\pm0.02$ & $0.18\pm0.02$ & $0.015\pm0.001$ & 14.9 & $2.3\pm0.1(11)$ & $3.7\pm0.5(-11)$\\ [1ex]
      & C$_2$H$(2-1)$ & $-26.20\pm0.02$ & $1.03\pm0.03$ & $0.18\pm0.02$ & $0.27\pm0.03$ & $7.53\pm1.04$ & $3.1\pm0.1$ & $2.1\pm0.3(15)$ & $3.5\pm0.6(-9)$\\ [1ex]
BLOB~1 & SO$(4_4-3_3)$ & $-26.76\pm0.02$ & $0.99\pm0.04$ & $0.22\pm0.03$ & $0.23\pm0.02$ & $0.030\pm0.001$ & 11.5 & $5.9\pm0.2(12)$ & $3.7\pm0.4(-10)$ \\ [1ex]
      & H$^{13}$CN$(2-1)$ & $-26.70\pm0.06$ & $1.31\pm0.13$ & $0.09\pm0.02$ & $0.16\pm0.02$ & $3.04\pm1.18$ & $3.0\pm0.1$ & $9.7\pm3.9(12)$ & $6.1\pm2.5(-10)$\\ [1ex]
      & H$^{13}$CO$^+(2-1)$ & $-26.70\pm0.02$ & $1.20\pm0.03$ & $0.31\pm0.03$ & $0.48\pm0.05$ & $0.122\pm0.001$ & 11.5 & $8.2\pm0.2(12)$ & $5.2\pm0.6(-10)$ \\ [1ex]
      & SiO$(4-3)$ & \ldots & \ldots & $<0.04$ & $<0.07$ & $<0.005$ & 11.5 & $<1.4(11)$ & $<9.3(-12)$ \\ [1ex]
      & HN$^{13}$C$(2-1)$ & $-26.72\pm0.03$ & $1.68\pm0.06$ & $0.21\pm0.02$ & $0.38\pm0.04$ & $0.029\pm0.001$ & 11.5 & $4.6\pm0.1(11)$ & $2.9\pm0.3(-11)$ \\ [1ex]
      & C$_2$H$(2-1)$ & $-26.70\pm0.02$ & $1.57\pm0.04$ & $0.21\pm0.03$ & $0.47\pm0.05$ & $7.90\pm0.80$ & $3.1\pm0.1$ & $3.3\pm0.3(15)$ & $3.2\pm0.5(-9)$\\ [1ex]
SMM~4 & SO$(4_4-3_3)$ & $-26.90\pm0.02$ & $1.03\pm0.01$ & $0.18\pm0.02$ & $0.19\pm0.02$ & $0.020\pm0.001$ & 13.2 & $3.9\pm0.1(12)$ & $1.5\pm0.2(-10)$\\ [1ex]
      & H$^{13}$CN$(2-1)$ & $-27.30\pm0.04$ & $1.57\pm0.12$ & $0.10\pm0.01$ & $0.18\pm0.02$ & $0.22\pm0.84$\tablefootmark{e} & $4.3\pm0.1$ & $8.2\pm0.6(11)$ & $3.3\pm0.4(-11)$ \\ [1ex]
      & H$^{13}$CO$^+(2-1)$ & $-27.40\pm0.01$ & $1.11\pm0.03$ & $0.46\pm0.05$ & $0.73\pm0.07$ & $0.149\pm0.001$ & 13.2 & $1.1\pm0.03(13)$ & $4.4\pm0.5(-10)$ \\ [1ex]
      & SiO$(4-3)$ & \ldots & \ldots & $<0.04$ & $<0.07$ & $<0.004$ & 13.2 & $<1.2(11)$ & $<5.0(-12)$\\ [1ex]
      & HN$^{13}$C$(2-1)$ & $-27.46\pm0.02$ & $1.52\pm0.04$ & $0.26\pm0.03$ & $0.42\pm0.04$ & $0.029\pm0.001$ & 13.2 & $5.0\pm0.1(11)$ & $2.0\pm0.2(-11)$\\ [1ex]
      & C$_2$H$(2-1)$ & $-27.60\pm0.01$ & $1.27\pm0.02$ & $0.55\pm0.06$ & $0.82\pm0.09$ & $3.17\pm0.27$ & $4.1\pm0.1$ & $1.1\pm0.1(15)$ & $1.9\pm0.3(-9)$\\ [1ex]
SMM~5 & SO$(4_4-3_3)$ & $-27.50\pm0.08$ & $1.45\pm0.17$ & $-0.05\pm0.01$ & $-0.08\pm0.01$ & $0.005\pm0.001$ & 14.0 & $1.4\pm0.2(12)$ & $2.6\pm0.5(-10)$ \\ [1ex]
      & H$^{13}$CN$(2-1)$ & $-27.50\pm0.05$ & $0.81\pm0.10$ & $-0.07\pm0.01$ & $-0.08\pm0.01$ & $2.90\pm1.59$ & $2.9\pm0.1$ & $5.8\pm3.2(12)$ & $1.1\pm0.6(-9)$ \\ [1ex]
      & H$^{13}$CO$^+(2-1)$ & $-27.45\pm0.02$ & $0.96\pm0.05$ & $-0.30\pm0.03$ & $-0.37\pm0.04$ & $0.089\pm0.001$ & 14.0 & $6.2\pm0.3(12)$ & $1.2\pm0.2(-9)$ \\ [1ex]
      & SiO$(4-3)$ & \ldots & \ldots & $<0.03$ & $<0.05$ & $<0.003$ & 14.0 & $<8.2(10)$ & $<1.7(-11)$\\ [1ex]
      & HN$^{13}$C$(2-1)$ & $-27.62\pm0.02$ & $0.88\pm0.05$ & $-0.18\pm0.02$ & $-0.17\pm0.02$ & $0.018\pm0.001$ & 14.0 & $2.0\pm0.1(11)$ & $3.9\pm0.6(-11)$\\ [1ex]
      & C$_2$H$(2-1)$ & $-27.70\pm0.01$ & $0.82\pm0.02$ & $0.40\pm0.04$ & $0.38\pm0.06$ & $2.84\pm0.65$ & $3.8\pm0.2$ & $6.2\pm1.4(14)$ & $3.5\pm0.9(-9)$\\ [1ex]
\hline 
\end{tabular} 
\tablefoot{Columns 3--10 give the local standard of rest (LSR) radial velocity ($v_{\rm LSR}$), full width at half maximum (FWHM; $\Delta v$), peak intensity ($T_{\rm MB}$), integrated intensity ($\int T_{\rm MB} {\rm d}v$), optical thickness ($\tau$), and excitation temperature ($T_{\rm ex}$) of the line, and the molecule's beam-averaged column density and fractional abundance with respect to H$_2$. The latter two quantities are given in the form $a\pm b(c)$, which stands for $(a\pm b) \times 10^c$.\tablefoottext{a}{A $3\sigma$ intensity upper limit is quoted for the non-detections.}\tablefoottext{b}{For the non-detections, the upper limit was calculated by multiplying the quoted $3\sigma$ intensity upper limit by the broadest linewidth among the detected lines in the case of SiO and C$_2$H, while for the H$^{13}$CN non-detections the width of the HN$^{13}$C line was used.}\tablefoottext{c}{For the H$^{13}$CN$(2-1)$ and C$_2$H$(2-1)$ lines, the quoted optical thicknesses are the total optical thicknesses derived through fitting the hyperfine structure of the lines. For the sources where H$^{13}$CN$(2-1)$ or C$_2$H$(2-1)$ were not detected, we adopted the average $T_{\rm ex}[{\rm H^{13}CN}(2-1)]$ and $T_{\rm ex}[{\rm C_2H}(2-1)]$ values ($3.7\pm0.4$~K and $3.5\pm0.1$~K, respectively) derived from the detected H$^{13}$CN and C$_2$H lines, and derived the optical thickness upper limits using those values. For all the other lines, the quoted value of optical thickness refers to the peak optical thickness. In case the line exhibits hyperfine structure, the latter value refers to the strongest hyperfine component.}\tablefoottext{d}{Apart from the H$^{13}$CN$(2-1)$ and C$_2$H$(2-1)$ lines, where $T_{\rm ex}$ could be derived from the optical thickness (Eq.~(\ref{eqn:Tex})) the transition was assumed to be thermalised at the dust temperature of the clump (see Table~\ref{table:sources}).}\tablefoottext{e}{The uncertainty of the derived optical thickness is larger than the nominal value. Hence, the uncertainty of $\tau$ was not propagated to the formal uncertainty of the corresponding $T_{\rm ex}$ value.} }
\end{minipage}}
\end{table*}

\addtocounter{table}{-1}

\begin{table*}
\caption{continued.}
{\scriptsize
\begin{minipage}{2\columnwidth}
\centering
\renewcommand{\footnoterule}{}
\label{table:otherparameters}
\begin{tabular}{c c c c c c c c c c}
\hline\hline 
Source & Transition & $v_{\rm LSR}$ & $\Delta v_{\rm LSR}$ & $T_{\rm MB}$\tablefootmark{a} & $\int T_{\rm MB} {\rm d}v$\tablefootmark{b} & $\tau$\tablefootmark{c} & $T_{\rm ex}$\tablefootmark{d} & $N$ & $x$\\
       &            & [km s$^{-1}$] & [km s$^{-1}$] & [K] & [K km s$^{-1}$] & & [K] & [cm$^{-2}$] & \\
\hline
IRAS 13037-6112 & SO$(4_4-3_3)$ & $-27.31\pm0.02$ & $0.96\pm0.04$ & $0.25\pm0.03$ & $0.25\pm0.03$ & $0.016\pm0.001$ & 20.0 & $3.1\pm0.1(12)$ & $3.1\pm0.4(-10)$ \\ [1ex]
      & H$^{13}$CN$(2-1)$ & $-27.40\pm0.03$ & $0.96\pm0.06$ & $0.10\pm0.02$ & $0.14\pm0.02$ & $5.67\pm1.14$ & $2.9\pm0.1$ & $1.3\pm0.3(13)$ & $1.3\pm0.3(-9)$ \\ [1ex]
      & H$^{13}$CO$^+(2-1)$ & $-27.30\pm0.01$ & $0.83\pm0.05$ & $0.45\pm0.05$ & $0.48\pm0.05$ & $0.083\pm0.001$ & 20.0 & $8.6\pm0.5(12)$ & $8.7\pm1.1(-10)$ \\ [1ex]
      & SiO$(4-3)$ & \ldots & \ldots & $<0.04$ & $<0.05$ & $<0.003$ & 20.0 & $<6.2(10)$ & $<6.6(-12)$\\ [1ex]
      & HN$^{13}$C$(2-1)$ & $-27.34\pm0.03$ & $0.91\pm0.07$ & $0.15\pm0.02$ & $0.15\pm0.02$ & $0.010\pm0.001$ & 20.0 & $1.9\pm0.1(11)$ & $1.9\pm0.2(-11)$ \\ [1ex]
      & C$_2$H$(2-1)$ & $-27.40\pm0.01$ & $0.89\pm0.02$ & $0.41\pm0.04$ & $0.44\pm0.05$ & $4.14\pm0.44$ & $3.7\pm0.1$ & $9.7\pm0.1(14)$ & $1.5\pm0.2(-9)$\\ [1ex]
SMM~6 & SO$(4_4-3_3)$ & $-26.73\pm0.03$ & $0.50\pm0.06$ & $-0.09\pm0.02$ & $-0.05\pm0.01$ & $0.009\pm0.001$ & 14.1 & $8.4\pm1.0(11)$ & $6.6\pm1.1(-11)$ \\ [1ex]
      & H$^{13}$CN$(2-1)$ & \ldots & \ldots & $<0.04$\tablefootmark{f} & $<0.06$\tablefootmark{f} & $<0.23$ & $3.7\pm0.4$ & $<7.7(11)$ & $<6.2(-11)$ \\ [1ex]
      & H$^{13}$CO$^+(2-1)$ & $-28.00\pm0.03$ & $0.44\pm0.08$ & $-0.16\pm0.03$ & $-0.10\pm0.01$ & $0.046\pm0.001$ & 14.1 & $1.5\pm0.3(12)$ & $1.2\pm0.2(-10)$ \\ [1ex]
      & SiO$(4-3)$ & \ldots & \ldots & $<0.04$ & $<0.06$ & $<0.004$ & 14.1 & $<9.7(10)$ & $<7.8(-12)$ \\ [1ex]
      & HN$^{13}$C$(2-1)$ & $-27.73\pm0.08$ & $1.25\pm0.18$ & $-0.08\pm0.01$ & $-0.11\pm0.02$ & $0.008\pm0.001$ & 14.1 & $1.3\pm0.1(11)$ & $1.0\pm0.1(-11)$\\ [1ex]
      & C$_2$H$(2-1)$ & $-28.00\pm0.01$ & $0.71\pm0.03$ & $-0.19\pm0.02$ & $-0.16\pm0.02$ & $3.35\pm1.01$ & $3.2\pm0.1$ & $6.3\pm1.9(14)$ & $6.3\pm2.0(-10)$\\ [1ex]
SMM~7 & SO$(4_4-3_3)$ & $-26.96\pm0.07$ & $0.51\pm0.14$ & $-0.04\pm0.01$ & $-0.02\pm0.01$ & $0.004\pm0.001$ & 14.5 & $3.7\pm1.0(11)$ & $4.0\pm1.2(-11)$\\ [1ex]
      & H$^{13}$CN$(2-1)$ & \ldots & \ldots & $<0.03$ & $<0.04$ & $<0.17$ & $3.7\pm0.4$ & $<5.1(11)$ & $<5.8(-11)$ \\ [1ex]
      & H$^{13}$CO$^+(2-1)$ & $-27.40\pm0.03$ & $0.62\pm0.08$ & $-0.08\pm0.02$ & $-0.05\pm0.01$ & $0.022\pm0.001$ & 14.5 & $1.1\pm0.1(12)$ & $1.2\pm0.2(-10)$ \\ [1ex]
      & SiO$(4-3)$ & \ldots & \ldots & $<0.04$ & $<0.05$ & $<0.004$ & 14.5 & $<8.5(10)$ & $<9.7(-12)$ \\ [1ex]
      & HN$^{13}$C$(2-1)$ & $-26.94\pm0.07$ & $1.14\pm0.14$ & $-0.04\pm0.01$ & $-0.05\pm0.01$ & $0.004\pm0.001$ & 14.5 & $5.9\pm0.9(10)$ & $6.5\pm1.3(-12)$\\ [1ex]
      & C$_2$H$(2-1)$ & \ldots & \ldots & $<0.04$ & $<0.05$ & $<0.32$ & $3.5\pm0.1$ & $<1.1(14)$ & $<2.3(-10)$ \\ [1ex]
IRAS 13039-6108 & SO$(4_4-3_3)$ & $-25.98\pm0.05$ & $0.91\pm0.12$ & $0.07\pm0.02$ & $0.07\pm0.01$ & $0.004\pm0.001$ & 22.2 & $7.6\pm1.0(11)$ & $9.0\pm1.5(-11)$ \\ [1ex]
      & H$^{13}$CN$(2-1)$ & $-26.00\pm0.16$ & $1.08\pm0.30$ & $0.05\pm0.02$ & $0.06\pm0.01$ & $1.31\pm0.66$ & $2.9\pm0.1$ & $3.5\pm2.0(12)$ & $4.1\pm2.4(-10)$ \\ [1ex]
      & H$^{13}$CO$^+(2-1)$ & $-26.10\pm0.01$ & $0.92\pm0.03$ & $0.26\pm0.03$ & $0.26\pm0.03$ & $0.042\pm0.001$ & 22.2 & $5.7\pm0.2(12)$ & $6.8\pm0.8(-10)$ \\ [1ex]
      & SiO$(4-3)$ & \ldots & \ldots & $<0.05$ & $<0.06$ & $<0.003$ & 22.2 & $<8.6(10)$ & $<1.1(-11)$\\ [1ex]
      & HN$^{13}$C$(2-1)$ & $-26.08\pm0.02$ & $0.75\pm0.04$ & $0.16\pm0.02$ & $0.13\pm0.01$ & $0.009\pm0.001$ & 22.2 & $1.7\pm0.1(11)$ & $2.0\pm0.2(-11)$ \\ [1ex]
      & C$_2$H$(2-1)$ & $-26.00\pm0.01$ & $1.11\pm0.03$ & $0.36\pm0.04$ & $0.52\pm0.06$ & $3.69\pm0.48$ & $3.6\pm0.1$ & $1.1\pm0.1(15)$ & $1.4\pm0.2(-9)$ \\ [1ex]
SMM~8 & SO$(4_4-3_3)$ & $-26.76\pm0.08$ & $0.73\pm0.24$ & $0.04\pm0.01$ & $0.03\pm0.01$ & $0.004\pm0.001$ & 13.9 & $5.6\pm1.8(11)$ & $6.5\pm2.3(-11)$ \\ [1ex]
      & H$^{13}$CN$(2-1)$ & $-26.70\pm0.07$ & $0.75\pm0.12$ & $0.05\pm0.01$ & $0.05\pm0.01$ & $4.20\pm2.74$ & $2.8\pm0.1$ & $7.9\pm5.3(12)$ & $9.2\pm6.2(-10)$ \\ [1ex]
      & H$^{13}$CO$^+(2-1)$ & $-26.70\pm0.01$ & $1.00\pm0.04$ & $0.29\pm0.03$ & $0.32\pm0.03$ & $0.086\pm0.001$ & 13.9 & $6.2\pm0.3(12)$ & $7.3\pm0.9(-10)$ \\ [1ex]
      & SiO$(4-3)$ & \ldots & \ldots & $<0.04$ & $<0.04$ & $<0.004$ & 13.9 & $<7.1(10)$ & $<8.8(-12)$\\ [1ex]
      & HN$^{13}$C$(2-1)$ & $-26.65\pm0.02$ & $0.92\pm0.05$ & $0.18\pm0.02$ & $0.18\pm0.02$ & $0.019\pm0.001$ & 13.9 & $2.1\pm0.1(11)$ & $2.5\pm0.3(-11)$ \\ [1ex]
      & C$_2$H$(2-1)$ & $-26.70\pm0.01$ & $0.91\pm0.03$ & $0.32\pm0.03$ & $0.38\pm0.05$ & $4.56\pm0.92$ & $3.4\pm0.1$ & $1.1\pm0.2(15)$ & $2.3\pm0.5(-9)$\\ [1ex]
SMM~9 & SO$(4_4-3_3)$ & $-26.92\pm0.05$ & $1.73\pm0.14$ & $0.04\pm0.01$ & $0.08\pm0.01$ & $0.003\pm0.001$ & 16.3 & $1.1\pm0.1(12)$ & $1.0\pm0.1(-10)$ \\ [1ex]
      & H$^{13}$CN$(2-1)$ & $-26.50\pm0.06$ & $1.23\pm0.20$ & $0.10\pm0.01$ & $0.14\pm0.02$ & $0.38\pm1.82$\tablefootmark{e} & $3.8\pm0.1$ & $1.1\pm0.2(12)$ & $1.0\pm0.2(-10)$ \\ [1ex]
      & H$^{13}$CO$^+(2-1)$ & $-26.60\pm0.01$ & $1.22\pm0.01$ & $0.33\pm0.04$ & $0.47\pm0.05$ & $0.079\pm0.001$ & 16.3 & $8.8\pm0.1(12)$ & $8.3\pm0.9(-10)$ \\ [1ex]
      & SiO$(4-3)$ & \ldots & \ldots & $<0.04$ & $<0.04$ & $<0.003$ & 16.3 & $<1.2(11)$ & $<1.1(-11)$ \\ [1ex]
      & HN$^{13}$C$(2-1)$ & $-26.61\pm0.01$ & $1.05\pm0.03$ & $0.21\pm0.02$ & $0.23\pm0.02$ & $0.018\pm0.001$ & 16.3 & $2.8\pm0.1(11)$ & $2.6\pm0.3(-11)$  \\ [1ex]
      & C$_2$H$(2-1)$ & $-26.60\pm0.01$ & $1.22\pm0.02$ & $0.47\pm0.05$ & $0.72\pm0.08$ & $3.46\pm0.32$ & $3.9\pm0.1$ & $1.1\pm0.2(15)$ & $2.7\pm0.4(-9)$\\ [1ex]
IRAS 13042-6105 & SO$(4_4-3_3)$ & $-26.42\pm0.12$ & $0.90\pm0.33$ & $0.04\pm0.01$ & $0.04\pm0.01$ & $0.004\pm0.001$ & 15.1 & $6.1\pm2.2(11)$ & $1.2\pm0.5(-10)$ \\ [1ex]
      & H$^{13}$CN$(2-1)$ & \ldots & \ldots & $<0.04$ & $<0.03$ & $<0.23$ & $3.7\pm0.4$ & $<4.1(11)$ & $<8.8(-11)$\\ [1ex]
      & H$^{13}$CO$^+(2-1)$ & $-26.60\pm0.03$ & $0.79\pm0.08$ & $0.18\pm0.02$ & $0.20\pm0.02$ & $0.048\pm0.001$ & 15.1 & $3.1\pm0.3(12)$ & $6.1\pm1.0(-10)$ \\ [1ex]
      & SiO$(4-3)$ & \ldots & \ldots & $<0.05$ & $<0.06$ & $<0.005$ & 15.1 & $<1.1(11)$ & $<2.1(-11)$\\ [1ex]
      & HN$^{13}$C$(2-1)$ & $-26.49\pm0.02$ & $0.72\pm0.05$ & $0.16\pm0.02$ & $0.12\pm0.01$ & $0.015\pm0.001$ & 15.1 & $1.4\pm0.3(11)$ & $2.9\pm0.6(-11)$ \\ [1ex]
      & C$_2$H$(2-1)$ & $-26.50\pm0.01$ & $0.79\pm0.03$ & $0.20\pm0.02$ & $0.20\pm0.02$ & $5.87\pm1.09$ & $3.2\pm0.1$ & $1.2\pm0.2(15)$ & $2.8\pm0.6(-9)$\\ [1ex]
\hline 
\end{tabular} 
\tablefoot{\tablefoottext{f}{The line appears in absorption, and hence the quoted intensity upper limits should be interpreted as absolute values.}  }
\end{minipage}}
\end{table*}


\begin{thebibliography}{}

\bibitem[Allen et al. 1980]{allen1980} Allen, T.~L., Goddard, J.~D., \& Schaefer, H.~F.\ 1980, \jcp, 73, 3255

\bibitem[Astropy Collaboration et al. 2013]{astropy2013} Astropy Collaboration, Robitaille, T.~P., Tollerud, E.~J., et al.\ 2013, \aap, 558, A33

\bibitem[2018]{astropy2018} Astropy Collaboration, Price-Whelan, A.~M., Sip{\H{o}}cz, B.~M., et al.\ 2018, \aj, 156, 123

\bibitem[Battersby et al. 2010]{battersby2010} Battersby, C., Bally, J., Jackson, J.~M., et al.\ 2010, \apj, 721, 222

\bibitem[Belitsky et al. 2018]{belitsky2018} Belitsky, V., Lapkin, I., Fredrixon, M., et al.\ 2018, \aap, 612, A23

\bibitem[Beltr{\'a}n et al. 2006]{beltran2006} Beltr{\'a}n, M.~T., Brand, J., Cesaroni, R., et al.\ 2006, \aap, 447, 221


\bibitem[Beuther \& Steinacker 2007]{beuther2007} Beuther, H., \& Steinacker, J.\ 2007, \apjl, 656, L85

\bibitem[Beuther et al. 2008]{beuther2008} Beuther, H., Semenov, D., Henning, T., et al.\ 2008, \apjl, 675, L33

\bibitem[Beuther et al. 2015]{beuther2015} Beuther, H., Ragan, S.~E., Johnston, K., et al.\ 2015, \aap, 584, A67

\bibitem[Busquet et al. 2016]{busquet2016} Busquet, G., Estalella, R., Palau, A., et al.\ 2016, \apj, 819, 139

\bibitem[Chambers et al. 2009]{chambers2009} Chambers, E.~T., Jackson, J.~M., Rathborne, J.~M., et al.\ 2009, \apjs, 181, 360



\bibitem[Colzi et al. 2018]{colzi2018} Colzi, L., Fontani, F., Caselli, P., et al.\ 2018, \aap, 609, A129

\bibitem[Cosentino et al. 2018]{cosentino2018} Cosentino, G., Jim{\'e}nez-Serra, I., Henshaw, J.~D., et al.\ 2018, \mnras, 474, 3760

\bibitem[Doty \& Neufeld 1997]{doty1997} Doty, S.~D., \& Neufeld, D.~A.\ 1997, \apj, 489, 122

\bibitem[Dumke \& Mac-Auliffe 2010]{dumke2010} Dumke, M., \& Mac-Auliffe, F.\ 2010, \procspie, 77371J

\bibitem[Egan et al. 1998]{egan1998} Egan, M.~P., Shipman, R.~F., Price, S.~D., et al.\ 1998, \apjl, 494, L199

\bibitem[Favre et al. 2014]{favre2014} Favre, C., Carvajal, M., Field, D., et al.\ 2014, \apjs, 215, 25

\bibitem[Finn et al. 2013]{finn2013} Finn, S.~C., Jackson, J.~M., Rathborne, J.~M., et al.\ 2013, \apj, 764, 102

\bibitem[Fixsen 2009]{fixsen2009} Fixsen, D.~J.\ 2009, \apj, 707, 916

\bibitem[Frerking et al. 1979]{frerking1979} Frerking, M.~A., Langer, W.~D., \& Wilson, R.~W.\ 1979, \apjl, 232, L65

\bibitem[Fuchs et al. 2004]{fuchs2004} Fuchs, U., Br{\"u}nken, S., Fuchs, G.~W., et al.\ 2004, Zeitschrift Naturforschung Teil A, 59, 861

\bibitem[Fuente et al. 1993]{fuente1993} Fuente, A., Martin-Pintado, J., Cernicharo, J., et al.\ 1993, \aap, 276, 473

\bibitem[Gerner et al. 2014]{gerner2014} Gerner, T., Beuther, H., Semenov, D., et al.\ 2014, \aap, 563, A97

\bibitem[Giannetti et al. 2013]{giannetti2013} Giannetti, A., Brand, J., S{\'a}nchez-Monge, {\'A}., et al.\ 2013, \aap, 556, A16

\bibitem[Giannetti et al. 2015]{giannetti2015} Giannetti, A., Wyrowski, F., Leurini, S., et al.\ 2015, \aap, 580, L7

\bibitem[Godard et al. 2010]{godard2010} Godard, B., Falgarone, E., Gerin, M., et al.\ 2010, \aap, 520, A20

\bibitem[Goldsmith et al. 1986]{goldsmith1986} Goldsmith, P.~F., Irvine, W.~M., Hjalmarson, A., et al.\ 1986, \apj, 310, 383

\bibitem[G{\"u}sten et al. 2006]{gusten2006} G{\"u}sten, R., Nyman, L. {\r{A}}., Schilke, P., et al.\ 2006, \aap, 454, L13

\bibitem[Hacar et al. 2020]{hacar2020} Hacar, A., Bosman, A.~D., \& van Dishoeck, E.~F.\ 2020, \aap, 635, A4

\bibitem[Heitsch et al. 2006]{heitsch2006} Heitsch, F., Slyz, A.~D., Devriendt, J.~E.~G., et al.\ 2006, \apj, 648, 1052

\bibitem[Helsel 2005]{helsel2005} Helsel, D.~R. 2005, Nondetects And Data Analysis: Statistics for Censored
Environmental Data (New York: John Wiley and Sons)

\bibitem[Henshaw et al. 2013]{henshaw2013} Henshaw, J.~D., Caselli, P., Fontani, F., et al.\ 2013, \mnras, 428, 3425

\bibitem[Henshaw et al. 2016]{henshaw2016} Henshaw, J.~D., Caselli, P., Fontani, F., et al.\ 2016, \mnras, 463, 146

\bibitem[Herbst 1978]{herbst1978} Herbst, E.\ 1978, \apj, 222, 508

\bibitem[Herbst \& Klemperer 1973]{herbst1973} Herbst, E., \& Klemperer, W.\ 1973, \apj, 185, 505

\bibitem[Herbst et al. 2000]{herbst2000} Herbst, E., Terzieva, R., \& Talbi, D.\ 2000, \mnras, 311, 869

\bibitem[Hily-Blant et al. 2010]{hilyblant2010} Hily-Blant, P., Walmsley, M., Pineau Des For{\^e}ts, G., et al.\ 2010, \aap, 513, A41

\bibitem[Hirota et al. 1998]{hirota1998} Hirota, T., Yamamoto, S., Mikami, H., et al.\ 1998, \apj, 503, 717

\bibitem[Hirota et al. 2003]{hirota2003} Hirota, T., Ikeda, M., \& Yamamoto, S.\ 2003, \apj, 594, 859

\bibitem[Hoq et al. 2017]{hoq2017} Hoq, S., Clemens, D.~P., Guzm{\'a}n, A.~E., et al.\ 2017, \apj, 836, 199

\bibitem[Hunter 2007]{hunter2007} Hunter, J.~D.\ 2007, Computing in Science and Engineering, 9, 90

\bibitem[Jackson et al. 2008]{jackson2008} Jackson, J.~M., Finn, S.~C., Rathborne, J.~M., et al.\ 2008, \apj, 680, 349

\bibitem[Jackson et al. 2010]{jackson2010} Jackson, J.~M., Finn, S.~C., Chambers, E.~T., et al.\ 2010, \apjl, 719, L185

\bibitem[Jim{\'e}nez-Serra et al. 2010]{jimenezserra2010} Jim{\'e}nez-Serra, I., Caselli, P., Tan, J.~C., et al.\ 2010, \mnras, 406, 187

\bibitem[Jin et al. 2015]{jin2015} Jin, M., Lee, J.-E., \& Kim, K.-T.\ 2015, \apjs, 219, 2

\bibitem[Juvela et al. 2018]{juvela2018} Juvela, M., Guillet, V., Liu, T., et al.\ 2018, \aap, 620, A26

\bibitem[Kaplan \& Meier 1958]{kaplan1958} Kaplan, E.~L., \& Meier, P.\ 1958, Journal of the American Statistical Association, 53:457–81

\bibitem[Kauffmann \& Pillai 2010]{kauffmann2010} Kauffmann, J., \& Pillai, T.\ 2010, \apjl, 723, L7

\bibitem[Kauffmann et al. 2008]{kauffmann2008} Kauffmann, J., Bertoldi, F., Bourke, T.~L., et al.\ 2008, \aap, 487, 993

\bibitem[Langer et al. 1984]{langer1984} Langer, W.~D., Graedel, T.~E., Frerking, M.~A., et al.\ 1984, \apj, 277, 581

\bibitem[Lee 2017]{lee2017} Lee, L. 2017, NADA: Nondetects and Data Analysis for Environmental Data. R package version 1.6-1. \url{https://CRAN.R-project.org/package=NADA}


\bibitem[Li et al. 2019]{li2019} Li, S., Wang, J., Fang, M., et al.\ 2019, \apj, 878, 29

\bibitem[Liu et al. 2013]{liu2013} Liu, X.-L., Wang, J.-J., \& Xu, J.-L.\ 2013, \mnras, 431, 27

\bibitem[Liszt 2017]{liszt2017} Liszt, H.~S.\ 2017, \apj, 835, 138

\bibitem[Loison et al. 2014]{loison2014} Loison, J.-C., Wakelam, V., \& Hickson, K.~M.\ 2014, \mnras, 443, 398

\bibitem[Mangum \& Shirley 2015]{mangum2015} Mangum, J.~G., \& Shirley, Y.~L.\ 2015, \pasp, 127, 266

\bibitem[Maret et al. 2011]{maret2011} Maret, S., Hily-Blant, P., Pety, J., et al.\ 2011, \aap, 526, A47

\bibitem[Mattern et al. 2018]{mattern2018} Mattern, M., Kainulainen, J., Zhang, M., et al.\ 2018, \aap, 616, A78

\bibitem[McKee \& Krumholz 2010]{mckee2010} McKee, C.~F., \& Krumholz, M.~R.\ 2010, \apj, 709, 308

\bibitem[McKinney 2010]{mckinney2010} McKinney, W.\ 2010, Proceedings of the 9th Python in Science Conference, 51

\bibitem[Miettinen 2012]{miettinen2012} Miettinen, O.\ 2012, \aap, 540, A104

\bibitem[Miettinen 2014]{miettinen2014} Miettinen, O.\ 2014, \aap, 562, A3

\bibitem[Miettinen 2018]{miettinen2018} Miettinen, O.\ 2018, \aap, 609, A123

\bibitem[Miettinen \& Harju 2010]{miettinenharju2010} Miettinen, O., \& Harju, J.\ 2010, \aap, 520, A102

\bibitem[Millar 2015]{millar2015} Millar, T.~J.\ 2015, Plasma Sources Science Technology, 24, 043001


\bibitem[Mladenovi{\'c} \& Roueff 2017]{mladenic2017} Mladenovi{\'c}, M., \& Roueff, E.\ 2017, \aap, 605, A22

\bibitem[Moore et al. 2018]{moore2018} Moore, D.~S., Notz, W.~I., \& Fligner, M.~A.\ 2018, \textit{The Basic Practice of Statistics} (Eight Edition), New York, NY: W. H. Freeman and Company 

\bibitem[Motte et al. 2018]{motte2018} Motte, F., Bontemps, S., \& Louvet, F.\ 2018, \araa, 56, 41

\bibitem[M{\"u}ller et al. 2005]{muller2005} M{\"u}ller, H.~S.~P., Schl{\"o}der, F., Stutzki, J., et al.\ 2005, Journal of Molecular Structure, 742, 215

\bibitem[Nagy et al. 2015]{nagy2015} Nagy, Z., Ossenkopf, V., van der Tak, F.~F.~S., et al.\ 2015, \aap, 578, A124

\bibitem[Ossenkopf \& Henning 1994]{ossenkopf1994} Ossenkopf, V., \& Henning, T.\ 1994, \aap, 291, 943

\bibitem[Padoan et al. 2020]{padoan2020} Padoan, P., Pan, L., Juvela, M., et al.\ 2020, \apj, \textit{submitted}, {\tt arXiv:1911.04465}

\bibitem[Padovani et al. 2009 ]{padovani2009} Padovani, M., Walmsley, C.~M., Tafalla, M., et al.\ 2009, \aap, 505, 1199

\bibitem[Padovani et al. 2011]{padovani2011} Padovani, M., Walmsley, C.~M., Tafalla, M., et al.\ 2011, \aap, 534, A77

\bibitem[Pearson \& Schaefer 1974]{pearson1974} Pearson, P.~K., \& Schaefer, H.~F.\ 1974, \apj, 192, 33

\bibitem[P\'erault et al. 1996]{perault1996} P\'erault, M., Omont, A., Simon, G., et al.\ 1996, \aap, 315, L165

\bibitem[Peretto \& Fuller 2009]{peretto2009} Peretto, N., \& Fuller, G.~A.\ 2009, \aap, 505, 405

\bibitem[Pickett et al. 1998]{pickett1998} Pickett, H.~M., Poynter, R.~L., Cohen, E.~A., et al.\ 1998, \jqsrt, 60, 883

\bibitem[Pilbratt et al. 2010]{pilbratt2010} Pilbratt, G.~L., Riedinger, J.~R., Passvogel, T., et al.\ 2010, \aap, 518, L1

\bibitem[Pillai et al. 2006]{pillai2006} Pillai, T., Wyrowski, F., Menten, K.~M., et al.\ 2006, \aap, 447, 929

\bibitem[R Core Team 2019]{R2019} R Core Team 2019. R: A language and environment for statistical computing. R Foundation for Statistical
Computing, Vienna, Austria. URL \url{http://www.R-project.org/}

\bibitem[Ragan et al. 2013]{ragan2013} Ragan, S.~E., Henning, T., \& Beuther, H.\ 2013, \aap, 559, A79

\bibitem[Rathborne et al. 2005]{rathborne2005} Rathborne, J.~M., Jackson, J.~M., Chambers, E.~T., et al.\ 2005, \apjl, 630, L181

\bibitem[Rathborne et al. 2006]{rathborne2006} Rathborne, J.~M., Jackson, J.~M., \& Simon, R.\ 2006, \apj, 641, 389

\bibitem[Rathborne et al. 2010]{rathborne2010} Rathborne, J.~M., Jackson, J.~M., Chambers, E.~T., et al.\ 2010, \apj, 715, 310

\bibitem[Rathborne et al. 2011]{rathborne2011} Rathborne, J.~M., Garay, G., Jackson, J.~M., et al.\ 2011, \apj, 741, 120


\bibitem[Rawlings et al. 2004]{rawlings2004} Rawlings, J.~M.~C., Redman, M.~P., Keto, E., et al.\ 2004, \mnras, 351, 1054

\bibitem[Reitblat 1980]{reitblat1980} Reitblat, A.~A.\ 1980, Soviet Astronomy Letters, 6, 406

\bibitem[Roberts et al. 2012]{roberts2012} Roberts, J.~F., Jim{\'e}nez-Serra, I., Gusdorf, A., et al.\ 2012, \aap, 544, A150

\bibitem[Ruze 1952]{ruze1952} Ruze, J.\ 1952, Il Nuovo Cimento, 9, 364

\bibitem[Sakai et al. 2010]{sakai2010} Sakai, T., Sakai, N., Hirota, T., et al.\ 2010, \apj, 714, 1658

\bibitem[Sakai et al. 2012]{saki2012} Sakai, T., Sakai, N., Furuya, K., et al.\ 2012, \apj, 747, 140

\bibitem[S{\'a}nchez-Monge et al. 2013]{sanchez2013} S{\'a}nchez-Monge, {\'A}., Beltr{\'a}n, M.~T., Cesaroni, R., et al.\ 2013, \aap, 550, A21

\bibitem[Sanhueza et al. 2012]{sanhueza2012} Sanhueza, P., Jackson, J.~M., Foster, J.~B., et al.\ 2012, \apj, 756, 60

\bibitem[Saral et al. 2018]{saral2018} Saral, G., Audard, M., \& Wang, Y.\ 2018, \aap, 620, A158

\bibitem[Sarrasin et al. 2010]{sarrasin2010} Sarrasin, E., Abdallah, D.~B., Wernli, M., et al.\ 2010, \mnras, 404, 518

\bibitem[Schilke et al. 1992]{schilke1992} Schilke, P., Walmsley, C.~M., Pineau Des Forets, G., et al.\ 1992, \aap, 256, 595

\bibitem[Schmid-Burgk et al. 2004]{schmid2004} Schmid-Burgk, J., Muders, D., M{\"u}ller, H.~S.~P., et al.\ 2004, \aap, 419, 949

\bibitem[Schuller et al. 2009]{schuller2009} Schuller, F., Menten, K.~M., Contreras, Y., et al.\ 2009, \aap, 504, 415

\bibitem[Simon et al. 2006]{simon2006} Simon, R., Jackson, J.~M., Rathborne, J.~M., et al.\ 2006, \apj, 639, 227

\bibitem[Siringo et al. 2009]{siringo2009} Siringo, G., Kreysa, E., Kov{\'a}cs, A., et al.\ 2009, \aap, 497, 945

\bibitem[Soam et al. 2019]{soam2019} Soam, A., Liu, T., Andersson, B.-G., et al.\ 2019, \apj, 883, 95

\bibitem[Svoboda et al. 2019]{svoboda2019} Svoboda, B.~E., Shirley, Y.~L., Traficante, A., et al.\ 2019, \apj, 886, 36

\bibitem[Tang et al. 2019]{tang2019} Tang, Y.-W., Koch, P.~M., Peretto, N., et al.\ 2019, \apj, 878, 10

\bibitem[Taniguchi et al. 2019]{taniguchi2019} Taniguchi, K., Saito, M., Sridharan, T.~K., et al.\ 2019, \apj, 872, 154

\bibitem[Turner 1991]{turner1991} Turner, B.~E.\ 1991, \apjs, 76, 617


\bibitem[Turner 2001]{turner2001} Turner, B.~E.\ 2001, \apjs, 136, 579

\bibitem[Turner et al. 2000]{turner2000} Turner, B.~E., Herbst, E., \& Terzieva, R.\ 2000, \apjs, 126, 427

\bibitem[van der Tak et al. 2009]{vandertak2009} van der Tak, F.~F.~S., M{\"u}ller, H.~S.~P., Harding, M.~E., et al.\ 2009, \aap, 507, 347

\bibitem[van der Walt et al. 2011]{vanderwawlt2011} van der Walt, S., Colbert, S.~C., \& Varoquaux, G.\ 2011, Computing in Science and Engineering, 13, 22

\bibitem[Vasyunina et al. 2009]{vasyunina2009} Vasyunina, T., Linz, H., Henning, T., et al.\ 2009, \aap, 499, 149

\bibitem[Vasyunina et al. 2011]{vasyunina2011} Vasyunina, T., Linz, H., Henning, T., et al.\ 2011, \aap, 527, A88

\bibitem[Vasyunina et al. 2014]{vasyunina2014} Vasyunina, T., Vasyunin, A.~I., Herbst, E., et al.\ 2014, \apj, 780, 85

\bibitem[Virtanen et al. 2020]{virtanen2020} Virtanen, P., Gommers, R., Oliphant, T.~E., et al.\ 2020, Nature Methods, \textit{in press}, 
{\tt arXiv:1907.10121}

\bibitem[Visser et al. 2009]{visser2009} Visser, R., van Dishoeck, E.~F., \& Black, J.~H.\ 2009, \aap, 503, 323

\bibitem[Walsh et al. 2010]{walsh2010} Walsh, A.~J., Thorwirth, S., Beuther, H., et al.\ 2010, \mnras, 404, 1396

\bibitem[Wang et al. 2006]{wang2006} Wang, Y., Zhang, Q., Rathborne, J.~M., et al.\ 2006, \apjl, 651, L125

\bibitem[Wang et al. 2011]{wang2011} Wang, K., Zhang, Q., Wu, Y., et al.\ 2011, \apj, 735, 64

\bibitem[Ward-Thompson et al. 2006]{wardthompson2006} Ward-Thompson, D., Nutter, D., Bontemps, S., et al.\ 2006, \mnras, 369, 1201

\bibitem[Waskom et al. 2017]{waskom2017} Waskom, M., Botvinnik, O., O'Kane, D., et al.\ 2017, {\tt mwaskom/seaborn: v0.8.1}, available from \url{https://doi.org/10.5281/zenodo.883859}

\bibitem[Watt et al. 1988]{watt1988} Watt, G.~D., White, G.~J., Millar, T.~J., et al.\ 1988, \aap, 195, 257

\bibitem[Wright et al. 2010]{wright2010} Wright, E.~L., Eisenhardt, P.~R.~M., Mainzer, A.~K., et al.\ 2010, \aj, 140, 1868

\bibitem[Zeng et al. 2017]{zeng2017} Zeng, S., Jim{\'e}nez-Serra, I., Cosentino, G., et al.\ 2017, \aap, 603, A22


\end{thebibliography}
\end{document}